\documentclass[11pt]{article}
\pdfoutput=1

\usepackage[utf8]{inputenc}
\usepackage{multirow}
\usepackage{amsmath, amsfonts, amssymb}
\usepackage{comment}
\usepackage{graphicx}
\usepackage{psfrag}
\usepackage{amsthm}
\usepackage[usenames,dvipsnames,svgnames,table]{xcolor}
\usepackage{enumerate}
\usepackage{arydshln}
\usepackage{soul}
\usepackage{array}
 \usepackage{slashed}
 \usepackage{nicefrac}
 \usepackage[most]{tcolorbox}
\usepackage[font=small,labelfont=bf]{caption} 
\usepackage{subcaption}
\usepackage{mathtools}
\usepackage{colortbl}
\usepackage{mathrsfs}
 \usepackage{a4wide}
  \usepackage{tikz}
  \usepackage{tikz-cd}
  \usetikzlibrary{shapes.geometric}
  \usepackage{tcolorbox}
\usepackage{diagbox}
\allowdisplaybreaks

  \usepackage{color}
  \definecolor{dark-gray}{gray}{0.20}
  \definecolor{gray}{gray}{0.30}
  \definecolor{light-gray}{gray}{0.80}
  \definecolor{dark-red}{rgb}{0.7,0,0}
  \definecolor{dark-green}{rgb}{0.1,0.4,0}
  \definecolor{dark-blue}{rgb}{0.3,0.3,0.7}
  \definecolor{light-blue}{rgb}{0.8,0.8,1}
      \definecolor{swamp}{RGB}{240, 199, 197}
       \definecolor{landscape}{RGB}{180, 250, 199}
          \definecolor{undecided}{RGB}{252, 252, 197}

  \usepackage{pifont}
\usepackage{newunicodechar} % To write the definition on the next line
\newunicodechar{ỳ}{\`y}

\usepackage{setspace}

\newcommand{\beq}{\begin{equation}}  \newcommand{\eeq}{\end{equation}}
\newcommand{\bal}{\begin{aligned}}   \newcommand{\eal}{\end{aligned}}
\newcommand{\be}{\begin{equation}}
\newcommand{\ee}{\end{equation}}

\def\be{\begin{equation}}
\def\ee{\end{equation}}
\def\bea{\begin{eqnarray}}
\def\eea{\end{eqnarray}}

\newcommand{\cO}{\mathcal{O}}
\newcommand{\cT}{\mathcal{T}}

\newcommand{\cC}{\mathcal{C}}

\DeclareMathOperator{\sign}{sign}
\newcommand{\abs}[1]{\lvert #1 \rvert}

\newcommand{\dd}{\mathrm{d}}

\def\simleq{\; \raise0.3ex\hbox{$<$\kern-0.75em
      \raise-1.1ex\hbox{$\sim$}}\; }
   \def\simgeq{\; \raise0.3ex\hbox{$>$\kern-0.75em
      \raise-1.1ex\hbox{$\sim$}}\; }

\numberwithin{equation}{section}

\usepackage{jheppub}
\usepackage{hyperref}
\usepackage{cleveref}

\hypersetup{
	colorlinks=true,
	linkcolor=dark-blue,
	citecolor=dark-red,
	urlcolor=dark-green,
	linktoc=page
}

\theoremstyle{remark}

\crefname{appendix}{Appendix}{Appendices}

\title{\centering EFT strings and dualities in 4d $\mathcal{N}=1$}

\author{Alessandra Grieco$^{1,2}$, Ignacio Ruiz$^{1,2}$}\affiliation{$^{1}$Instituto de F\'{i}sica Te\'{o}rica IFT-UAM/CSIC,
C/ Nicol\'{a}s Cabrera 13-15, Campus de Cantoblanco, 28049 Madrid, Spain}
\affiliation{$^2$Departamento de F\'{i}sica Te\'{o}rica, Universidad Aut\'{o}noma de Madrid, Cantoblanco, 28049 Madrid, Spain}
\author{and Irene Valenzuela$^{1,2,3}$}\affiliation{$^3$CERN, Theoretical Physics Department, 1211 Meyrin, Switzerland}

\preprint{IFT-UAM/CSIC-25-39\\ \vspace*{-0.1cm} 
\hfill CERN-TH-2025-086}

\abstract{We investigate the global structure of the states becoming light at perturbative limits of 4d $\mathcal{N}=1$ string and M-theory compactifications, identifying the different duality frames that emerge asymptotically and how they fit together in moduli space. These limits are characterized by the presence of EFT strings—a special class of axionic BPS strings whose tension, derived from the IR Kähler potential, vanishes in Planck units at infinite field distance. An intriguing integer scaling relation, $m \sim \mathcal{T}^w$ with $w = \{1,2,3\}$ in Planck units, connects the tension $\mathcal{T}$ of these strings to the mass scale $m$ of the leading tower of states along the string flow. We show that this relation also holds for the subleading towers below the species scale that generate the tower convex hull, implying that their associated $\vec\zeta = -\vec{\nabla} \log m$ vectors lie in a lattice generated by those of the EFT strings. This reveals a striking UV/IR interplay and offers organizing principles for the parametric hierarchies among the relevant UV scales in a given perturbative limit and the web of dualities governing 4d string vacua. }

\begin{document}

\hypersetup{pageanchor=false}
\makeatletter
\let\old@fpheader\@fpheader

\makeatother

\maketitle

\hypersetup{
    pdftitle={EFTstrings},
    pdfauthor={Alessandra Grieco, Ignacio Ruiz, Irene Valenzuela},
    pdfsubject={EFT strings and Duality frames}
}

\newcommand{\remove}[1]{\textcolor{red}{\sout{#1}}}

\section{Introduction}

The Distance Conjecture \cite{Ooguri:2006in} is now a central pillar of the Swampland program \cite{Vafa:2005ui,Brennan:2017rbf,Palti:2019pca,vanBeest:2021lhn,Grana:2021zvf,Harlow:2022ich,Agmon:2022thq}, which aims to determine the constraints that effective quantum field theories must satisfy in order to admit a UV completion within a consistent theory of quantum gravity. These constraints, formulated in terms of a series of interconnected conjectures, represent universal features observed across the string theory landscape and continue to accumulate growing evidence over time. 

The Distance Conjecture is one such swampland constraint, which predicts the existence of an infinite tower of states becoming exponentially light as one approaches any infinite distance boundary of the field space of a gravitational theory. The mass scale of the tower is given by
\beq\label{eq.SDC}
m\sim m_0 e^{-\alpha \Delta\phi}\to 0\quad \text{ as }\;\Delta\phi\rightarrow \infty
\eeq
where $\Delta\phi$ is the geodesic distance in field space, measured in the Einstein frame. Although surprising from a low energy perspective, this behavior is arguably one of the most characteristic features of the string theory landscape. Infinite distance boundaries in the moduli space of string theory compactifications are typically associated with either having extra dimensions opening up -- leading to towers of Kaluza-Klein modes becoming light --  or with weak coupling limits, in which towers of higher spin fields become massless in Planck units. The existence of these towers underlies the rich network of dualities observed in string theory, which has been one of the most profound insights in the field.

This typical feature of string theory was conjectured to be a universal quantum gravity consistency constraint in \cite{Ooguri:2006in}, and since the Swampland program's revival in 2015, it has received considerable attention with a plethora of works testing its validity across the most remote corners of the string landscape (see, e.g. \cite{Baume:2016psm,Klaewer:2016kiy, Blumenhagen:2017cxt, Grimm:2018ohb,Heidenreich:2018kpg, Blumenhagen:2018nts, Grimm:2018cpv, Buratti:2018xjt, Corvilain:2018lgw, Joshi:2019nzi,  Erkinger:2019umg, Marchesano:2019ifh, Font:2019cxq,  Lee:2019wij,Gendler:2020dfp, Lanza:2020qmt, Klaewer:2020lfg, Lee:2021qkx,Lee:2021usk,Alvarez-Garcia:2023gdd,Alvarez-Garcia:2023qqj,Rudelius:2023mjy,Aoufia:2024awo,Friedrich:2025gvs}). Given the amount of evidence gathered in recent years, the Distance conjecture has become one of the most firmly established and quantitatively tested swampland conjectures in string theory nowadays, with multiple ramifications and connections to other swampland proposals \cite{Lee:2018spm,Ooguri:2018wrx,Lust:2019zwm,Gendler:2020dfp,Hamada:2021yxy,Rudelius:2022gbz,Casas:2024oak}. It has also served to reveal universal geometric structures underlying stringy moduli spaces, leading to fruitful connections with mathematics (see e.g. \cite{Grimm:2018ohb,Grimm:2018cpv,Grimm:2019bey,Grimm:2019ixq,Lee:2019tst,Lee:2019wij,Lee:2020gvu,Klawer:2021ltm,Alvarez-Garcia:2023gdd,Friedrich:2025gvs}). Furthermore, the conjecture is starting to yield promising results in the context of AdS/CFT, where it is also providing guidance in uncovering universal features of conformal manifolds \cite{Baume:2020dqd,Perlmutter:2020buo,Baume:2023msm,Ooguri:2024ofs,Calderon-Infante:2024oed}. Several works are even on the track to providing a general proof of the conjecture in AdS using conformal field theory techniques. \\

Although the conjecture has always been widely accepted in its qualitative features, it initially lacked a more precise quantitative formulation. For instance, what are the possible values of the exponential rate $\alpha$ of the tower in \eqref{eq.SDC}? What is the microscopic nature of the tower? What is the exact energy scale at which the effective field theory will necessarily break down and quantum gravity description must take over (also known as quantum gravity scale or species scale)? Over time, the conjecture has undergone significant development, with all the gathered string-theoretic evidence helping to refine it into a more concrete and quantitative proposal. 
The most important of these refinements is the Emergent String Conjecture \cite{Lee:2019xtm,Lee:2019wij}, which states that in Einstein gravity theories\footnote{There are examples in non-Einstein AdS spaces where the leading tower of states are higher spin modes associated to non-critical strings \cite{Calderon-Infante:2024oed}.}, the nature of the tower is, in some dual frame, either a Kaluza-Klein tower or a tower of oscillator modes of a critical string, rather than e.g. excitation modes of a light membrane, see \cite{Alvarez-Garcia:2021pxo}. This significantly constrains the possible values of the exponential rate of the tower, suggesting in particular a lower bound $\alpha\geq \frac{1}{\sqrt{d-2}}$ in $d$ space-time dimensions, as proposed in \cite{Etheredge:2022opl}. The more we study the conjecture, the more we uncover universal constraints pointing toward an even deeper underlying structure. Examples of this are the universal pattern relating the tower's mass scale and the species scale \cite{Castellano:2023stg,Castellano:2023jjt} or the taxonomy rules obeyed among towers of states \cite{Etheredge:2024tok} in large types of limits.\\

Another universal intriguing relation associated with the Distance conjecture is the integer scaling relation observed in the context of 4d $\mathcal{N}=1$ string theory compactifications in \cite{Lanza:2021udy}. In this work, it was shown that infinite distance limits in these compactifications are associated with the emergence of approximate axionic shift symmetries, and consequently, with the existence of axionic BPS strings becoming tensionless in Planck units at infinite distance. These objects -- dubbed EFT strings \cite{Lanza:2021udy,Lanza:2022zyg,Grimm:2022sbl,Marchesano:2022avb,Martucci:2022krl,Wiesner:2022qys,Cota:2022yjw,Marchesano:2022axe,Martucci:2024trp}-- are particularly interesting because their backreaction profile induced in the EFT, also interpretable as an RG flow, is in one-to-one correspondence with infinite distance paths in moduli space. This provides a bottom-up tool for understanding the asymptotic geometry of moduli spaces and also offers a potential explanation for the Distance Conjecture itself.

However, the key feature for this paper is the \textbf{Integer Scaling Conjecture}, which relates the tension $\cT_{\bf e}$ of the EFT string of charge ${\bf e}$ to the mass $m_*$ of the leading tower of states:
\beq
\label{scal}
m_*^2\simeq M_{\rm Pl,4}^2\, A\, \left(\dfrac{\cT_{\bf e}}{M_{\rm Pl,4}^2}\right)^w \quad \text {with } w\in \mathbb{Z}_{>0}\\
\quad\text{ as }s^i=s^i_0+e^i\sigma\text{ with }\sigma\rightarrow \infty \ ,
\eeq
with $A$ being some coefficient depending on the non-flowing scalars. This relation was discovered by checking string theory examples, and the underlying reason remains a mystery -- particularly why the scaling weight $w$ only takes integer values $w=1,2,3$ in all known string theory examples. Since the  EFT string tension can be read from the K\"ahler potential, the relation \eqref{scal} provides a purely bottom-up method for inferring the tower’s mass, up to the integer $w$. Moreover, it was argued in \cite{Lanza:2021udy} that this relation remains valid even when supersymmetry is broken at low energies or when a scalar potential is introduced -- making these results directly applicable to realistic models of our universe.\\

The goal of this paper is to explore this fascinating constraint in greater depth. We will extend the analysis of \cite{Lanza:2021udy} by testing \eqref{scal} along more general infinite distance trajectories and for \emph{all} towers of states (not just the leading one) in a wide range of 4d $\mathcal{N}=1$ string compactifications. To this end, we will use the techniques recently developed in \cite{Calderon-Infante:2020dhm,Etheredge:2022opl,Etheredge:2024tok,Calderon-Infante:2023ler} to perform a global analysis of tower behavior along various trajectories by computing their scalar charge-to-mass ratio vectors $\vec\zeta=-\vec\partial_\phi \log m$ and constructing the convex hull of these vectors. 

This tower convex hull encodes how different limits and towers fit together globally in the moduli space. It also provides information about the microscopic interpretation of the various perturbative descriptions that emerge in asymptotic corners of the moduli space, and the dualities connecting them. For each string theory setup studied in \cite{Lanza:2021udy}, we will classify and plot the convex hull of the towers of states, the species scale and the EFT strings, identifying the dual theory emerging in each case. Here we will not aim to provide rigorous proofs for the population of the towers of states, which is still even a topic of current research in the highly supersymmetric cases \cite{Hattab:2025aok,Monnee:2025msf,Baines:2025upi}, but simply study the scaling behaviour of the masses and use this to determine the implied structure of dualities across moduli space. 
This way, we will provide a new way to describe systematically the dualities arising in 4d $\mathcal{N}=1$ string theory compactifications, since the convex hull of the towers directly inform us on how different duality frames should fit together in the moduli space. Since many of these $\mathcal{N}=1$ dualities are largely unknown,  we hope this can provide interesting targets for future research to yield further evidence for the dualities implied by the Distance conjecture in this work. Along the way, we will also learn new features about the sliding behavior of $\zeta-$vectors as we move in the moduli space, as well as uncover new bounds for $w$ in \eqref{scal}.\\

To our surprise and delight, we will find that the Integer Scaling Relation \eqref{scal} is even more general than previously anticipated: it holds for any multi-field trajectory associated with an EFT string flow (including non-elementary ones) and applies to all (sub)leading towers of states below the species scale that generate the tower convex hull. Taking this into account, the Integer Scaling Relation implies that the $\zeta$-vectors of all such towers lie on the lattice generated by the EFT string $\zeta$-vectors. This suggests a deeper connection between EFT strings and the UV towers of states, potentially tied to the existence of dualities and approximate global symmetries, which is yet to be understood.\\

In an accompanying paper \cite{bottomUpTBA}, we will reverse the logic and ask: How much information can be inferred about the towers of states given only the K\"ahler potential $K$ -- and therefore, the EFT string tensions -- if assuming \eqref{scal}? Clearly, once $K$ is fixed, the relation \eqref{scal} restricts the mass of the possible towers of states to a discrete set. Combined with the Emergent String Conjecture (which constrains the allowed values of the exponential rate of the tower), this leads to a finite set of viable tower convex hulls. We will show in \cite{bottomUpTBA} that this set precisely matches those arising from known string theory compactifications explored in this current paper -- thus realizing yet another instance of string universality.\\

The outline of the paper goes as follows. We start by reviewing the relevant aspects of perturbative regimes in 4d $\mathcal{N}=1$ theories, the role of EFT strings, and the construction of convex hulls of towers of states in Section \ref{s.REVIEW}. In Section \ref{ss.summary}, we reformulate the Integer Scaling Relation \eqref{scal} as a local condition on the $\zeta$-vectors, and proceed to verify this condition for all (sub)leading towers of states in a wide range of string theory compactifications throughout the paper. For convenience, the main results are summarized in Section~\ref{ss.summary}, allowing the reader to grasp the key ideas without delving into the full technical details of the string constructions.  For readers interested in the global description of the different towers and duality frames emerging asymptotically in diverse 4d $\mathcal{N}=1$  string theory setups, we provide detailed analyses -- accompanied by illustrative figures -- in the later sections.  Section \ref{s.het} focuses on Heterotic $E8\times E8$ on Calabi-Yau threefolds,  Section \ref{s.more topdown} covers Type II on orientifolds -- including Type IIA, Type I and F-theory setups--, and Section \ref{s.Mth} addresses M-theory on Joyce manifolds. We conclude with a summary and outlook in Section \ref{s.conc}.

\section{Review of EFT strings and Summary of new results\label{s.REVIEW}}

\subsection{Review of perturbative regimes in 4d $\mathcal{N}=1$ and EFT strings}

Consider a 4d $\mathcal{N}=1$ effective field theory coupled to gravity. The scalar fields can be complexified as $z^\alpha=a^\alpha+i s^\alpha$ where $\alpha $ runs over the number of chiral multiplets of the theory. The effective action for the scalars is given by
\be
S=M^2_{\rm Pl,4}\int \left(\frac{1}{2}\mathcal{R}*1-K_{\alpha\bar \beta}\,\dd z^\alpha\wedge*\dd\bar z^{\bar \beta}-V*1 \right)\, ,
\label{effaction}
\ee 
The field space metric and the potential can be derived from a K\"ahler potential $K$ and superpotential $W$ as follows, $K_{\alpha\bar \beta}=\partial_{z^\alpha}\partial_{\bar z^{\bar \beta}} K$ and $V=e^K(K^{\alpha\bar \beta}D_\alpha WD_{\bar \beta}W-3 |W|^2)$. 

We are interested in perturbative regimes of this EFT, where $K$ and $W$ can be expanded in terms of some perturbative leading contribution plus exponentially suppressed non-perturbative corrections. This can occur whenever the field space metric develops an approximate axionic symmetry, as we review in the following. Following \cite{Lanza:2021udy}, we also review how these perturbative regimes correspond to infinite distance in the field space where certain strings (dubbed EFT strings) become tensionless.

For simplicity, we are going to consider $W=0$, since this will not be relevant for the purposes of this paper (see \cite{Lanza:2021udy}, though, for the analysis with $W\neq 0$). The leading non-perturbative corrections are expected to come from $\frac12$-BPS instantons, whose contribution to the K\"ahler potential is given by  $\cO(e^{2\pi i m_iz^i})$ where $m_i\in \mathbb{Z}$ are the instanton charges. Since these instanton corrections are suppressed by  $e^{-2\pi \, m_i s^i}$, we can reach these EFT perturbative regimes by taking limits in which linear combinations of saxions $s^i=\textrm{Im}(z^i)$ become very large, with $\{z^i\}_i\subseteq\{z^\alpha\}_\alpha$.\footnote{Notice that $\{z^i\}_i\subseteq\{z^\alpha\}_\alpha$ can be a subset of the complete set of chiral scalars, as some might be kept at finite fixed value.} To leading order, the K\"ahler potential in these perturbative regimes is given by
\beq
\label{KP}
K=-\log \left[P(s)+\cO(e^{2\pi i m_iz^i})\right]
\eeq
in all known 4d $\mathcal{N}=1$ EFTs arising from string theory compactifications,  where $P(s)$ is some homogeneous function of positive integral degree on the saxions. These perturbative regimes are, therefore, characterized by having approximate axionic shift symmetries in the field space metric, since the field space metric is independent to leading order of the real scalars $a^i=\text{Re}\,z^i$, which play the role of axions in the EFT under which the instantons are electrically charged.

For each region  with perturbative axionic symmetries in the field space metric, the breaking of the said symmetry is measured by a set of instanton charges $\mathcal{C}_{\rm I}\subset M_{\mathbb{Z}}$. 
The perturbative region is then identified by requiring that 
  \be\label{asymptregion}
  \langle {\bf m}, {\bf s}\rangle  \equiv m_is^i\gg 1\quad~~~ \forall\, {\bf m}\in \cC_{\rm I}\, ,
  \ee
  so that the instanton corrections are exponentially suppressed.
Here ${\bf s}$ collectively denotes the saxions $s^i$.   This means that the perturbative regime can be identified as the {\em deep interior} of a saxionic cone $\Delta$, which is defined as follows:
  \be
\Delta \equiv \{{\bf s}\in N_{\mathbb{R}}\ | \ \langle {\bf m}, {\bf s} \rangle > 0\,,\ \forall {\bf m}\in \cC_{\rm I}\}\, .
\label{defDelta}
\ee
Notice that trajectories in the field space where the saxions become parametrically large brings us deeper in the interior of the saxionic cone. Due to the leading form of $K$ in \eqref{KP}, the limit $s^i\rightarrow \infty$ is at infinite distance in the field space, as long as we stay within the saxionic cone so that we can still trust \eqref{KP}.\\

On a different note, the emergence of approximate axionic symmetries in the field space metric allows for the existence of axionic $\frac12$-BPS strings, that are magnetically charged under the axions. They correspond to supersymmetric string-like solutions of \eqref{effaction} with $V=0$ that imply the following non-trivial (harmonic) profile of the scalars near the string core,
\begin{subequations}
\label{solsplit}
\bea\label{imt}
\label{axionmon}
a^i &= &a^i_0\frac{\theta}{2\pi}\,e^i\, ,\\
s^i & = & s_{0}^i+e^i \sigma \ , \quad  \sigma\equiv-\frac1{2\pi}\log \left(\frac{r}{r_0}\right)\ .
\label{saxionflow}
\eea
\end{subequations}

The first equation is the monodromy that the axion undergoes when performing a rotation $\theta\rightarrow \theta+2\pi$ around a string of charges $e^i\in \mathbb{Z}$. As approaching the string core (located at $r=0$), the backreaction of the string in the space-time geometry implies a non-trivial profile for the saxion given by \eqref{saxionflow}, that goes from a reference value $s_0^i\equiv s^i(r_0)$ to $s^i\rightarrow  e^i \cdot \infty$ at the string core. Hence, the saxions are driven to the boundaries of the field space at the string core. This particular path is denoted as the \emph{string flow}. Note that these strings are fundamental in the sense that they cannot be resolved into a smooth solitonic object
within a four-dimensional EFT.

The BPS string tension is field-dependent and given by
\beq
\label{TEFT}
\cT_{\bf e}=M_p^2e_il^i \quad , \ l_i=-\frac12\frac{\partial K}{\partial s^i}
\eeq
where recall that $e_i$ are the string charges and $l_i$ are known as the \emph{dual saxions}. BPS strings satisfy that they have a positive tension at any point of the saxionic cone $\Delta$. 

A subset of these BPS strings were denoted in \cite{Lanza:2020qmt,Lanza:2021udy} as \emph{EFT strings} (see also \cite{Lanza:2022zyg,Grimm:2022sbl,Marchesano:2022avb,Martucci:2022krl,Wiesner:2022qys,Marchesano:2022axe,Cota:2022yjw,Martucci:2024trp}) and satisfy the additional property of remaining weakly coupled along their entire string flow \eqref{saxionflow}, as we show next. Consider a flow that starts in the perturbative region \eqref{asymptregion}, that is $\langle {\bf m},{\bf s}_0\rangle\gg 1$ for any ${\bf m}\in\cC_{\rm I}$. For  the string backreaction to remain within the perturbative region  \eqref{asymptregion}
as we approach the core of the string, we need that all instanton corrections remain exponentially suppressed. In other words, we need that 
\be
e^{-2\pi \langle {\bf m}, {\bf s} \rangle} = e^{-2\pi (\langle {\bf m}, {\bf s}_0 \rangle+\sigma \langle {\bf m},{\bf e}\rangle )} \, \ll 1 \quad \text{ as } \sigma \rightarrow \infty
\label{inststr}
\ee  
which is satisfied if $\langle {\bf m},{\bf e}\rangle \geq 0$ for any ${\bf m}\in\cC_{\rm I}$. 
Such a charge ${\bf e}$  represents then an EFT string.\footnote{Not all BPS strings are EFT strings. If instead $\langle{\bf m}, {\bf e}\rangle <0$ for some ${\bf m} \in \cC_{\rm I}$, the associated non-perturbative effects grow along the string flow of charges ${\bf e}$, until they reach $\cO(1)$ contributions at a finite radial distance. Hence, non-EFT strings become strongly coupled along the saxionic flow and, as shown in \cite{Lanza:2021udy}, get tensionless at finite distance in field space.}.
Therefore, we may identify the set of EFT strings associated with a given perturbative regime with the following lattice cone
 \be
 \cC^{\text{\tiny EFT}}_{\rm S}=\overline\Delta\cap N_{\mathbb{Z}}
 \label{CSEFT}\, ,
 \ee

where $\overline\Delta$ is the closure of $\Delta$. Therefore, the EFT string charges act as generators of the saxionic cone parameterizing the perturbative regime. 

Moreover, a K\"ahler potential of the form \eqref{KP} implies that the tension \eqref{TEFT} vanishes asymptotically along the string flow as
\be\label{univtensionflow}
\cT_{\bf e}\sim \frac{M^2_{\rm P}}{\sigma}\rightarrow 0\, ,\quad~~~ \text{for $\sigma\rightarrow\infty$}\, .
\ee

There exists, thus, a map between the backreacted string flows driven by EFT strings in a given EFT, and infinite distance paths in the moduli space of such EFT (see also \cite{Lanza:2020qmt} for the interpretation of these string backreacted solutions in terms of RG flows). Remarkably, one can then use physical properties of the EFT strings to explain some of the asymptotic geometric features observed in the moduli space of string theory compactifications (like for example the logarithmic behavior of $K$ in \eqref{KP} \cite{Lanza:2020qmt}).

The above map (if universal) implies that there will always be an EFT string becoming tensionless in Planck units along any infinite distance/perturbative limit of a 4d EFT weakly coupled to gravity. Therefore, one can use the existence of these objects (which can be motivated by the absence of global symmetries \cite{Lanza:2021udy}) to provide a bottom-up explanation for the EFT breakdown predicted by the Swampland Distance Conjecture \cite{Ooguri:2006in}. In particular, the energy scale of the string tension, which  can be directly computed from the K\"ahler potential as in \eqref{TEFT}, provides an upper bound for the EFT cut-off, so that $\Lambda_{\rm cut-off}\lesssim  \cT^{1/2}$. As shown in \cite{Lanza:2020qmt}, imposing that the EFT string satisfies the WGC, implies that this cut-off must behave exponentially in the moduli space distance, as predicted by the Distance Conjecture.\\

The microscopic nature of these EFT strings was identified in a plethora of 4d $\mathcal{N}=1$ string compactifications in \cite{Lanza:2021udy}. It is important to remark that even if they are present at every infinite distance limit of these theories, they do not necessarily correspond to the leading tower of states becoming light. For instance, there can be lighter towers of Kaluza-Klein states or other string modes with $m_*< \mathcal{T}^{1/2}$. Nevertheless, a very intriguing relation between the leading tower of states and the EFT string tension was found in \cite{Lanza:2021udy}. All analyzed string theory examples satisfied that
\begin{align}
\label{scalingweight}
\begin{array}{c}
m_*^2\simeq M_{\rm Pl,4}^2\left(\dfrac{\cT_{\bf e}}{M_{\rm Pl,4}^2}\right)^w \quad \text {with } w=1,2,3\\
\text{ as }s^i=s^i_0+e^i\sigma\text{ with }\sigma\rightarrow \infty \ .
\end{array}
\end{align}
Hence, the mass of the leading tower scales as the EFT string tension up to some integer value, along the infinite distance trajectory mapping the EFT string flow.
 This relation was simply denoted as \emph{Conjecture 2: cut-off asymptotics} in  \cite{Lanza:2021udy}, but we will refer to it us the \textbf{Integer Scaling Conjecture} for convenience, following \cite{Lanza:2022zyg}.

The main open question left was what fixes the value of the scaling weight $w$ and why it is upper bounded by 3. Recall than $\cT$ can be derived from bottom-up from $K$. Hence, if we are able to fix $w$, we could derive the scale of the leading tower of states from bottom-up!

Moreover, the above integer scaling relation was only studied for the leading tower along elementary string flows\footnote{Elementary string flows are those that cannot be decomposed as superposition of other BPS strings. They correspond to single-field trajectories as they only have one non-vanishing string charge, and therefore, only one saxion is driven to infinite distance.} and certain non-elementary ones.  But what happens for more general infinite distance trajectories along which more than one saxion grow at possibly different rates? And what about other subleading towers of states? In this paper, we will perform a systematic analysis of the fate of \eqref{scalingweight} for all (sub)leading lights towers of states along all multi-field trajectories of the saxionic cone to provide an answer to these questions.

\subsection{Review of convex hulls of towers of states\label{ss.review towers}}

When transitioning from analyzing the scaling of the towers along a single infinite distance limit to examining the global structure of the towers across different limits, convex hulls of $\zeta$-vectors have proven to be extremely useful. Let us review next how to construct these convex hulls for the towers of states.

First of all, one defines the  $\zeta$-vector of each tower with characteristic mass $m$ as follows,
\beq\label{eq.zeta def}
\vec\zeta =- \vec\nabla \log \frac{m(\vec{\varphi})}{M_{{\rm Pl},d}}\,,
\eeq
where the derivative is taken with respect to the different light scalars $\{\varphi^i\}_i$. The vector $\vec{\zeta}$ is sometimes denoted as the \emph{scalar charge-to-mass ratio vector} \cite{Lee:2018spm,Gonzalo:2019gjp,DallAgata:2020ino,Andriot:2020lea,Benakli:2020pkm}, by analogy to the WGC \cite{Palti:2017elp}.\footnote{To better understand this analogy, consider scalar field $\psi$ with mass $m(\varphi)$, and expand its Lagrangian $\mathcal{L}\supset\frac{1}{2}m(\varphi)^2\psi^2=(\frac12 m_0^2+m_0\varphi\partial_\varphi m(\varphi))\psi^2+...$, where now $\partial_\varphi m$ can be understood as a scalar Yukawa charge induced by the modulus $\varphi$ on $\psi$.} 

\begin{figure}[htp!] 	
 \vspace{-0.6cm}
 	\centering
			\includegraphics[width=0.45\textwidth]{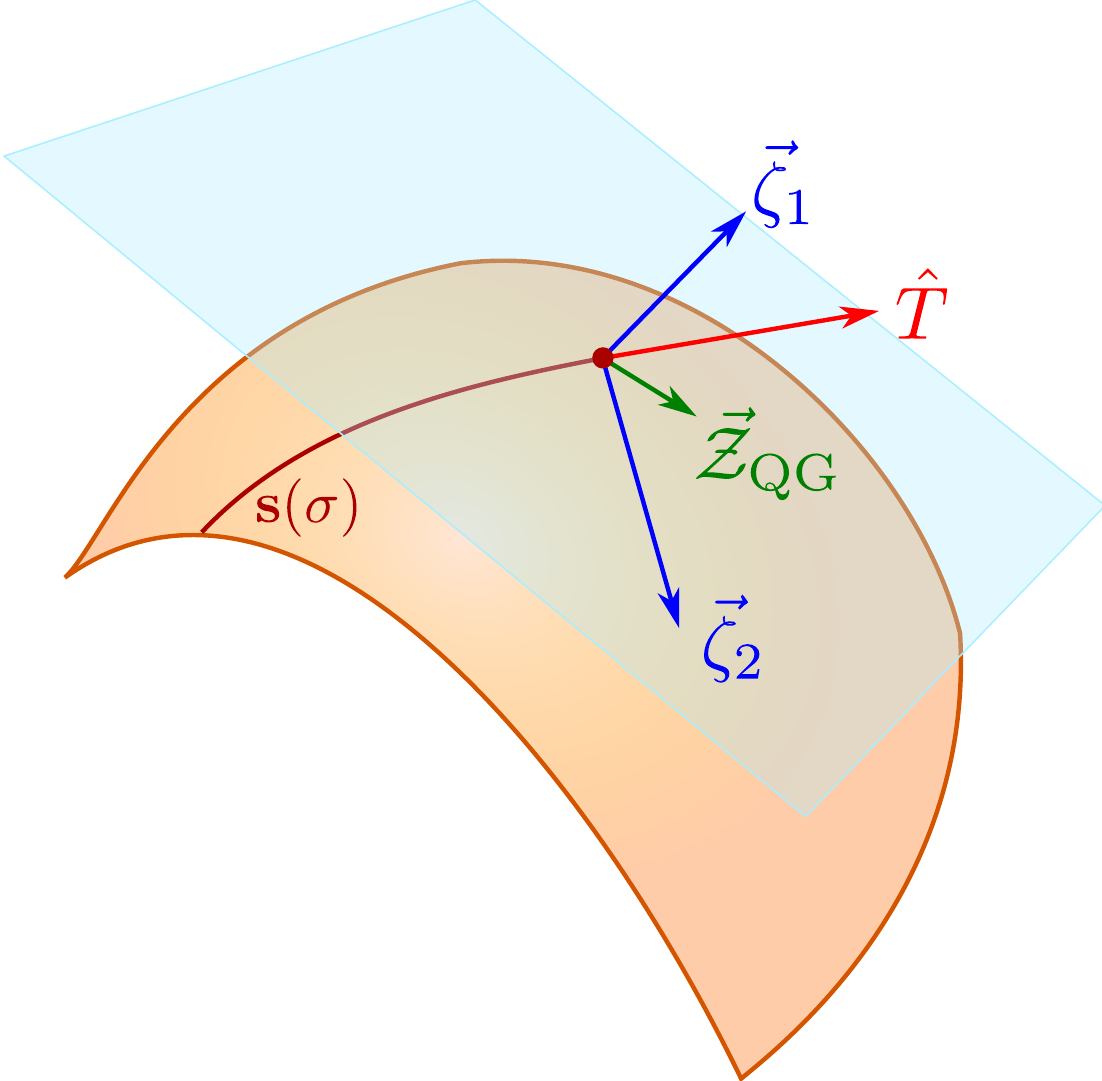}
		\caption{The $\zeta$-vectors of towers of states (and species scale vector $\vec{\mathcal{Z}}_{\rm QG}$) represented in the tangent space of the field space at a given point of a trajectory $\mathbf{s}(\sigma)$ with asymptotic tangent vector $\hat T$.}\label{fig:tangent-space}
	
\end{figure}

Consider now a particular infinite distance trajectory with asymptotic direction given by the unit tangent vector $\hat T$.  The exponential rate $\alpha$ of a tower of states along such trajectory is given by $\alpha=\vec\zeta \cdot \hat T$, where the scalar product is taken with respect to the field space metric.\footnote{From \eqref{effaction}, the saxionic (i.e., associated to non-compact scalars) part of the field space metric is given by
\begin{equation}  \label{e.Kahler metric}
	\mathsf{G}_{ij}=\frac{1}{2} \frac{\partial^2 K}{\partial s^i\partial s^j} \;,
\end{equation} with $i$ and $j$ running over the scalars taken to large values.}
In the presence of multiple towers, we can plot the $\zeta$-vectors of all the towers of states becoming light along such trajectory, which will lie in the tangent space of the moduli space (see Figure \ref{fig:tangent-space}). The leading (lightest) tower will be that with the largest projection of $\vec\zeta$. Taking this into account, it becomes useful to draw the convex hull of all the  $\zeta$-vectors, where the generators will precisely be the leading towers of states along different directions. As first described in \cite{Calderon-Infante:2020dhm} and further used in \cite{Etheredge:2022opl,Etheredge:2023odp,Calderon-Infante:2023ler,Etheredge:2023usk,Castellano:2023jjt,Etheredge:2024tok,Etheredge:2024amg}, the Distance conjecture will be satisfied with a minimal exponential rate $\alpha_0$ if the convex hull of the towers includes the ball of radius $\alpha_0$. \\

Moreover, as noted in \cite{Etheredge:2024tok}, the geometry of these vectors is highly constrained by the microscopic nature of the towers. In particular, assuming the Emergent String Conjecture (recursively upon decompactification) and certain condition on the regularity of the limit (see \cite{Etheredge:2024tok}), the light towers becoming light at the same time satisfy the following taxonomic rule:
\beq\label{e.taxonomy}
\vec\zeta_i\cdot \vec\zeta_j=\frac{1}{d-2}+\frac1n\delta_{ij}
\eeq
where $d$ is the number of space-time dimensions and $n$ the number of extra dimensions decompactifying in the limit (for an emergent string limit, one should formally set $n=\infty$).
The convex hull of the towers forms a frame simplex whose geometry is rigid under variations of the direction of the infinite distance limit, as long as the above assumptions remain valid.\\

Analogously, one can also define a similar vector for the species scale \cite{Calderon-Infante:2023ler,Castellano:2023jjt,vandeHeisteeg:2023ubh,vandeHeisteeg:2023uxj,Castellano:2023stg},
\beq\label{eq.species}
\vec{\mathcal{Z}}_{\rm QG} =- \vec\nabla \log \frac{\Lambda_{\rm QG}}{M_{{\rm Pl},d}}
\eeq
which will give us information about the cut-off scale at which the EFT breaks down due to quantum gravitational effects \cite{Arkani-Hamed:2005zuc,Distler:2005hi,Dimopoulos:2005ac,Dvali:2007hz,Dvali:2010vm}, i.e. no local QFT description weakly coupled to Einstein gravity is longer possible above $\Lambda_{\rm QG}$. Note that since $\Lambda_{\rm QG}$ is a uniquely defined function over all moduli space, $\vec{\mathcal{Z}}_{\rm QG}$ defines a global vector field. We will be interested in the different asymptotic expressions of $\vec{\mathcal{Z}}_{\rm QG}$, which determine the microscopic interpretation of the dual theory emerging at infinite distance.

Remarkably, the following universal (asymptotic) pattern relating the species scale and the \textit{leading} tower of states was observed in \cite{Castellano:2023stg,Castellano:2023jjt}:
\beq\label{eq.PATTERN}
\vec\zeta \cdot \vec{\mathcal{Z}}_{\rm QG}=\frac1{d-2}
\eeq
which is satisfied in all known string theory examples for the moment. The above relation only depends on the dimension of space-time, and it works regardless of whether the leading tower is a string or Kaluza-Klein tower. It can also be derived from bottom-up using the same assumptions that were used to derive \eqref{e.taxonomy} in \cite{Etheredge:2024tok}, so it becomes part of the taxonomic rules resulting from the Emergent String Conjecture.\\

When moving in the moduli space of a string theory compactification, we can typically find different regimes that are better described by different perturbative descriptions related by dualities (thereby, known as duality frames). Each duality frame will be characterized by a particular microscopic interpretation of $\Lambda_{\rm QG}$, which identifies a particular $\vec{\mathcal{Z}}_{\rm QG}$. The $\zeta$-vectors of the leading towers are located at the interfaces between different duality frames. From the geometry of these vectors (i.e., their length and angles), one can infer the microscopic interpretation of the dual description emerging at the infinite distance limit. Moreover, the above taxonomic rules provide information about how different duality frames can fit together in the moduli space.

\begin{figure}[htp!] 	
 	\centering
			\includegraphics[width=0.45\textwidth]{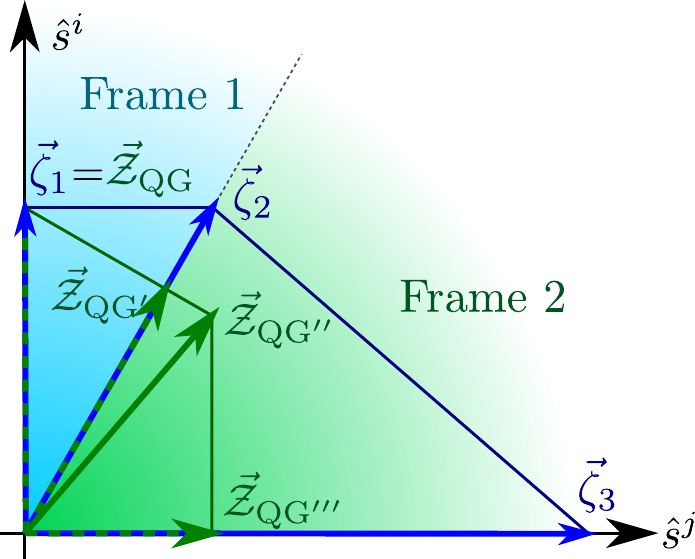}
		\caption{Example of an arrangement of $\zeta$-vectors for towers (blue) and $\mathcal{Z}$-vectors for the possible asymptotic expression of the species scale $\Lambda_{\rm QG}$ (green), in terms of two canonically normalized saxions, $\hat{s}^i$ and $\hat{s}^j$. We also draw the convex hull of the $\zeta$-vectors for the towers and for the species scale. The blue and green regions correspond to two duality frames associated to different perturbative regimes of the theory. The microscopic nature of the emergent dual theories at infinite distance is determined by the nature of the species scale and the light towers located at the boundaries of each region. }\label{fig:FramesEJ}
\end{figure}

For illustration purposes, we provide an example of the convex hulls for both the towers of states and the  species scales in Figure \ref{fig:FramesEJ}. 
The different duality frames are represented with different colors. Each duality frame corresponds to the region in which a given $\vec{\mathcal{Z}}_{\rm QG}$ dominates, so that the interfaces are precisely the loci along which different asymptotic expressions of the species scale (in green) decay at the same rate, i.e. they project to the same vector. In each duality frame, we draw the convex hull of the $\zeta$-vectors of all the light towers (in blue). These convex hulls can then be combined to build a polytope. This tower polytope (or tower convex hull) will be characteristic of each possible UV completion of a 4d perturbative regime, and we will see that the generators satisfy the integer scaling relation \eqref{scalingweight}. This example is not made up, but actually correspond to one of the convex hulls obtained in heterotic string theory in Section \ref{s.het}, where we will describe the microscopic nature of the towers in detail. In that example, the axes will be associated to the 4d dilaton and the overall compactification volume, canonically normalized. For clarity reasons, similar figures from now on will feature points rather than vectors depicting the different $\zeta$- and $\mathcal{Z}$-vectors.

While expressions \eqref{e.taxonomy} and \eqref{eq.PATTERN} (as well as \eqref{lattice} later on) are covariant in moduli space coordinates, the asymptotic expressions for the $\zeta$- and $\mathcal{Z}$-vectors simplify greatly when working in local \emph{flat coordinates} $\{\hat{\varphi}^a\}_a$ (i.e. canonically normalized scalars in Einstein frame), for which the moduli space metric takes the form $\mathsf{G}_{ab}=\delta_{ab}$. The expressions of the different vectors are then given by
	\begin{equation}
	\zeta^i=-\mathsf{e}^{ia}\partial_{\varphi^a}\log\frac{m}{M_{{\rm Pl,}d}}\;,
	\end{equation}
(analogously for $\vec{\mathcal{Z}}_{\rm QG}$), where $\mathsf{e}^{ia}$ are the inverse vielbein associated to $\mathsf{G}_{ab}$ (so that $\mathsf{e}^{ia}\mathsf{e}^{jb}\delta_{ij}=\mathsf{G}^{ab}$), defined up to an $O(n)$ transformation (with $n$ the number of moduli). In general, these flat coordinates cannot be defined globally, unless the chosen slice of moduli space is (Riemann) flat. However, they can be obtained locally as we move along an asymptotic trajectory, without any loss of generality (see Section \ref{sec.no homo} for more details on this). Therefore, Figure \ref{fig:FramesEJ} (and similar plots in this paper) have to be interpreted as  local representations of the $\zeta$-vectors evaluated at a given point of the perturbative regime.  As mentioned above, the figure remains unchanged even if we move in the moduli space (at least) as long as we stay in the interior of the same duality regime. The vectors plotted in all these figures will always be given in terms of these flat coordinates, so that the axes correspond to the canonically normalized saxions denoted as $\hat s^i$, with
\begin{equation}
	S\supset \frac{M_{\rm Pl,4}^2}{2}\int\dd^4 x\sqrt{-g}\left\{\mathcal{R}-\delta_{ij}\partial_\mu\hat{s}^i\partial^\mu\hat{s}^j\right\}\ .
\end{equation}

The plan of this paper is to apply this machinery of convex hulls to 4d $\mathcal{N}=1$ compactifications, to analyze how the towers of states and EFT strings fit together in a more global way in the perturbative regime. The pattern \eqref{eq.PATTERN} was already checked to hold in these compactifications in \cite{Castellano:2023jjt}, but EFT strings have not been considered yet in this context. We will see that they indeed play an essential role to determine the possible duality frames and highly constrain the structure of the UV towers of states.\\

\subsection{Summary of new results and general lessons}

\subsubsection{EFT strings as lattice generators\label{ss.summary}}

We begin by defining a $\zeta$-vector for the EFT strings as follows,
\beq
	\label{defstringvectors}
\vec\zeta_{\cT} =- \vec\nabla \log \frac{\cT(\vec\varphi)^{1/2}}{M_{\rm Pl,4}}
\eeq
so that it provides information of the exponential rate of the energy scale $\cT^{1/2}$. Simple manipulation shows that for a given EFT string charge $\mathbf{e}$, one has $\vec{\zeta}_{\mathcal{T}_{\mathbf{e}}}\propto\mathbf{e}$, where precisely $\mathbf{e}=\partial_\sigma \mathbf{s}(\sigma)$ is a tangent vector (not necessarily of unit norm) of the flow trajectory  \eqref{saxionflow}.\footnote{To see this, simply note that
	\begin{equation}
		{\zeta}^{i}_{\mathcal{T}_{\mathbf{e}}}=-\tfrac{1}{2}\mathsf{G}^{ij}\partial_j\log\left(-\tfrac{1}{2}e^a\partial_a K\right)=-(\partial^2K)^{ij}\frac{-\tfrac{1}{2}e^a\partial_i\partial_a K}{-\frac{1}{2}e^a\partial_a K}=\frac{M_{\rm Pl,4}^2}{2\mathcal{T}_{\mathbf{e}}}e^i\;.
\end{equation}}

The convex hull of the $\zeta$-vectors  of all EFT string in a given perturbative regime forms a hyperplane. This was already noted in \cite{Lanza:2020qmt}, since for codimension-2 objects in four dimensions, the scalar charge-to-mass-ratio vector $\vec\zeta$ is equal to the usual gauge charge-to-mass-ratio vector $\vec Q/ \cT$. Since the tension is linear in the quantized charges, the convex hull turns out to span a hyperplane for mutually BPS objects. The generators of this hyperplane will be the \textbf{elementary EFT strings} (namely, those with only one non-vanishing charge), which will be located at the boundaries of the saxionic cone. As for \textbf{non-elementary EFT strings}, with $e^i\neq 0$ for $i\in I$ (such that $I$ has at least two elements so the string has more than one non-vanishing charge), they are located in the simplex $\Delta_I$ generated by the $I$ elementary strings, $\Delta_I=\{\vec{\zeta}_{\mathcal{T}_{e^i}}\}_{i\in I}$. The precise location of $\vec{\zeta}_{\mathcal{T}_{\mathbf{e}}}$ for the non-elementary strings will depend on the point of moduli space. If we evaluate the vector $\vec{\zeta}_{\mathcal{T}_{\mathbf{e}}}$ along the string flow \eqref{saxionflow}, then  it gets perpendicular to such simplex. In that case, the closest point from the origin to the simplex will set its length along the flow, given by
\footnote{\label{fn.perp} To see this, notice that from \eqref{TEFT}, along a given EFT string flow \eqref{saxionflow} associated to a charge $\mathbf{e}$, the tension of EFT strings with said charge, as well as elementary EFT strings, scale as 
	\begin{equation}
		\mathcal{T}_{\mathbf{e}}=M_{\rm Pl,4}^2\frac{{\rm deg}_\sigma P(s)|_{\rm flow}}{2\sigma}\;,\quad \mathcal{T}_{e^{i_0}}=M_{\rm Pl,4}^2\frac{{\rm deg}_\sigma P(s^{i_0})|_{\rm flow}}{2\sigma}\;,
	\end{equation}
	where for elementary EFT strings we only consider the degree with respect to  $s^{i_0}\sim e^{i_0}\sigma$. This way (not summing over $i_0$),
	\begin{equation}
		\left.\vec{\zeta}_{\mathcal{T}_\mathbf{e}}\cdot\vec{\zeta}_{\mathcal{T}_{e^{i_0}}}\right|_{\rm flow}=\left.\zeta^{j}_{\mathcal{T}_\mathbf{e}}\zeta^{i_0}_{\mathcal{T}_{e^{i_0}}}\mathsf{G}_{ji_0}\right|_{\rm flow}=\left.\frac{M_{\rm Pl,4}^4e^je^{i_0}\partial_j\partial_{i_0}K}{8 \mathcal{T}_{\mathbf{e}} \mathcal{T}_{e^{i_0}}} \right|_{\rm flow}=\left.\frac{-\sigma^2\partial_\sigma^2\log P(s^{i_0})}{2{\rm deg}_\sigma P(s){\rm deg}_\sigma P(s^{i_0})}\right|_{\rm flow}= \frac{1}{2{\rm deg}_\sigma P(s)|_{\rm flow}}
	\end{equation}
	if $i_0\in I$ and 0 it $i_0\not\in I$. Now, since the simplex $\Delta_I$ is generated by the elementary EFT strings $\vec{\zeta}_{\mathcal{T}_{e^i}}$, with $i\in I$, forming its vertices, points in $\Delta_I$ will be \textit{convex} linear combinations of $\{\vec{\zeta}_{\mathcal{T}_{e^i}}\}_{i\in I}$. However, as shown above, $\left.\vec{\zeta}_{\mathcal{T}_\mathbf{e}}\cdot\vec{\zeta}_{\mathcal{T}_{e^{i}}}\right|_{\rm flow}$ takes the same value for all $i\in I$, so that actually said product is the same for all points in $\Delta_I$ (including $\left.\vec{\zeta}_{\mathcal{T}_\mathbf{e}}\right|_{\rm flow}$). This means that \textbf{(a)} $\left.\vec{\zeta}_{\mathcal{T}_\mathbf{e}}\right|_{\rm flow}$ is perpendicular to $\Delta_{I}$, and \textbf{(b)} the distance from $\Delta_I$ to the origin is precisely ${\rm dist}(\mathbf{0},\Delta_I)=(2{\rm deg}_\sigma P(s)|_{\rm flow})^{-1/2}$.}
\begin{equation} \label{eftstring_degp}
	|\vec{\zeta}_{\mathcal{T}_\mathbf{e}}|^2_{\rm flow}=\frac{1}{2{\rm deg}_\sigma P(s)|_{\rm flow}}\;,
\end{equation}
with $P(s)$ defined as in \eqref{KP}. In this paper, we will always plot the vector for non-elementary strings evaluated along the flow $\vec{\zeta}_{\mathcal{T}_{\mathbf{e}}}|_{\rm flow}$ rather than evaluated in a generic point of moduli space (for elementary strings, the location is always the same, though). \\

Notice that the integer scaling relation in \eqref{scalingweight} only applies when moving along a particular trajectory corresponding to the saxionic string flow. It is useful then to rewrite it in terms of the $\zeta$-vectors as follows,
\beq
\label{lattice}
\boxed{\vec\zeta_{*} \cdot \vec\zeta_{\cT_{\mathbf{e}}}=w|\vec\zeta_{\cT_{\mathbf{e}}}|^2 \ , \quad \text{ with } w=1,2,3}
\eeq
which is now valid locally at any point of the perturbative regime. Here $\vec\zeta_\ast$ refers to the $\zeta$-vector of the leading tower of states, and $w$ is the scaling weight associated to the tower $m_\ast$ and EFT string with charge ${\mathbf{e}}$. In this paper, we have checked systematically whether this integer scaling relation also holds for other subleading towers of states along both elementary and non-elementary string flows. The result is that \emph{it holds for any string flow} (including non-elementary ones), \emph{and for the leading tower + all subleading towers of states at or below $\Lambda_{\rm QG}$ which are the generators of the convex hull} (i.e., that become the leading tower in \emph{some} trajectory of the perturbative regime, but can be subleading if moving along the string flow).\footnote{
Oscillator modes of non-BPS states will be located in the middle of facets, so they do not generate the convex hull and do not generically satisfy the integer scaling (moreover, they are at the species scale or higher). However, they will satisfy the integer scaling if we are moving along a string flow in which they are the leading tower.}
Hence, the integer scaling relation holds even more generally than what was originally anticipated in \cite{Lanza:2021udy}. Moreover, in all examples checked in this paper, it still happens that $w$ only takes three possible  values ($w=1,2,3$) for all towers in all string flows. This is an extremely strong condition that highly constraints the possible structure of the towers.\\

Equations \eqref{eftstring_degp} and \eqref{lattice} can be used to argue in more generality for the \textbf{convexity} on the scaling weight $w$ observed in \cite{Lanza:2021udy}. In a nutshell\footnote{To show this, consider \eqref{lattice} for an EFT string with charge $\mathbf{e}+\mathbf{e}'$, and consider that along its flow, the vector is perpendicular to the plane spanned by $\vec\zeta_{\cT_{\mathbf{e}}}$ and $\vec\zeta_{\cT_{\mathbf{e}'}}$ (also along their flows if they are not elementary):
\begin{equation}
w_{\mathbf{e}+\mathbf{e}'}=\frac{\left.\vec{\zeta}_0\cdot\vec{\zeta}_{\mathcal{T}_{\mathbf{e}+\mathbf{e}'}}\right|_{\rm flow}}{|\vec{\zeta}_{\mathcal{T}_{\mathbf{e}+\mathbf{e}'}}|^2_{\rm flow}}=\frac{\vec{\zeta}_0}{|\vec{\zeta}_{\mathcal{T}_{\mathbf{e}+\mathbf{e}'}}|^2_{\rm flow}}\cdot\frac{{\rm deg}_\sigma P(s)|_{\rm flow,\mathbf{e}}\vec\zeta_{\cT_{\mathbf{e}}}+{\rm deg}_\sigma P(s)|_{\rm flow,\mathbf{e}'}\vec\zeta_{\cT_{\mathbf{e}'}}}{{\rm deg}_\sigma P(s)|_{\rm flow,\mathbf{e}}+{\rm deg}_\sigma P(s)|_{\rm flow,\mathbf{e}'}}\;,
\end{equation}
where ${\rm deg}_\sigma P(s)|_{\rm flow,\mathbf{e}}$ is the degree with respect to $\sigma$ of $P(s)$ along the EFT string flow associated to the charge $\mathbf{e}$. Now, using \eqref{lattice} and \eqref{eftstring_degp}, we have
\begin{equation}
w_{\mathbf{e}+\mathbf{e}'}=\frac{{\rm deg}_\sigma P(s)|_{\rm flow,\mathbf{e}+\mathbf{e}'}}{{\rm deg}_\sigma P(s)|_{\rm flow,\mathbf{e}}+{\rm deg}_\sigma P(s)|_{\rm flow,\mathbf{e}'}}(w_{\mathbf{e}}+w_{\mathbf{e}'})\leq w_{\mathbf{e}}+w_{\mathbf{e}'}\,,
\end{equation}
since ${\rm deg}_\sigma P(s)|_{\rm flow,\mathbf{e}+\mathbf{e}'}\leq {\rm deg}_\sigma P(s)|_{\rm flow,\mathbf{e}}+{\rm deg}_\sigma P(s)|_{\rm flow,\mathbf{e}'}$ by definition.
}, given some tower with $\zeta$-vector $\vec{\zeta}_0$ and two EFT string charges $\mathbf{e}$ and $\mathbf{e}'$, then the scaling weight of said tower along an EFT string flow with charge $\mathbf{e}+\mathbf{e}'$ is less or equal than the sum of the scaling weight associated to each of the individual towers, i.e.,
\begin{equation}
w_{\mathbf{e}+\mathbf{e}'}\leq w_{\mathbf{e}}+w_{\mathbf{e}'}\;,
\end{equation}
where
\begin{equation}
\vec\zeta_{*} \cdot \vec\zeta_{\cT_{\mathbf{e}}}=w_{\mathbf{e}}|\vec\zeta_{\cT_{\mathbf{e}}}|^2\,\quad \vec\zeta_{*} \cdot \vec\zeta_{\cT_{\mathbf{e}'}}=w_{\mathbf{e}'}|\vec\zeta_{\cT_{\mathbf{e}'}}|^2\,,\quad \vec\zeta_{*} \cdot \vec\zeta_{\cT_{\mathbf{e}+\mathbf{e}'}}=w_{\mathbf{e}+\mathbf{e}'}|\vec\zeta_{\cT_{\mathbf{e}+\mathbf{e}'}}|^2\;.
\end{equation}

Taking the above results into account, the integer scaling relation \eqref{lattice} can be reformulated as follows:
\begin{equation*}
	\tcbhighmath[boxrule=1pt,drop fuzzy shadow=black]{\begin{matrix}\text{The $\zeta$-vectors of all the towers of states (generating the tower convex hull)}\\\text{ necessarily lie at points of the lattice generated by EFT string vectors.}\end{matrix}}
\end{equation*}

In this paper, we will check and illustrate this statement in a plethora of 4d $\mathcal{N}=1$ string compactifications, including Heterotic on Calabi-Yau threefolds, Type IIA on orientifolds, F-theory on Calabi-Yau fourfolds and M-theory on Joyce manifolds. We will study the arrangement of the towers of states, the species scale and the EFT strings along infinite distance limits of the moduli space, assuming $V=0$ for simplicity. See also \cite{Etheredge:2024amg,muldrowTBA} for other instances in the literature of a lattice structure relating the mass scale of towers and extended objects.

Turning on a scalar potential does not change the asymptotic exponential rates of the towers and strings, so the integer scaling relation \eqref{lattice} will be unchanged. However, the presence of the potential can obstruct certain infinite distance limits. In that sense, the reader should keep in mind that, in this paper, we are exploring the behavior along all \emph{a priori possible} infinite distance limits, but some of these limits might be obstructed by a diverging potential or quantum corrections in certain cases. Hence, the perturbative limits accessible by the EFT in concrete compactifications might only be a subset of the infinite distance limits explored in this paper.\\

Finally, equipped with this information, we can return to the key question of whether it is possible to determine the value of $w$ from bottom-up. 
Applying the condition \eqref{lattice} to the leading tower along a single EFT string flow is clearly insufficient to fix the tower’s mass scale, as it still allows for some freedom in choosing among $w=1,2,3$. However, imposing \eqref{lattice} along \emph{all} EFT string flows of a given perturbative regime is a much stronger constraint. In an accompanying paper \cite{bottomUpTBA} we will see that this condition, when combined with the Emergent String Conjecture, seems \emph{sufficient} to reconstruct from bottom-up the specific tower convex hulls that arise in known string theory compactifications. This enables us to extract  UV information of the towers of states directly from the IR K\"ahler potential.

\subsubsection{Bounds on the scaling weight\label{sec.bounds}}

Before delving into specific string theory compactifications, we want to provide some bounds on $w$ that can be derived from analyzing the interplay between the species scale $\Lambda_{\rm QG}$ and the EFT strings.

Recall that $\Lambda_{\rm QG}$ is defined as the scale at which quantum gravitational effect become important, and a local EFT description weakly coupled to gravity is no longer possible. The Emergent String Conjecture \cite{Lee:2019wij} implies then that $\Lambda_{\rm QG}$ is asymptotically given by either a higher dimensional Planck mass or the mass scale of the oscillator modes of a critical string, $m_{\rm str}$. If moving along an infinite distance limit corresponding to an EFT string flow, we will have $\Lambda_{\rm QG}\sim \sqrt{\mathcal{T}_{\mathbf{e}}}$ whenever the EFT string can be uplifted to a critical superstring at weak string coupling (so that \eqref{scalingweight} is realized with $w=1$). Otherwise, the EFT string should not be quantizable and their would-be excitation masses should be above the species scale\footnote{We also find instances in which the EFT string scales parametrically as a higher dimensional Planck scale, the latter playing the role of the species scale. In that case, the EFT string does not uplift to a critical string and the flow has $w>1$.} \cite{Cota:2022yjw, Martucci:2024trp}. This translates to the upper bound on the species scale proposed in \cite{Martucci:2024trp},
\begin{equation} \label{MleqT1}
	\Lambda_{\rm QG} \lesssim \sqrt{2\pi \mathcal{T}_{\mathbf{e}}}, 
\end{equation}
where in the presence of several EFT strings, $\mathcal{T}_{\mathbf{e}}$ is the  tension of the dominant (i.e., the lightest). Note that this will depend on the asymptotic limit and will in general vary as we move in different moduli space directions.  In this paper, we will provide further evidence for \eqref{MleqT1}, since the bound will be satisfied in all explored string theory examples.

Even though the bound above has only been argued along EFT string flows, it can be extended to any infinite distance limit as follows. First, observe that it is sufficient for condition \eqref{MleqT1} to hold at the boundary of the saxionic cone $\Delta$ in order for it to hold throughout its interior. This follows from the convexity of the polytope generated by constructing the convex hull of the various asymptotic $\mathcal{Z}$-vectors, and the fact that all EFT string $\zeta$-vectors lie on the same hyperplane. Next, notice that the $\zeta$-vectors of elementary EFT strings lie along the boundaries of the saxionic cone. Therefore, satisfying \eqref{MleqT1} along the EFT string flows implies that such hierarchy is preserved along any infinite distance limit. This will be confirmed in all string theory examples that we investigate in this paper. \\

Let us now use this bound to constrain the possible values of $w$. The condition \eqref{MleqT1} can be recast as the following constraint on the species scale $\mathcal{Z}$-vectors,
\begin{equation}
	\frac{\vec{\mathcal{Z}}_{\rm QG}\cdot \vec{\zeta}_{\mathcal{T}_{\mathbf{e}}}}{\abs{\vec{\zeta}_{\mathcal{T}_{\mathbf{e}}}}^2}	\geq 1 \ .
\end{equation}

Combining this with the integer scaling relation \eqref{scalingweight}, we get the following bound
\begin{equation} \label{cond_eftstr_Z}
\frac{n}{n+2} =\frac{\vec{\mathcal{Z}}_{\rm QG}\cdot \vec{\zeta}_{\mathcal{T}_{\mathbf{e}}}}{\vec{\zeta}_*\cdot \vec{\zeta}_{\mathcal{T}_{\mathbf{e}}}}\geq \frac{1}{w} %\leq 1.
\end{equation}
where in the first equality we have used the pattern \eqref{eq.PATTERN} for the species scale, and set the norm of the leading tower to be $\abs{\vec{\zeta}_*} = \sqrt{\frac{n+2}{2n}}$, where $n$ is the number of decompactifying dimensions\footnote{Here we are assuming that warping effects get diluted in the limit, since otherwise they can change the value of the exponential rate of a KK tower as in \cite{Etheredge:2023odp}.} ($n\to\infty$ for emergent strings). Here we are replacing the particular asymptotic behavior of the species scale associated to the leading tower entering in \eqref{cond_eftstr_Z}, which is collinear to the tower and therefore has $|\vec{\mathcal{Z}}_{\rm QG}|= \sqrt{\frac{n}{2(2+n)}}$ due to  \eqref{eq.PATTERN}. Notice that this bound remains valid for all string flows, even when the tower vector is not collinear with the string vector, since the angle dependence cancels out.\\

The constraint \eqref{cond_eftstr_Z} yields the following possible values of $n$ for each $w$ of a string flow. For $w=1$, the condition is only satisfied for $n \to \infty$, corresponding to identifying the leading tower with an emergent string tower $\zeta$-vector with modulus $|\vec{\zeta_{\rm osc}}| = \sqrt{\frac{1}{2}}$. This results in $\sqrt{\mathcal{T}}\sim \Lambda_{\rm QG}\sim m_{\rm osc}$, and  the EFT string uplifts to a critical emergent BPS string. 
	For $w=2$, the condition is satisfied for $n\geq 2$ (including $n\to\infty$).	
		Finally, for $w=3$, the condition \eqref{cond_eftstr_Z} is satisfied for all $n$. 

Analogously, \eqref{cond_eftstr_Z} can also give us precious information on the value of the scaling weight depending on the number of dimension which are decompactifying in each frame. Indeed, it turns out that if $n=1$, we need necessarily that $w\geq 3$.  In the top-down constructions along the following sections, this will be associated to M-theory limits in which a circle or the Ho\v{r}ava-Witten interval decompactify in strong coupling limits of type IIA/heterotic $E_8\times E_8$, respectively. If $n$ is bigger than 1 \textit{but finite} we can have in principle any scaling weight $\omega >1 $, but in practice only $w =\{2,3\}$ are found in string examples. Finally, for $n \to \infty$, where we recover the modulus of an emergent string tower vector, we get that a priori any value of $w$ is allowed. Whenever the EFT string uplifts to a critical string, then $w=1$, while cases with $w>1$ correspond to emergent string limits in which the critical string is non-BPS, so it does not correspond to an EFT string. In those latter cases, we have therefore two different strings; the EFT BPS string which uplifts to some wrapping brane, and the non-BPS critical emergent string becoming light faster (so $w>1$). \\

Although a priori limits with $n>1$ may have any $w> 1$, we can show that $w$ is actually upper bounded by $w\leq 3$ for $n>1$ (or $w\leq 2$ for $n=\infty$) if the number of extra dimensions is $n_{\rm{total}}\leq 7$. 
To do this, we just need to combine the integer scaling relation \eqref{lattice} with the result for the norm of the EFT string vector in \eqref{eftstring_degp}, obtaining that
\beq
\label{boundP}
w=\sqrt{\frac{n+2}{n}}\sqrt{{\rm deg_\sigma}P}\,{\rm cos}(\theta)\leq \sqrt{\frac{n+2}{n}}\sqrt{n_{\rm{total}}}
\eeq
where ${\rm deg_\sigma}P$ is the degree of the polynomial of the K\"ahler potential (see \eqref{KP}) along the string flow. Here, we have set again the norm of the tower vector to be $\abs{\vec{\zeta}_*} = \sqrt{\frac{n+2}{2n}}$, while $\theta$ is the angle between the tower vector $\vec\zeta$  and the EFT string vector $\vec\zeta_{\mathcal{T}_{\mathbf{e}}}$. In the last inequality we have used that the degree of $P(s)$ cannot be bigger than the total number of internal dimensions $n_{\rm{total}}$.
The above bound is therefore again valid for any string flow, even if the tower vector and the EFT string vector are not aligned. For an emergent string tower (i.e., $n=\infty$) this translates to
\begin{equation}
	w = {\rm cos}(\theta)\sqrt{{\rm deg_\sigma}P}\leq \sqrt{n_{\rm{total}}}
\end{equation}
 so $w\leq 2$ for emergent string towers arising from superstring theory and M-theory compactifications.

More generally, \eqref{boundP} places an upper bound on $w$ in terms of the degree of the polynomial $P(s)$ of the K\"ahler potential, which is in turn upper bounded by the total number of extra dimensions. For six or seven (total) extra dimensions, one recovers $w\leq 4$ for $n=1$ or $w\leq 3$ for $n>1$ but finite. Hence, $w=4$ is in principle also allowed by \eqref{boundP} if it is associated to decompactifying only one extra dimension while keeping ${\rm deg_\sigma}P=\{6,7\}$. However, this case is not realized in any of the top-down examples explored in this paper, where we always find $w\leq 3$. In fact, using  \eqref{boundP}, one can check that it would be inconsistent with having an integer value of the polynomial $P(s)$ in the case that the vectors were collinear (i.e. if assuming $\theta=0$)\footnote{Analogously, the cases excluded by \eqref{cond_eftstr_Z} (such as $n = 1$ with $w = 2$, or $w = 1$ with finite $n$) are also pathological when associated with flows where $\theta = 0$, as they would similarly lead to a non-integer degree for the polynomial $P(s)$, according to \eqref{boundP}.}. Therefore, if such decompactification limit exists, it cannot correspond to an elementary EFT string flow, along which all relevant vectors are collinear. This explains (although it does not prove) why we only find $w=1,2,3$ in all explicit top-down examples, even if M-theory compactifications can have ${\rm deg_\sigma}P$ up to 7. \\

In Table \ref{tab.bounds}, we summarize the possible values of $(w, n)$ allowed by the bounds discussed above. Entries marked in blue are excluded by the condition \eqref{cond_eftstr_Z}, while those in red are ruled out by \eqref{boundP} assuming a total of $n_{\rm total} = 7$ extra dimensions. The values in orange are compatible with \eqref{boundP} only if $n_{\rm total} = 7$, but not for $n_{\rm total} = 6$, and hence must arise from M-theory compactifications rather than from weakly coupled string theory. The intersection of both bounds restricts $(w, n)$ to a finite set of values, highlighted in green.
If we further require that ${\rm deg}_\sigma P$ is an integer (as occurs in all known string theory examples) then only the values marked with an ``$\times$" are consistent with \eqref{boundP} for elementary EFT string flows, where the tower and string vectors are aligned ($\theta = 0$). Combining all these constraints leaves only a small set of viable values, all of which are realized in explicit string constructions.
As we will see in concrete examples, non-elementary EFT string flows -- where the tower vector is not aligned with that of the string -- permit additional values of $(w, n)$ beyond those marked with an “$\times$”. Such cases are indeed realized in certain M-theory limits discussed in Section~\ref{s.Mth}. In contrast, the bounds \eqref{cond_eftstr_Z} and \eqref{boundP} apply more generally to any (elementary or non-elementary) string flow and are satisfied in all top-down constructions considered in this work.

\begin{table}[h]
	\centering
	\begin{tabular}{|c|ccccccccc|}
	\hline
	\backslashbox{$w$~}{$n$~~} &1&2&3&4&5&6&7&\dots&$\infty$\\\hline
	1&\cellcolor{blue!20!}&\cellcolor{blue!20!}&\cellcolor{blue!20!}&\cellcolor{blue!20!}&\cellcolor{blue!20!}&\cellcolor{blue!20!}&\cellcolor{blue!20!}&\cellcolor{blue!20!}&\cellcolor{green!20!}$\times$\\
	2&\cellcolor{blue!20!}&\cellcolor{green!20!}$\times$&\cellcolor{green!20!}&\cellcolor{green!20!}&\cellcolor{green!20!}&\cellcolor{green!20!}$\times$&\cellcolor{green!20!}&\cellcolor{green!20!}&\cellcolor{green!20!}$\times$\\
	3&\cellcolor{green!20!}$\times$&\cellcolor{green!20!}&\cellcolor{green!20!}&\cellcolor{green!20!}&\cellcolor{orange!20!}&\cellcolor{orange!20!}&\cellcolor{orange!20!}$\times$&\cellcolor{red!20!}&\cellcolor{red!20!}$\times$\\
	4&\cellcolor{green!20!}&\cellcolor{red!20!}$\times$&\cellcolor{red!20!}&\cellcolor{red!20!}&\cellcolor{red!20!}&\cellcolor{red!20!}$\times$&\cellcolor{red!20!}&\cellcolor{red!20!}&\cellcolor{red!20!}$\times$\\
	5&\cellcolor{red!20!}&\cellcolor{red!20!}&\cellcolor{red!20!}$\times$&\cellcolor{red!20!}&\cellcolor{red!20!}&\cellcolor{red!20!}&\cellcolor{red!20!}&\cellcolor{red!20!}&\cellcolor{red!20!}$\times$\\\hline	
	\end{tabular}
	\caption{Allowed values for $w$ (the scaling weight of the string flow) and $n$ (the number of decompactifying dimensions, with $n = \infty$ corresponding to an emergent string limit) for EFT string flows. Values in blue are excluded by \eqref{cond_eftstr_Z}, while those in red are ruled out by \eqref{boundP} assuming a total of $n_{\text{total}} = 7$ extra dimensions. Values in orange are allowed by \eqref{boundP} only if $n_{\text{total}} = 7$, but not if it is 6. Elementary string flows are further constrained if one requires the polynomial degree of $P(s)$ to be an integer along the flow, in which case only the values marked with an “$\times$” are allowed.\label{tab.bounds}}
\end{table}

\section{Heterotic $E_8\times E_8$ compactifications\label{s.het}}
We will start the study of 4d $\mathcal{N}=1$ tower arrangements and EFT string limits through compactifications of $E_8\times E_8$ heterotic string theory on a Calabi Yau 3-fold $X$. We will focus in the large volume/weak string coupling perturbative regime, so that the saxionic cone is parameterized by the 4d dilaton and the K\"ahler moduli.

\subsection{Asymptotic geometry and possible towers of states}

Starting from an integral basis $\{\omega_a\}_{a=1}^{h^{1,1}}$ of $H^{1,1}(X;\mathbb{Z})$, dual to a set of divisors (4-cycles) $\{[\omega_a]\}_{a=1}^{h^{1,1}}$, the K\"ahler form can be expanded as
\begin{equation}\label{e.kahler sax}
	J= s^a [\omega_a],
\end{equation}
with saxions $\{s^a\}_{a=1}^{h^{1,1}}\subset\mathbb{R}_{>0}$, and $J$ taking values in the K\"ahler cone $\mathcal{K}(X)$. For a choice of integral basis $\{[\omega_a]\}_{a=1}^{h^{1,1}}$, the \emph{triple intersection numbers} are then given by
\begin{equation}
	\kappa_{abc}=\int_X\omega_a\wedge \omega_b\wedge\omega_c\;.
\end{equation}
The 3-fold volume is then given in string units by
\begin{equation}
\label{hV}
	\mathcal{V}_X= \int_XJ\wedge J\wedge J=\frac{1}{3!}\kappa_{abc}s^as^bs^c\;,
\end{equation}
while the divisors $[\omega_a]$ and intersection curves $[\omega_a]\cap[\omega_b]$ read
\begin{subequations}\label{e. volumes}
\begin{align}
	\mathcal{V}_a &= \int_X \omega_a \wedge J \wedge J = \frac{1}{2}\kappa_{abc}s^bs^c \label{e. vol divisor}\\
	\mathcal{V}_{ab}&=\int_X \omega_a \wedge \omega_b \wedge J  = \kappa_{abc}s^c\;, \label{e. vol curve}
	\end{align}
\end{subequations}
also in string units. We can further define the universal saxion $s^0$ as
\begin{equation}
\label{Hets0}
	s^0 = e^{-2\Phi}\mathcal{V}_X = e^{-2\phi_4},
\end{equation}
with $\Phi$ and $\phi_4$ the 10- and 4-dimensional dilatons, respectively, so that $g_s=e^{\Phi}$. 

The saxionic cone is $\Delta= \mathbb{R}_{>0}\oplus \mathcal{K}(X)$ with $ \mathbb{R}_{>0}$ parameterized by $s^0$ and $\mathcal{K}(X)$ being the K\"ahler cone. In this perturbative regime, the K\"ahler potential takes the following leading form, modulo non-perturbative corrections that can be safely ignored,
\begin{equation}\label{e.kahler pot def}
	K = -\log s^0-\log \mathcal{V}_X\;,
\end{equation}
from which we can derive the saxionic moduli space metric \eqref{e.Kahler metric}
\begin{equation}  
	\mathsf{G}_{ij}=\frac{1}{2} \frac{\partial^2 K}{\partial s^i\partial s^j},  \ i,j=0,\dots, h^{2,1}\;.
\end{equation}
Since in general $\mathcal{V}_X$ is not a homogeneous function of the K\"ahler saxions, $\mathsf{G}_{ij}$ will neither be diagonal nor flat for an arbitrary set of intersection numbers $\kappa_{abc}$. However, as it will be explained in further detail in Section \ref{sec.no homo}, we will still be able to divide the asymptotic regime into growth sectors where $\mathcal{V}_X$ simplifies and is well approximated by a single monomial to leading order.

The 10-dimensional string frame metric is given by
\begin{equation}\label{e. metric het}
	\mathrm{d}s^2_{10}= e^{2A}\dd s_4^2+\ell_{\rm s}^2\dd s^2_X,
\end{equation}
where $\ell_{\rm s} = \sqrt{\alpha'}$ is the sting length, $\dd s_4^2$ is the 4d Einstein frame metric and $A$ is a warping factor. Through dimensional reduction and a pertinent conformal rescaling, the explicit expression for $A$ can be found to be
\begin{equation} \label{4dframe}
	e^{2A} = \frac{\ell_{\rm s}^2M^2_{\rm Pl,4}e^{2\Phi}}{4\pi \mathcal{V}_X} = \frac{\ell_{\rm s}^2M^2_{\rm Pl,4}}{4\pi s^0}.
\end{equation}
Having summarized the relevant aspects of the asymptotic geometry in this perturbative regime of the heterotic moduli space, we can now move on to the set of possible towers of states becoming light in the different asymptotic limits, as well as the duality frames that are associated with them. 
To start with, we have two types of EFT strings \cite{Lanza:2021udy}:
\begin{itemize}
	\item \textbf{Fundamental heterotic string}: 
	Using \eqref{4dframe}, we can write the fundamental string tension in 4d Planck units as follows,
	\begin{equation}\label{e.m osc het}
	\mathcal{T}_{\rm F1} =\frac{2\pi e^{2A}}{\ell_{\rm s}^2}= \frac{M_{\rm Pl,4}^2}{2s^0}=M_{\rm Pl,4}^2e^0 l_0\;.
		%\frac{m_{\rm osc}}{M_{\rm Pl,4}} \sim (s^0)^{-1/2}\;.
	\end{equation}
	as expected by \eqref{TEFT} since $l_0 = -\frac{1}{2}\frac{\partial K}{\partial s^0} = \frac{1}{2s^0}$. The oscillator modes of the string will have a characteristic mass scale given by
	\begin{equation}\label{eq.mosc}
		\frac{m_{\rm osc}}{M_{\rm Pl,4}}=\frac{\sqrt{\mathcal{T}_{\rm F1}}}{M_{\rm Pl,4}}\sim (s^0)^{-1/2}\;.
	\end{equation}
	\item \textbf{NS5 brane wrapping a 4-cycle}:  The tension of an EFT string arising from a NS5 brane wrapping an effective divisor $D=e^a[\omega_a]$ is given by 
	\begin{equation}
	\label{TNS5}
		\mathcal{T}_{\text{NS5},\mathbf{e}}= \frac{\pi e^{2A}}{\ell_{\rm s}^2}\int_{D}e^{-2\Phi}J\wedge J = M_{\rm Pl,4}^2e^al_a\;,
	\end{equation}
	where the dual saxions read
	\begin{equation}\label{eq.dualsaxions}
		  l_a=  -\frac{1}{2}\frac{\partial K}{\partial s^a} = \frac{1}{2\mathcal{V}_X}\kappa_{abc}s^bs^c=\frac{\mathcal{V}_a}{\mathcal{V}_X}\;.
	\end{equation}
	The string becomes tensionless and weakly coupled when the volume $\mathcal{V}_a$ of the divisor shrinks in comparison to the overall volume. If the asymptotic geometry of this divisor is  $\mathbb{T}^4$ or K3 (these are the only possibilities, see \cite{Oguiso1993,Grimm:2019bey}), the EFT string is dual to the fundamental Type IIA string or the fundamental heterotic string in a dual  $E_8\times E_8$ theory, respectively (see e.g.\cite{Lee:2019xtm}).	
\end{itemize}
	
\noindent	In addition to the EFT strings, we also have different types of towers of states becoming light in this perturbative regime:
	\begin{itemize}
	\item \textbf{KK towers}: As we take different decompactification limits, we can get towers of KK modes with a characteristic mass given by
	\begin{equation}
	\label{KKtowers}
		\frac{m_{{\rm KK},\mathcal{C}}}{M_{\rm Pl,4}} = (s^0)^{-1/2}\mathcal{V}_\mathcal{C}^{-1/{\rm dim}_{\mathbb{R}}\mathcal{C}}\;.
	\end{equation}
Depending on whether $\mathcal{V}_{\mathcal{C}}$ is associated to a curve, a divisor or the overall volume, with volumes given by \eqref{hV} and \eqref{e. volumes}, we get the following types of KK towers:
	\begin{subequations}\label{eq.KK het}
	\begin{align}
	\frac{m_{{\rm KK},X}}{M_{\rm Pl,4}}&\sim (s^0)^{-1/2}(\kappa_{abc}s^as^bs^c)^{-1/6}\sim (s^0)^{-1/2}\mathcal{V}_X^{-1/6}\label{eq.KKX het}\\
	 \frac{m_{{\rm KK},a}}{M_{\rm Pl,4}}&\sim (s^0)^{-1/2}(\kappa_{abc}s^b s^c)^{-1/4}\label{eq.KK divisor}\\
	  \frac{m_{{\rm KK},ab}}{M_{\rm Pl,4}}&\sim (s^0)^{-1/2}(\kappa_{abc}s^c)^{-1/2}\;,\label{eq.KK curve}
	\end{align}
	\end{subequations}
	respectively decompactifying 6, 4 and 2 internal dimensions.
	\item \textbf{M-theory tower}: The above 4d perturbative regime also contains regions of strong coupling for the 10d dilaton as follows: take e.g. $\mathcal{V}_X\rightarrow \infty$  while keeping $s^0 = e^{-2\Phi}\mathcal{V}_X$ fixed, so that the 10d dilaton $g_s = e^{\Phi}\sim \sqrt{\mathcal{V}_X}\to\infty$. In heterotic $E_8 \times E_8$ string theory, the strong coupling limit is resolved through H\v{o}rava-Witten theory, i.e. M-theory compactified on an interval $S^1/\mathbb{Z}_2$ \cite{Horava:1995qa,Horava:1996ma}. Thus, we have a new KK scale associated to the length of the M-theory interval which in 4d Planck units is given by 
	\begin{equation}\label{eq. Mth het}
		m_{{\rm KK,M-th}}\sim \frac{M_{\rm Pl,10}}{\mathbf{R}_{10}^{9/8}}\sim \frac{M_{\rm Pl,4}} {{\mathcal{V}}_X^{1/2}}\;,
	\end{equation}
	where $\mathbf{R}_{10}=e^{\frac{2}{3}\Phi}$ is the radius of the H\v{o}rava-Witten interval in $M_{\rm Pl,11}$ units, and we have used that
$M_{\rm Pl,4}=M_{\rm Pl,10}V_X^{1/2}=M_{\rm Pl,10} (s^0)^{3/8}\mathcal{V}_X^{1/8}$.
	\item \textbf{Tower of wrapped M2-branes}: In regions where the 10d dilaton grows but some 2-cycle shrinks, we can also get towers of particle states from wrapping M2-branes of increasing winding number. In particular, take this 2-cycle to be an elliptic curve of volume $\mathcal{V}_\mathbb{E}$ in string units. The mass scale of the towers is given by
	\begin{align}\label{eq.M2}
		m_{\rm M2}=M_{\rm Pl,11}^3\,\ell_s^{2}\mathcal{V}_{\mathbb{E}}=M_{\rm Pl,4}\frac{\mathcal{V}_{\mathbb{E}}}{\sqrt{\mathcal{V}_X}}\;.
	\end{align}
	where we have used $M_{\rm Pl,11}=e^{-\Phi/12}M_{\rm Pl,10}=(s^0)^{-1/3}\mathcal{V}_X^{-1/6}M_{\rm Pl,4}$.
	
	\item Finally, we can also have \textbf{bound states} of the above towers of states. For instance, when decompactifying two different spaces $\mathcal{C}_n$ and ${\mathcal{C}_m}$, we will get KK towers scaling as 
	 \begin{equation}\label{e.bounded}
	 \frac{m_{{\rm KK},\mathcal{C}_n\times\mathcal{C}_m}}{M_{\rm Pl,4}}=\left[\left(\frac{m_{{\rm KK},\mathcal{C}_n}}{M_{\rm Pl,4}}\right)^n \left(\frac{m_{{\rm KK},\mathcal{C}_m}}{M_{\rm Pl,4}}\right)^m\right]^{\frac{1}{n+m}}\;.
	 \end{equation}
	 They will only become important when decompactifying the two spaces at the same rate. Moreover, it will also be useful to consider bound states of the M-theory KK tower and the tower of wrapping M2-branes, yielding
	\begin{equation}\label{eq. M-th+M2}
		\frac{m_{\rm KK, M-th+M2}}{M_{\rm Pl,4}}\sim \mathcal{V}_{\mathcal{B}}^{-1/2}\;,
	\end{equation}
\end{itemize}

In the above derivations of the scalings we have been implicitly working in the large momentum limit, where by Weyl's law the KK mass is parametrically set by the internal volume, and additional corrections (as the towers are non-BPS) become subleading asymptotically. A similar logic can be applied to particles resulting from wrapped branes which will be dual to KK towers in a dual frame, so that their mass is given by the brane tension times the wrapped internal volume to leading order in the infinite distance limit.

In addition to the scaling of the towers of states, we will also be interested in the result for the species scale $\Lambda_{\rm QG}$ (see \ref{eq.species}) in each limit.  Since all the limits will correspond to either decompactification or perturbative string limits (as predicted by the Emergent String Conjecture \cite{Lee:2019xtm,Lee:2019wij}), the species scale will be either a string scale $\Lambda_{\rm QG}\approx m_{\rm osc}\sim \sqrt{\mathcal{T}_{\rm F1}}$ or a higher dimensional Planck mass with
\begin{subequations}
\begin{align}\label{eq.Mpl}
	\frac{M_{{\rm Pl},4+n}}{M_{\rm Pl,4}}&\sim (s^0)^{-\frac{n}{2(n+2)}}\mathcal{V}_{\mathcal{C}_n}^{\frac{1}{n+2}}\\
	\frac{M_{\rm Pl,11}}{M_{\rm Pl,4}}&\sim (s^0)^{-1/3}\mathcal{V}_X^{-1/6}\label{Mp11Mth}
	\\
\frac{M_{\rm Pl,5}}{M_{\rm Pl,4}}&\sim \mathcal{V}_X^{-1/6}\label{MpMth}
	\end{align}
\end{subequations}
with $\mathcal{V}_{\mathcal{C}_n}$ the volume of the decompactifying $n$-cycle in string units. Hence, \eqref{eq.Mpl} is the higher dimensional Planck scale associated to the decompactification signaled by the heterotic KK towers in \eqref{KKtowers}, while \eqref{MpMth} is the Planck scale associated with decompactifying the M-theory circle (KK tower in \eqref{eq. Mth het}). We can also have combinations in which we decompactify both some cycle ${\mathcal{C}_n}$ and the M-theory circle, yielding
\begin{equation}
\frac{M_{{\rm Pl,}5+n}}{M_{\rm Pl,4}}\sim(s^0)^{-\frac{n}{2(n+3)}}\mathcal{V}_{\mathcal{C}_n}^{-\frac{1}{n+3}}\mathcal{V}_X^{-\frac{1}{2(n+3)}}\;,
\end{equation}
which generalizes \eqref{Mp11Mth} and \eqref{MpMth}. In general, the $M_{{\rm Pl},4+k}$ can be given in terms of the associated KK-$k$ tower scale by \cite{Castellano:2023jjt}
\begin{equation}\label{eq.species from KK}
\frac{M_{{\rm Pl},4+k}}{M_{\rm Pl,4}}\sim\left(\frac{m_{{\rm KK},k}}{M_{\rm Pl,4}}\right)^{\frac{n}{2+n}}
\end{equation}

In the presence of multiple towers becoming light, the species scale will be in practice the lowest scale of all the above possibilities.

\subsection{Top-down examples}

In the following we will investigate how the different light towers and EFT strings fit globally within the perturbative regime,  which will result into a description of the different duality regimes that can arise asymptotically. To do this, we are going to divide the perturbative regime of the moduli space in different regions, in which the overall volume will be asymptotically dominated by a single homogeneous monomial in the K\"ahler saxions, either $\mathcal{V}_X\sim (s^1)^3$, $\mathcal{V}_X\sim (s^1)^2s^2$ or $\mathcal{V}_X\sim s^1s^2s^3$. They will then serve as building blocks to describe the general case of a heterotic Calabi-Yau compactification. We will describe how to \emph{glue} these regimes in Section \ref{sec.no homo}.  Notice that addtional structure of the Calabi-Yau compactification (such as gauge bundles, etc) will play no role in our discussion; even if they are essential to determine the specific dual theory, we will only be interested in the coarse-grained information about the microscopic nature of the asymptotic limit (i.e. whether it is a perturbative limit, how many dimensions decompactify, etc.) that can be read from the asymptotic arrangement of the towers. We will also make extensive use along this section of the known result \cite{Oguiso1993,Grimm:2019bey} that a CY threefold only admits three types of fibration structures, where the fiber is topologically either a $\mathbb{T}^2$, $\mathbb{T}^4$, or K3.

\subsubsection{Case 1: $\mathcal{V}_X\sim (s^1)^3$\label{s.het vols3}}
The first and simplest example has a moduli space parameterized only by two scalars, namely the universal saxion $s^0$ and a single K\"ahler saxion, with $\mathcal{V}_X=(s^1)^3$. From \eqref{e.Kahler metric}, the moduli space metric is found to be diagonal and flat, given by
\begin{equation}
	\mathsf{G}_{00} = \frac{1}{2(s^0)^2}, \, \mathsf{G}_{11} = \frac{3}{2(s^1)^2}, \, \mathsf{G}_{01}=\mathsf{G}_{10}=0\;.
\end{equation}
The dual saxions can be computed from \eqref{eq.dualsaxions}, $l_0=\frac{1}{2}(s^0)^{-1}$ and $l_1=\frac{3}{2}(s^1)^{-1}$, in such a way that the tension of the elementary EFT strings is given by
\beq
\mathcal{T}_{e^0}=\mathcal{T}_{\rm F1}=M_{\rm Pl,4}^2e^0 l_0=\frac{M_{\rm Pl,4}^2}{2s^0}\ , \quad  \mathcal{T}_{e^1}=\mathcal{T}_{\rm NS5}=M_{\rm Pl,4}^2e^1 l_1=\frac{3M_{\rm Pl,4}^2}{2s^1}
\eeq
where we have used \eqref{e.m osc het} and \eqref{TNS5}. We denote the elementary string charged in the $i$-th direction as $\mathcal{T}_{e^i}$.
The $\zeta$-vectors associated to the mass scale of the EFT strings (see \eqref{eq.mosc} and \eqref{TNS5}) are given by\footnote{Note that the 1 or 0 subindices indicate the presence or absence of charge component in $\mathbf{e}$, with all EFT strings from each family having the same scaling. This is, the EFT strings with $\mathbf{e}=(1,1)$ and $(5,9)$ have the same asymptotic dependence on the scalars modulo numerical factors.}
\begin{equation}\label{eq.EFT 1}
\vec{\zeta}_{\mathcal{T}_{e^0}}=\left(\frac{1}{\sqrt{2}},0\right)\;,\quad\vec{\zeta}_{\mathcal{T}_{e^1}}=\left(0,\frac{1}{\sqrt{6}}\right)\;,\left.\quad\vec{\zeta}_{\mathcal{T}_{(e^0,e^1)}}\right|_{\rm flow}=\left(\frac{1}{4\sqrt{2}},\frac{1}{4}\sqrt{\frac{3}{2}}\right)\;,
\end{equation}

As discussed in Section \ref{ss.summary}, the $\zeta$-vector for non-elementary strings generally depends on the values of the charges $\mathbf{e}$ and the location in moduli space, interpolating along the simplex generated by the $\zeta$-vectors of the corresponding elementary strings. However, as shown in Footnote \ref{fn.perp}, the expression of  $\vec{\zeta}_{\mathcal{T}_{\mathbf{e}}}$ along the \emph{EFT flow} generated by the string $\mathcal{T}_{\mathbf{e}}$ is fixed and  given by the point of said simplex closest to the origin, namely  $\vec{\zeta}_{\mathcal{T}_{\mathbf{e}}}|_{\rm flow}$ above.

 The towers of particle states becoming light are the tower of oscillator modes of the fundamental heterotic, the M-theory KK tower and the KK tower associated to the overall volume of the Calabi-Yau $X$, with masses given by \eqref{eq.mosc}, \eqref{eq.KKX het} and \eqref{eq. Mth het} respectively. The $\zeta$-vectors of these towers read
\begin{equation}
	\vec{\zeta}_{\rm osc}=\left(\frac{1}{\sqrt{2}},0\right)\;,\quad\vec{\zeta}_{{\rm KK},X}=\left(\frac{1}{\sqrt{2}},\frac{1}{\sqrt{6}}\right)\;,\quad\vec{\zeta}_{\rm KK, M-th}=\left(0,\sqrt{\frac{3}{2}}\right)\;.
\end{equation}
We can also have bound states with $\vec{\zeta}_{\rm KK,7}=\frac{1}{7}\left(\vec{\zeta}_{\rm KK, M-th}+6\vec{\zeta}_{{\rm KK},X}\right)=\left(\frac{3\sqrt{2}}{7},\frac{3}{7}\sqrt{\frac{3}{2}}\right)$ associated to the decompactification of both the M-theory interval and the whole 3-fold. Hence, depending on the infinite distance limit taken, we will be decompactifying 1, 6 or 7 dimensions.

The role of the species scale will then be played by the fundamental string scale, the 5d Planck scale, the 10d Planck scale or the 11d Planck scale. The $\mathcal{Z}$-vectors of these scales are given by
\begin{equation}
	\vec{\mathcal{Z}}_{\rm str}=\left(\frac{1}{\sqrt{2}},0\right),\quad \vec{\mathcal{Z}}_{\rm Pl_5}=\left(0,\frac{1}{\sqrt{6}}\right),\quad \vec{\mathcal{Z}}_{\rm Pl_{10}}=\left(\frac{3}{4\sqrt{2}},\frac{1}{4}\sqrt{\frac32}\right),\quad \vec{\mathcal{Z}}_{\rm Pl_{11}}=\left(\frac{\sqrt{2}}{3},\frac{1}{\sqrt{6}}\right)\,.
\end{equation}

\begin{figure}[h] 	
 	\centering
			\includegraphics[width=0.55\textwidth]{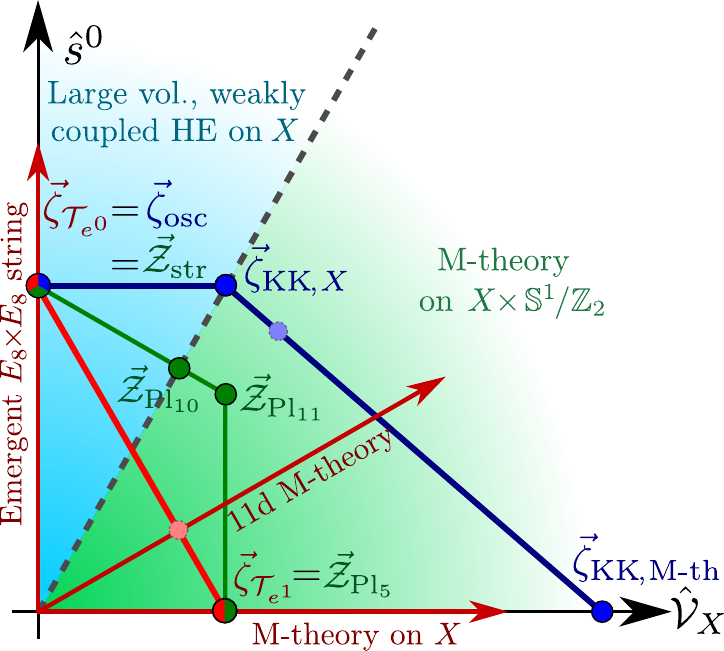}
		\caption{Arrangement of $\zeta$-vectors for towers (blue) and EFT strings (red), as well as $\mathcal{Z}$-vectors for the possible asymptotic expression of the species scale $\Lambda_{\rm QG}$ (green), in terms of the canonically normalized fields associated to the universal saxion $\hat{s}^0$ and the overall volume in string units $\hat{\mathcal{V}}_X\equiv\hat{s}^1$. The dots represent the different $\vec\zeta$ or $\vec{\mathcal{Z}}$ vectors and the lines correspond to the convex hull of the vectors. We also show the different 4d perturbative descriptions, as well as the emergent dual theories along the different EFT string flow (darker red arrows). The $\zeta$-vectors of bounded KK states and non-elementary EFT string are not labeled, and depicted in lighter colors only for EFT string flows or if they are co-leading in the limit.}\label{fig:het2d}
\end{figure}

All the above $\zeta$-vectors of the towers of states, the EFT strings and the species scale are plotted in Figure \ref{fig:het2d}. We will use the following color code in this paper: blue for the towers of particle states, red for the EFT strings and green for the species scale. We can see that the convex hull of the towers is always outside the convex hull of the species scale, which is in turn outside of the convex hull of the EFT strings. As described at the end of Section \ref{ss.summary}, this is consistent with having a hierarchy of scales of the form $m_*\leq \Lambda_{\rm QG}\leq \mathcal{T}^{1/2}$, since the convex hull is a sense a measure of the exponential decay rate of the corresponding mass scale.\\

The elementary string flows, as defined in \eqref{saxionflow}, are given by $s^0\rightarrow \infty$ and $s^1\rightarrow \infty$. One can check that indeed the integer scaling $\vec{\zeta}_{\mathcal{T}_{\mathbf{e}}} \cdot \vec{\zeta}_* = w \  \abs{\vec{\zeta}_{\mathcal{T}_{\mathbf{e}}}}^2$ in \eqref{lattice} (with $w \in \{0,1,2,3\}$) holds, not only for the leading towers, but also for all subleading towers generating the convex hull along the different (both elementary and non-elementary) EFT string limits $\mathbf{e}=(e^0,0)$, $(0,e^1)$ and $(e^0,e^1)$.
The identification of the leading tower, the species scale and the value of $w$ (for the leading tower) along each string flow are summarized in Table \ref{tab.bigs3}. For completeness, the scaling weights of the (sub)leading towers (namely, those with a characteristic mass scale in between the leading tower and the species scale) are shown in Table \ref{tab.w s3}. The leading tower in each limit is highlighted in bold face, so that its integer value of $w$ fixes the scaling weight of the string flow, which is also shown in Table  \ref{tab.bigs3}. The EFT string limits are represented as red arrows in Figure \ref{fig:het2d}.

\begin{table}
	\begin{center}
			\begin{tabular}{|c|c|c|c|c|c|}
			\hline
			$\mathbf{e}$ &Lead. tower(s) $m_\star$ & $w$ & $\Lambda_{\text{QG}}$ & Sublead. towers & Emergent dual theory \\
			\hline 
			\rowcolor{blue!10!}$(e^0,0)$ &  $m_{\text{osc}}$, $m_{\text{KK},X}$& 1 & $m_{\rm str}$ & $\setminus$ & Emergent $E_8\times E_8$ string\\
			\hline
		\rowcolor{red!10!}	$(0,e^1)$ &$m_\text{KK,M-th}$ & 3& $M_\text{Pl,5}$&$m_{\text{KK},X}$& M-theory on $X$ \\
			\hline
			\rowcolor{red!10!}$(e^0,e^1)$ & $m_\text{KK,M-th}$  & 3&$M_\text{Pl,11}$ & $m_{\text{KK},X}$& 11d M-theory\\
			\hline
		\end{tabular}
		\caption{Description of the different EFT string limits driven by the string charges ${\mathbf{e}}$  for $E_8\times E_8$ heterotic theory on a CY 3-fold with $\mathcal{V}_X\sim (s^1)^3$. We provide the identification of the leading tower(s) $m_\ast$, its scaling weight $w$, the asymptotic species scale $\Lambda_{\rm QG}$, subleading towers with $m\lesssim \Lambda_{\rm QG}$ and the microscopic interpretation of the dual theory emerging at the infinite distance limit. \label{tab.bigs3}} 
	\end{center}
\end{table}

\begin{table}[h]
	\centering
	\begin{tabular}{|c|ccc|}
	\hline
	$\mathbf{e}$ &$\vec{\zeta}_{\rm osc}$&$\vec{\zeta}_{{\rm KK},X}$&$\vec{\zeta}_{\rm KK,M-th}$\\\hline
	$(e^0,0)$&\textbf{1}&\textbf{1}&0\\
	$(0,e^1)$&0&1&\textbf{3}\\
	$(e^0,e^1)$&1&2&\textbf{3}\\\hline	
	\end{tabular}
	\caption{Integer scaling weight $w$ of the different towers of states along each EFT string limit, for $E_8\times E_8$ heterotic theory on a CY 3-fold with $\mathcal{V}_X\sim (s^1)^3$. Each tower with vector $\vec{\zeta}$ satisfies  $\vec{\zeta}_{\mathcal{T}_{\mathbf{e}}} \cdot \vec{\zeta} = w \  \abs{\vec{\zeta}_{\mathcal{T}_{\mathbf{e}}}}^2$ along the flow driven by an EFT string with vector $\vec{\zeta}_{\mathcal{T}_{\mathbf{e}}}$ and charge ${\mathbf{e}}$. The leading tower(s) for each EFT string limit are highlighted in bold face. \label{tab.w s3}}
\end{table}

Finally, we can comment on the different 4d dual descriptions that emerge in different corners of this perturbative region of the moduli space (also depicted in Figure \ref{fig:het2d}). To do this, we need to consider general infinite distance trajectories within the saxionic cone, and not only EFT string limits. In this case, the perturbative regime is divided in two regions corresponding to two different duality frames (blue and green regions in Figure \ref{fig:het2d}). Each of these two regions will have a different microscopic interpretation of the species scale:
\begin{itemize}
\item  For $s^0\gg (s^1)^3$ with $s^0\rightarrow \infty$ (blue region in Figure \ref{fig:het2d}), we have $g_s^2 \sim (s^0)^{-1}(s^1)^3\rightarrow 0$, using \eqref{Hets0}. Hence, this region is described by \textbf{weakly coupled heterotic $E_8\times E_8$ string theory on the CY$_3$}, with the species scale being the heterotic string scale. Depending on the scaling of $s^1$ we can reach the following theories at infinite distance:
\begin{itemize}
	 \item For $s^0\gg s^1\sim s^1_0$, we have a weak coupling limit in which $g_s\rightarrow 0$ while the volume of $X$ in string units remains fixed. The lightest tower is that of heterotic string oscillators, $m_{\rm osc}$, which also sets the QG cutoff, $\Lambda_{\rm QG}=m_{\rm str}$. As usual for these limits, there is also a KK tower whose mass  $m_{{\rm KK},X}$ decays at the same rate as $m_{\rm osc}$, since the KK copies of the string modes also become light. It is, therefore, an \textbf{emergent $E_8\times E_8$ heterotic string} limit.  
	 \item For $s^0\gg (s^1)^3\to\infty$, the string mass $m_{\rm s}$ is becoming light in $M_{ \rm Pl,4}$ units, while $\mathcal{V}_X$ becomes large with respect to $\ell_{\rm s}$. Here, while the species scale is still given by $\Lambda_{\rm QG}=m_{\rm str}$, we have a light tower given by $m_{{\rm KK},X}$. The limit corresponds to a \textbf{weakly coupled SUGRA description of 10d $E_8\times E_8$ heterotic string theory}. 
	 \item For $s^0\sim (s^1)^3\to\infty$, we have that $g_s\sim $ constant, so that   the string scale remains constant in 10d Planck units, $m_{\rm str}\sim M_{\rm Pl,10}$. In this precise limit the species scale is given by $M_{\rm Pl,10}$ and the lightest tower is $m_{{\rm KK},X}$. Thus our limit corresponds to a decompactification to a 10d theory with fixed string coupling, better described as a \textbf{10d Heterotic $E_8\times E_8$ SUGRA}. As one can see in Figure \ref{fig:het2d}, moving in such direction, the $m_{\rm KK,M-th}$ and $m_{\rm osc}$ towers become light at the same rate as $M_{\rm Pl,10}$, precisely corresponding to the towers present in the 10d moduli space of the heterotic $E_8\times E_8$ theory.
	 \end{itemize}
	 \item  For $(s^1)^3\gg s^0\rightarrow \infty$  (green region in Figure \ref{fig:het2d}), we have $g_s^2 \sim (s^0)^{-1}(s^1)^3\rightarrow \infty$. Hence,  we go to strong string coupling, opening up the M-theory Ho\v{r}ava-Witten interval, so that  this asymptotic region of the moduli space is better described by \textbf{M-theory on the CY$_3$} and the species scale is given by the 11d Planck scale. Depending on the concrete scaling of $s^0$ we can reach the following theories at infinite distance:
	 \begin{itemize}
	 \item For $(s^1)^3\gg s^0\to\infty$, the overall $X$ volume is also growing. Hence, even if the M-theory KK tower $m_{\rm KK,M-th}$ is the leading one, we also have the overall $m_{{\rm KK},X}$ tower becoming light, which results in the species scale being given by $\Lambda_{\rm QG}=M_{\rm Pl,11}$. We are thus decompactifying to \textbf{11d M-theory}. When moving exactly along the direction of $\vec{\mathcal{Z}}_{\rm Pl_{11}}$ (i.e. in the limit $(s^1)^2\sim s^0$, one can check that indeed there are no light towers lying at the $M_{\rm Pl,11}$ scale, as it is expected from the absence of moduli in 11d M-theory.
	 \item For $s^1\gg s^0\sim s^0_0$ with $s^1\to\infty$, $M_{\rm Pl,5}$, $M_{\rm Pl,11}$ and $m_{{\rm KK},X}$ all scale at the same rate, see \eqref{eq.Mpl} and \eqref{eq.KKX het}. Hence, the leading tower is still given by $m_{\rm KK,M-th}$ but the species scale is given by $\Lambda_{\rm QG}=M_{\rm Pl,5}$, since along this limit $m_{{\rm KK},X}\sim M_{\rm Pl,5}\to 0$, and does these KK states do not contribute to $\Lambda_{\rm QG}$. Therefore, this signals the decompactification to a 5d theory given by \textbf{M-theory on $X$}. Again, the fact that $m_{{\rm KK,}X}$ lies at the $M_{\rm Pl,5}$ scale signals the existence of this tower in the resulting 5d theory.
\end{itemize}
\end{itemize}

\subsubsection{Case 2: $\mathcal{V}_X\sim (s^1)^2 s^2$ \label{ss.het 2sec}}
We now consider a second, more involved case. Consider an overall volume which behaves to leading order as $\mathcal{V}_X\sim (s^1)^2 s^2$, so the field metric (again flat) is given by
\begin{equation}
	{\mathsf{G}}_{ij}= \text{diag}\left(\frac{1}{2(s^0)^2}, \ \frac{1}{(s^1)^2}, \ \frac{1}{2(s^2)^2}\right)\;,
\end{equation}
Here, the only non-vanishing intersection number is $\kappa_{112}$, so that we have the following nef curves and divisors
\begin{align}
\begin{array}{rl}
\text{Divisor }[\omega_2]\text{ with }\mathcal{V}_2\sim (s^1)^2\, ,& \quad \text{Divisor }[\omega_1]\text{ with }\mathcal{V}_1\sim s^1s^2\\
\text{Curve }[\omega_1]\cap[\omega_1]\text{ with }\mathcal{V}_{11}\sim s^2\, ,&\quad
\text{Curve }[\omega_1]\cap[\omega_2]\text{ with }\mathcal{V}_{12}\sim s^1
\end{array}
\end{align}
where we have used \eqref{e. volumes} to compute the volumes.

Using the above field metric and the result for the EFT string tensions in \eqref{e.m osc het} and \eqref{TNS5}, we can compute the $\zeta$-vectors for the elementary (and non-elementary) EFT strings, obtaining
\begin{equation}\label{e.EFT string s12 s2}
\begin{array}{c}
\vec{\zeta}_{\mathcal{T}_{e^0}}=\left(\frac{1}{\sqrt{2}},0,0\right)\;,\quad
\vec{\zeta}_{\mathcal{T}_{e^1}}=\left(0,\frac{1}{2},0\right)\;,\quad
\vec{\zeta}_{\mathcal{T}_{e^2}}=\left(0,0,\frac{1}{\sqrt{2}}\right)\;\\
\left.\vec{\zeta}_{\mathcal{T}_{(e^0,e^1,0)}}\right|_{\rm flow}=\left(\frac{1}{3\sqrt{2}},\frac{1}{3},0\right)\;,\quad
\left.\vec{\zeta}_{\mathcal{T}_{(e^0,0,e^1)}}\right|_{\rm flow}=\left(\frac{1}{2\sqrt{2}},0,\frac{1}{2\sqrt{2}}\right)\\
\left.\vec{\zeta}_{\mathcal{T}_{(0,e^1,e^2)}}\right|_{\rm flow}=\left(0,\frac{1}{3},\frac{1}{3\sqrt{2}}\right)\;,\quad
\left.\vec{\zeta}_{\mathcal{T}_{(e^0,e^1,e^2)}}\right|_{\rm flow}=\left(\frac{1}{4\sqrt{2}},\frac{1}{4},\frac{1}{4\sqrt{2}}\right)\;,
\end{array}
\end{equation}
where again for non elementary EFT strings the $\zeta$-vector is only given along the associated EFT string flow directions.
Notice that all the EFT strings vectors are located in the plane spanned by the $\zeta$-vector of the three elementary EFT strings. Two of these elementary strings ($\vec{\zeta}_{\mathcal{T}_{e^0}}$ and $\vec{\zeta}_{\mathcal{T}_{e^2}}$) have the right value of $|\vec{\zeta}|$ to correspond to a fundamental critical string (see \eqref{e.taxonomy}). Indeed, the first one is the fundamental heterotic string, while we will see that the third one (which originates from an NS5 wrapping the divisor $[\omega_2]$) corresponds to another fundamental heterotic string in a dual frame. 

Using the expression for the masses of the possible towers at the beginning of the Section, we can also compute the $\zeta$-vectors of all the relevant light towers of states. In addition to towers of string oscillator modes, we also have the following KK towers, particularized to the case of interest with $\mathcal{V}_X\sim (s^1)^2 s^2$:
\begin{equation}
\begin{array}{c}
	\vec{\zeta}_{{\rm KK},X}=\left(\frac{1}{\sqrt{2}},\frac{1}{3},\frac{1}{3\sqrt{2}}\right)\;,\quad\vec{\zeta}_{\rm KK,M-th}=\left(0,1,\frac{1}{\sqrt{2}}\right)\\
		\vec{\zeta}_{{\rm KK},[\omega_2]}=\left(\frac{1}{\sqrt{2}},\frac{1}{2},0\right)\;,\quad\vec{\zeta}_{{\rm KK},[\omega_1]\cap[\omega_1]}=\left(\frac{1}{\sqrt{2}},0,\frac{1}{\sqrt{2}}\right)\;,
\end{array}
	\end{equation}
	Here, we have used \eqref{eq.KKX het} and \eqref{eq. Mth het} for the overall KK tower $\vec{\zeta}_{{\rm KK,X}}$ and the M-theory KK tower $\vec{\zeta}_{\rm KK,M-th}$. Similarly, the other two KK towers are obtained from using \eqref{eq.KK divisor} and \eqref{eq.KK curve} and are associated to the divisor $[\omega_2]$ and the curve $[\omega_1]\cap[\omega_1]$ respectively. Notice that the overall KK tower can also be obtained from considering bound states of the KK towers associated to the above divisor and curve on which $X$ factorizes.\footnote{Similarly to the above, we could also compute the $\zeta$-vectors associated to the divisor $[\omega_1]$ and curve $[\omega_1]\cap[\omega_2]$, with
\begin{equation}
	\vec{\zeta}_{{\rm KK},[\omega_1]}=\left(\frac{1}{\sqrt{2}},\frac{1}{4},\frac{1}{2\sqrt{2}}\right)\;,\quad\vec{\zeta}_{{\rm KK},[\omega_1]\cap[\omega_2]}=\left(\frac{1}{\sqrt{2}},\frac{1}{2},0\right)\;.
\end{equation}	
See that $\vec{\zeta}_{{\rm KK},[\omega_1]\cap[\omega_2]}=\vec{\zeta}_{{\rm KK},[\omega_2]}$, as $\mathcal{V}_{12}\sim \sqrt{\mathcal{V}_2}$, while $\vec{\zeta}_{{\rm KK},[\omega_1]}=\frac{1}{2}(\vec{\zeta}_{{\rm KK},[\omega_1]\cap[\omega_1]}+\vec{\zeta}_{{\rm KK},[\omega_1]\cap[\omega_2]})$, and thus can be seen as a bound state of the KK towers of two curves. Since these $\zeta$-vectors will not generate the tower polytope, for clarity's sake we will not consider them further. 
	\label{fn.bounded}}

	Finally, we can also have M2-branes wrapping the curve $[\omega_1]\cap[\omega_1]$ yielding 
		\begin{equation}
		\vec{\zeta}_{{\rm M2}}=\left(0,1,-\frac{1}{\sqrt{2}}\right)
	\end{equation}
	where we have used \eqref{eq.M2}. We will see that this is dual to a KK tower in a dual F-theory frame. Notice, though, that the $\zeta$-vector of this tower is outside the K\"ahler cone, so we will never move in a direction that follows the gradient of $\vec{\zeta}_{{\rm M2}}$. However, we can consider bound states of this tower with the KK M-theory tower above, yielding
\begin{equation} 
\vec{\zeta}_{\rm KK,M-th+M2}=\left(0,1,0\right)
\end{equation}
which is inside the K\"ahler cone and will be the leading tower if we take the infinite distance limit $s^1\rightarrow \infty$. We can also have further bound states combining the KK towers above using \eqref{e.bounded}, leading to other types of subleading towers.\\

Similarly, for the quantum gravity cut-off, we will have the following options depending on the concrete infinite distance limit followed:
\begin{equation}
		\begin{array}{c}
		\vec{\mathcal{Z}}_{\rm str}=\left(\frac{1}{\sqrt{2}},0,0\right)\,,\quad \vec{\mathcal{Z}}_{\rm Pl,{8}}=\left(\frac{\sqrt{2}}{3},\frac{1}{3},0\right)\,,\quad \vec{\mathcal{Z}}_{\rm Pl,{10}}=\left(\frac{3}{4 \sqrt{2}},\frac{1}{4},\frac{1}{4 \sqrt{2}}\right)\,,\quad \vec{\mathcal{Z}}_{\rm Pl,6}=\left(\frac{1}{2 \sqrt{2}},0,\frac{1}{2 \sqrt{2}}\right)\\\vec{\mathcal{Z}}_{\rm Pl,{10}}=\left(\frac{1}{2 \sqrt{2}},\frac{1}{2},0\right)\,,\quad \vec{\mathcal{Z}}_{\rm Pl,9}=\left(\frac{2 \sqrt{2}}{7},\frac{3}{7},\frac{1}{7 \sqrt{2}}\right)\,,\quad				\vec{\mathcal{Z}}_{\rm Pl,{11}}=\left(\frac{\sqrt{2}}{3},\frac{1}{3},\frac{1}{3\sqrt{2}}\right)\\
		 			\vec{\mathcal{Z}}_{\rm Pl,7}=\left(\frac{\sqrt{2}}{5},\frac{1}{5},\frac{3}{5 \sqrt{2}}\right)	\,,\quad\vec{\mathcal{Z}}_{\rm Pl,6}=\left(0,\frac{1}{2},0\right)\,,\quad\vec{\mathcal{Z}}_{\rm Pl,5}=\left(0,\frac{1}{3},\frac{1}{3\sqrt{2}}\right)\,,\quad\vec{\mathcal{Z}}_{\rm str}=\left(0,0,\frac{1}{\sqrt{2}}\right)
		\end{array}
\end{equation}
To avoid confusion, we want to remark that, even if there are several vectors associated to a 10d Planck scale or to a string scale, they actually refer to scales in different duality frames.\\

\begin{figure}[h]
\begin{center}
\begin{subfigure}[b]{0.49\textwidth}
\captionsetup{width=.95\linewidth}
\center
\includegraphics[width=\textwidth]{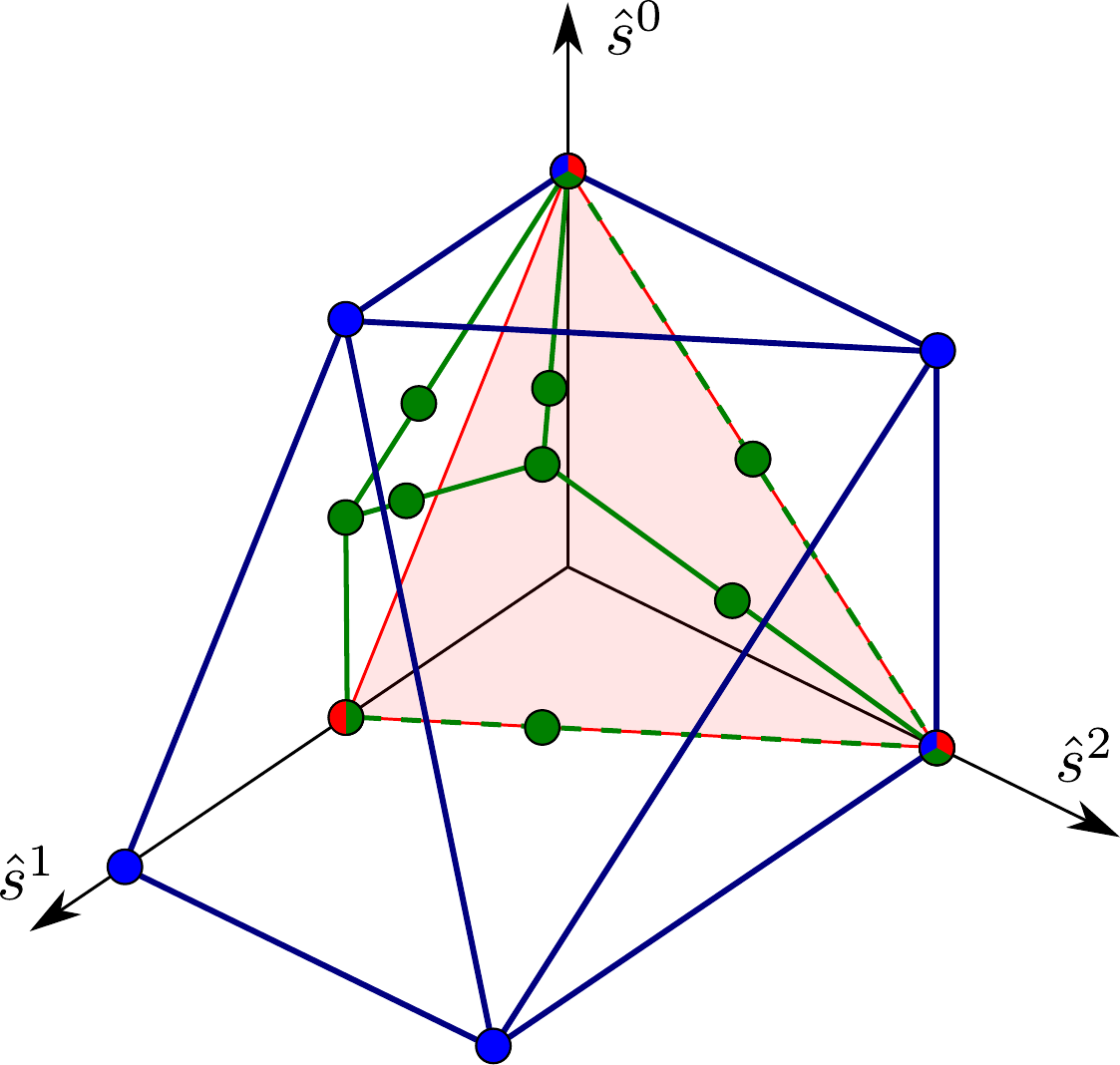}
\caption{\hspace{-0.3em} Arrangement of $\zeta$-vectors for the leading towers (in blue), elementary EFT strings (red) and $\mathcal{Z}$-vectors for species scale (green).} \label{f.towers2-1}
\end{subfigure}\begin{subfigure}[b]{0.45\textwidth}
\captionsetup{width=.95\linewidth}
\center
\includegraphics[width=\textwidth]{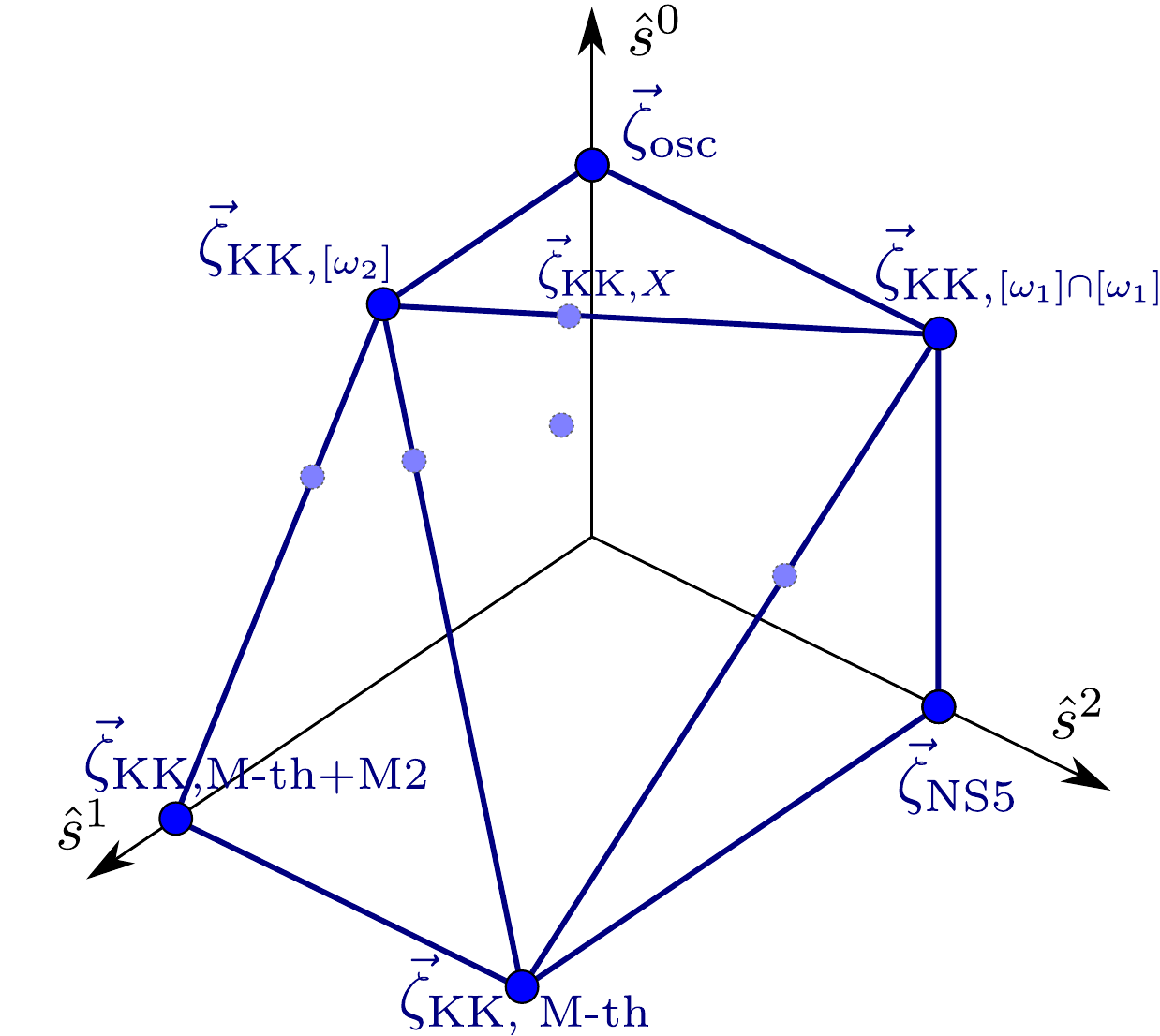}
\caption{\hspace{-0.32em} Convex hull of the $\zeta$-vectors for towers of states. Bounded states are pictured are smaller dots only when they are co-leading.} \label{f.towers2-2}
\begin{minipage}{.1cm}
            \vfill
            \end{minipage}
\end{subfigure}
\hfill
\begin{subfigure}[b]{0.45\textwidth}
\captionsetup{width=.95\linewidth}
\center
\includegraphics[width=\textwidth]{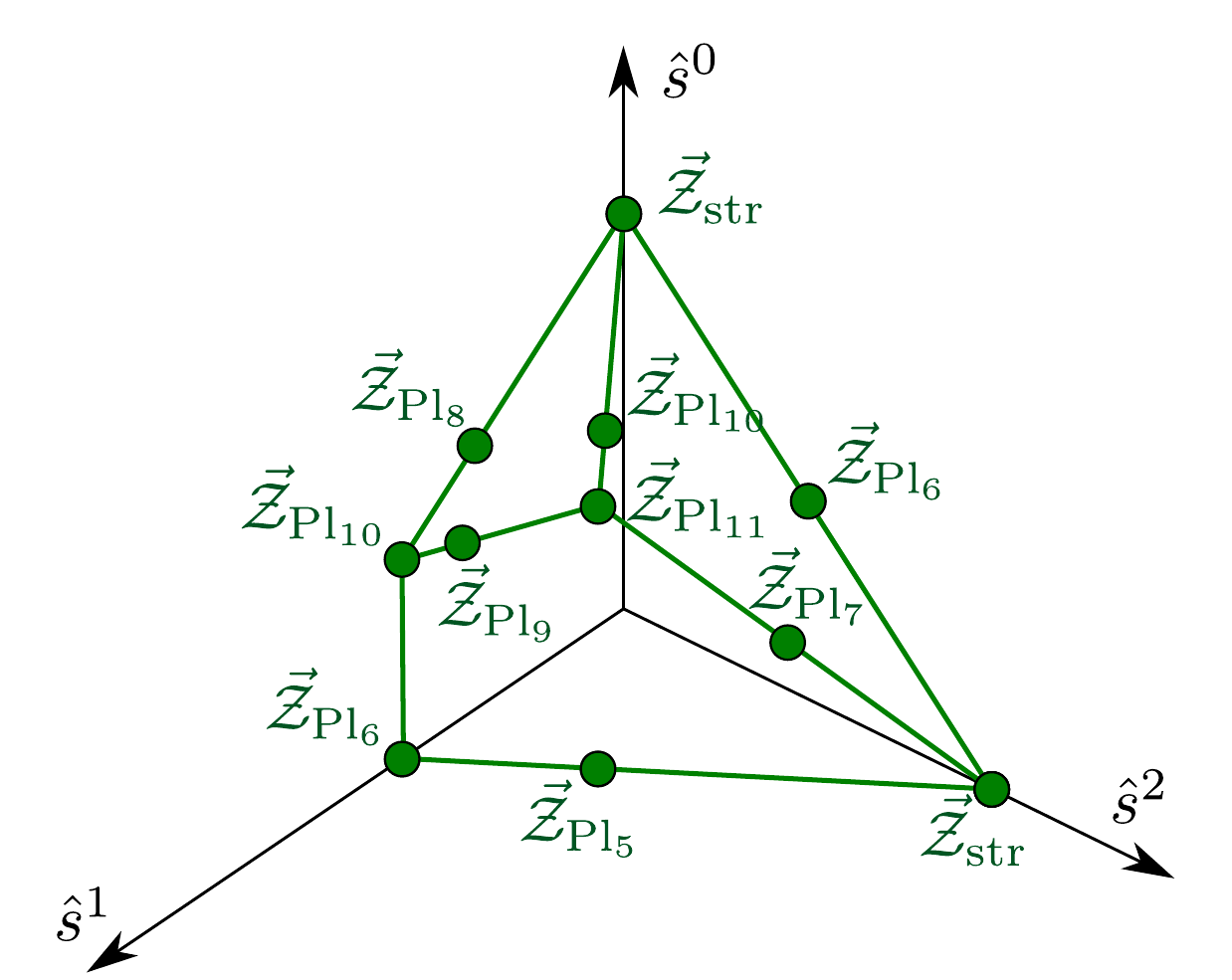}
\caption{\hspace{-0.3em} Convex hull of the $\mathcal{Z}$-vectors that will play the role of the species scale along different directions.} \label{f.towers2-3}
\begin{minipage}{.1cm}
            \vfill
            \end{minipage}
\end{subfigure}\begin{subfigure}[b]{0.45\textwidth}
\captionsetup{width=.95\linewidth}
\center
\includegraphics[width=\textwidth]{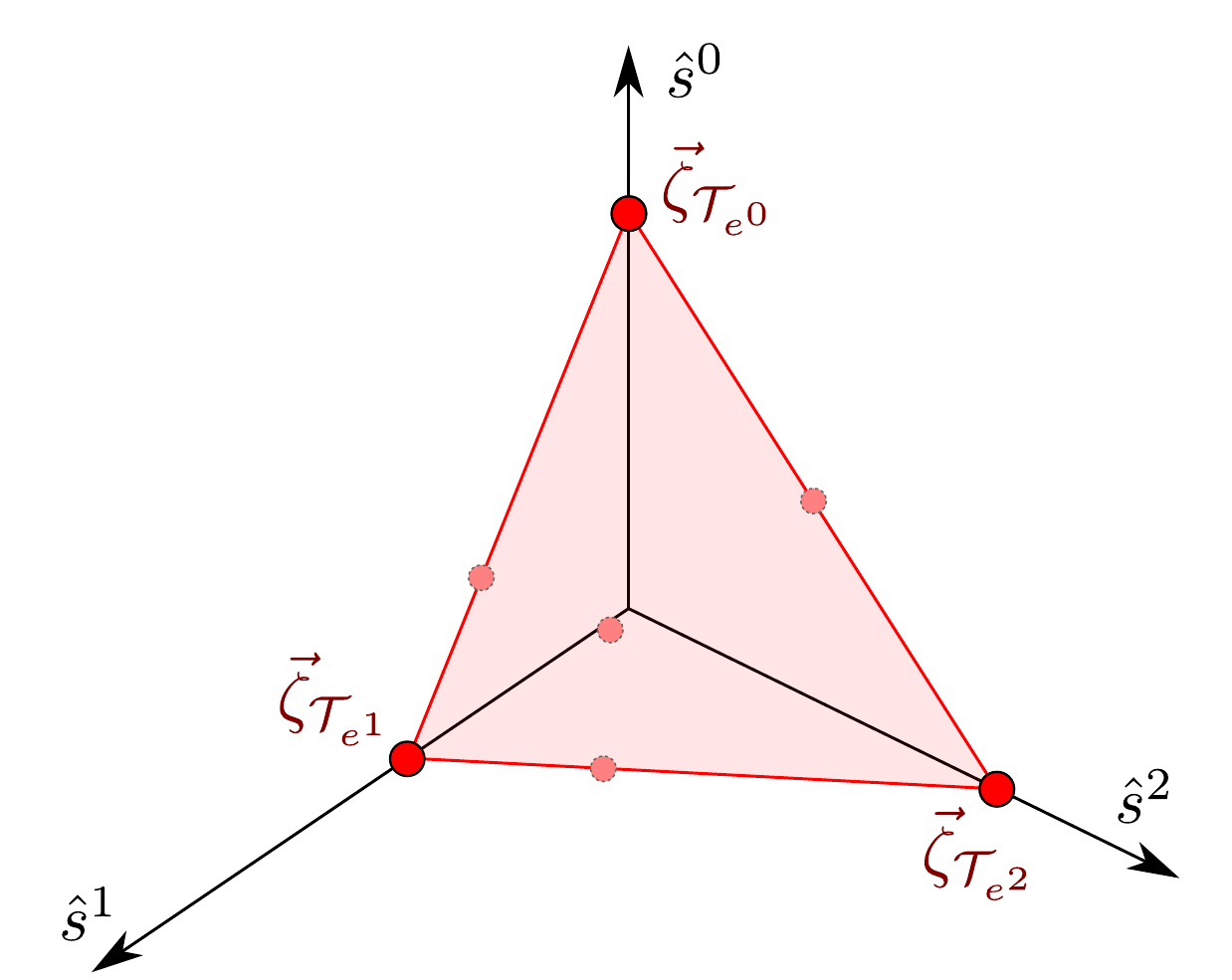}
\caption{\hspace{-0.3em} Convex hull of the $\zeta$-vectors for EFT strings, generated by the elementary ones. The non-elementary strings are shown as smaller dots only along their EFT string flow.} \label{f.towerse2-4}
\end{subfigure}
\caption{Convex hull of towers (blue), possible species scales (green) and EFT strings (red) for heterotic $E_8\times E_8$ on a Calabi-Yau 3-fold with $\mathcal{V}\sim (s^1)^2s^2$, in terms of the canonically normalized saxions $\hat{s}^0$, $\hat{s}^1$ and $\hat{s}^2$. The figures are contained within each other, preserving the hierarchy $m_\ast\leq \Lambda_{\rm QG}\leq \sqrt{\mathcal{T}}$.}
			\label{f.towers2}
	\end{center}
\end{figure}

All the above $\zeta$-vectors for the towers and EFT strings, alongside the $\mathcal{Z}$-vectors for the species scale, are plotted in Figure \ref{f.towers2}. The polytope of the EFT strings is contained in the species polytope, which is contained in the tower polytope. All towers of states lie in points of the lattice generated by the  $\zeta$-vectors of the EFT strings, so that they all satisfy the integer scaling relation \eqref{lattice} with $w={1,2,3}$ when moving along a string flow, see Table \ref{tab.w s2s}. 
In Table \ref{t.dual frames 2} we provide the identification of the leading (and subleading) towers, the value of $w$ for the leading tower, the species scale, and asymptotic theory obtained along each EFT string limit.

 \begin{figure}[h]
	\begin{center}
			\includegraphics[width=0.65\textwidth]{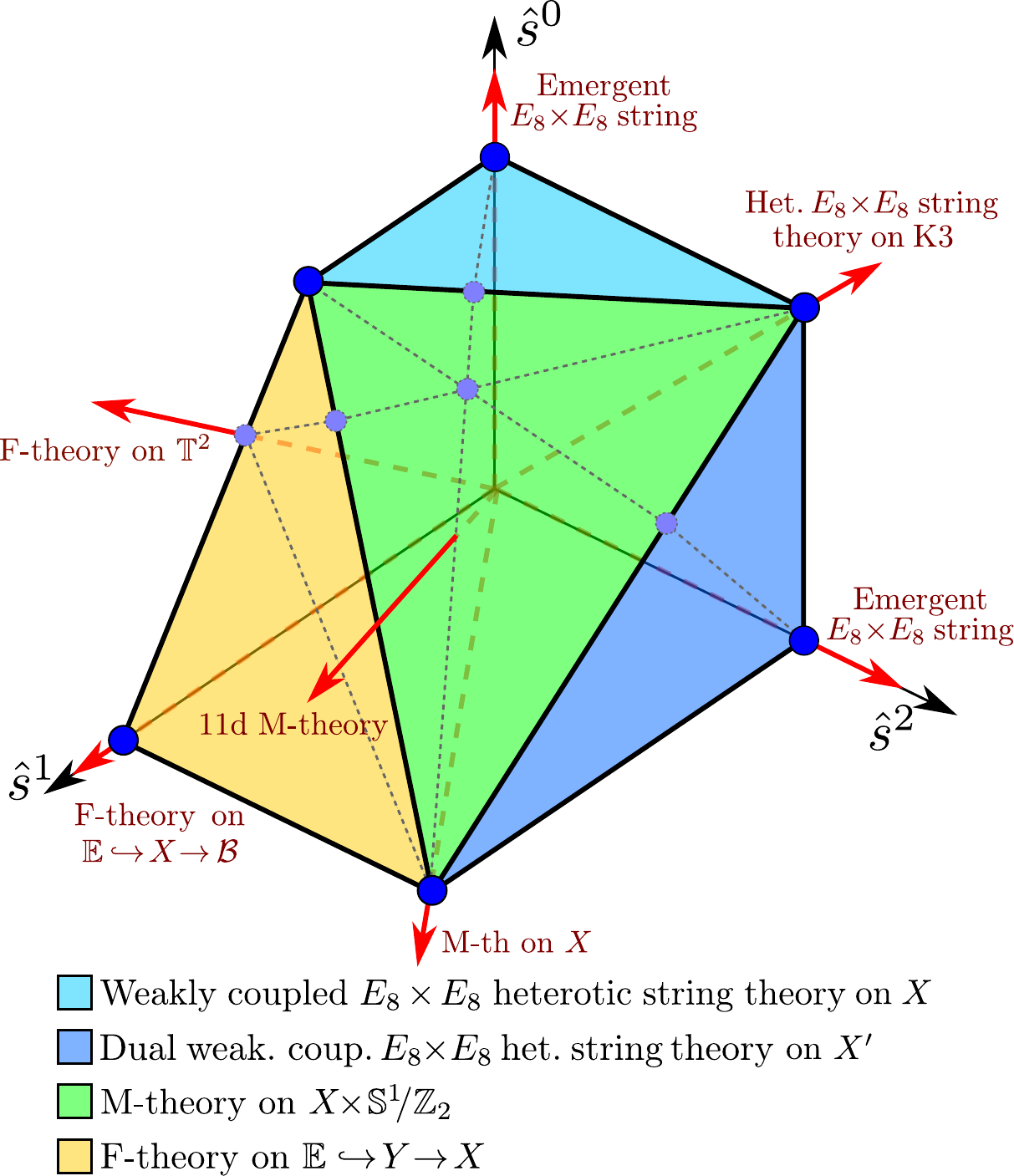}
		\caption{Arrangement of duality frames for  heterotic $E_8\times E_8$ on a Calabi-Yau 3-fold with $\mathcal{V}\sim (s^1)^2s^2$, in terms of the canonically normalized saxions $\hat{s}^0$, $\hat{s}^1$ and $\hat{s}^2$. The different leading towers are depicted as blue dots, but for clarity have not been labeled (see Figure \ref{f.towers2}). The different labels for each EFT string limit (red arrows) correspond to the dual theory emerging at infinite distance. Each region with a different color is described by a different perturbative description (denoted as duality frames) connected by dualities. The dashed lines sub-dividing each duality frame mark the regions where   internal cycles decompactify at different rates, setting the hierarchies between the subleading towers, though the final asymptotic higher-dimensional theory is the same.}
		 
			\label{f.dual2}
	\end{center}
\end{figure}

\begin{table}
	\begin{center}
	\resizebox{\textwidth}{!}{
			\begin{tabular}{|c|c|c|c|c|c|}
			\hline
			$\mathbf{e}$ & Leading tower(s) $m_*$ & $w$ & $\Lambda_{\text{QG}}$ & Subleading towers & Emergent dual theory \\
			\hline 
			\rowcolor{blue!10!}$(e^0,0,0)$ &  \begin{tabular}{@{}c@{}}$m_{\text{osc}}$, $m_{\text{KK},[\omega_2]}$,\\ $m_{\text{KK},[\omega_1]\cap[\omega_1]}$, ($m_{\text{KK},X}$)\end{tabular} & 1 & $m_{\rm str}$ & $\setminus$ & \begin{tabular}{@{}c@{}}Emergent $E_8\times E_8$\\  heterotic string\end{tabular}\\
			\hline
			\rowcolor{green!10!}$(0,e^1,0)$ & $m_{\text{KK,M-th+M2}}$, $m_\text{KK,M-th}$ & 2 & $M_\text{Pl,6}$ & $m_{\text{KK}, [\omega_2]}$& \begin{tabular}{@{}c@{}}F-theory on\\$\mathbb{E}\hookrightarrow X\to\mathcal{B}$  \end{tabular}\\
			\hline
			\rowcolor{blue!10!}$(0,0,e^2)$ & $m_{\text{NS5}}$, $m_{\text{KK}, [\omega_1]\cap[\omega_1]}$, $m_{\text{KK,M-th}}$& 1 & $m_{\rm str}$ & $\setminus$ &\begin{tabular}{@{}c@{}}Emergent $E_8\times E_8$\\  heterotic string\end{tabular} \\
			\hline
			\rowcolor{green!10!}$(e^0,e^1,0)$ &\begin{tabular}{@{}c@{}}$m_{\text{KK,M-th+M2}+[\omega_2]}$, $m_{{\rm KK},[\omega_2]}$, \\ $m_\text{KK,M-th+M2}$, ($m_\text{KK,M-th}$) \end{tabular}  & 2 & $M_{\text{Pl,10}}$& $\setminus$& F-theory on $\mathbb{T}^2$\\
			\hline
			\rowcolor{green!10!}$(e^0,0,e^1)$ & $m_{\text{KK},[\omega_1]\cap[\omega_1]}$& 2& $M_{\rm Pl,6}$&  
\begin{tabular}{@{}c@{}}$m_\text{osc}$, $m_\text{NS5}$,\\  $m_\text{KK,M-th}$, $m_{{\rm KK},[\omega_2]}$\end{tabular}& $E_8\times E_8$ on K3 \\
			\hline
		\rowcolor{red!10!}	$(0,e^1,e^2)$ &$m_\text{KK,M-th}$ & 3& $M_\text{Pl,5}$&\begin{tabular}{@{}c@{}}$m_\text{KK,M-th+M2}$, $m_\text{NS5}$,\\  $m_{\text{KK},[\omega_2]}$, $m_{\text{KK},[\omega_1]\cap[\omega_1]}$\end{tabular}  & M-theory on $X$ \\
			\hline
			\rowcolor{red!10!}$(e^0,e^1,e^2)$ & $m_\text{KK,M-th}$  & 3&$M_\text{Pl,11}$ &\begin{tabular}{@{}c@{}}$m_\text{KK,M-th+M2}$, \\  $m_{\text{KK},[\omega_2]}$, $m_{\text{KK},[\omega_1]\cap[\omega_1]}$\end{tabular}   & \begin{tabular}{@{}c@{}}11d \\ M-theory \end{tabular} \\
			\hline
		\end{tabular}}
		\caption{Description of the different EFT string limits driven by the string charges ${\mathbf{e}}$  for $E_8\times E_8$ heterotic theory on a CY 3-fold with $\mathcal{V}_X\sim (s^1)^2s^2$. We provide the identification of the leading tower(s) $m_\ast$, their scaling weight $w$, the asymptotic species scale $\Lambda_{\rm QG}$, subleading towers with $m\lesssim \Lambda_{\rm QG}$ and the microscopic interpretation of the dual theory emerging at the infinite distance limit. \label{t.dual frames 2}}
	\end{center}
\end{table}

\begin{table}[h]
	\centering
	\begin{tabular}{|c|cccccc|}
	\hline
	$\mathbf{e}$ &$\vec{\zeta}_{\rm osc}$&$\vec{\zeta}_{\rm NS5}$&$\vec{\zeta}_{{\rm KK},[\omega_2]}$&$\vec{\zeta}_{{\rm KK},[\omega_1]\cap[\omega_1]}$&$\vec{\zeta}_{\rm KK,Mth+M2}$&$\vec{\zeta}_{\rm KK,Mth}$\\\hline
	$(e^0,0,0)$&$\mathbf{1}$&0&$\mathbf{1}$&$\mathbf{1}$&0&0\\
	$(0,e^1,0)$&0&0&1&0&$\mathbf{2}$&$\mathbf{2}$\\
	$(0,0,e^2)$&0&$\mathbf{1}$&0&$\mathbf{1}$&0&$\mathbf{1}$\\
	$(e^0,e^1,0)$&1&0&$\mathbf{2}$&1&$\mathbf{2}$&$\mathbf{2}$\\
	$(e^0,0,e^2)$&1&1&1&$\mathbf{2}$&0&1\\
	$(0,e^1,e^2)$&0&1&1&1&2&$\mathbf{3}$\\
	$(e^0,e^1,e^2)$&1&1&2&2&2&$\mathbf{3}$\\\hline
	\end{tabular}
	\caption{Integer scaling weight $w$ of the different towers of states along each EFT string limit, for $E_8\times E_8$ heterotic theory on a CY 3-fold with $\mathcal{V}_X\sim (s^1)^2s^2$. Each tower with vector $\vec{\zeta}$ satisfies  $\vec{\zeta}_{\mathcal{T}_{\mathbf{e}}} \cdot \vec{\zeta} = w \  \abs{\vec{\zeta}_{\mathcal{T}_{\mathbf{e}}}}^2$ along the flow driven by an EFT string with vector $\vec{\zeta}_{\mathcal{T}_{\mathbf{e}}}$ and charge ${\mathbf{e}}$. The leading tower(s) for each EFT string limit are highlighted in bold face. \label{tab.w s2s}}
\end{table}

Let us now discuss the different duality frames, depicted in Figure \ref{f.dual2}. The perturbative regime is divided in four regions described by different duality frames, depending on the nature of $\Lambda_{\rm QG}$:
\begin{itemize}
	\item For $s^0\gg \mathcal{V}_X\sim (s^1)^{2}s^2$ with $s^0\to \infty$ (light blue region in Figure \ref{f.dual2}), the theory is weakly coupled in terms of the 10d string dilaton. Hence, this region is described by \textbf{heterotic $E_8\times E_8$ theory on CY$_3$}, with the species scale given by the fundamental heterotic string scale, $\Lambda_{\rm QG}=m_{\rm str}$. Depending on the relative growth between $s^1$ and $s^2$, we will have different subregions yielding a different asymptotic theory upon taking the infinite distance limit: 
	\begin{itemize}
		\item For $\mathcal{V}_X$ constant, we will have an \textbf{$E_8\times E_8$ heterotic emergent string limit}, with the leading tower being also $m_{\rm osc}$.
		\item If only one of the two saxions $s^1$ or $s^2$ grow while the other one remains fixed, we will have partial decompactifications, resulting in \textbf{weakly coupled $E_8\times E_8$ heterotic string theory on $\mathbb{T}^2$} (for $s^1\rightarrow \infty$) or \textbf{on K3} (for $s^2\to\infty$).
				\item For $s^1,s^2\rightarrow \infty$, the whole $X$ will decompactify, going to \textbf{10d weakly coupled $E_8\times E_8$ heterotic string theory}.
	\end{itemize}
For $s^0\ll \mathcal{V}_X$, $e^{2\Phi}=(s^0)^{-1}\mathcal{V}_X\to\infty$, entering a strongly coupled regime. Depending on the relative scaling of the saxions controlling the internal volume, we will find different dual 4d perturbative descriptions. Unlike in the $\mathcal{V}_X\sim (s^1)^2$ case from Section \ref{s.het vols3}, here it will be possible to have cycles relatively shrinking with respect to the overall volume, on which extended objects can be wrapped and provide new light degrees of freedom.
	\item For $s^2\gg s^0 s^1$ with $s^2\to\infty$ (darker blue region in Figure \ref{f.dual2}), the volume  $\mathcal{V}_{11}\sim s^2$ of the curve $[\omega_1]\cap[\omega_1]$ grows, which also makes the overall volume grow at least at the same rate, $\mathcal{V}_X\gtrsim \mathcal{V}_{11}\rightarrow \infty$. This implies that the dual divisor $[\omega_2]$, with $\mathcal{V}_2\sim (s^1)^2$ (relatively) shrinks with respect to the total volume, so the CY$_3$ develops a fibration structure in this limit given by a K3 surface fibered over a $\mathbb{P}^1$ base \cite{Lee:2019wij}.\footnote{Note that a $\mathbb{T}^4$ fibration would imply the existence of two curves in $[\omega_2]$, which in principle should be able to grow independently, yielding $\mathcal{V}_2\sim s^a s^b$ with $a\neq b$, rather than $\mathcal{V}_2\sim (s^a)^2$.} Therefore, NS5-branes wrapping this shrinking divisor are going to yield strings that become tensionless in 4d Planck units, $m_{\rm osc}\sim \mathcal{T}^{1/2}_{\rm NS5}\sim (s^2)^{-1/2} M_{\rm Pl,4} $. Since the topology of the divisor $[\omega_2]$ is that of K3, the resulting string is the fundamental heterotic string of a dual $E_8\times E_8$ theory, whose string scale sets the species scale, $\Lambda_{\rm QG}\sim \mathcal{T}^{1/2}_{\rm NS5}$. This region is therefore better described by this dual \textbf{heterotic $E_8\times E_8$ theory on a CY$_3$} $\text{K3}'\hookrightarrow X'\to \mathbb{P}^1$, different from the original $X$, and whose fibration structure can be inferred from the limiting regimes of this duality frame as follows. 
		
	In this new duality frame, $s^2$ acts as the new universal dilaton and $s^0$ measures the volume of the $\mathbb{P}^1$ base in the new string units. As for the fiber, K3$'\simeq \mathbb{T}^4/\mathbb{Z}_2$, in the $s^1\gg \sqrt{s^0}$ regime, it factorizes as K3$'\simeq \mathbb{S}^1/\mathbb{Z}_2\times Y_3$, with the interval $\mathbb{S}^1/\mathbb{Z}_2$ having volume $s^1$ in the new string units (compare \eqref{KKtowers} with \eqref{eq. Mth het}), and $Y_3\simeq \mathbb{S}^1\times \mathbb{T}^2$ being a real 3-manifold whose volume is not controlled by the scalars $\{s^i\}_{i=0}^2$,\footnote{Note that this is not against the expected \emph{K\"ahler} asymptotic limits described in \cite{Lee:2019wij} since $\{s^i\}_{i=0}^2$ do not need to map to K\"ahler saxions of the new Calabi-Yau $X'$, more generally being a combination of K\"ahler and complex structure.}. In summary,
	\begin{equation}
\text{K3}'\hookrightarrow X'\to \mathbb{P}^1\xRightarrow{\sqrt{s^0}\ll s^1\to\infty}\mathbb{S}^1/\mathbb{Z}_2\times\mathbb{S}^1\times(\mathbb{T}^2\hookrightarrow \widehat{\text{K3}}\to\mathbb{P}^1)\,,
	\end{equation}
	with $\widehat{\text{K3}}$ a dual K3 manifold \cite{Kachru:1995wm}, different from the other two K3s. As above, depending on the relative scaling of $s^0$ and $s^1$, we can reach different asymptotic dual theories at infinite distance along the interfaces:
	\begin{itemize}
		 \item For $s^0$ and $s^1$ fixed, as $s^2\to\infty$ we have an \textbf{emergent (dual) $E_8\times E_8$ heterotic string limit}, with $m_{\rm osc}=\mathcal{T}^{1/2}_{\rm NS5}$, where the KK modes of the internal cycles $\mathbb{P}^1$ and $\mathbb{S}^1/\mathbb{Z}_2$ scale at the same rate as the string scale.
		 \item For $s^0\sim s^2\to\infty$ and $s^1$ fixed, only the curve  $[\omega_1]\cap[\omega_1]$ decompactifies and we reach \textbf{weakly coupled (dual) $E_8\times E_8$ heterotic string theory on K3$'$}. This is dual to the original $E_8\times E_8$ heterotic theory on the original K3. 
		 \item Conversely, for $s^2\sim s^1\to\infty$ with $s^0$ fixed, we get the (dual) \textbf{$E_8\times E_8$ heterotic string theory on }$\widehat{\text{K3}}\times\mathbb{S}^1$ \cite{Kachru:1995wm}. This is also dual to M-theory on the original K3 $\hookrightarrow X\to \mathbb{P}^1$, as it will become clear below.
		 \item For both $s^0, s^1\to\infty$ while $s^2\gg s^0 s^1$, we decompactify to the (dual) \textbf{$E_8\times E_8$ heterotic string theory on} $Y_3$.
		 	\end{itemize} 
	
	\item For $s^1\gg\sqrt{s^0}s^2$ (yellow region in Figure \ref{f.dual2}), our 3-fold adopts an elliptic fibration $\mathbb{E}\hookrightarrow X\to\mathcal{B}$ with the fiber (being the curve $[\omega_1]\cap[\omega_1]$ of volume $\mathcal{V}_{11}\sim s^2$) shrinking relatively to the base (which is the divisor $[\omega_2]$ of volume $\mathcal{V}_2\sim (s^1)^2$) and, therefore, shrinking also relative to the overall volume $\mathcal{V}_X$. Moreover, since $g_s=e^{\Phi}=(s^0)^{-1/2}\mathcal{V}_X^{1/2}\to\infty$, the M-theory KK tower in \eqref{eq. Mth het} will become light. In addition to this tower, we can also find a tower of M2-branes wrapped on $\mathbb{E}$ with mass given in \eqref{eq.M2}\footnote{Note that the gradient flow of this tower, with $m_{\rm M2}\sim M_{\rm Pl,4}\sim \mathcal{V}_{\mathbb{E}}\mathcal{V}_X^{-1/2}\sim (s^1)^{-1}(s^2)^{1/2} $ points outside the K\"ahler cone, towards non-perturbative regimes with $s^2\to 0$.}, and their bound states with $m_{\rm KK,M-th}$. It is well known that M-theory on a shrinking elliptic curve is dual to type IIB on a circle getting large. Therefore, this asymptotic region of the moduli space is better described by \textbf{F-theory}\footnote{The complex structure $\tau$ of the shrinking curve is mapped to the axio-dilaton $\tau=C_0+ie^{-\Phi}$ of the dual type IIB theory \cite{Weigand:2018rez}. Since we are only moving in the moduli space of the K\"ahler sector of the original heterotic frame, $\tau$ remains fixed, and we are not capturing weakly coupled limits of the dual type IIB string theory. Hence, this region of the moduli space is better described by F-theory. } \textbf{on} $(\mathbb{E}\hookrightarrow X''\to\mathcal{B})\times \mathbb{S}^1\times \mathbb{S}^1/\mathbb{Z}_2$, where $\mathcal{B}$ is the base of the original CY $\mathbb{E}\hookrightarrow X\to\mathcal{B}$, the first $\mathbb{S}^1$ is the type IIB circle becoming large (whose inverse radius can be related to the volume of the shrinking curve $\mathbb{E}$) and the $\mathbb{S}^1/\mathbb{Z}_2$ is the original Ho\v{r}ava-Witten interval. Since we are generically at the strong coupling regime of Type IIB, the species scale is given by $M_{\rm Pl,10}$.
	
	The EFT string $\vec\zeta_{\mathcal{T}_{0,1,0}}$ is not a fundamental string, but maps to a D3-brane wrapping the curve $[\omega_1]\cap[\omega_2]\subset\mathcal{B}=[\omega_2]$ in the dual F-theory frame.\footnote{In the original heterotic frame, this corresponds to an NS5 brane wrapping the divisor $[\omega_1]$. In the M-theory picture this corresponds to a M5 brane wrapping the same divisor. In the limit where the fiber $[\omega_1]\cap[\omega_1]\subset[\omega_1]$ shrinks, the duality with type IIB/F-theory maps this M5 to a D4-brane in type IIA, corresponding to a  type IIB/F-theory D3-brane under T-duality, wrapping the curve $[\omega_1]\cap[\omega_2]$.}  This is consistent with the fact that $w>1$ for all towers along this string flow.

	For a generic limit in this region (i.e. when all saxions grow), we will decompactify to F-theory on the elliptic fiber. Additionally, we can find the following dual asymptotic theories when moving along the interfaces of this duality frame:
	\begin{itemize}
	\item If $s^1\to\infty$ with $s^0$ and $s^2$ fixed (corresponding to the EFT string limit for $\mathbf{e}=(0,e^1,0)$), $\mathcal{B}$ does not grow, and we simply obtain \textbf{F-theory on $\mathbb{E}\hookrightarrow X''\to\mathcal{B}$} and the species scale is given by $M_{\rm Pl,6}$. 
	\item For $s^2\sim(s^1)^3\sim(s^0)^4\to\infty$ or $s^0\sim (s^1)^{2/3}\to\infty$ (with fixed $s^2$) we have partial decompactifications to \textbf{F-theory on $\mathbb{E}\hookrightarrow\mathbb{T}^3\to\mathbb{S}^1$} ($\Lambda_{\rm QG}=M_{\rm Pl,9}$) or \textbf{F-theory on} $\mathbb{E}\hookrightarrow\text{K3}\to\mathbb{P}^1$ ($\Lambda_{\rm QG}=M_{\rm Pl,8}$), respectively. Note that the latter is also dual to heterotic string theory on a torus, see Figure \ref{f.dual2}.
	\end{itemize}
	
	\item Finally, for $(s^1)^2s^2\gg s^0\gg s^2/s^1,\, (s^1/s^2)^2$ with $s^0,\, s^1,\,s^2\to\infty$ (green region in Figure \ref{f.dual2}), the string coupling diverges, $e^{\Phi}=(s^0)^{-1/2}\mathcal{V}_X^{\frac{1}{2}}\to\infty$, leading to the opening of the Ho\v{r}ava-Witten interval, at the same time as the overall volume $X$ decompactifies (without any internal cycle relatively shrinking with respect to the overall volume). Hence, the species scale is given by $M_{\rm Pl,11}$ and the region is indeed described by a \textbf{compactification of M-theory on $X\times \mathbb{S}^1/\mathbb{Z}_2$}. The leading KK tower will depend on the relative scaling between the saxions. For a limit in the interior of this region, we will then simply decompactify to \textbf{11d M-theory}. If instead we move along the boundaries of this region, we obtain additional asymptotic dual descriptions:

	\begin{itemize}
		\item For $s^1\sim s^2\to\infty$ and fixed $s^0\sim s^0_0$, the overall $X$ volume scales at the same rate as $M_{\rm Pl,5}$ and $M_{\rm Pl,11}$, so that we simply decompactify the interval $\mathbb{S}^1/\mathbb{Z}_2$ to obtain \textbf{M-theory on $X$}.
		\item For $s^0\sim (s^1)^2 s^2$, we find ourselves back to the interface with weakly coupled, large volume $E_8\times E_8$ heterotic on $X$. When both $s^1$ and $s^2$ are sent to infinity, we decompactify to \textbf{M-theory on $\mathbb{S}^1/\mathbb{Z}_2$} of fixed size. When either $s^1$ or $s^2$ are left finite, the asymptotic limit corresponds to either \textbf{M-theory on $\mathbb{T}^2\times \mathbb{S}^1/\mathbb{Z}_2$} ($\Lambda_{\rm QG}=M_{\rm Pl,8}$) or \textbf{M-theory  on} K3$\times \mathbb{S}^1/\mathbb{Z}_2$ ($\Lambda_{\rm QG}=M_{\rm Pl,6}$).
	\end{itemize}

\end{itemize}

\subsubsection{Case 3: $V_X\sim s^1s^2s^3$\label{s.het s1s2s3}}
Consider now a CY volume of the kind $\mathcal{V}_X \sim s^1s^2s^3$ at leading order. Looking at its symmetries, we expect a tower structure which is invariant under an $S_3$ transformation interchanging the three K\"ahler saxions. This is also clear from the shape of the moduli space metric, again flat and diagonal, given by
\begin{equation}
	\mathsf{G}_{ij}=\frac{1}{2} \text{diag}\left(\frac{1}{(s^0)^2}, \ \frac{1}{(s^1)^2}, \ \frac{1}{(s^2)^2}, \frac{1}{(s^3)^2} \right).
\end{equation}
Since the only non-vanishing intersection number is $\kappa_{123}$, the nef curves and divisors are 
\begin{equation}
	\text{Divisors }\{[\omega_a]\}^3_{a=1} \text{ with } \mathcal{V}_{[\omega_a]}\sim  s^bs^c, \ a\neq b \neq c 
\end{equation}
\begin{equation}
 \text{Curves }\{[\omega_a]\cap[\omega_b]\}_{a \neq b}^3 \text{ with }  \mathcal{V}_{[\omega_a]\cap {[\omega_b]}} \sim s^c, \ a \neq b \neq c 
\end{equation}

 The EFT string charges are  $\mathbf{e}= \{e ^0,e^1,e^2,e^3\}\in\mathbb{Z}_{\geq 0}^4$ so we have $2^4 -1 =  15$ EFT string flow families, depending on which charge components are non-vanishing.
 The elementarily-charged EFT strings are
\begin{equation}\label{e.EFT strings s1 s2 s3}
	\begin{array}{c}
		\vec{\zeta}_{\mathcal{T}_{e^0}}=\left(\frac{1}{\sqrt{2}},0,0,0\right)\;,\quad
		\vec{\zeta}_{\mathcal{T}_{e^1}}=\left(0,\frac{1}{\sqrt{2}},0,0\right)\;,\\
		\vec{\zeta}_{\mathcal{T}_{e^2}}=\left(0,0,\frac{1}{\sqrt{2}},0\right)\;,\quad 
		\vec{\zeta}_{\mathcal{T}_{e^3}}=\left(0,0,0,\frac{1}{\sqrt{2}}\right);\;\quad 
	\end{array}
\end{equation}
while the non-elementary EFT string vectors, evaluated along the string flow, are given by
\begin{equation}\label{e.non-el EFT strings s1 s2 s3}
	\begin{array}{c}
		\vec{\zeta}_{\mathcal{T}_{e^ie^j}}\big|_{\rm flow}=\frac{\delta^i_{m}+\delta^j_m}{2\sqrt{2}}\left(\delta^m_0,\delta^m_1,\delta^m_2,\delta^m_3\right), \ i\neq j\\
		\vec{\zeta}_{\mathcal{T}_{e^ie^je^k}}\big|_{\rm flow}=\frac{\delta^i_{m}+\delta^j_{m}+\delta^k_{m}}{3\sqrt{2}}\left(\delta^m_0,\delta^m_1,\delta^m_2,\delta^m_3\right) \ i\neq j \neq k\;,\\
		\vec{\zeta}_{\mathcal{T}_{(e^0,e^1,e^2,e^3)}}\big|_{\rm flow}=\frac{1}{4\sqrt{2}}\left(1,1,1,1\right)\;\\
	\end{array}
\end{equation}

Because of the invariance of the setup under permutations of the saxions, there are really only seven  inequivalent types of string flows, represented in Table \ref{tab.EFT lim s1s2s3} and distinguished by the number of growing saxionic directions (including both the K\"ahler saxions and the 4d dilaton). $\mathcal{T}_{e_0}$ corresponds, as in the previous cases, to the tension of the fundamental heterotic string, while the three other elementary EFT strings have the microscopical interpretation of NS5 branes wrapped on (relatively) shrinking $[\omega_i]$ divisors.

As in the previous cases, in addition to string oscillation modes we have KK towers whose $\zeta$-vectors are given by
 \begin{equation}
 	\begin{array}{c}
 		\vec{\zeta}_{\text{KK,M-th}} = \left(0, \frac{1}{\sqrt{2}}, \frac{1}{\sqrt{2}},\frac{1}{\sqrt{2}}\right)\;, \vec{\zeta}_{{\rm KK},X}=\left(\frac{1}{\sqrt{2}},\frac{1}{3 \sqrt{2}},\frac{1}{3 \sqrt{2}},\frac{1}{3 \sqrt{2}}\right)\;,\\  \vec{\zeta}_{{\rm KK},[\omega_a]\cap[\omega_b]}=\frac{1}{\sqrt{2}}\left(1,1-\delta_{1a}-\delta_{1b},1-\delta_{2a}-\delta_{2b},1-\delta_{3a}-\delta_{3b}\right)\;.
 	\end{array}
 \end{equation}
These correspond respectively to the KK modes of the M-theory interval decompactification, the overall volume KK tower and KK modes associated with the three intersection curves $[\omega_a]\cap [\omega_b]$. Note that since the only nonvanishing intersection number is $\kappa_{123}$, the divisors $[\omega_i]$ do not self-intersect.
 
 Finally, from wrapping M2-branes over the intersection curves we get, as in the previous case, additional towers of particles. We can use \eqref{eq.M2} to compute their $\zeta$-vectors,
 \begin{equation}
 	\vec{\zeta}_{\rm M2,ab} =-\frac{1}{\sqrt{2}} \left(0,1-\delta_{1a}-\delta_{1b},1-\delta_{2a}-\delta_{2b},1-\delta_{3a}-\delta_{3b}\right), \ a\neq b.
 \end{equation}
 which are outside of the K\"{a}hler cone, as in the previous subsection. Nevertheless,
the bound states of these with the M-theory tower yield $\zeta$-vectors which are inside the K\"{a}hler cone  and will play a relevant role for the tower polytope, with
 \begin{equation}
 	\vec{\zeta}_{\rm KK,M-th+M2,ab}=\frac{1}{\sqrt{2}}\left(0,\delta_{1a}+\delta_{1b},\delta_{2a}+\delta_{2b},\delta_{3a}+\delta_{3b}\right)\;,
 \end{equation}
 where again $a\neq b$.
All the other subleading towers can be realized as bound states of those above.
 
 Analogously, depending on the direction we move in moduli space one of the following quantum gravity cut-offs will dominate in some duality frame\footnote{In some particular infinite distance limits, namely along the interfaces between different duality frames, we can also have partial decompactifications where the species scale will be given by other higher dimensional Planck scales that can be computed using \eqref{eq.species from KK}.},
\begin{equation}
	\begin{array}{c}
		\vec{\mathcal{Z}}_{\text{str}}= \left(\frac{1}{\sqrt{2}},0,0,0\right)\;,
		\vec{\mathcal{Z}}_{\text{str},a}= \frac{1}{\sqrt{2}}\left(0,\delta_{1a},\delta_{2a},\delta_{3a}\right)\;, \\
			\vec{\mathcal{Z}}_{\text{Pl},6,ab}=\left(\frac{1}{2\sqrt{2}},\frac{1-\delta_{1a}-\delta_{1,b}}{2\sqrt{2}},\frac{1-\delta_{2a}-\delta_{2b}}{2\sqrt{2}},\frac{1-\delta_{3a}-\delta_{3b}}{2\sqrt{2}}\right)\;,\vec{\mathcal{Z}}_{\text{Pl},{10}}=\left(\frac{3}{4\sqrt{2}},\frac{1}{4\sqrt{2}},\frac{1}{4\sqrt{2}},\frac{1}{4\sqrt{2}}\right)\;,\\
			%\vec{\mathcal{Z}}_{\text{Pl}_8,a}=\frac{1}{3\sqrt{2}}\left(2,1-\delta_{1a},1-\delta_{2a},1-\delta_{3a}\right)\;,	
			
			\vec{\mathcal{Z}}_{\text{Pl},{5}} = \left(0,\frac{1}{3\sqrt{2}},\frac{1}{3\sqrt{2}},\frac{1}{3\sqrt{2}}\right)\;,
			\vec{\mathcal{Z}}_{\text{Pl},{11}} = \left(\frac{\sqrt{2}}{3},\frac{1}{3\sqrt{2}},\frac{1}{3\sqrt{2}},\frac{1}{3\sqrt{2}}\right)\;.
			
	\end{array}
\end{equation}
	which are associated either to some string scale or to a higher dimensional Planck scale. \\

 As in the previous examples, we observe that all the towers of states lie on the points of a square lattice generated by $\zeta$-vectors corresponding to the elementarily-charged EFT strings. Thanks to this lattice structure, moving along a string flow the scaling weight of all the towers is integer, as we summarize in Table \ref{tab.w sss}. The information of the leading and subleading towers, the quantum gravity cut-off and the interpretation of the theory emerging at the infinite distance limit, along each saxionic flow, is summarized in Table \ref{tab.EFT lim s1s2s3}.

 \begin{table}[htp!]
 	\centering
 	\begin{tabular}{|c|ccccc|}
 		\hline
 		$\mathbf{e}$ &$\vec{\zeta}_{\rm osc}$&$\vec{\zeta}_{\rm NS5,23}$&$\vec{\zeta}_{{\rm KK},[\omega_2]\cap[\omega_3]}$&$\vec{\zeta}_{\rm KK,Mth+M2,23}$&$\vec{\zeta}_{\rm KK,Mth}$\\\hline
 		($e^0$,0,0,0)&\textbf{1}&0&\textbf{1}&0&0\\
 		(0,$e^1$,0,0)&0&\textbf{1}&\textbf{1}&0&\textbf{1} \\
 		($e^0$,$e^1$,0,0)&1&1&\textbf{2}&0&1\\
 		(0,$e^1$,$e^2$,0)&0&1&1&1&\textbf{2}\\
 		($e^0$,$e^1$,$e^2$,0)&1&1&\textbf{2}&1&\textbf{2}\\
 		(0,$e^1$,$e^2$,$e^3$) &0&1&1&2&\textbf{3}\\
 		($e^0$,$e^1$,$e^2$,$e^3$)&1&1&2&2&\textbf{3}\\\hline
 	\end{tabular}
 	\caption{Integer scaling weight $w$ of the different towers of states along each EFT string limit, for $E_8\times E_8$ heterotic theory on a CY 3-fold with $\mathcal{V}_X\sim s^1s^2s^3$.  Each tower with vector $\vec{\zeta}$ satisfies  $\vec{\zeta}_{\mathcal{T}_{\mathbf{e}}} \cdot \vec{\zeta} = w  \abs{\vec{\zeta}_{\mathcal{T}_{\mathbf{e}}}}^2$ along the flow driven by an EFT string with vector $\vec{\zeta}_{\mathcal{T}_{\mathbf{e}}}$ and charge ${\mathbf{e}}$. The leading tower(s) for each EFT string limit are highlighted in bold face. We consider only a representative of each set of equivalent limits since the CY$_3$ volume is $S_3$-symmetric. \label{tab.w sss}}
 \end{table}

\begin{table}[htp!]
	\begin{center}
	\resizebox{\textwidth}{!}{
			\begin{tabular}{|c|c|c|c|c|c|}
			\hline
			$\mathbf{e}$ &Leading tower(s) $m_\star$ & $w$ & $\Lambda_{\text{QG}}$ & Subleading towers & Emergent dual theory \\
			\hline 
		\rowcolor{blue!10!}	($e^0$,0,0,0) & \begin{tabular}{@{}c@{}}	$m_{\rm osc}$, $m_{{\rm KK},[\omega_i]\cap[\omega_j]}$, $(m_{{\rm KK},X})$ \end{tabular}	&1	&$m_{\rm str}$	&$\setminus$	& \begin{tabular}{@{}c@{}}Emergent $E_8\times E_8$\\  heterotic string\end{tabular} 	\\\hline
			
			\rowcolor{blue!10!}(0,$e^1$,0,0) &	\begin{tabular}{@{}c@{}} $m_{{\rm NS5},1}$, $m_{{\rm KK},[\omega_2]\cap[\omega_3]}$, $m_{\rm KK,M-th}$	\end{tabular}&1	&$m_{\rm str}$	&$\setminus$	&\begin{tabular}{@{}c@{}}Emergent \\  Type IIA string\end{tabular}	\\\hline
			
			\rowcolor{green!10!}($e^0$,$e^1$,0,0) &	\begin{tabular}{@{}c@{}} $m_{{\rm KK},[\omega_2]\cap[\omega_3]}$	\end{tabular}&2	&$M_{\rm Pl,6}$	&\begin{tabular}{@{}c@{}} $m_{\rm osc}$, $m_{\rm NS5,1}$, $m_{\rm KK,M-th}$,\\ $m_{{\rm KK},[\omega_1]\cap[\omega_2]}$, $m_{{\rm KK},[\omega_1]\cap[\omega_3]}$				\end{tabular}	& $E_8\times E_8$ on $\mathbb{T}^4$	\\\hline
			
			\rowcolor{green!10!}(0,$e^1$,$e^2$,0) &	\begin{tabular}{@{}c@{}} $m_{\rm KK,M-th+M2,12}$, $m_{\rm KK,M-th}$	\end{tabular}&2	&$M_{\rm Pl,6}$	&\begin{tabular}{@{}c@{}} $m_{{\rm KK},[\omega_1]\cap[\omega_3]}$, $m_{{\rm KK},[\omega_2]\cap[\omega_3]}$\\
			$m_{\rm NS5,1}$, $m_{\rm NS5,2}$	\end{tabular}	& \begin{tabular}{@{}c@{}}F-theory on\\$\mathbb{E}\hookrightarrow X\to\mathbb{T}^4$  \end{tabular}	\\\hline
			
			\rowcolor{green!10!}	($e^0$,$e^1$,$e^2$,0) &	\begin{tabular}{@{}c@{}} $m_{\rm KK,M-th+M2+[\omega_3]}$, $m_{\rm KK,[\omega_1]\cap[\omega_3]}$,\\ $m_{\rm KK,[\omega_2]\cap[\omega_3]}$, $m_{\rm KK,M-th+M2}$, ($m_{\rm KK,M-th}$)	\end{tabular}&	2&$M_{\rm Pl,10}$	&\begin{tabular}{@{}c@{}}	$\setminus$\end{tabular}	&F-theory on $\mathbb{T}^2$\\\hline
			\rowcolor{red!10!}	(0,$e^1$,$e^2$,$e^3$) &	\begin{tabular}{@{}c@{}}$m_{\rm KK,M-th}$	\end{tabular}&	3&$M_{\rm Pl,5}$	&	\begin{tabular}{@{}c@{}} All but $m_{\rm osc}$\\
			($m_{\rm KK,M-th+II}\ll M_{\rm Pl,5}$)	\end{tabular}& M-theory on $X$	\\ \hline
			\rowcolor{red!10!}($e^0$,$e^1$,$e^2$,$e^3$) &\begin{tabular}{@{}c@{}} $m_{\rm KK, M-th}$	\end{tabular}&	3&$M_{\rm Pl,11}$	& All but $m_{\rm osc}$, $\ll M_{\rm Pl,5}$	& \begin{tabular}{@{}c@{}}11d \\ M-theory \end{tabular}	\\\hline
			
		\end{tabular}
		}
		\caption{Description of the different EFT string limits driven by the string charges ${\mathbf{e}}$  for $E_8\times E_8$ heterotic theory on a CY 3-fold with $\mathcal{V}_X\sim s^1s^2s^3$. We provide the identification of the leading tower(s) $m_\ast$, their scaling weight $w$, the asymptotic species scale $\Lambda_{\rm QG}$, subleading towers with $m\lesssim \Lambda_{\rm QG}$ and the microscopic interpretation of the dual theory emerging at the infinite distance limit. Notice that we only depict a representative of each set of equivalent limits under the $S_3$ action on the K\"ahler saxions.\label{tab.EFT lim s1s2s3}} 
	\end{center}
\end{table} 

 By combining the information given by the leading towers and the quantum gravity cutoff, we can reconstruct the  microscopic interpretation of the different duality frames associated to each saxionic growth sector. This way, the saxionic cone gets divided in the different perturbative regimes:
 
 \begin{itemize}
 	\item For $s^0\gg \mathcal{V}_X$ with $s^0\to \infty$, we reach a weak coupling regime which can be described in terms of \textbf{heterotic $E_8\times E_8$ string theory on CY$_3$}, the species scale being $\Lambda_{\rm QG}= m_{\rm str}$. Depending on the relative growth of the three saxions we reach asymptotically the following theories:
 	\begin{itemize}
 		
 \item If $\mathcal{V}_X$ remains constant we go to an \textbf{emergent $E_8\times E_8$ heterotic string limit}, with leading tower given by $m_{\rm osc}$.
  \item If the three K\"ahler saxions $s^1,s^2,s^3$ also grow, we decompactify to \textbf{10d heterotic string theory}. The leading tower will depend on which saxion grows faster, being the KK tower associated to that direction.  We can also have partial decompactifications to \textbf{6d $E_8\times E_8$ heterotic string theory on $\mathbb{T}^4$ or }K3 (if only one saxion grows while the other two remain fixed) and \textbf{8d  $E_8\times E_8$ heterotic string theory on $\mathbb{T}^2$} (if only two saxions grow while the other one remains fixed). 
 \end{itemize}
 	\item For $s^a\gg s^0\sqrt{s^bs^c}$,  the volume of the divisor $[\omega_a]$ will shrink relatively to the overall volume. Then, we will get a string becoming tensionless in Planck units coming from a NS5-brane wrapping such divisor. This string will have a dual description in terms of a Type IIA string if such divisor is topologically a $\mathbb{T}^4$, as we expect from the asymptotic form of the volume\footnote{This works as long as the perturbative corrections remain subleading along any direction of this region of the moduli space. Otherwise, we could also have a K3 fibration rather than a $\mathbb{T}^4$, reaching a dual heterotic frame.} $\mathcal{V}_{[\omega_a]}\sim s^bs^c$ and will set the species scale. Therefore, we expect this whole region to be described by \textbf{Type IIA string theory on a six-dimensional compact space} $\mathcal{X}_6'$,  where now the universal saxion is given by $s^a$. 
However, further investigation is required to properly describe this internal space (see e.g. \cite{Lee:2019wij} for dual descriptions of M-theory limits with vanishing $\mathbb{T}^4$ fibers with 8 supercharges).
 	Again, depending on the relative scaling of the saxions, we will reach different theories asymptotically:
 	\begin{itemize}
 			\item If $s^0$, $s^b$, $s^c$ are fixed and since $s^a\to\infty$, we go towards an \textbf{emergent type IIA string limit}, with both the leading tower and the species scale given by $\mathcal{T}_{e^a}^{1/2}$.
 			\item We can get a rudimentary understating of the new space $\mathcal{X}_6'$ by studying the KK towers that become light at the same rate as the string. First, $m_{{\rm KK},[\omega_b]\cap[\omega_c]}\sim M_{\rm Pl,4}(s^0)^{-1/2}(s^a)^{-1/2}$ can be understood in this new duality frame as associated to a curve $\mathcal{C}\subset \mathcal{X}_6'$ with volume $s^0$ in the new string units, while $m_{\rm KK,M-th}\sim M_{\rm Pl,4}(s^a)^{-1/2}(s^b s^c)^{-1/2}$ comes from an interval $\mathbb{S}^1/\mathbb{Z}_2$ with length $\sqrt{s^b s^c}$. This means that in the interfaces with other duality frames, we are able to recover decompactification limits to \textbf{weakly coupled 6d heterotic $E_8\times E_8$} (for $s^0\sim s^a\to\infty$ and fixed $s^b$ and $s^c$) and \textbf{5d M-theory on $X$} (by $s^a\sim s^b\sim s^c\to\infty$ and fixed $s^0$).
 	\end{itemize}
 
 	\item For $s^as^b\gg s^0(s^c)^2$, one recovers an F-theory description over a elliptic fibration with $[\omega_a]\cap[\omega_b]=\mathbb{E}\hookrightarrow X\to\mathcal{B}=[\omega_c]$ base, with the complete $X$ decompactifying to \textbf{10d F-theory on $\mathbb{T}^2$}. In analogy with Section \ref{ss.het 2sec}, can have further \textbf{partial decompactifications to 6d} ($s^as^b\to\infty$ with fixed $s^c$ and $s^0$), \textbf{8d} (with $s^0\sim (s^a s^b)^{1/3}\to\infty$ and fixed $s^c$). The fundamental IIB string is again not accessible from the K\"ahler sector, as the IIB dilaton is part of the complex-structure moduli which remains fixed.
 	\item Finally, for $s^1s^2s^3 \gg s^0 \gg s^a/\sqrt{s^bs^c}, s^as^b/(s^c)^2$, the heterotic string coupling diverges at the same time as the overall volume, with no internal cycles shrinking. Hence this region, where the species scale is given by $M_{\rm Pl,11}$, is described by \textbf{M-theory on $X\times \mathbb{S}^1/\mathbb{Z}_2$}. 
 	\begin{itemize}
 			\item For $s^1 \sim s^2 \sim s^3 \to \infty$ and fixed $s^0$ the perturbative description is just \textbf{M-theory on $X$}, with the $\mathbb{S}^1/\mathbb{Z}_2$ interval decompactifying.
 
 \item	Along the interfaces where $s^0 \sim s^1s^2s^3$ we can have partial decompactifications. If the three saxions are sent to infinity at the same time, the species scale is given by $M_{\rm Pl,10}$ and the dual asymptotic description is \textbf{M-theory on $\mathbb{S}^1/\mathbb{Z}_2$}. If only $s^c$ is fixed while  $s^a,  s^b \to \infty$, the species scale is $M_{\rm Pl,8}$ and we get asymptotically \textbf{M-theory on $\mathbb{S}^1/\mathbb{Z}_2 \times \mathbb{T}^2$}. Finally, if only $s^a \to \infty$ while $s^b$ and $s^c$ are fixed, $M_{\rm Pl,6}$ is the species scale and we approach \textbf{M-theory on $\mathbb{S}^1/\mathbb{Z}_2 \times \mathbb{T}^4$}.
		\end{itemize}

	Since the saxionic cone is 4-dimensional, it is not easy to depict in a picture. In Figure \ref{f.slicess1s2s3} we depict two different 3-dimensional slices, keeping fixed one of the saxions. Note that equivalent slices can be obtained upon an $S_3$ transformation on $\{s^i\}_{i=1}^3$.
    \begin{figure}[hpt!]
\begin{center}
\begin{subfigure}[b]{0.5\textwidth}
\captionsetup{width=.95\linewidth}
\center
\includegraphics[width=\textwidth]{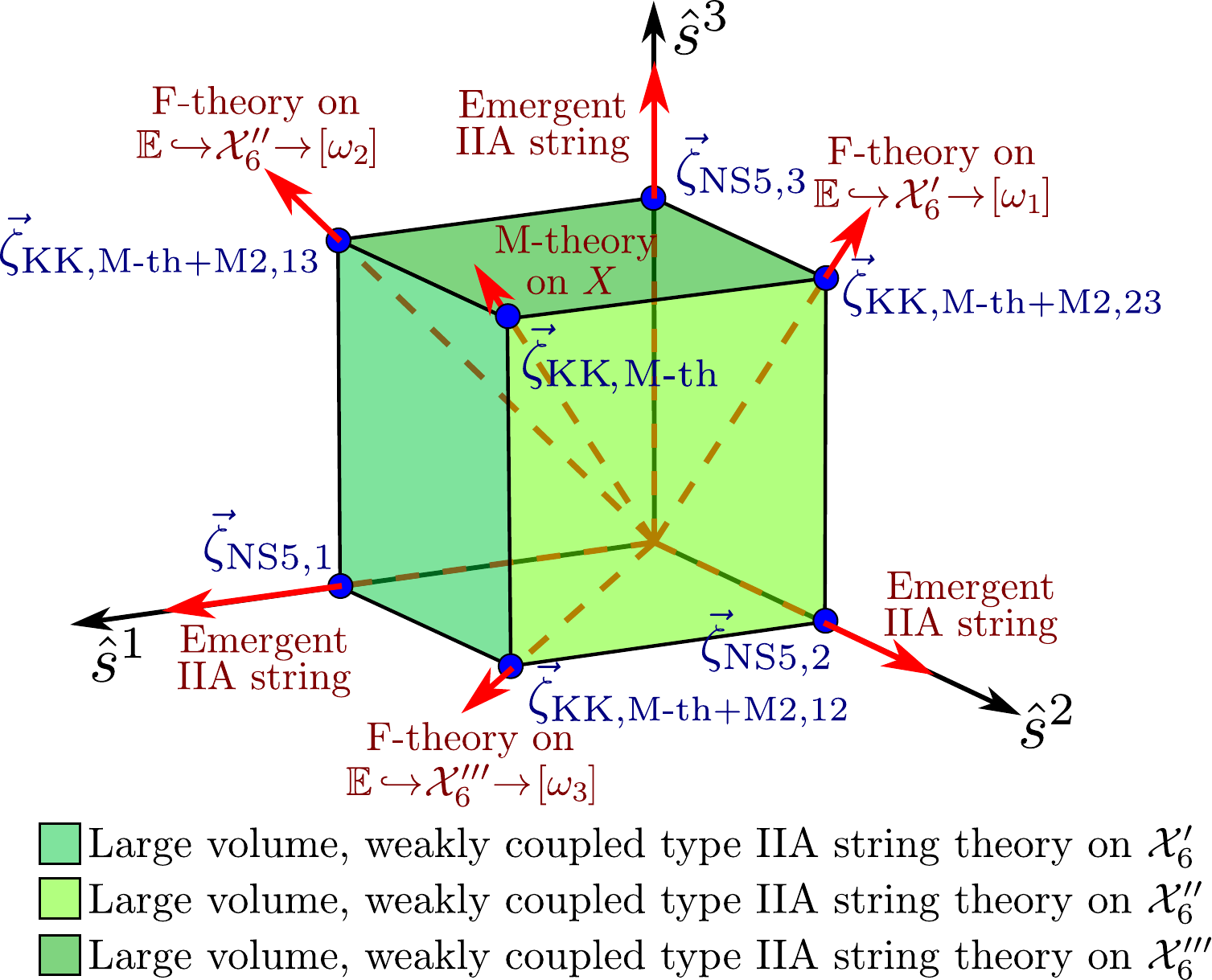}
\caption{\hspace{-0.3em} Tower convex hull (blue) and duality frames for the slice with $s^0$ fixed.} \label{f.slicess1s2s3-1}
\end{subfigure}
\begin{subfigure}[b]{0.44\textwidth}
\captionsetup{width=.95\linewidth}
\center
\includegraphics[width=\textwidth]{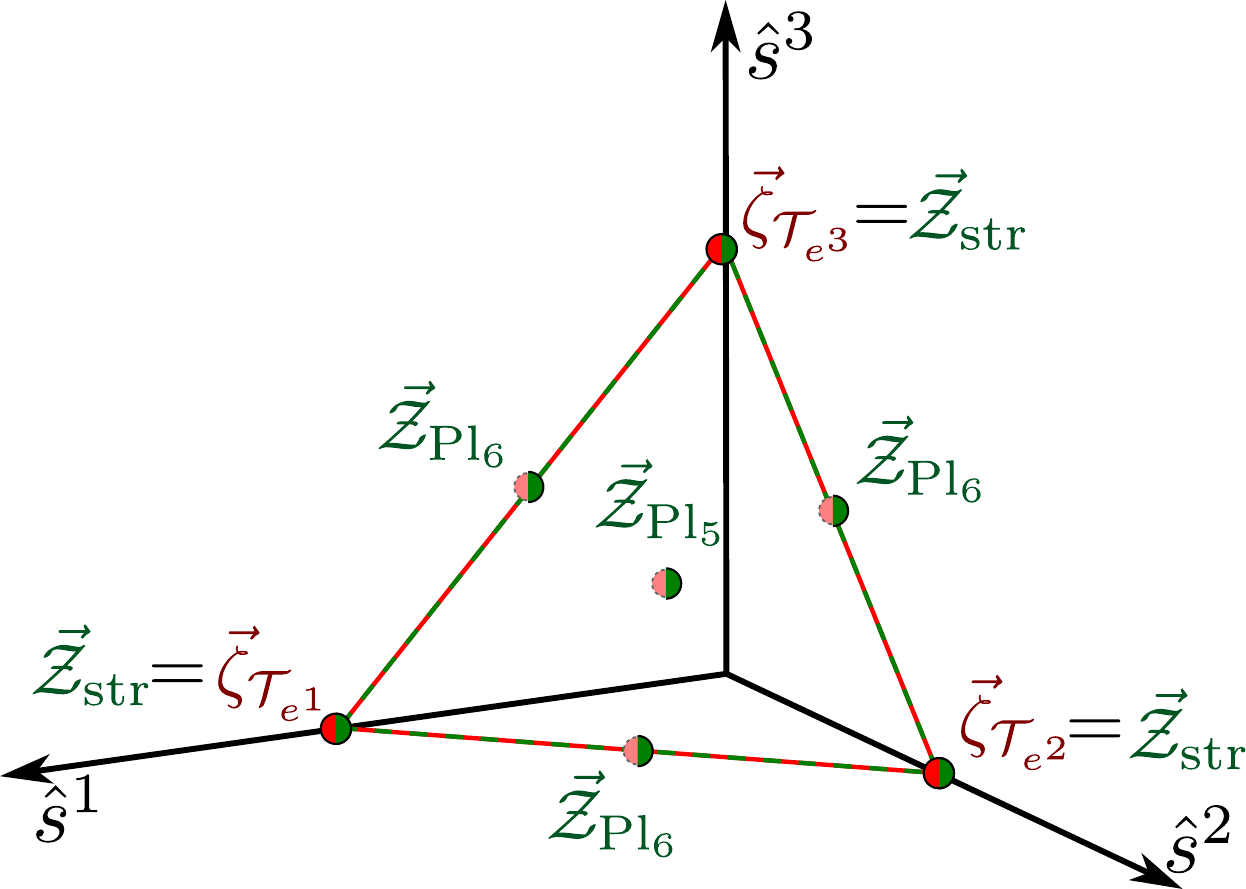}
\caption{\hspace{-0.32em} Species scale (green) and EFT strings (red) for the slice with $s^0$ fixed. } \label{f.slicess1s2s3-1sp}
\end{subfigure}
\begin{subfigure}[b]{0.5\textwidth}
\captionsetup{width=.95\linewidth}
\center
\includegraphics[width=\textwidth]{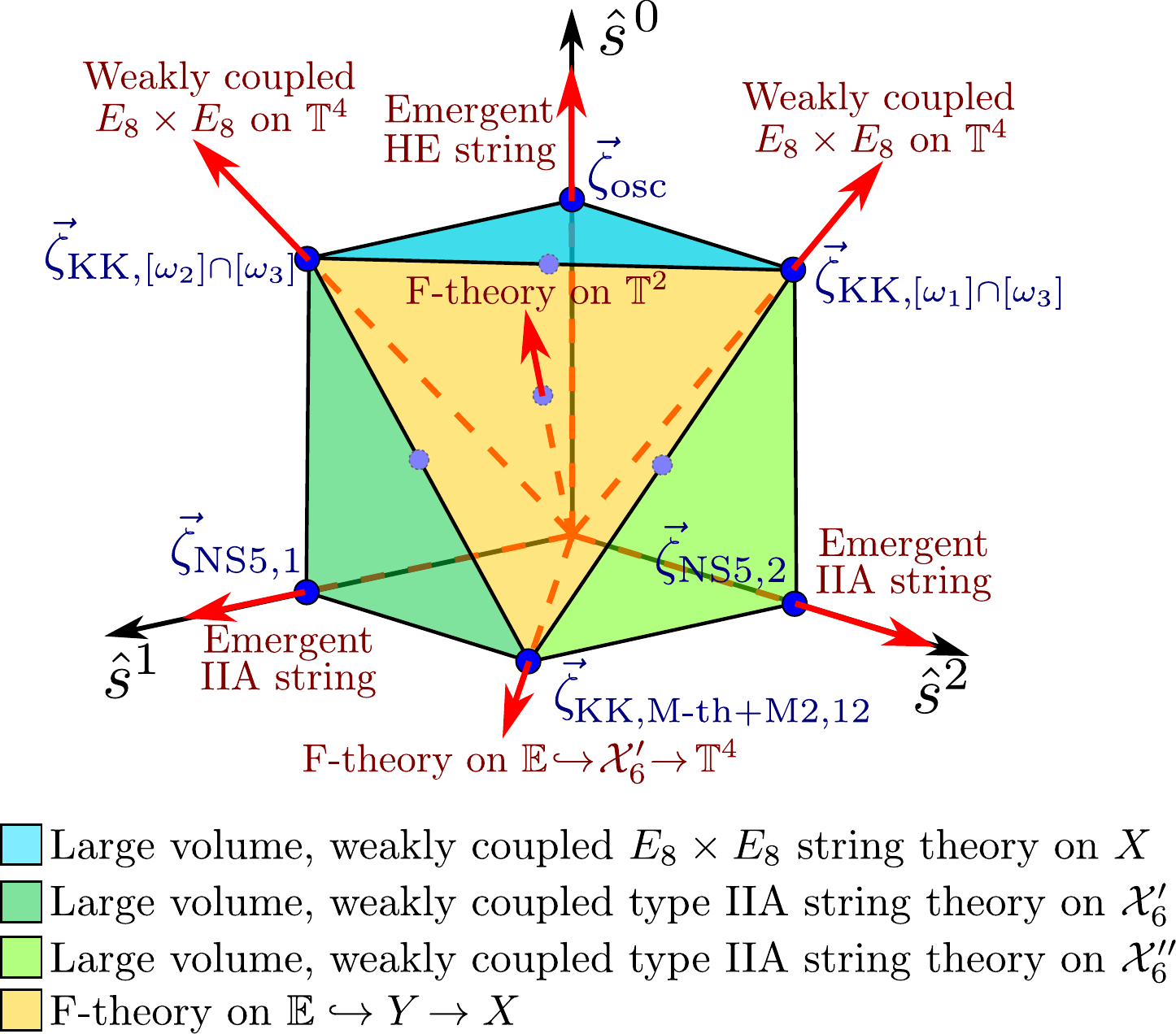}
\caption{\hspace{-0.3em} Tower convex hull (blue) and duality frames for the slice with $s^3$ fixed.} \label{f.slicess1s2s3-2}
\end{subfigure}
\begin{subfigure}[b]{0.44\textwidth}
\captionsetup{width=.95\linewidth}
\center
\includegraphics[width=\textwidth]{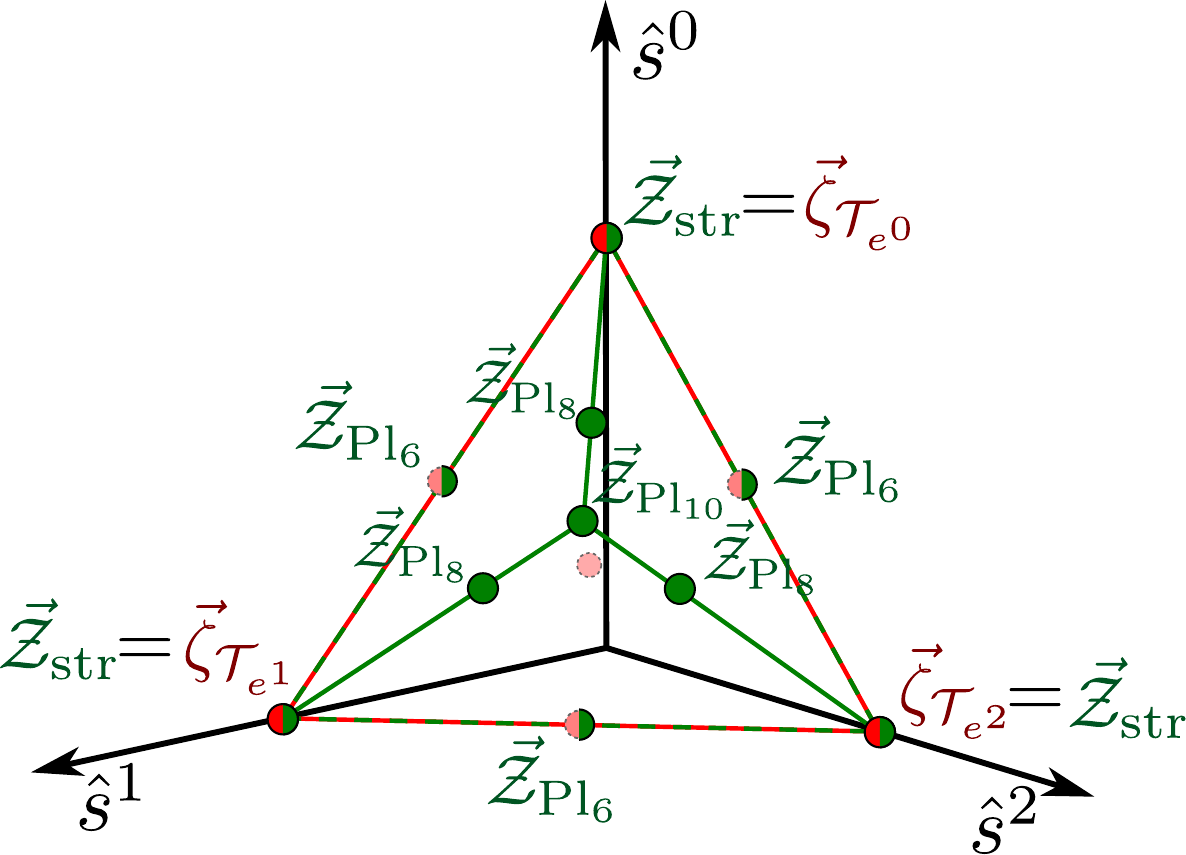}
\caption{\hspace{-0.32em} Species scale (green) and EFT strings (red)  for the slice with $s^3$ fixed. } \label{f.slicess1s2s3-2sp}
\end{subfigure}
\caption{Arrangements for the $\zeta$-vectors of towers and EFT strings, as well as $\mathcal{Z}$-vectors for asymptotic behavior of $\Lambda_{\rm QG}$, for two slices for the saxionic cone of heterotic $E_8\times E_8$ compactified on a CY 3-fold with $\mathcal{V}_X\sim s^1s^2 s^3$. All quantities are depicted in terms of the canonically normalized saxions. The different colored facets represent the different perturbative duality frames of each theory. Since the six-dimensional space $\mathcal{X}_6'$ on which the dual type IIA description is compactified needs not be the exact same manifold for the different $s^i\to\infty$ ($i=1$, 2 or 3) limits, we denote them by $\mathcal{X}_6'$, $\mathcal{X}_6''$ and $\mathcal{X}_6'''$, respectively.
\label{f.slicess1s2s3}}
\end{center}
\end{figure}
 \end{itemize}

\subsubsection{Non-homogeneous $V_X$ and growth sectors\label{sec.no homo}}

In the cases we have just studied, we have considered compactifications on Calabi-Yau  where the asymptotic form of their internal volumes (in string units) $ \mathcal{V}_X = \frac{1}{3!}\kappa_{abc}s^as^bs^c$ could be expressed as an homogeneous function of the saxions. In other words, there is a single non-vanishing intersection number. This possibility only occurs for at most 3 K\"ahler saxions, with $ \mathcal{V}_X\sim (s^1)^3,\,(s^1)^2s^2$ and $s^1s^2s^3$ being precisely the cases studied in the previous subsections. However, a generic 3-fold $X_6$ will have more than 3 K\"ahler moduli and several monomials in the asymptotic expression of $ \mathcal{V}_X$. As a result, the metric $\mathsf{G}_{ij}$ \eqref{e.Kahler metric} will not be diagonal and, save for very specific cases, will have non-vanishing curvature, so that the  previous expressions and global classification of the duality frames, towers and limits will not immediately hold. 

In spite of this, our results above will still serve as building blocks to construct the more general case. Given a compactification with $k$ K\"ahler saxions $\{s^i\}_{i=1}^k$, we can still divide our perturbative region (defined by the saxionic cone) in \emph{growth sectors} \cite{Grimm:2018cpv,Corvilain:2018lgw} 
\begin{equation}
\mathcal{R}_{i_1,\dots,i_k}=\{s^{i_1}\gg s^{i_2}\gg \dots \gg s^{i_k}\}\;.
\end{equation}
 Note that this covers all asymptotic regions of the K\"ahler moduli space, except for asymptotic direction loci where several saxions grow at the same rate.\footnote{The space of infinite distance limits is given by the equivalence classes of all \emph{parallel asymptotic geodesics}, being a manifold with one dimension less than the number of non-compact scalars of our moduli space, see Appendix C from \cite{Castellano:2023jjt} for more details. Considering the inherited Lebesgue measure, the set of asymptotic directions over which several saxions grow at the same rate can be shown to have measure zero.} Then, for an asymptotic direction in a given growth sector $\mathcal{R}_{i_1,\dots,i_k}$, $ \mathcal{V}_X$  will be dominated by a single cubic monomial (see also \cite{Grimm:2022sbl} for the interplay between growth sectors and EFT strings). Hence, to describe the asymptotic structure of the towers and duality frames in these limits,  we can simply borrow the results of the cases studied above. If the asymptotic moduli space is flat (which will not always be the case), we can then glue the different building blocks to obtain a polytope representing the full perturbative regime. However, this gluing is not trivial as it can involve some \emph{jumping} of the $\zeta$-vectors of the towers, as we will describe below.
\vspace{0.25cm}

But first, let us identify explicitly the building blocks that will describe the structure of the towers within each possible growth sector. In Figure \ref{f.growth} we show again the structure of the towers obtained in Sections \ref{s.het vols3} and \ref{ss.het 2sec} for $\mathcal{V}_X\sim (s^1)^3$ and $\mathcal{V}_X\sim (s^1)^2s^2$, also taking into account the universal saxion $s^0$ direction. Notice that these polytopes were constructed under the assumption that this form of the volume is valid through the entire saxionic cone. Rather, if it is only valid in a given growth sector, we have to consider only the part of the figure associated to that region of moduli space. For instance, consider Figure \ref{f.growth2} representing the towers for  $V_X\sim (s^1)^2s^2$. We have divided the polytope in two parts, the orange and green regions, to illustrate the towers that are relevant in each of the two growth sectors, $s^1\gg s^2$ or $s^2\gg s^1$. If this asymptotic form of the volume is only valid for $s^1\gg s^2$, then we will have to consider only the green part of the Figure. This green region is enough to capture all towers that become light below the species scale for any trajectory within that growth sector. 

Similarly, we have illustrated in Figure \ref{f.growth3} the part of the tower polytope that is relevant for a growth sector given by $s^2\gg s^1\gg s^3$ and $\mathcal{V}_X\sim s^1s^2s^3$. Since the volume is symmetric under permutations of the saxions, all growth sectors have the same structure of towers in this case. It is important to emphasize that the tower polytope will be the same regardless of which infinite distance trajectory we take, as long as we stay in the \emph{strict interior} of a given growth sector. In this sense, we say that the tower polytope is \emph{rigid} within each growth sector.\\

Even if for a given growth sector $\mathcal{R}_{i_1,\dots,i_k}$ the  volume $\mathcal{V}_X$ is asymptotically a monomial, and the moduli space metric $\mathsf{G}_{ij}$ becomes diagonal to leading order (asymptotically), this does not translate in the moduli space becoming (asymptotically) flat along said limits. In the computation of the Riemann tensor and the related curvature tensors and invariants, inverse metrics and their derivatives come into play, in such a way that the asymptotically vanishing subleading terms in $\mathsf{G}_{ij}$ contribute to the moduli space curvature in a non-vanishing way. We will show an explicit example of this behavior later this section. As explained in Section \ref{ss.review towers}, the different $\zeta$-vectors live in the tangent space at a given point of moduli space. Moreover, in a given growth sector the different towers below the species scale are the same (since there is no change in the hierarchy at which the different saxions scale), and as we explained above, the tower polytope is rigid inside a given growth sector. Therefore, even if the non-vanishing moduli space curvature prevents the identification of the field and tangent spaces, the $\zeta$-vector arrangement will remain invariant inside a growth sector.

\begin{figure}[h]
\begin{center}
\includegraphics[width=.5\textwidth]{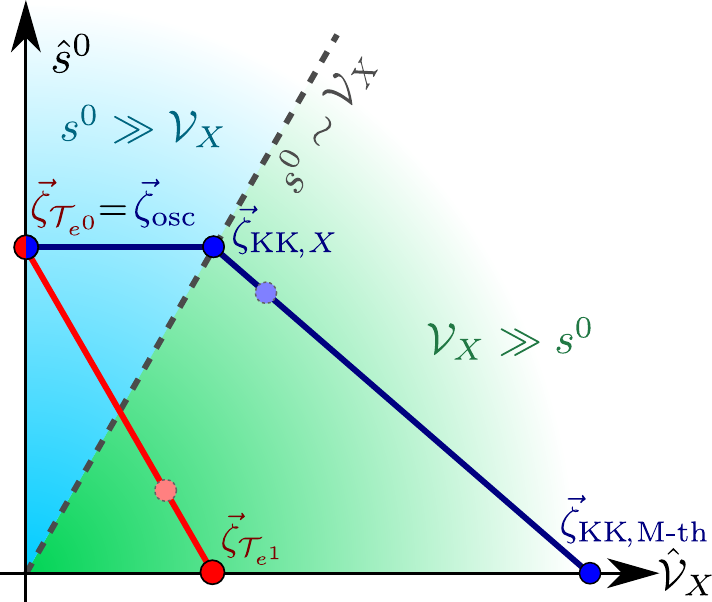}
\caption{Growth sectors $s^0\gg \mathcal{V}_X$ and $\mathcal{V}_X\gg s^0$ for Calabi-Yau compactification of heterotic $E_8\times E_8$, where $s^0$ is the universal dilaton and $\mathcal{V}_X$ the overall volume in string units. Note that precisely the correspond to the duality frames in Figure \ref{fig:het2d}. All quantities are given in terms of the canonically normalized moduli.
\label{f.growth1}}
\end{center}
\end{figure}

    \begin{figure}[h]
\begin{center}
\begin{subfigure}[b]{0.45\textwidth}
\captionsetup{width=.95\linewidth}
\center
\includegraphics[width=\textwidth]{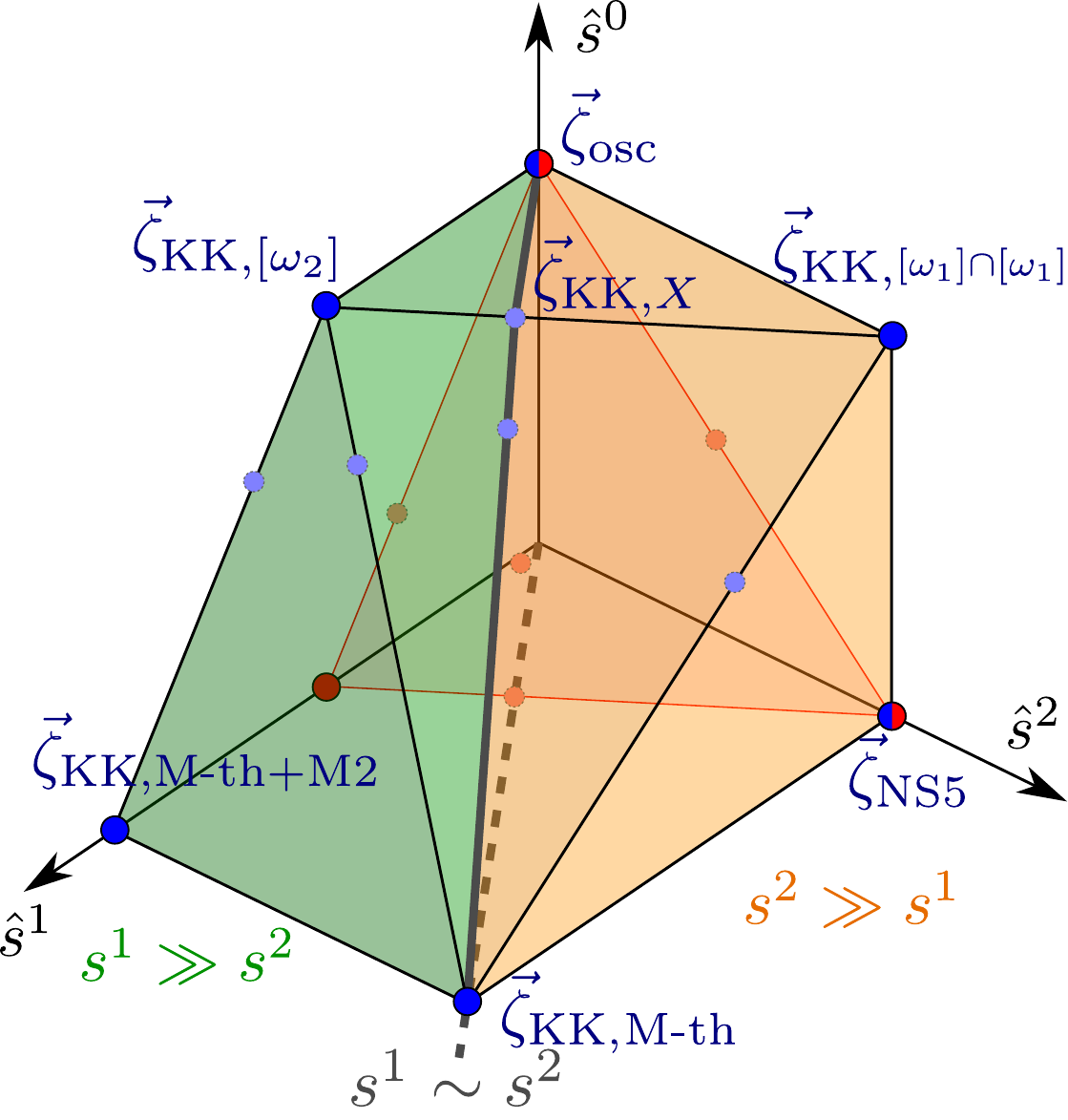}
\caption{\hspace{-0.32em} Growth sectors $s^1\gg s^2$ and $s^2\gg s^1$ for $\mathcal{V}\sim (s^1)^2s^2$.} \label{f.growth2}
\end{subfigure}
\begin{subfigure}[b]{0.54\textwidth}
\captionsetup{width=.95\linewidth}
\center
\includegraphics[width=\textwidth]{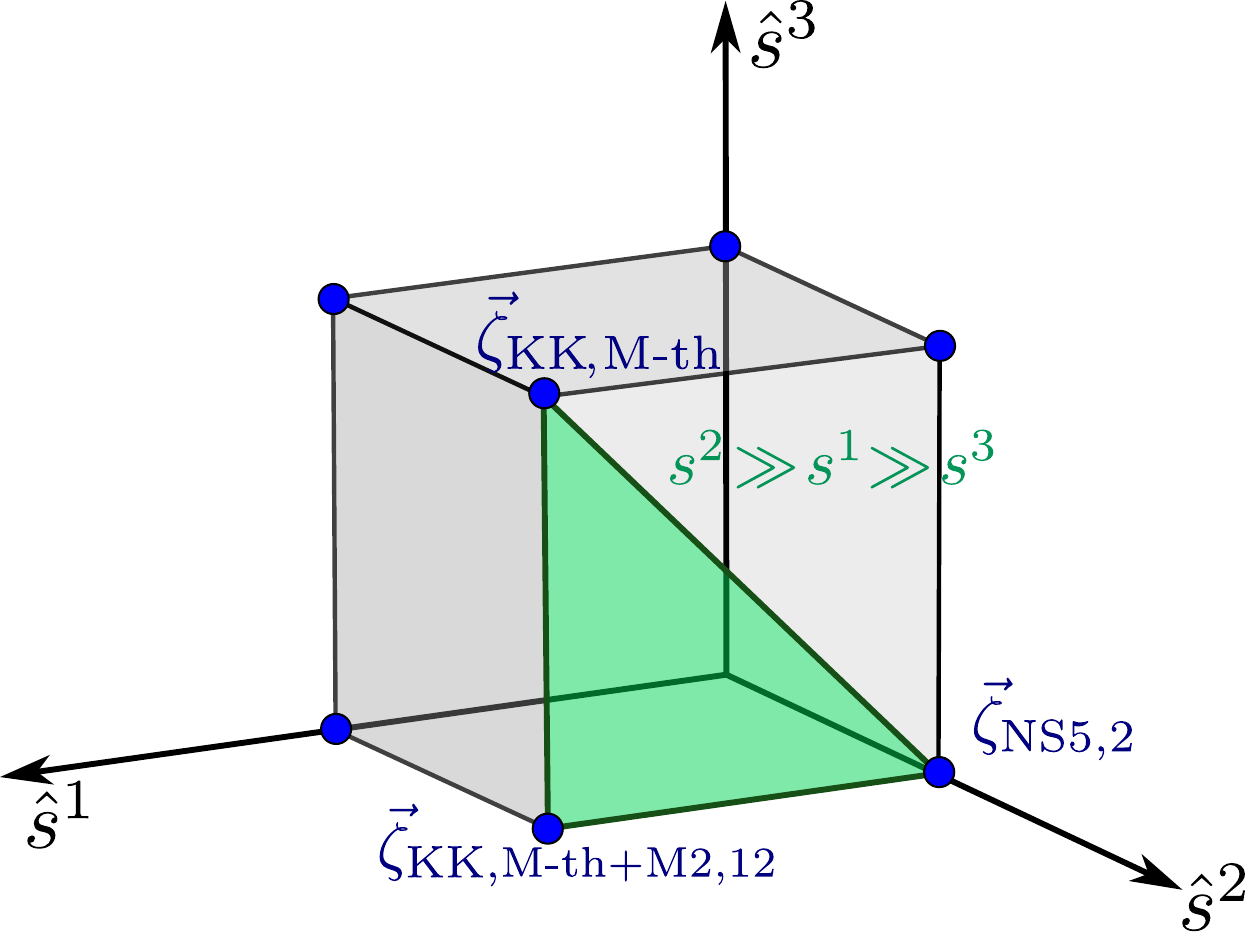}
\caption{\hspace{-0.32em} Growth sectors $s^2\gg s^1\gg s^3$ (fixed $s^0$) for $\mathcal{V}\sim s^1s^2s^3$. The rest growth sectors in the $\{s^1,s^2,s^3\}$ slice can be obtained with a $S_3$ transformation. } \label{f.growth3}
\end{subfigure}
\caption{Illustration of growth sectors for heterotic $E_8\times E_8$ compactified on CY 3-folds with volumes $\mathcal{V}_X\sim (s^1)^2s^2$ and $s^1s^2s^3$ in string units. All quantities are given in terms of the canonically normalized moduli.   
\label{f.growth}}
\end{center}
\end{figure}

When defining the growth sectors, it is sufficient to consider only  the relative growth between the K\"ahler saxions, without including the 4d string coupling/universal saxion $s^0$.  The reason is that the latter factorizes in the definition of the K\"ahler potential, $K=-\log s^0-\log  \mathcal{V}_X$ (see \eqref{e.kahler pot def}), so the asymptotic form of the K\"ahler potential will be the same regardless of whether $s^0$ grows faster or slower than the others. This implies that the tower polytope remains also rigid under changing the $s^0$ component of the infinite distance limit, as long as we stay within the same growth sector for the  K\"ahler  moduli. Moreover, if we take a 2d subspace spanned by $s^0$ and $ \mathcal{V}_X$, the convex hulls of the towers and strings will always take the same form, as shown in Figure \ref{f.growth1}. The weak coupling regime $s^0\gg V_X$ corresponds to the region $\hat{s}^0>\frac{1}{3}\hat{\mathcal{V}}_X$ in terms of the canonically normalized fields in the figure.
\vspace{0.25cm}

One caveat is in order, though. As explained in \cite{Witten:1996mz,Cvetic:2024wsj} for certain compactifications of $E_8\times E_8$ string theory on $X$ with standard embedding, the regime $ \mathcal{V}_X\gg s^0$ results in strong coupling limits that are obstructed by quantum corrections: the coupling in one of the $E_8$ gauge groups diverges at finite value of the moduli, classically resulting in a finite distance boundary in the saxionic cone. From the M-theory point of view, this maps to the warp factor over the Ho\v{r}ava-Witten interval diverging at one of the 9-branes, which would appear as 5d End of the World brane in the lower-dimensional theory. In those cases, the boundary of the saxionic cone would actually be located at $s^0\sim  \mathcal{V}_X$, since non-perturbative instantonic corrections become important beyond that. In other words, only the regime $s^0\gg \mathcal{V}_X$ is a priori part of the 4d perturbative regime, and only the blue part of Figure \ref{f.growth1} should be considered. 
In \cite{Cvetic:2024wsj} it was shown that, after non-perturbative effects are taken into account, the warp factor divergence can be regulated, resolving this way the strong coupling singularity and extending the field space beyond this classical boundary. However, it is not clear whether  the endpoint of those trajectories will be at finite distance or infinite distance, nor what is the dual description of the new region.\footnote{In analogy with similar transitions in \cite{Atiyah:2000zz,Acharya:2000gb}, in \cite{Cvetic:2024wsj} it was conjectured that the new 4d $\mathcal{N}=1$ region might be described by M-theory compactified on curved 7-manifold with fluxes.} Hence, we will not discuss this case here any further.

\vspace{0.5cm}

Let us illustrate the above discussion using the following example. Take the overall volume as follows,
\begin{equation}\label{e.potej}
\mathcal{V}_X=3\kappa_{112}(s^1)^2s^2+3\kappa_{113}(s^1)^2s^3+\kappa_{123}s^1s^2s^3\;.
\end{equation}
The asymptotic regions of the K\"ahler moduli space are non flat, with Ricci scalar\footnote{While for this example $\mathcal{R}$ is negative everywhere, there are examples in the literature for which $\mathcal{R}$ is positive or even diverges to $+\infty$ for specific directions and trajectories of measure zero \cite{trenner2010asymptoticcurvaturemodulispaces,Marchesano:2023thx,Marchesano:2024tod,Castellano:2024gwi}. This would not affect our approach or conclusions.}
\begin{equation}
	\mathcal{R}_{\mathsf{G}}=-\frac{108\kappa_{112}\kappa_{113}(s^1)^2\left[18\kappa_{112}\kappa_{113}(s^1)^2+3\kappa_{123}s^1(\kappa_{112}s^2+\kappa_{113}s^3)+\kappa_{123}s^2s^3\right]}{\left[36\kappa_{112}\kappa_{113}(s^1)^2+3\kappa_{123}s^1(\kappa_{112}s^2+\kappa_{113}s^3)+\kappa_{123}s^2s^3\right]^2}<0\;,
	%\quad\frac{12 X^3 \left(2 x^3+x^2 (y+z)+x y z\right)}{\left(4 x^3+x^2 (y+z)+x y z\right)^2}<0\;.
\end{equation}
We show in Table \ref{tab.example} the different tower polytopes for the different growth sectors. As commented above, the non-zero curvature does not prevent us from depicting together the $\zeta$-vectors for the different light towers in a given growth sector, with the tower polytope being rigid and thus being the same for all the limits inside each growth sector. However, the curvature prevents us from performing a global identification of the tangent space with the saxionic moduli space, and thus from gluing the tower polytopes of Table \ref{tab.example} together in a single figure.
Notice that the tower of oscillator modes from the fundamental heterotic string depends only on $s^0$ through \eqref{eq.mosc}, and thus is the same and present in all growth sectors.  \\
\begin{table}[htp!]
\begin{center}
	\resizebox{\textwidth}{!}{
	\begin{tabular}{|c|c|c|c|c|}\hline
	Growth sector & $\mathcal{V}_X$&${\rm dim}_\infty\mathcal{M}$& $\mathcal{R}$  & Tower structure \\\hline
	$s^1\gg s^2\gtrsim s^3$ & $(s^1)^2s^2$& 1+2&$\left[-\frac{3}{2},0\right)$& \begin{minipage}{.275\textwidth}
      \includegraphics[width=\linewidth]{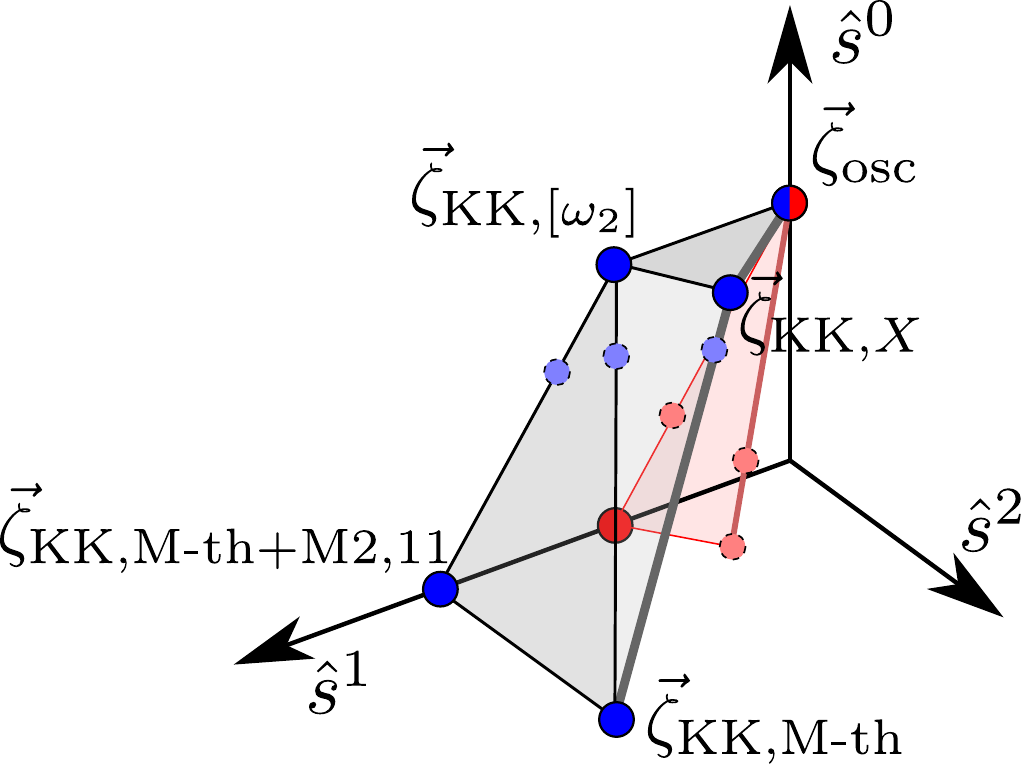}
    \end{minipage} \\\hline
	$s^2\gg s^1\gtrsim s^3$ &$(s^1)^2s^2$&1+2& $0^-$& \begin{minipage}{.2\textwidth}
      \includegraphics[width=\linewidth]{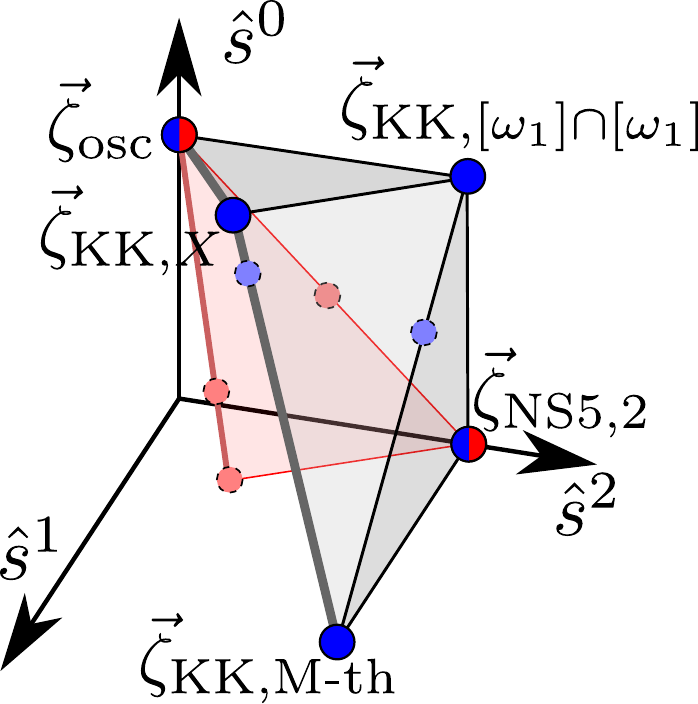}
    \end{minipage}\\\hline
	$s^2\gg s^3\gg s^1$ & $s^1s^2s^3$&1+3& $0^-$& \begin{minipage}{.5\textwidth}
      \includegraphics[width=\linewidth]{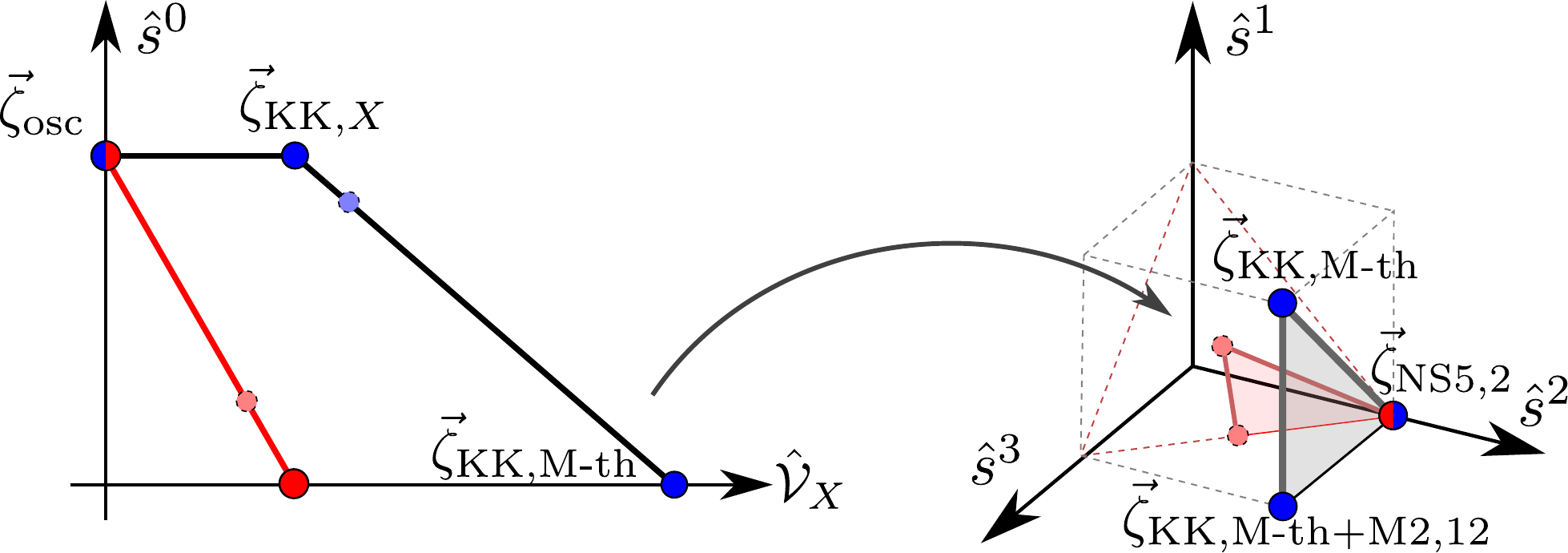}
    \end{minipage} \\\hline
	$s^1\sim s^2\gtrsim s^3$ & $(s^1)^3$&1+1& $\left[-\frac{3}{2},0\right)$& \begin{minipage}{.275\textwidth}
      \includegraphics[width=\linewidth]{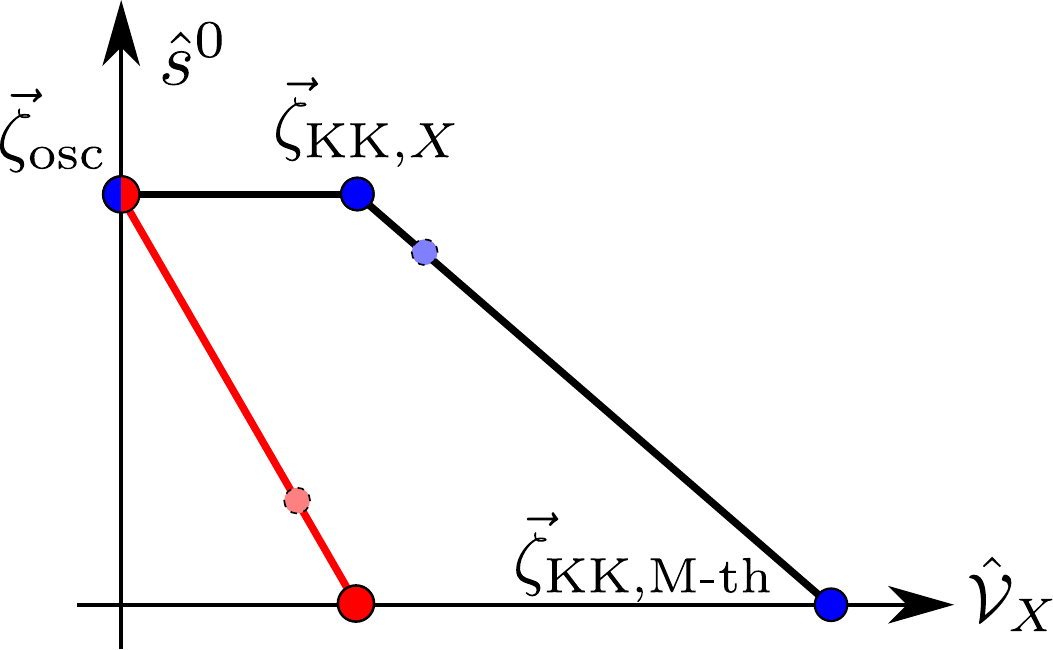}
    \end{minipage} \\\hline
	$s^2\sim s^3\gg s^1$ & $s^1(s^2)^2$&1+2& $0^-$& \begin{minipage}{.295\textwidth}
      \includegraphics[width=\linewidth]{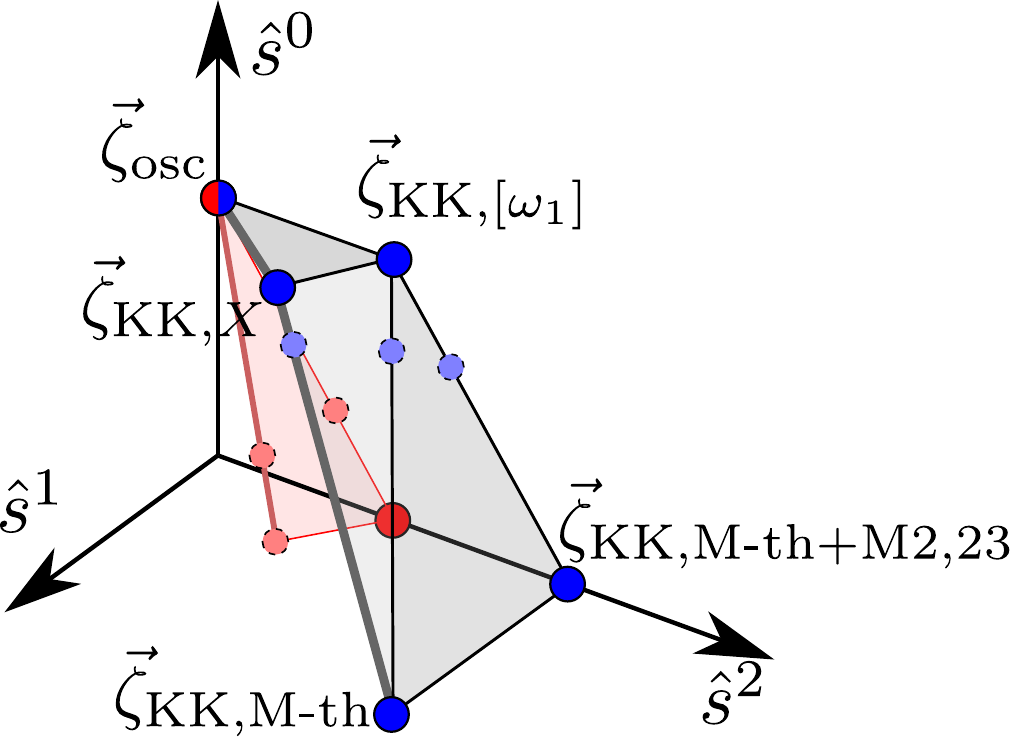}
    \end{minipage} \\\hline
	\end{tabular}
	}
	\end{center}
	\caption{Different growth sectors for the volume \eqref{e.potej}. The asymptotic structure of $\mathcal{V}_X$ and tower structure are present, as well as the asymptotic dimensionality of the saxionic moduli space (including $s^0$), and its scalar curvature $\mathcal{R}$ (for the case in which an interval is presented, the $\mathcal{R}=-\frac{3}{2}$ is only achieved in the strict $\gtrsim\to\gg$ regime). Because of the $s^2\leftrightarrow s^3$ symmetry in $\mathcal{V}_X$ (modulo numeric prefactors), all the above regimes have a counterpart under such saxion interchange. In the $s^2\gg s^3\gg s^1$ growth sector, with $\mathcal{V}_X\sim s^1 s^2 s^3$, we sketch two slices of the four-dimensional figure. Note that the first and last tower arrangements in the table are equivalent.\label{tab.example}}
\end{table}

The above discussion only applies if we move along an infinite distance limit in the strict interior of a given growth sector. So what happens if we move instead along a trajectory in the \emph{interface} between two or more growth sectors? In that case, two or more saxions will grow at the same rate (e.g. $s^1\sim s^2$ or $s^1\sim s^2\sim s^3$) and several terms of the volume expression $\mathcal{V}_X$ might compete. This may lead to changes in the tower arrangement, as we explain next. 

Consider a trajectory parameterized by $s^j=\alpha s^i$ with $s^i\rightarrow \infty$, up to subleading terms,  with $\alpha\to 0,\,\infty$ corresponding to $s^i\gg s^j$ and $s^j\gg s^i$, respectively. If $\alpha$ is finite, both saxions grow at the same parametric rate and the trajectory will lie in the interface between the two growth sectors. Notice that the trajectories that differ by a finite value of $\alpha$ share the same asymptotic unit tangent vector, so the direction is the same. We can then regard the coefficient $\alpha$ as an \textbf{impact parameter} of our trajectory.  As we vary $\alpha$, the $\zeta$-vectors of the towers may change in such a way that the tower polytope of one growth sector gets continuously\footnote{This is necessary to guarantee that the light states do not suddenly disappear from the spectrum when moving in a perturbative corner of the moduli space, and it is realized in all examples studied in this paper.} deformed into the tower polytope of the other growth sector. In all examples considered, this deformation occurs in a very particular way: the sliding of the $\zeta$-vectors occurs such that the projection of $\zeta$ over the tangent vector of the asymptotic trajectory remains constant. In other words, when comparing parallel trajectories differing in the impact parameter, only the component of $\zeta$ that is \emph{perpendicular} to the trajectory may vary, as explained in Appendix B of \cite{Etheredge:2024tok}. This implies that the exponential rate of the tower along trajectories differing by the value of $\alpha$ is the same, so that the exponential rate only depends on the direction of the asymptotic infinite distance limit and not on the impact parameter. Therefore, as we vary the direction of the infinite distance limit, the continuous sliding of the towers is actually seen as a discrete \emph{jumping} of the vectors, as first described in \cite{Etheredge:2023odp}.

The same behavior occurs along the interface between growth sectors with different tower content. As we vary $\alpha$, the towers which are only present at one of the two growth sectors can start \emph{sliding} out (in) from (to) infinity, perpendicular to the growth sectors interface (where our trajectory is contained). 
The exponential decay rate, given by the projection $\lambda=\vec{\zeta}\cdot\hat T$ where $\hat T$ is the tangent vector of the trajectory, remains constant as we vary the impact parameter. However, as we take the strict limit $\alpha\to 0,\, \infty$, which corresponds to slightly varying the direction of the trajectory and entering the new growth sector, $|\vec\zeta|$ formally diverges and such tower is no longer present in the new duality frame. Less dramatically, it can occur that individual $\zeta$ vectors slide a finite distance between their original position to that of another vector. This is what occurs for the KK modes associated to decompactifications of submanifolds $\mathcal{C}\subseteq \mathcal{D}$. While in some growth sector $\mathcal{C}$ might grow much faster than the overall $\mathcal{D}$, and thus we have different $\vec{\zeta}_{\rm KK,\mathcal{C}}$ and $\vec{\zeta}_{\rm KK,\mathcal{D}}$, when moving into another growth sector this might no longer be the case, such that $\mathcal{C}$ and $\mathcal{D}$ scale in the same way. This results in the sliding $\vec{\zeta}_{\rm KK,\mathcal{C}}\to\vec{\zeta}_{\rm KK,\mathcal{D}}$.
 
It is also interesting to notice that not all the towers slide. As described in Section \ref{s.het}, we basically have two types of towers. First we have those whose mass expression depends only on $s^0$ and/or $\mathcal{V}_X$, which are defined globally always and factorize. These towers are located in the $\{s^0,\mathcal{V}_X\}$ slice, and their $\zeta$-vectors do not change when comparing different growth sectors of the moduli space. The second set of towers are those whose mass depends on the volume of internal cycles (in other words, on individual K\"ahler saxions rather than the whole $\mathcal{V}_X$). From the geometric point of view, only in some specific limits a given fibration structure emerges such that a given KK/wrapped extended object effectively  exists. In the 4d language, the overall $\mathcal{V}_X=\frac{1}{6}\kappa_{ijk}s^i s^j s^k$ gets dominated by a term associated to that particular fibration that becomes subleading and effectively null when moving away from those specific limits. This translates in the $\zeta$-vectors of the corresponding towers experiencing sliding from one expression to another (when the tower exists but the cycle structure has changed) or coming in/out from/to infinity, with their norm diverging, when the corresponding fibration structure ceases to exist.\\

In order to illustrate this sliding phenomenon better, consider the following example. Take the Calabi-Yau 3-fold $\mathbb{P}^{(1,1,1,6,9)}[18]$ \cite{Candelas:1993dm,Hosono:1994ax}, with volume
\begin{equation}
\label{VXex2}
	\mathcal{V}_X=\frac{3}{2}(s^1)^3+\frac{3}{2}(s^1)^2s^2+\frac{1}{2}s^1(s^2)^2\;,
\end{equation}
depending only on the saxions $s^1$ and $s^2$. This example can also be recovered as a particular slice of the moduli space of \eqref{e.potej}.\footnote{The sliding in the example with volume $\mathcal{V}_X$ in \eqref{e.potej}, illustrated in Table \eqref{tab.example}, has three saxionic directions, so that the sliding arrangements are more involved. For the slice $s^2\sim s^3$ one obtains the same behavior as in the example shown now, while for more general directions we can have up to the three saxions competing, translating in co-dimension 2 sliding loci, which makes the $\zeta$-vector of the moving towers quite involved (one has to consider the curvature of the moduli space too). The example shown in \eqref{VXex2} is simpler, and has been chosen since it corresponds to the particular slice $s^2\sim s^3$ of \eqref{e.potej}.} One can explicitly check that the associated metric is flat (this is always the case for two K\"ahler moduli), and that there are two growth sectors, $\mathcal{R}_{12}$ ($s^1\gg s^2$) and $\mathcal{R}_{12}$ ($s^2\gg s^1$), respectively with dimensionalities 1+1 and 1+2 (taking also $s^0$ into account). The dominant term of $\mathcal{V}_X$ is given by $\mathcal{V}_X\sim (s^1)^3$ in $\mathcal{R}_{12}$  and $\mathcal{V}_X\sim s^1(s^2)^2$ in $\mathcal{R}_{21}$. The tower arrangement associated with $\mathcal{R}_{12}$ is that of Figure \ref{f.growth1} while the one associated with $\mathcal{R}_{21}$ is the green part of Figure \ref{f.growth2} (after a relabeling of $s^1$ and $s^2$). Along the interface where $s^1\sim s^2$, the volume takes the form of  $\mathcal{V}_X\sim (s^1)^3$. On one hand, the vectors of the string oscillator modes \eqref{eq.mosc}, and the KK towers associated to X  \eqref{eq.KKX het} and the M-theory interval \eqref{eq. Mth het},  are present in both growth sectors and only depend on $s^0$ and the overall $\mathcal{V}_X$, so do not vary. On the other hand, the KK tower associated to the divisor $[\omega_1]$, with volume $\mathcal{V}_1=\frac{9}{2}(s^1)^2+3s^1s^2+\frac{1}{2}(s^2)^2$, which are present in one growth sector but not the other (since the fibration structure is different in $\mathcal{R}_{12}$ and $\mathcal{R}_{21}$), will vary depending on the value of the impact parameter to the trajectory lying in the interface.  We can explicitly compute their $\zeta$-vectors along a trajectory $s^2\sim \alpha s^1$ to obtain
\begin{subequations}\label{eq.sliding}
\begin{align}
	\vec{\zeta}_{\rm KK,[\omega_1]}=\left(\frac{1}{\sqrt{2}},\frac{1}{\sqrt{6}},\hat{f}(\alpha)\right)\,,\quad \text{with}\quad \hat{f}(\alpha)=\tfrac{1}{2\sqrt{3}}\left(\tfrac{\alpha}{3+\alpha}\right)^{3/2}\approx\left\{\begin{array}{ll}
	0&\text{for }\alpha\to 0\\
	\frac{1}{2\sqrt{3}}&\text{for }\alpha\to\infty
	\end{array}\right.\\
	\vec{\zeta}_{\rm KK,M-th+M2}=\left(0,\sqrt{\frac{2}{3}},\hat{g}(\alpha)\right)\,,\quad \text{with}\quad \hat{g}(\alpha)=\tfrac{3+2\alpha}{2\sqrt{3}\sqrt{\alpha(\alpha+3)}}\approx\left\{\begin{array}{ll}
	\infty&\text{for }\alpha\to 0\\
	\frac{1}{\sqrt{3}}&\text{for }\alpha\to\infty
	\end{array}\right.
	\end{align}
\end{subequations}

It is possible to understand the above behavior intuitively as follows. In the $\mathcal{R}_{21}$ growth sector (i.e., $\alpha\to\infty$) the volumes of $X$ and the $[\omega_1]$ divisor scale like $\mathcal{V}_X\sim \frac{1}{2}s^1(s^2)^2$ and $\mathcal{V}_{[\omega_1]}\sim\frac{1}{2}(s^2)^2$, so that $[\omega_1]$ grows much faster than $X$, and the $[\omega_2]\cap[\omega_2]\simeq\mathbb{T}^2$ fiber relatively shrinks with respect to the overall volume, $\mathcal{V}_{[\omega_2]\cap[\omega_2]}=s^1$. Because of this, we find KK towers associated to $[\omega_1]$ and the bounded states between the Ho{\v{r}}ava-Witten interval KK and wrapped M2 branes on $[\omega_2]\cap[\omega_2]\simeq\mathbb{T}^2$ both becoming light. However, this is not the case in the $\mathcal{R}_{12}$ growth sector (where $\alpha\to 0$), where the whole Calabi-Yau $X$ decompactifies homogeneously, $\mathcal{V}_X\sim \frac{3}{2}(s^1)^3$. There is no independent KK tower associated to $[\omega_1]$, and the $\vec{\zeta}_{\rm KK,[\omega_1]}$ vector slides to coincide with $\vec{\zeta}_{{\rm KK},X}=\left(\frac{1}{\sqrt{2}},\frac{1}{\sqrt{6}},0\right)$, since their KK scales are the same. On the other hand, there is no $\mathbb{T}^2$ fiber becoming small here, and thus there is no light tower of M-th+M2 bounded states. This causes the $\vec{\zeta}_{\rm KK,M-th+M2}$ vector to slide out to infinity, as it is no longer present in the $\mathcal{R}_{12}$ growth sector.

%Similar expressions, with the same dependence $\vec{\zeta}(\alpha)=\vec{\zeta}_\infty+(0,0,\hat{f}(\alpha))$ can be obtained for the rest of towers and mass scales of EFT strings that experience sliding, namely all those that have an explicit dependence on $s^2$ rather that the volume $V_X$ as a whole, so are present in only one of the two growth sectors. 

In the interface between the two growth sectors (which overlaps with $\mathcal{R}_{12}$ as the later has lower dimensionality than $\mathcal{R}_{21}$) all the sliding towers, are found to be at the species scale or lighter. However, they are all heavier than the leading tower, which does not slide.

In summary, since the leading towers when moving along the interface are located along the plane $s^0-\mathcal{V}_X$, they do not slide, and for instance the pattern relating the leading tower and the species scale \eqref{eq.PATTERN} still holds when moving along the interfaces (i.e. along irregular limits using the language of \cite{Etheredge:2024tok}). The subleading towers will, though, slide perpendicularly to the tangent vector of the trajectory, so the taxonomy rules in \eqref{e.taxonomy} between the leading and subleading towers will not be preserved.\\

 \begin{figure}[h] 	
 	\centering
			\includegraphics[width=0.5\textwidth]{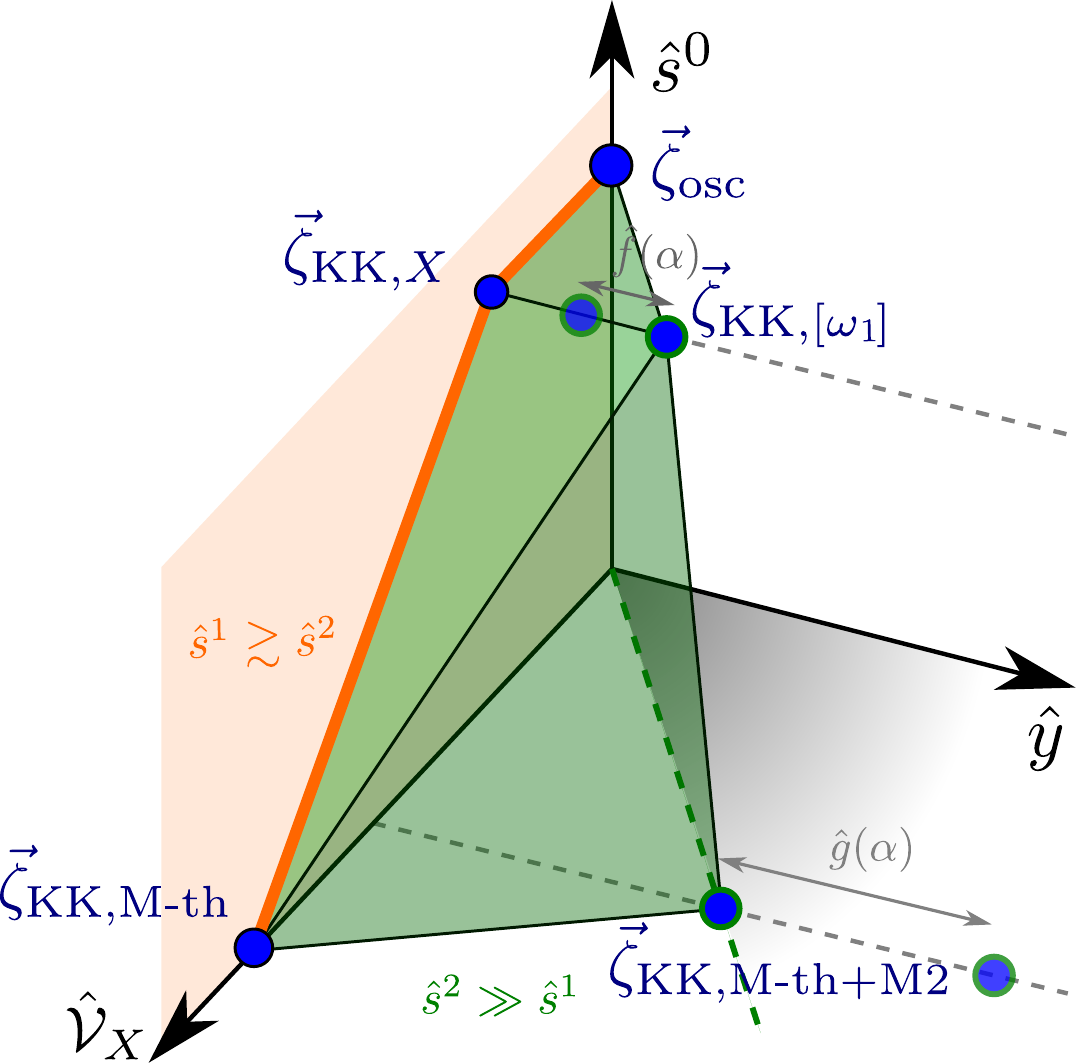}
		\caption{Sliding of the towers for $E_8\times E_8$ Heterotic string theory compactified on $\mathbb{P}^{(1,1,1,6,9)}[18]$, in terms of the canonically normalized coordinates, which in the saxionic plane are given by the volume $\hat{\mathcal{V}}_X$ and an additional direction $\hat{y}$ accounting for the contribution of $s^2$ to $V_X$. The growth sectors $s^1\gg s^2$ (in orange) and $\hat{s}^2\gg s^2$ (in green) are depicted. The towers in the orange plane do not slide, while those in the right do slide according to \eqref{eq.sliding}, i.e., perpendicular to the infinite distance trajectory (which is parallel to the orange plane).  The region shaded in gray is not part of the K\"ahler cone in the $\{\hat{s}^0,\hat{\mathcal{V}}_X,\hat{y}\}$ coordinates.\label{f.EXEX}}
\end{figure}

Something similar happens to the EFT strings, with 
\begin{equation}
\vec{\zeta}_{\mathcal{T}_{e^1}}=\left(0,\frac{1}{\sqrt{6}},-2\hat{f}(\alpha)\right)\;, \vec{\zeta}_{\mathcal{T}_{e^2}}=\left(0,\frac{1}{\sqrt{6}},\frac{1}{6\hat{g}(\alpha)}\right)\;,
\end{equation}
with $\hat{f}(\alpha)$ and $\hat{g}(\alpha)$ as defined in \eqref{eq.sliding}. It is interesting to notice that in this case the $\zeta$-vectors of EFT strings remain finite, interpolating between the arrangement for $\mathcal{V}_X\sim (s^1)^3$ ($\alpha\to 0$, with both vectors coinciding) and $\mathcal{V}_X\sim s^1(s^2)$ ($\alpha\to\infty$), described in the previous sections. Since the EFT string flows occur in the deep regime of the two growth sectors, for such trajectories there is no sliding, and the Integer Scaling Conjecture is satisfied.
 
The features of the sliding observed in these examples match with those already described in \cite{Etheredge:2023odp,Etheredge:2024tok}, in the sense that the sliding is always perpendicular to the direction of the trajectory. The sliding observed here is, though, milder in the sense that it only affects subleading (and not the leading) towers. It would be interesting to perform a more systematic analysis of the sliding in generic volume forms $\mathcal{V}_X=\frac{1}{6}\kappa_{abc} s^a s^b s^c$, to understand all the possible ways in which the $\zeta$-vectors can move.

\section{Other string theory constructions\label{s.more topdown}}
After the detailed study of top down constructions of 4d $\mathcal{N}=1$ effective theories from compactifications of heterotic $E_8\times E_8$ string theory on Calabi-Yau threefolds in Section \ref{s.het}, we will study additional top-down string theory constructions discussed in \cite{Lanza:2021udy}. We will see that, in certain cases, the arrangement of towers and duality frames is completely analogous to that of heterotic $E_8\times E_8$, so we will not enter into such detail, and focus mainly on the polytopes that are different. The case of 4d compactifications of M-theory is left for Section \ref{s.Mth}.

\subsection{Type IIA on CY orientifold\label{sIIA}}
Consider first type IIA string theory compactified on a Calabi-Yau orientifold resulting from the involution of $X$ under the O6 action $\mathfrak{i}:X\to X$, acting on the (string frame) K\"ahler form $J$ and (3,0)-form $\Omega$ by $\mathfrak{i}^\ast J=-J$ and $\mathfrak{i}^\ast\Omega=\bar{\Omega}$, with the O6-planes calibrated under ${\rm Re}\Omega$. The complexified version of these forms can be expanded in terms of odd and even cohomology bases $\{[D_a^+]\}_{a=1}^{b_2(X)_-}$ and $\{[\Sigma_\alpha^-]\}_{\alpha=1}^{b_3(X)_+}$ (recall that Poincar\'e duality pairs even/odd cycles with odd/even cohomology classes, given that $\mathfrak{i}$ inverts orientation), thus introducing two sets of chiral fields
\begin{subequations}
	\begin{align}
		B_2+iJ&=t^b[D^+_b]=(a^b+is^b)[D^+_b]\in H^2(X;\mathbb{R})_-\\
		C_3+ie^{-\Phi}{\rm Re}\Omega&=\hat{t}^\alpha[\Sigma^-_\alpha]=(\hat{a}^\alpha+i\hat{s}^\alpha)[\Sigma_\alpha^-]\in H^3(X;\mathbb{R})_+\;,
	\end{align}
\end{subequations}
where $\Phi$ is the 10d dilaton. The K\"ahler $s^b$ and complex structure $\hat s^\alpha$ saxions take values on a factorized saxionic cone $\Delta=\Delta_{\rm K}\times \hat{\Delta}$, independently generated. As in Section \ref{s.het}, we will focus on large volume/weak coupling perturbative regimes. The string frame volume $\mathcal{V}_X$ and Hitchin function $\mathcal{H}$ are given by
\begin{subequations}
	\begin{align}
	\mathcal{V}_X&=\frac{1}{3!}\int_XJ\wedge J\wedge J=\frac{1}{3}\kappa_{abc}s^as^bs^c\\
	\mathcal{H}&=\frac{i}{8}\int_Xe^{-2\Phi}\Omega\wedge\bar{\Omega}\;,
	\end{align}
\end{subequations}
where $\kappa_{abc}$ are the $X$ intersection numbers and $\mathcal{V}_X$ and $\mathcal{H}$ only depend on $s^a$ and $\hat{s}^\alpha$, respectively. Using the normalization
\begin{equation}
	\frac{1}{3!}J\wedge J\wedge J=\frac{i}{8}\Omega\wedge\bar{\Omega}\;,
\end{equation}
we obtain\footnote{Since we are working in the small coupling/large volume perturbative regime, internal warping \cite{Giddings:2001yu} and backreaction can be neglected, see \cite{Shiu:2008ry,Frey:2008xw,Martucci:2009sf,Martucci:2014ska,Martucci:2016pzt} for more details on how they could be incorporated.}
\begin{equation}
	e^{2\Phi}=\frac{\mathcal{V}_X}{\mathcal{H}}\;.
\end{equation} 
The 10d string frame metric and reduction to 4d Einstein frame are equivalent to \eqref{e. metric het} and \eqref{4dframe}, upon replacing $s^0\to \mathcal{H}$.

Since the complex structure and K\"ahler saxions are factorized, in analogy with the $E_8\times E_8$ heterotic case, the role of the universal saxion will be taken by the Hitchin function $\mathcal{H}>0$, in such a way that we can take $\Delta\equiv \mathcal{K}(X)\times \mathbb{R}_{>0}$, and the EFT K\"ahler potential is given by \cite{Grimm:2004ua}
\begin{equation}\label{e. kahler IIA}
	K=-2\log\mathcal{H}-\log\mathcal{V}_X\;.
\end{equation}
The moduli space metric $\mathsf{G}_{ij}$ is computed as in \eqref{e.Kahler metric}. Again, we will restrict our analysis to specific growth sectors where $\mathcal{V}_X$ can be approximated by its leading term, so $\mathsf{G}_{ij}$ is diagonal up to subleading terms.

Similar to the heterotic $E_8\times E_8$ case, we will find the following types of EFT strings:
\begin{itemize}
	\item\textbf{Fundamental type IIA string}: The tensions of the fundamental string in 4d Planck units is given by
	\begin{equation}
		\mathcal{T}_{\rm F1}=\frac{2\pi e^{2A}}{\ell_{\rm s}^2}=\frac{M_{\rm Pl,4}^2}{2\mathcal{H}}=M_{\rm Pl,4}^2e^0 l_0,
	\end{equation}		
	with $l_0=\mathcal{H}^{-1}$ being the dual saxion associated to the Hitchin function in \eqref{e. kahler IIA}. As expected, this expression is completely analogous to \eqref{e.m osc het} in heterotic $E_8\times E_8$ constructions. 
	\item\textbf{NS5-branes wrapping 4-cycles}: By wrapping NS5-branes along \emph{even} (nef and effective, see \cite{Katz:2020ewz}) divisors $D^+=e^aD^+_a$, an EFT string with tension 
	\begin{equation}
	\mathcal{T}_{{\rm NS5},\mathbf{e}}=\frac{\pi e^{2A}}{2\ell_{\rm s}^2}\int_{D^+}e^{-2\Phi}J\wedge J=M_{\rm Pl,4}^2e^al_a\;,
\end{equation}
arises, again analogous to \eqref{TNS5}, with the dual saxions, measuring the relative volume of the $D^+_a$ divisors with respect to the overall $X$, defined in the usual way.	
\end{itemize}
Associated to  the complex structure sector (in addition to the $\mathcal{H}$ direction), there will be additional EFT strings coming from D4-branes wrapping SLag 3-cycles (calibrated by $e^{-\Phi}{\rm Im}\Omega$), with analogous expression for the tension in terms of the dual saxions $\hat{l}_\alpha=\frac{1}{2}\partial_{\hat{s}^\alpha}K$, \cite{Lanza:2021udy}. Since the understanding of the EFT strings and light towers in this sector is more limited than in the K\"ahler sector, we only focus on the latter in this paper.

\vspace{0.25cm}
We are now ready to study the different towers, strings, and species polytopes. However, the arrangement of the towers and dualities (as well as the integer scaling of the towers with respect to the EFT strings), is almost translated \emph{verbatim} from the $E_8\times E_8$ heterotic case in Section \ref{s.het}, with the only difference that now in the $\mathcal{H}\to\infty$ direction the fundamental strings are type IIA, and that in the M-theory limits the growing direction is a $\mathbb{S}^1$ circle rather than the Ho\v{r}ava-Witten interval. Therefore, all the figures will be equivalent and we can borrow the results for the integer scaling in  the tables of the previous section. Moreover, the possible obstruction arising at strong coupling limits in heterotic $E_8\times E_8$ compactifications discussed in Section \ref{sec.no homo} does not obviously occur here, so the entire saxionic cone can in principle be covered. For the dual theories emerging in the infinite distance limits,  the pertinent manifold can be obtained from our Calabi-Yau orientifold through an analogous analysis to that in Section \ref{s.het}.

\subsection{Heterotic $SO(32)$/type I on CY threefold\label{s.SO32}}
We now turn to heterotic $SO(32)$ string theory compactifications on a Calabi-Yau threefold. As we will see, since the strong coupling limit differs from the M-theory decompactification of the $E_8\times E_8$ theory, the tower and duality arrangement will differ from Section \ref{s.het}.

The saxionic cone for this 4d $\mathcal{N}=1$ theory is the same as in Section \ref{s.het}, again parameterized by the universal and K\"ahler saxions, and with identical expressions for the moduli space metric and volumes of cycles and divisors. Furthermore, the EFT strings will again be obtained from either the fundamental heterotic string or NS5 branes wrapping effective divisors, with tensions given by \eqref{e.m osc het} and \eqref{TNS5} and, therefore, with identical $\zeta$-vectors.

\vspace{0.25cm}

Regarding the towers of light states, we find the following options for this perturbative regime of $SO(32)$ heterotic string theory compactified on a Calabi-Yau $X$:
\begin{itemize}
	\item \textbf{Heterotic $SO(32)$ string and KK towers}: The perturbative description of heterotic $SO(32)$ CY-compactifications is analogous to that of $E_8\times E_8$, and thus the fundamental string and  KK towers will have the same expressions as in \eqref{eq.mosc} and \eqref{eq.KK het}, with the same $\zeta$-vectors from Section \ref{s.het}. The vector of the fundamental heterotic string will be simply labeled by $\vec\zeta_{\rm osc}$.
	\item\textbf{Type I string}: For $s^0\ll \mathcal{V}_{X}$ we go to strong coupling, with the 10d heterotic dilaton diverging, $e^{\Phi/4}=\left(\mathcal{V}_X/s^0\right)^{1/2}\to\infty$. Unlike the $E_8\times E_8$ case, it is well known \cite{Witten:1995ex}, that Heterotic $SO(32)$ string theory is S-dual to type I string theory on $X$, with
	\begin{equation}\label{eq. type I string}
	\ell_{\rm s, I}=\ell_{\rm s,het}\sqrt{g_{\rm s,het}}=\ell_{\rm s,het}e^{\Phi/2}\Longrightarrow\frac{m_{\rm osc,I}}{M_{\rm Pl,4}}=(s^0)^{-1/4}\mathcal{V}_X^{-1/4}\;,
\end{equation}
	which in the basis $\{s^0,\mathcal{V}_X\}$ results in
	\begin{equation}
		\vec{\zeta}_{\rm osc,I}=\left(\frac{1}{2\sqrt{2}},\frac{1}{2}\sqrt{\frac{3}{2}}\right)\;.
	\end{equation}
\end{itemize}
Type I string theory can be interpreted as a type IIB orientifold (with O9-planes and D9-branes). In certain toroidal-like cases, when the volume of some even cycle $\mathcal{C}$ becomes small in comparison to the type I string scale $\ell_{\rm s,I}$, one can perform an \emph{even} number of T-dualities along the shrinking submanifold $\mathcal{C}$, resulting in a new description in a type IIB CY compactification with O-planes of different dimensionality. Since the string scale is not affected by T-dualities, under such transformation $\mathcal{V}_{\mathcal{C}}\to\frac{1}{\mathcal{V}_{\mathcal{C}}}$, the $\zeta$-vector associated with the new cycle $\mathcal{C}'$ would be given by the reflection of $\vec{\zeta}_{\rm KK,\mathcal{C}}$ with respect to $\vec{\zeta}_{\rm osc,I}$, i.e.,
\begin{equation}
	\vec{\zeta}_{\rm KK,\mathcal{C}'}=2\vec{\zeta}_{\rm osc,I}-\vec{\zeta}_{\rm KK,\mathcal{C}}\;,
\end{equation}
This can result in additional KK towers in the saxionic cone.\\

The integer scaling of the (sub)leading towers along the different string flows is shown in Tables \ref{tab.w type I s3} and \ref{tab.w typeI s2s} for different expressions of the overall volume, $\mathcal{V}_X\sim (s^1)^3$ and $(s^1)^2s^2$, respectively. As it can be seen in both tables, the integer scaling does not generally hold for the type I string (with vector $\vec{\zeta}_{\rm osc,I}$). This is not in contradiction with the integer scaling \ref{lattice}, since we only expect it to hold for the $\zeta$-vectors generating the tower polytope inside the saxionic cone. Unlike the heterotic $SO(32)$ string, which has $\vec{\zeta}_{\rm osc}$ located at the boundary of the saxionic cone, and thus acts as generator of the tower polytope (respecting \ref{lattice}), $\vec{\zeta}_{\rm osc,I}$ is located in the middle of a polytope facet, and is not needed to generate the $\zeta$-vector convex hull. Only if we move along a trajectory which is parallel to the direction set by $\vec{\zeta}_{\rm osc,I}$, the string tower is co-leading with the KK towers located at the vertices of its facet, thus scaling with the same exponential rate. In that particular limit, the integer scaling is then also fulfilled for the Type I string tower, since the $\zeta$-vectors of the KK vertices fulfill \eqref{lattice} for every EFT string limit. 

In Table \ref{flowsTypeI} we summarize the microscopic identification of the (sub)leading towers, the integer scaling of the leading one, the species scale and the dual theory emerging asymptotically along the different EFT string flows for the case $\mathcal{V}_X\sim (s^1)^2s^2$. Analogously, one can obtain the limits for $\mathcal{V}_X\sim s^1s^2s^3$, while the case $\mathcal{V}_X\sim (s^1)^3$ only involves limits that are equivalent to the first and last row of Table \ref{flowsTypeI}.  

\begin{figure}[htp!]
\begin{center}
\begin{subfigure}[b]{0.49\textwidth}
\captionsetup{width=.95\linewidth}
\center
\includegraphics[width=\textwidth]{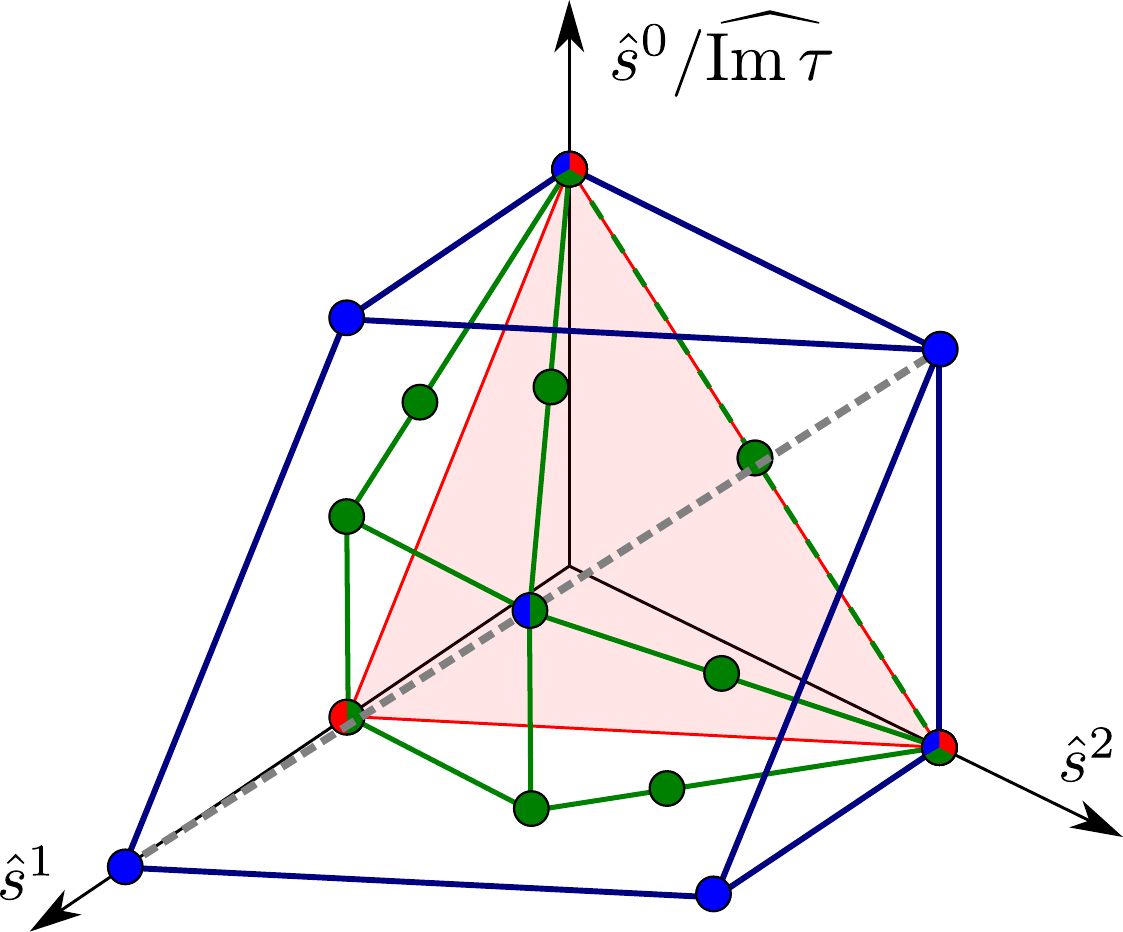}
\caption{\hspace{-0.3em} Arrangement of $\zeta$-vectors for towers of states (in blue), elementary EFT strings (red) and $\mathcal{Z}$-vectors for species scale (green).} \label{f.towers2-1BIS}
\end{subfigure}
\hfill
\begin{subfigure}[b]{0.45\textwidth}
\center
\includegraphics[width=\textwidth]{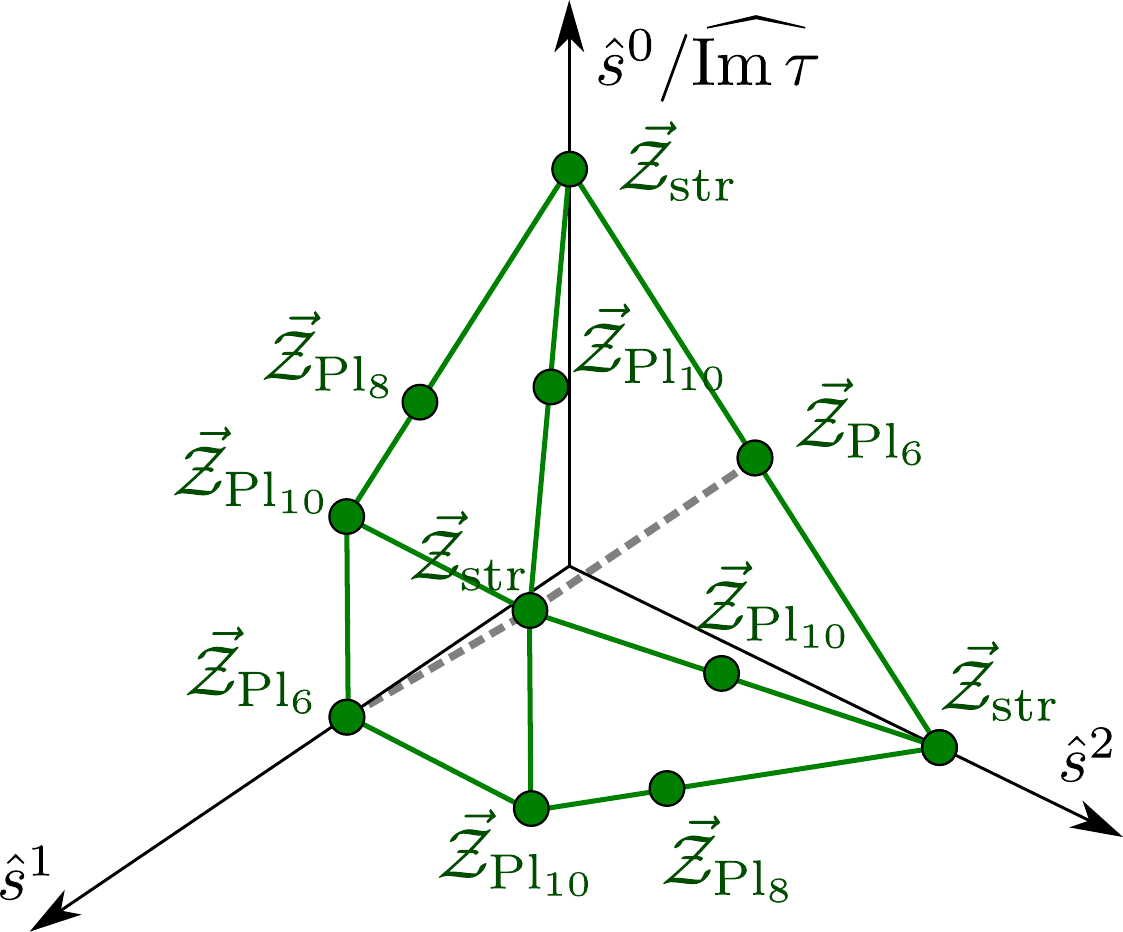}
\caption{\hspace{-0.3em} $\mathcal{Z}$-vectors for the asymptotic behavior of the specie scale along different directions.} \label{f.towers2-3BIS}
\begin{minipage}{.1cm}
            \vfill
            \end{minipage}
\end{subfigure}
\caption{Convex hull of towers (blue), possible species scales (green) and EFT strings (red), for heterotic $SO(32)$/type I on a Calabi-Yau 3-fold with $\mathcal{V}_X\sim (s^1)^2s^2$ or type IIB on Calabi-Yau orientifold with $V_X\sim v_1^2v_2$, in terms of the canonically normalized saxions $\hat{s}^0$, $\hat{s}^1$ and $\hat{s}^2$. By drawing the full polytope, we are assuming that we can perform the required T-dualities to transition through the diverse duality frames completing the saxion cone. A perturbative description is known for heterotic $SO(32)$/type I and type IIB/F-theory for the upper and lower part of the dashed gray line, respectively. The hierarchy $m_\ast\leq \Lambda_{\rm QG}\leq \sqrt{\mathcal{T}}$ is preserved. The microscopic interpretation of the different towers is depicted in Figures \ref{f.typeI2} and \ref{f.II2B} (respectively), while the EFT strings are the same as in Figure \ref{f.towerse2-4}.}
			\label{f.towers2BIS}
	\end{center}
\end{figure}

\begin{figure}[h!]
\begin{center}
\begin{subfigure}[b]{0.39\textwidth}
\captionsetup{width=.95\linewidth}
\center
\includegraphics[width=\textwidth]{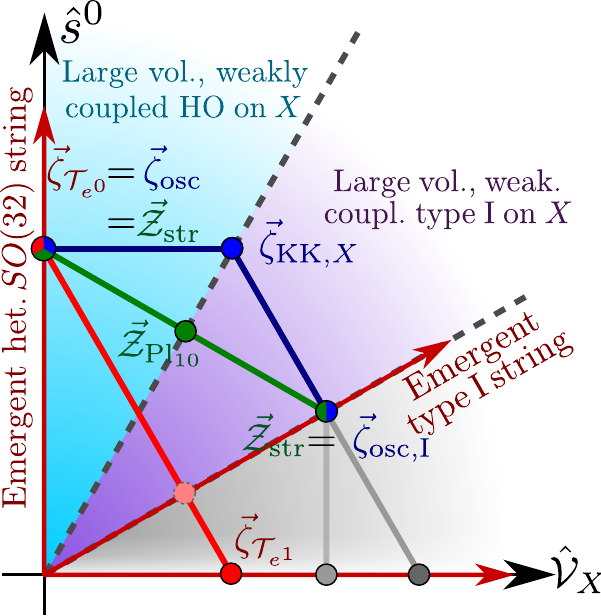}
\caption{\hspace{-0.3em} Convex hull of $\zeta$-vectors for towers (blue) and EFT strings (red), as well as $\mathcal{Z}$-vectors for  the species scale (green) and the corresponding duality frames, in terms of $s^0$ and overall volume $\mathcal{V}_X$.} \label{f.typeI1}
\end{subfigure}\begin{subfigure}[b]{0.6\textwidth}
\captionsetup{width=.95\linewidth}
\center
\includegraphics[width=\textwidth]{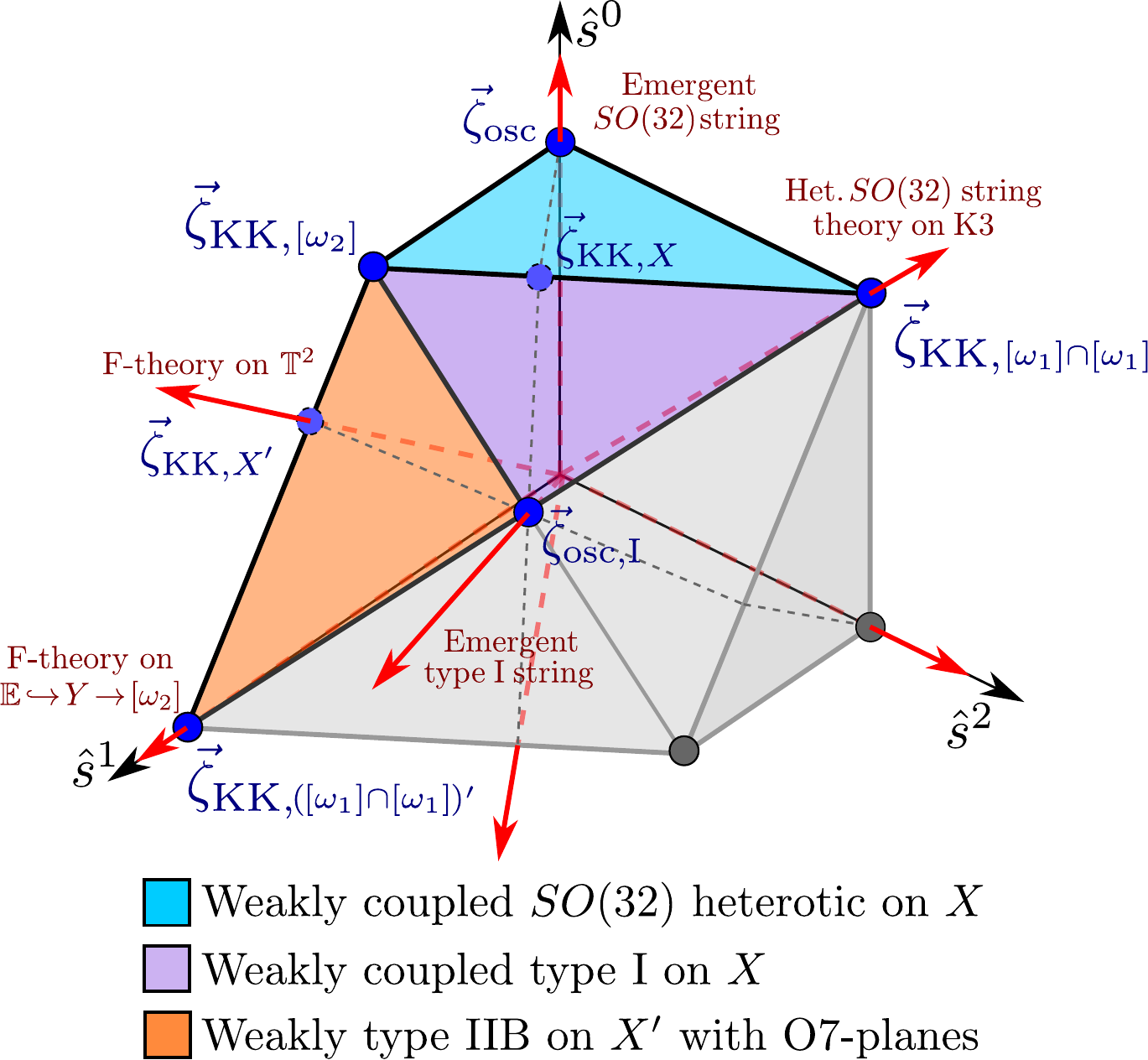}
\caption{\hspace{-0.32em} Arrangement of $\zeta$-vectors and duality frames for $\mathcal{V}_X\sim(s^1)^2s^2$.} \label{f.typeI2}
\end{subfigure}
\caption{Arrangement of duality frames and towers for heterotic $SO(32)$/type I compactifications on Calabi-Yau 3-folds with overall volume $\mathcal{V}_X$ (\ref{f.typeI1}) and $\mathcal{V}_X\sim (s^1)^2s^2$ (\ref{f.typeI2}), in terms of canonically normalized moduli. Depicted in gray are the regions which would result through a series of dualities assuming a toroidal-like fibration structure (but not necessarily in a general CY). Just for reference, the $\zeta$-vectors of the hypothetical towers also appear in gray.  We show as red arrows the  EFT string flows and the dual asymptotic descriptions emerging at infinite distance.
\label{f.typeI}}
\end{center}
\end{figure}

The convex hulls of the $\zeta$-vectors for the towers of states, EFT strings and species scale are represented in Figure \ref{f.towers2BIS} for compactifications on a CY 3-fold with volume of the form $\mathcal{V}_X\sim (s^1)^2s^2$. 
In Figure \ref{f.typeI}, we show the different duality descriptions (in different colors) arising at the different regions of the perturbative moduli space, together with the dual theory emerging at infinite distance along the different string flows (red arrows) for the cases of $\mathcal{V}\sim (s^1)^3$ and $\mathcal{V}\sim (s^1)^2 s^2$.\footnote{Note that Figure \ref{f.typeI1} replicates the arrangement of towers and 1-branes (upon considering both the EFT strings and the non-BPS type I string) in Figure (4d) of \cite{Etheredge:2024amg}. In said reference the convex hulls of states were found for the 2-dimensional slice of 4d maximal supergravity moduli space, spanned by the 10d dilaton and overall volume, obtained from compactification of type IIB on $\mathbb{T}^6$. Figure \ref{f.IIB 1}, corresponding to a type IIB orientifold compactification, has an analogous tower and 1-brane arrangement, replicating the figure of \cite{Etheredge:2024amg}. Note that the amount of SUSY is lowered when going to 4d $\mathcal{N}=1$, and thus some branes and towers are no longer BPS.\label{fn.braneTax}}
\vspace{0.25cm}

\begin{table}[h]
	\centering
	\begin{tabular}{|c|ccc|}
	\hline
	$\mathbf{e}$ &$\vec{\zeta}_{\rm osc}$&$\vec{\zeta}_{{\rm KK},X}$&$\vec{\zeta}_{\rm osc,I}$\\\hline
	$(e^0,0)$&\textbf{1}&\textbf{1}&$\color{gray}\frac{1}{2}$\\
	$(e^0,e^1)$&1&\textbf{2}&\textbf{2}\\\hline	
	\end{tabular}
	\caption{Scaling weight $w$ for the different EFT strings $\mathcal{T}_{\mathbf{e}}$ and tower $\vec{\zeta}$, such that $\vec{\zeta}_{\mathcal{T}_{\mathbf{e}}} \cdot \vec{\zeta} = w\abs{\vec{\zeta}_{\mathcal{T}_{\mathbf{e}}}}^2$, with the leading tower(s) highlighted for each EFT string limit in a 4d $\mathcal{N}=1$ compactification obtained from $SO(32)$ heterotic theory on a CY 3-fold with $\mathcal{V}_X\sim (s^1)^3$. Only those EFT string flow moving in a known perturbative description are considered.\label{tab.w type I s3}}
\end{table}

\begin{table}[h]
	\centering
	\begin{tabular}{|c|ccccc|}
	\hline
	$\mathbf{e}$ &$\vec{\zeta}_{\rm osc}$&$\vec{\zeta}_{{\rm KK},[\omega_2]}$&$\vec{\zeta}_{{\rm KK},[\omega_1]\cap[\omega_1]}$&$\vec{\zeta}_{{\rm KK},([\omega_1]\cap[\omega_1])'}$&$\vec{\zeta}_{\rm osc,I}$\\\hline
	$(e^0,0,0)$&\textbf{1}	&\textbf{1}	&\textbf{1}	&0	&$\color{gray}\frac{1}{2}$	\\
	$(0,e^1,0)$&0	&1	&0	&2	&1	\\
	$(e^0,e^1,0)$&1	&\textbf{2}	&1	&\textbf{2}	&$\color{gray}\frac{3}{2}$	\\
	$(e^0,0,e^2)$&1	&1	&\textbf{2}	&0	&1	\\
	$(e^0,e^1,e^2)$&1	&\textbf{2}	&\textbf{2}	&\textbf{2}	&\textbf{2}	\\
	\hline\end{tabular}
	\caption{Integer scaling weight $w$ of the different towers of states along each EFT string limit, for $SO(32)$ heterotic theory on a CY 3-fold with $\mathcal{V}_X\sim (s^1)^2s^2$.  Each tower with vector $\vec{\zeta}$ satisfies  $\vec{\zeta}_{\mathcal{T}_{\mathbf{e}}} \cdot \vec{\zeta} = w\  \abs{\vec{\zeta}_{\mathcal{T}_{\mathbf{e}}}}^2$ along the flow driven by an EFT string with vector $\vec{\zeta}_{\mathcal{T}_{\mathbf{e}}}$ and charge ${\mathbf{e}}$. The leading tower(s) for each EFT string limit are highlighted in bold face.  \label{tab.w typeI s2s}}
\end{table}

\begin{table}
	\begin{center}
	\resizebox{\textwidth}{!}{
			\begin{tabular}{|c|c|c|c|c|c|}
			\hline
			$\mathbf{e}$ &Leading tower(s) $m_*$ & $w$ & $\Lambda_{\text{QG}}$ & Sublead. towers & Emergent dual theory \\
			\hline 
			\rowcolor{blue!10!}$(e^0,0,0)$ &  \begin{tabular}{@{}c@{}}$m_{\text{osc}}$, $m_{\text{KK},[\omega_2]}$,\\ $m_{\text{KK},[\omega_1]\cap[\omega_1]},\;(m_{{\rm KK},X})$\end{tabular} & 1 & $m_{\rm str}$ & $\setminus$ & \begin{tabular}{@{}c@{}}Emergent $SO(32)$\\  heterotic string\end{tabular}\\
			\hline
			\rowcolor{green!10!}$(0,e^1,0)$ & $m_{\text{KK},([\omega_1]\cap[\omega_1])'}$ & 2 & $M_\text{Pl,6}$ & $m_{\text{KK}, [\omega_2]}$& \begin{tabular}{@{}c@{}}F-theory on\\$\mathbb{E}\hookrightarrow Y\to[\omega_2]$  \end{tabular}\\
			\hline
			\rowcolor{green!10!}$(e^0,e^1,0)$ &\begin{tabular}{@{}c@{}}$m_{\text{KK},[\omega_1]\cap[\omega_1]}$, $m_{\text{KK}, [\omega_2]}$\end{tabular}  & 2 & $M_{\text{Pl,10}}$& $m_{\text{KK},[\omega_1]\cap[\omega_1]}$, $m_{\text{osc}}$ & F-theory on $\mathbb{T}^2$\\
			\hline
			\rowcolor{green!10!}$(e^0,0,e^2)$ & $m_{\text{KK},[\omega_1]\cap[\omega_1]}$& 2& $M_{\rm Pl,6}$&  
\begin{tabular}{@{}c@{}}$m_\text{osc}$, $m_\text{osc,I}$,\\   $m_{{\rm KK},[\omega_2]}$\end{tabular}& \begin{tabular}{@{}c@{}}Heterotic $SO(32)$\\  string theory on K3\end{tabular} \\
			\hline
			\rowcolor{green!10!}$(e^0,e^1,e^2)$ & \begin{tabular}{@{}c@{}}$m_\text{osc,I}$,   $m_{\text{KK},[\omega_2]}$,\\ $m_{\text{KK},[\omega_1]\cap[\omega_1]}$, $m_{\text{KK},([\omega_1]\cap[\omega_1])'}$\end{tabular} & 2&$m_{\rm str}$ & $\setminus$  & \begin{tabular}{@{}c@{}}Emergent Type I string \end{tabular} \\
			\hline
		\end{tabular}}
		\caption{Description of the different EFT string limits driven by the string charges ${\mathbf{e}}$  for $SO(32)$ heterotic theory on a CY 3-fold with $\mathcal{V}_X\sim (s^1)^2s^2$. We provide the identification of the leading tower(s) $m_\ast$, their scaling weight $w$, the asymptotic species scale $\Lambda_{\rm QG}$, subleading towers with $m\lesssim \Lambda_{\rm QG}$ and the microscopic interpretation of the dual theory emerging at the infinite distance limit. \label{flowsTypeI}}
	\end{center}
	\end{table}

As in the previous cases, the saxionic cone gets divided in different asymptotic regions associated to different perturbative descriptions in 4d:
\begin{itemize}
	\item The region $s^0\gg \mathcal{V}_X$ is analogous to the $E_8\times E_8$ counterpart described in Section \ref{s.het}, with the $\Lambda_{\rm QG}$ given by the heterotic string scale and the theory described by \textbf{large volume, weakly coupled $SO(32)$ heterotic string theory compactified on the Calabi-Yau $X$}. The particular limits along interfaces of this duality frames are exactly analogous as in the $E_8\times E_8$ cases from Section \ref{s.het}.
	\item If $s^0\ll \mathcal{V}_X$, the heterotic $SO(32)$ 10d dilaton diverges and this region is better described by \textbf{type I string theory on the same Calabi-Yau $X$}, with the species scale given by the type I string scale. Note that in order for this new description to be valid, we need the different internal cycles of $X$ to be large in type I string units. From \eqref{eq. type I string}, this translates to
	\begin{equation}
		\mathcal{V}_{\mathcal{C}_n}\gg\mathcal{V}_X^{n/4}(s^0)^{-n/4}\,,
	\end{equation}
	where $\mathcal{V}_{\mathcal{C}_n}$ are the volume of the cycles $\mathcal{C}_n\subseteq X$ in \emph{heterotic} string units, as in \eqref{hV} and \eqref{e. volumes}. From the type I point of view,\footnote{Under S-duality we can map the heterotic K\"ahler form and universal saxion to type I variables, $e^{-\Phi}J=s^a[D_a]$ and $s^0=e^{-\Phi}\mathcal{V}_X=\frac{1}{3!}e^{2\Phi}\kappa_{abc}s^a s^b s^c$, giving an analogous description, see \cite{Lanza:2021udy}.} the fundamental heterotic string corresponds to a D1-string, and the EFT strings are given by wrapped D5-branes. Since the S-duality does not affect the geometry of the Calabi-Yau, the KK towers are the same as in the weakly-coupled/large volume heterotic description, with expressions \eqref{eq.KK het}.
	\item If $\mathcal{V}_X\sim (s^1)^2s^2$ (analogously for $\mathcal{V}_X\sim s^1s^2s^3$), in the limit $s^1\to\infty$ our compact manifold adopts a fibration structure $[\omega_1]\cap[\omega_2]\hookrightarrow X\to[\omega_1]$, with the volume of $[\omega_1]\cap[\omega_2]$ shrinking relatively to the overall $X$, as  the base $[\omega_1]$ grows with $s^1$. From the classification of asymptotic Calabi-Yau threefold fibrations \cite{Oguiso1993,Grimm:2019bey}, topologically the complex dimension-one fiber must be a 2-torus, $[\omega_1]\cap[\omega_2]\simeq \mathbb{T}^2$. This allows us to bypass the small volume limit 
	\begin{equation}
	s^2\sim\mathcal{V}_{[\omega_1]\cap[\omega_2]}\ll\mathcal{V}_X^{1/2}(s^0)^{-1/2}\sim (s^0)^{-1/2}s^1(s^2)^{1/2},
	\end{equation}
	by performing a double T-duality on the $\mathbb{T}^2$ fiber, which allows us to describe the new region as \textbf{large volume, weakly coupled type IIB on a new Calabi-Yau $X'$ with O7-planes}, due to the effect of T-dualities on boundary conditions. Since the string scale is not affected by T-dualities, in this new duality frame the species scale is still given by the type I string.
	
	As can be seen in Figure \ref{f.towers2-3BIS}, along asymptotic limits of this duality frame with $s^2$ fixed (or limits with fixed $s^a$ for the case $\mathcal{V}_X\sim s^1s^2s^3$), the Type I string scale $m_{\rm str}$ becomes light at the same rate as $M_{\rm Pl,6}$ (for $s^1\to\infty$ and fixed $s^0$), $M_{\rm Pl,8}$ (for $s^1\sim \sqrt{s^0}\to\infty$) or $M_{\rm Pl,10}$ (for any case in between). These mass scales set the species scale along those limits. This implies that the emergent dual theory at infinite distance becomes, respectively, F-theory on $\mathbb{E}\hookrightarrow Y\to[\omega_2]$, F-theory on $\mathbb{E}\hookrightarrow Y'\to [\omega_1]\cap[\omega_1]$, and F-theory on $\mathbb{T}^2$, since in this last limit the higher-dimensional type IIB dilaton remains fixed, see Figure~\ref{f.typeI2}. Here $Y$ is an elliptically fibered 3-fold, with $[\omega_2]$ as base (we do not assume $Y$ is the same manifold as $X$, since the actual fibration details might not be the same in both cases) and $Y'$ is an elliptically fibered 2-fold with $[\omega_1]\cap[\omega_1]\simeq \mathbb{P}^1$ as base. From the duality between heterotic string theory on $\mathbb{T}^2$ and F-theory on K3 \cite{Vafa:1996xn}, we conclude that $Y'$ must be a K3 manifold.
\end{itemize}

Notice that, unlike in heterotic $E_8\times E_8$ and type IIA, the whole K\"ahler cone cannot be immediately covered by a web of dualities to different perturbative regimes. The reason is that a divisor gets of string size when crossing the interface $\mathcal{V}_{\mathcal{C}_n}\sim\mathcal{V}_X^{n/4}(s^0)^{-n/4}$ (this is, $s^0\sim s^a$ in the cases $\mathcal{V}_X\sim s^a(s^b)^2$ or $\mathcal{V}_X\sim s^as^bs^c$, with $a\neq b,c$), corresponding precisely to the region colored in gray in Figure \ref{f.typeI}.  To follow the fate of trajectories in that gray region, we would need to transition to a dual frame as the Type I description breaks down. If the compact space was toroidal-like (such as some  instances), we could simply perform 4 T-dualities and obtain a different Type I description. This will occur whenever the geometry the CY is such that it develops  a fibration $\mathbb{T}^4\hookrightarrow X\to\mathcal{B}$ as $s^0\ll s^2$, yielding indeed Type IIB on a CY with a different arrangement of O-planes. The towers if gray in Figure \ref{f.typeI} are the ones we would obtained in such toroidal-like fibration.
However, the dualities can be more complicated in a general CY, and the limits might be even obstructed by quantum corrections in certain cases (since we only have 4 supercharges) along the lines of \cite{Klaewer:2020lfg}. Hence, we leave a more detailed analysis of this region for future work.

For the sake of brevity, we do not include a table showing the integer scaling of the different towers and EFT string tensions as in Tables \ref{tab.w type I s3} and \ref{tab.w typeI s2s} for the case with $\mathcal{V}_X\sim s^1 s^2s^3$. One can check, though, that the integer scaling relation \eqref{lattice} also holds since limits with $s^a\to\infty$ and $s^as^b\to\infty$ are analogous to those with $s^2\to \infty$ and $(s^1)^2$ in Table \ref{tab.w typeI s2s}. This asymptotic form of the volume $\mathcal{V}_X\sim s^1 s^2s^3$ precisely suggests the toroidal fibration structure that in principle would allow us to perform T-dualities to cover the entire K\"ahler cones with different perturbative descriptions. In this case, the tower convex hull would have a $S_4$ symmetry acting on $\{s^0,s^1,s^2,s^3\}$.

\subsection{F-theory/Type IIB on CY orientifold\label{s.Fth}}

We finally consider Type IIB (or more generally, F-theory)  compactifications down to 4d $\mathcal{N}=1$. Consider then F-theory compactified on an elliptically fibered Calabi-Yau 4-fold $\mathbb{E}\hookrightarrow Y\to X$, with the base $X$ being a K\"ahler manifold. Starting from the 10d type IIB Einstein frame action, the dimensional reduction is performed akin to \eqref{e. metric het}, with
\begin{equation}
		\dd s_{10}^{\rm (E)2}=e^{2A}\dd s_4^2+\ell_{\rm s}^2\dd s_X^2\;,\quad\text{with }\;e^{2A}=\frac{\ell_{\rm s}{M_{\rm Pl,4}}^2}{4\pi V_X}\;,
\end{equation}
where now $V_X$ denotes the base volume in 10d Planck units. For more details on the 4d $\mathcal{N}=1$ EFT, see \cite{Denef:2008wq,Grimm:2010ks,Lanza:2021udy}. Furthermore, as it was already the case in the type IIA compactifications, we will not consider possible warping effects \cite{Giddings:2001yu,Shiu:2008ry,Frey:2008xw,Martucci:2009sf,Martucci:2014ska,Martucci:2016pzt} as we expect them to be diluted in the particular asymptotic limits under consideration.

As in the previous cases, let us focus on the K\"ahler sector $\mathcal{K}(X)$, whose moduli $\{v_a\}_{a=1}^{b_4(X)}$ now appear from the expansion of the (Einstein frame) K\"ahler form $J=v_a[D^a]$, with $\{D^a\}_{a=1}^{b_4(X)}$ a basis of divisors for $H_4(X;\mathbb{Z})$. The volume of said divisors can be parameterized through the EFT saxions
\begin{equation}\label{eq.saxion kahler}
	s^a=\frac{1}{2}\int_{D^a}J\wedge J=\frac{1}{2}\kappa^{abc}v_bv_c\;,\quad\text{for }\;a=1,\dots,b_4(X)\;,
\end{equation}
generating the \emph{saxionic cone} $\Delta$ \eqref{defDelta}. The above relation can be inverted so that the K\"ahler potential for chiral fields $\{t^b=a^b+i s^b\}_{b=1}^{b_4(X)}$ reads,
\begin{equation}
	K=-2\log\int_XJ\wedge J\wedge J=-2\log\left(\kappa^{abc}v_a v_b v_c\right)
\end{equation}
which is, in turn, a function of $\{s^a\}_{a=1}^{b_4(X)}$ via \eqref{eq.saxion kahler}. The K\"ahler base $X$ volume in 10d Planck units is given by
\begin{equation}
V_X=\frac{1}{3}\kappa^{abc}v_av_bv_c\;,
\end{equation}
as one would expect, so that up to a numerical constant which can be ignored in asymptotic limits, $K\sim-2\log V_X$. The moduli space metric for the K\"ahler sector is then obtained analogously to \eqref{e.Kahler metric}. 

For the basis of 2-cycles $C_a\in H_2(X;\mathbb{Z})$ dual to $\{D^a\}_{a=1}^{b_4(X)}$ (this is $C_a\cdot D^b=\delta_a^b$), one can compute the dual saxions
\begin{equation}
\label{dualsF}
	l_a=\frac{\int_{C_a}J}{2V_X}=\frac{3v_a}{\kappa^{ijk}v_iv_jv_k}\;,
\end{equation} 
giving the relative volume of $C_a$ with respect to $V_X$ in  $M_{\rm Pl,10}$ units. In \cite{Cota:2022yjw} a general study of EFT string limits, together with light towers and the species scale, was carried out for the K\"ahler space of the $X$ base. In this work, we will consider more general limits including the type IIB dilaton/complex structure of the fiber, which results in additional duality frames, EFT strings and light towers. Our results are compatible with \cite{Cota:2022yjw} when restricted to directions along the K\"ahler cone $\mathcal{K}(X)$.\\

The axio-dilaton $\tau=C_0+ie^{-\Phi}$ of the Type IIB description corresponds to the complex structure of the elliptical fiber of $\mathbb{E}\hookrightarrow Y\to X$. Since in the asymptotic regime the dilaton and above K\"ahler directions factorize, we can extend the 4d perturbative regime by considering a saxionic cone including both the K\"ahler and dilaton directions, namely $\mathcal{K}(X)\times\mathbb{R}_{>0}$. This is the 4d perturbative regime that we will be interested in from now on.
For future reference, the moduli space metric for the dilaton is given by  $\mathsf{G}_{{\rm Im\,}\tau{\rm Im\,}\tau}=\frac{1}{2({\rm Im\,}\tau)^2}$.\\

As discussed in more detail in \cite{Lanza:2021udy}, 4d EFT strings are obtained from two sources:
\begin{itemize}
	\item \textbf{D3-branes wrapping an effective curve $C=e^aC_a$}: Using $\mathcal{T}_{{\rm Dp}}=(2\pi)^{p}(\alpha')^{-\frac{p+1}{2}}e^{-\Phi}$, where $\Phi$ is the 10d type IIB dilaton, one obtains
	\begin{equation}
	\mathcal{T}_{{\rm D3},\mathbf{e}}=\frac{1}{2}M_{\rm Pl,10}^2\int_CJ=M_{\rm Pl,4}^2e^al_a\,,
	\end{equation}	
	for the string arising from wrapping D3's, with the dual saxions given in \eqref{dualsF}, as expected from the general expression \eqref{TEFT}. 
	The elementary EFT string vectors are given by
	\begin{subequations}
\begin{align}
		\text{For }V_X\sim v_1^3\sim (s^1)^{3/2} \ :\ & \ \vec{\zeta}_{\mathcal{T}_{e^1}}=\left(0,\frac{1}{\sqrt{6}}\right)\label{eq.VOlIIBEFT1}
	\\
		\text{For }V_X\sim v_1^2v_2\sim s^1\sqrt{s^2} \ :\ &\  \vec{\zeta}_{\mathcal{T}_{e^1}}=\left(0,\frac{1}{2},0\right) \ , \ \vec{\zeta}_{\mathcal{T}_{e^2}}=\left(0,0,\frac{1}{\sqrt{2}}\right)\label{eq.VOlIIBEFT2}
	\\
		\text{For }V_X\sim v_1v_2v_3\sim\sqrt{s^1 s^2 s^3} \ :\ &\left\{ \begin{array}{c}
		\vec{\zeta}_{\mathcal{T}_{e^1}}=\left(0,\frac{1}{\sqrt{2}},0,0\right)\,,\;\vec{\zeta}_{\mathcal{T}_{e^2}}=\left(0,0,\frac{1}{\sqrt{2}},0\right)\,,\\
		\vec{\zeta}_{\mathcal{T}_{e^3}}=\left(0,0,0,\frac{1}{\sqrt{2}}\right)
		\end{array}\right.
		\end{align}
\end{subequations}
in terms of the canonically normalized moduli $\{\widehat{{\rm Im}\,\tau},\hat{s}^i\}$. Note that several of them have the proper length of $\frac{1}{\sqrt{2}}$ to correspond to critical strings. This is because in those cases, the EFT string flow corresponds to a limit in which a curve $C_a$ shrinks to zero size with respect to the overall volume, so that the $J$-class limits of \cite{Lee:2019tst,Klaewer:2020lfg} are realized ($J$-class A for volume with leading term $V_X\sim\kappa^{112}v_1^2v_2$ and $J$-class B for $V_X\sim\kappa^{123}v_1v_2v_3$).
In those cases, the EFT string will be indeed dual to a heterotic or a type IIB critical string, depending on whether the $C_a$ fiber is rational ($C_a\simeq \mathbb{P}^1$) or elliptic ($C_a\simeq\mathbb{T}^2$). 

The other cases correspond to EFT string limits which are not of $J$-class; the volume of the (relatively) shrinking curve $C$ does not become small fast enough, so that the EFT string tension is heavier than some KK scale, and its oscillation modes do not correspond to light modes of the emergent dual theory at infinite distance. In the new dual frame, the EFT string ($\vec{\zeta}_{\mathcal{T}_{e^1}}$ in \eqref{eq.VOlIIBEFT1} and \eqref{eq.VOlIIBEFT2}) is interpreted as a D3 brane wrapped along a growing curve. For $V_X\sim v_1^3$, the D3 brane has constant tension in $M_{\rm Pl,10}$, but since it wraps the curve $C_1\subset X\simeq C_1\times C_1\times C_1$, the string tension $\mathcal{T}_{e^1}\sim M_{\rm Pl,10}^{2}V_{C_1}$ diverges in $ M_{\rm Pl,10}$ units. On the other hand, for $V_X\sim v_1^2v_2$, the D3 brane wraps $C_1\subset D^2\simeq C_1\times C_1$ (whose volume co-scales with $D^2$), so that the resulting string tension scales as $\mathcal{T}_{e^1}\sim M_{\rm Pl,10}^{2}V_{C_1}\sim M_{\rm Pl,6}^2$. See \cite{Cota:2022yjw} for a more detailed analysis on this type of EFT strings and the associated limits.

	\item \textbf{D7-brane wrapping the entire internal space $X$}:\footnote{As noted in  \cite{Collinucci:2008pf}, in this orientifold backgrounds D(-1)-branes contribute by  $e^{i\pi\tau}$ to the path-integral, so that the shift symmetry is $\tau\simeq\tau+2$ rather than $\tau\simeq \tau+1$. This means that in order to generate the monodromy, an \emph{even} number of D7-branes needs to be wrapped along the internal space. We refer to said paper for more details on this.} It gives rise to an EFT string with tension (assuming constant $\Phi$ over $X$)
		\begin{equation}
\mathcal{T}_{{\rm EFT,D7}}\sim M_{\rm Pl,4}^2e^{\Phi}=\frac{M_{\rm Pl,4}}{{\rm Im\,}\tau}\;,
\end{equation}
where  $\tau=C_0+ie^{-\Phi}$ is the IIB axio-dilaton. The $\zeta$-vector of this EFT string is given by $\vec{\zeta}_{\mathcal{T}_{\rm D7}}=(\frac{1}{\sqrt{2}},0,\dots,0)$ where the non-vanishing component is along the $\widehat{{\rm Im\,}\tau}$ direction (with the weak coupling limit being ${{\rm Im\,}\tau}\to\infty$).
\end{itemize}

The EFT string arrangement over the saxionic cone $\Delta$ is the same as in $\mathcal{V}_X\sim (s^1)^3$, $(s^1)^2s^2$ and $s^1s^2 s^3$ from Sections \ref{s.het} and \ref{s.SO32}, since they have equivalent K\"ahler potentials upon identifying ${\rm Im}\,\tau\equiv s^0$ and the previous $s^i$ saxions (measuring volumes in string units) with those given in \eqref{eq.saxion kahler} for the current set-up (measuring volumes in 10d Planck units).

 Notice that the fundamental type IIB string, which has a mass dependence
 \begin{equation}\label{eq. massIIB}
 \frac{m_{\rm osc}}{M_{\rm Pl,4}}\sim V_X^{-1/2}({\rm Im\,}\tau)^{-1/4}\;
 \end{equation}
 does not yield an EFT string since it is not BPS.
 The presence of O7-planes and D7 branes on $X$ allow this string to break, so it is analogous to the situation of the type I string in Section \ref{s.SO32}. \\

 The scaling weight of the different towers along the EFT string flows is shown in Tables \ref{tab.w type IIB s3} and \ref{tab.w IIB s2s} for the case of $V_X=v_1^3$ and $V_X=v_1^2v_2$, respectively. Notice that the KK towers (below the species scale) always satisfy the integer scaling relation as they are generators of the tower convex hull. However, the (non-BPS) Type IIB string lies in the middle of a facet of the tower convex hull and it is not a generator, so its scaling weight is not necessarily an integer in general. For instance, it has $w=\frac{3}{2}$ along the EFT string flow with $\mathbf{e}=(0,e^1)$ for $V_X\sim v_1^3$ (see Table \ref{tab.w type IIB s3}) or the flow $\mathbf{e}=(0,e^1,e^2)$ for $V_X\sim v_1^2v_2$ (see Table \ref{tab.w IIB s2s}). Nevertheless, the scaling weight of this Type IIB string is an integer whenever its tower of oscillator modes is the leading tower, which occurs for instance along the EFT string flow with $\mathbf{e}=(e^0,e^1,...,e^n)$, yielding $w=2$.\\
 
 \begin{table}[h]
	\centering
	\begin{tabular}{|c|cc|}
	\hline
	$\mathbf{e}$ &$\vec{\zeta}_{\rm osc}$&$\vec{\zeta}_{{\rm KK},X}$\\\hline
	$(0,e^1)$&$\color{gray}\nicefrac32$&\textbf{2}\\
	$(e^0,e^1)$&\textbf{2}&\textbf{2}\\\hline	
	\end{tabular}
	\caption{Integer scaling weight $w$ of the different towers of states along each EFT string limit, for type IIB string theory on a CY$_3$ orientifold with $V_X\sim v_1^3$.  Each tower with vector $\vec{\zeta}_*$ satisfies  $\vec{\zeta}_{\mathcal{T}_{\mathbf{e}}} \cdot \vec{\zeta}_* = w \  \abs{\vec{\zeta}_{\mathcal{T}_{\mathbf{e}}}}^2$ along the flow driven by an EFT string with vector $\vec{\zeta}_{\mathcal{T}_{\mathbf{e}}}$ and charge ${\mathbf{e}}$. The leading tower(s) for each EFT string limit are highlighted in bold face.\label{tab.w type IIB s3}}
\end{table}

\begin{table}[h]
	\centering
	\begin{tabular}{|c|ccccc|}
	\hline
	$\mathbf{e}$ &$\vec{\zeta}_{\rm osc}$&$\vec{\zeta}_{{\rm KK},D^2}$&$\vec{\zeta}_{{\rm KK},C_2}$&$\vec{\zeta}_{{\rm KK},\mathbb{T}^2}$&$\vec{\zeta}_{\rm D3}$\\\hline	
	$(0,e^1,0)$&1	&1	&\textbf{2}	&2	&0	\\
	$(0,0,e^2)$&$\color{gray} \nicefrac12$	&0	&\textbf{1}	&\textbf{1}	&\textbf{1}	\\	
	$(0,e^1,e^2)$&$\color{gray} \nicefrac32$	&\textbf{2}	&\textbf{2}	&	1&1	\\
	$(e^0,0,e^2)$&1	&1	&0	&\textbf{2}	&1	\\
	$(e^0,e^1,e^2)$&\textbf{2}	&\textbf{2}	&\textbf{2}	&\textbf{2}	&1	\\
	\hline\end{tabular}
	\caption{Integer scaling weight $w$ of the different towers of states along each EFT string limit, for type IIB string theory on a CY$_3$ orientifold with $V_X\sim v_1^2v_2$.  Each tower with vector $\vec{\zeta}$ satisfies  $\vec{\zeta}_{\mathcal{T}_{\mathbf{e}}} \cdot \vec{\zeta} = w \  \abs{\vec{\zeta}_{\mathcal{T}_{\mathbf{e}}}}^2$ along the flow driven by an EFT string with vector $\vec{\zeta}_{\mathcal{T}_{\mathbf{e}}}$ and charge ${\mathbf{e}}$. The leading tower(s) for each EFT string limit are highlighted in bold face.\label{tab.w IIB s2s}}
\end{table}

\begin{figure}[h]
\begin{center}
\begin{subfigure}[b]{0.45\textwidth}
\center
\includegraphics[width=\textwidth]{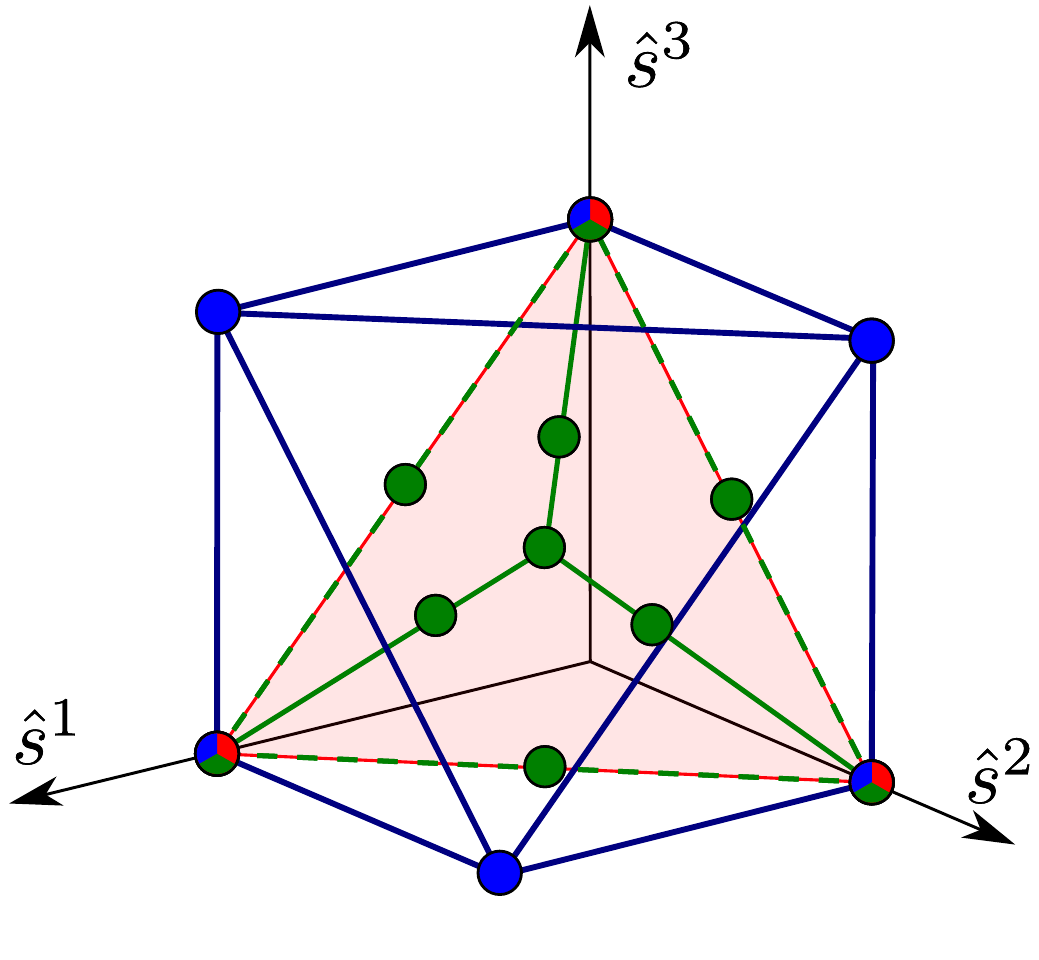}
\caption{\hspace{-0.3em} Arrangement of $\zeta$-vectors for towers (in blue), EFT strings (red) and $\mathcal{Z}$-vectors for species scale (green).} \label{f.towerse-1}
\end{subfigure}\begin{subfigure}[b]{0.45\textwidth}
\center
\includegraphics[width=\textwidth]{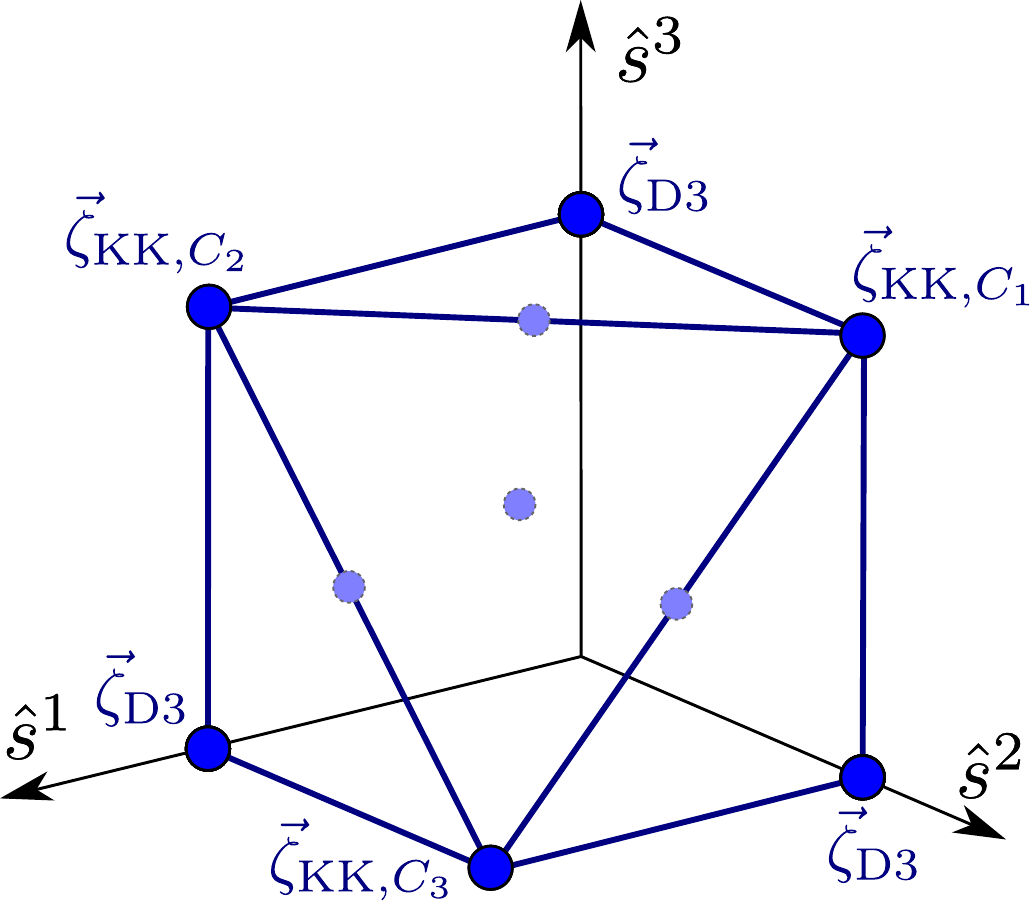}
\caption{\hspace{-0.32em} Convex hull of the towers of states.} \label{f.towerse-2}
\end{subfigure}
\hfill
\begin{subfigure}[b]{0.45\textwidth}
\center
\includegraphics[width=\textwidth]{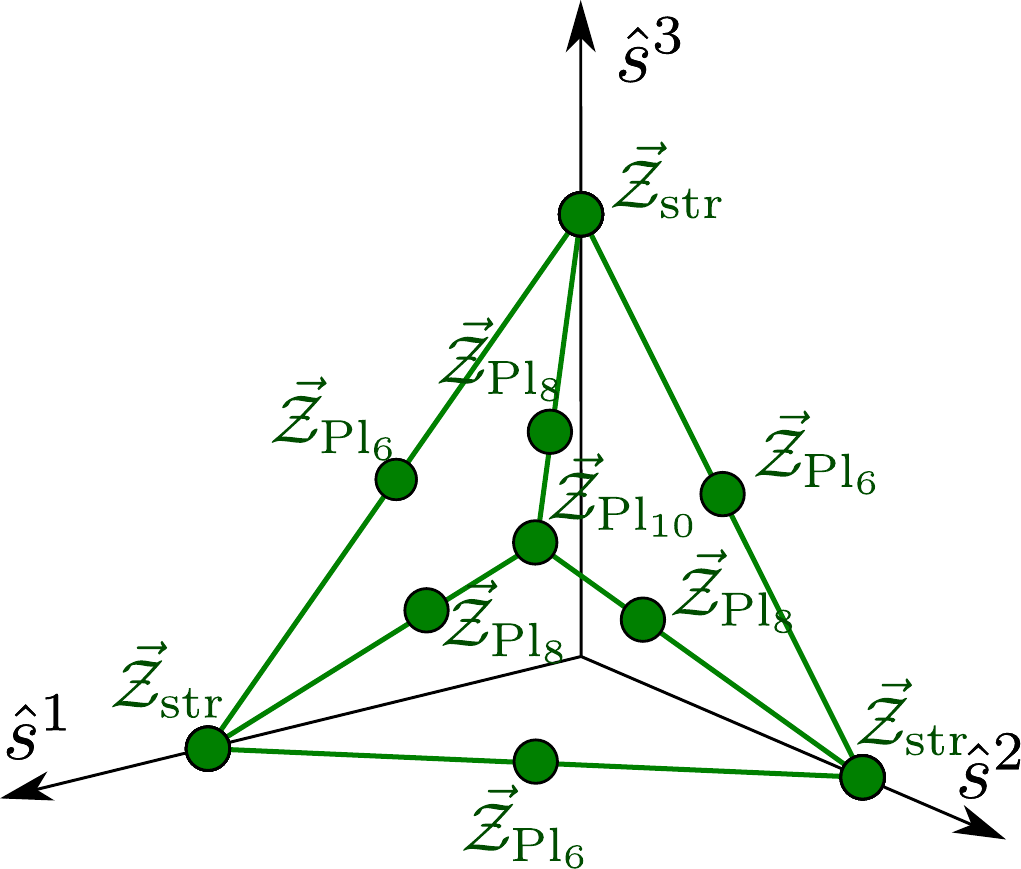}
\caption{\hspace{-0.3em} Asymptotic behavior of species scale.} \label{f.towerse-3}
\end{subfigure}\begin{subfigure}[b]{0.45\textwidth}
\center
\includegraphics[width=\textwidth]{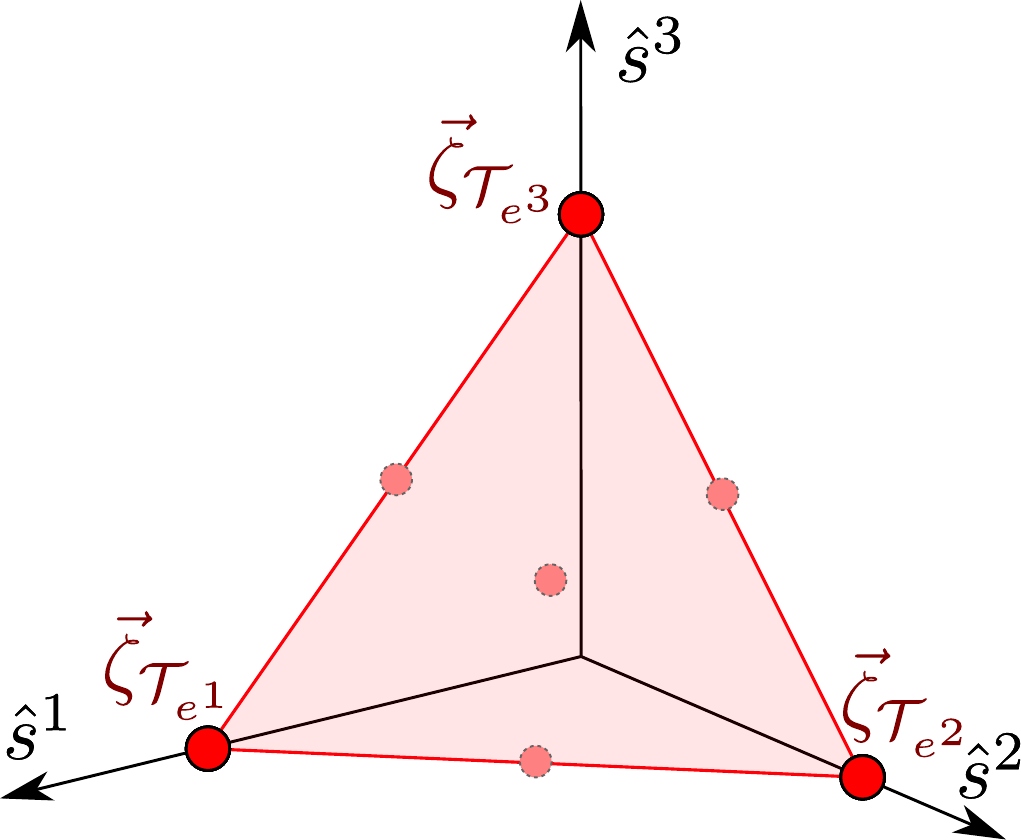}
\caption{\hspace{-0.3em} Convex hull of EFT strings.} \label{f.towerse-4}
\end{subfigure}
\caption{Arrangement of towers (blue), possible species scales (green) and EFT strings (red), for F-theory on a CY 4-fold $\mathbb{E}\hookrightarrow Y\to X$ with volume base $V_X\sim \sqrt{s^1s^2s^3}$ in 10d Planck units. For simplicity we do not label the vectors associated to bounded towers/non elementary EFT strings (only depicted along their EFT string flow). All quantities are depicted in terms of the canonically normalized saxions. The hierarchy $m_\ast\leq \Lambda_{\rm QG}\leq \sqrt{\mathcal{T}}$ is preserved.}
			\label{f.towers3}
	\end{center}
\end{figure}

We can now describe the different perturbative descriptions, or duality frames, emerging at the different regions of moduli space:
\begin{itemize}
	\item For ${\rm Im\,}\tau\gg 1$ and large internal cycles in string units, i.e., $V_{\mathcal{C}_n}\gg (\ell_{\rm s}M_{\rm Pl,10})^n\sim({\rm Im}\,\tau)^{n/4}$, the theory is described by \textbf{weakly coupled, large volume type IIB on orientifold $X$}. The species scale $\Lambda_{\rm QG}$ is given by the type IIB string scale given in \eqref{eq. massIIB}, and there are additional KK towers coming from the decompactification of the different cycles $\mathcal{C}_n\subseteq X$, with 
	\begin{equation}
		\frac{m_{{\rm KK},\mathcal{C}_n}}{M_{\rm Pl,4}}=V_{\mathcal{C}_n}^{-1/n}V_X^{-1/2}\;.
	\end{equation}
	The different $\zeta$-vectors generating the tower simplex are depicted in Table \ref{t.KK IIB} for volumes of the form $V_X\sim v_1^3$, $v_1^2v_2$ and $v_1v_2v_3$. These are depicted in Figures \ref{f.IIB 1}, \ref{f.II2B} and \ref{f.towerse-2} (for the later only the slice with fixed ${\rm Im}\,\tau$, since it is not possible to represent the 4-dimensional cone). 
	
	For a generic limit in this region, we decompactify to 10d Type IIB. However, asymptotic limits along the boundary of this duality frame can lead to different dual emergent theories:
	\begin{itemize}
		\item For overall decompactifications, where the base volume $X$ grows homogeneously, $V_X\to\infty$, while keeping ${\rm Im}\,\tau$ fixed, we simply decompactify to a 10d theory given by \textbf{F-theory on $\mathbb{T}^2$}.
		\item As commented before, for volumes with leading term $V_X\sim v_1^2v_2$, sending the associated saxion $s^1\to\infty$ does not result in a critical string (as the EFT string wrapping $C_1$ is too heavy), but rather a decompactification of the curve $C_2$ dual to the divisor $D^2$. This way, sending $s^1\to\infty$ while keeping $s^2$ and ${\rm Im}\,\tau$ fixed, results in the emergent dual theory given by \textbf{F-theory on an elliptically fibered 6-dimensional space $\mathbb{E}\hookrightarrow Y\to D^2$}.
	\end{itemize}

	\end{itemize}
	\begin{table}[hpt!]
    \centering
    \resizebox{\textwidth}{!}{
    \begin{tabular}{|c||c|c|c|}\hline
       $V_X$ &$v_1^3\sim(s^1)^{3/2}$&$v_1^2v_2\sim s^1\sqrt{s^2}$	&$v_1v_2v_3\sim \sqrt{s^1s^2s^3}$	\\\hline \hline
      EFT saxions& $s^1\sim v_1^2$	& $\begin{array}{c}
      s^1\sim v_1v_2,\,s^2\sim v_1^2
      \end{array}$	& $s^a=v_bv_c,\;a\neq b,c$	\\\hline
      Region described &${\rm Im\,}\tau\gg 0,\, s^1\gg{\rm Im\,}\tau$ & ${\rm Im\,}\tau\gg 0, s^1\gg{\rm Im\,}\tau,\, s^2\gg{\rm Im\,}\tau$& ${\rm Im\,}\tau\gg 0,\,s^as^b\gg s^c{\rm Im\,}\tau$ \\\hline
     \begin{tabular}{@{}c@{}} Towers $\zeta$-vectors\\generating\\ convex hull\end{tabular}&$\begin{array}{c}\vec{\zeta}_{\rm osc}=\left(\frac{1}{2\sqrt{2}},\frac{1}{2}\sqrt{\frac{3}{2}}\right)\\\vec{\zeta}_{{\rm KK},X}=\left(0,\sqrt{\tfrac{2}{3}}\right)\end{array}$	&$\begin{array}{c}\vec{\zeta}_{\rm osc}=\left(\frac{1}{2\sqrt{2}},\frac{1}{2},\frac{1}{2\sqrt{2}}\right)\\
      \vec{\zeta}_{{\rm KK},C_2}=(0,1,0)\\\vec{\zeta}_{{\rm KK},C_1}=\left(0,\frac{1}{2},\frac{1}{\sqrt{2}}\right)
      \end{array}$	&$\begin{array}{c}\vec{\zeta}_{\rm osc}=\left(\frac{1}{2\sqrt{2}},\frac{1}{2\sqrt{2}},\frac{1}{2\sqrt{2}},\frac{1}{2\sqrt{2}}\right)\\\left\{\vec{\zeta}_{{\rm KK},C_i}=\left(0,\tfrac{1-\delta^i_1}{\sqrt{2}},\tfrac{1-\delta^i_2}{\sqrt{2}},\tfrac{1-\delta^i_3}{\sqrt{2}}\right)\right\}_{i=1}^3\end{array}$	\\\hline
      \begin{tabular}{@{}c@{}} Divisors in\\terms of curves\end{tabular} &\begin{tabular}{@{}c@{}} No independent\\internal cycles.\end{tabular}	& $D^1\simeq C_1\times C_2,\,D^2\simeq C_1\times C_1$& $D^a\simeq C_b\times C_c,\, a\neq b,c$	\\\hline
    \end{tabular}}
    \caption{$\zeta$-vectors of the towers generating the convex hull (or frame simplex) associated to the perturbative regime described by weakly coupled, large volume type IIB on orientifold $X$. We show the result for different asymptotic behaviors of $V_X$, in terms of canonical coordinates $(\widehat{{\rm Im}\,\tau},\hat{s}^1,\dots,\hat{s}^n)$. Note that $\vec{\zeta}_{\rm osc}$ is always associated to the type IIB string. We also include the dictionary between the EFT saxions $s^a$ and K\"ahler moduli $v_b$, as well as the region of moduli space covered by this duality frame. In the last row, the relations between different internal cycles leading to possible KK bounded states are presented.}
    \label{t.KK IIB}
\end{table}

	The above region is only part of the entire saxionic cone, and is depicted in Figures \ref{f.IIB 2-IIB} and \ref{f.IIB 2-het} in light orange. When $V_{\mathcal{C}_n}\ll (\ell_{\rm s}M_{\rm Pl,10})^n\sim({\rm Im}\,\tau)^{n/4}$ (while keeping both large), some volume cycle becomes small in string units and the above Type IIB perturbative description breaks down, so a different duality frame  takes over. Let us analyze this depending on whether the cycle becoming of string size is a divisor or a curve of $X$. \\
	
	On one hand, if the shrinking cycle is a divisor, the chain of dualities is difficult to follow for a general Calabi-Yau fourfold, and the limit might be even obstructed by quantum corrections in certain cases or suffer from strong warping effects that do not get diluted asymptotically as in \cite{Etheredge:2023odp}. If the divisor is toroidal-like, we could perform 4 T-dualities to obtain the new duality description, and continue this way.
	Assuming this toroidal-like fibration structure, the arrangement of towers has been represented in Figures \ref{f.IIB 2-IIB} and \ref{f.IIB 2-het} in gray color, since it might be different in a general CY. Notice that the EFT string arising from a wrapping D7-brane points precisely in the direction of small overall volume (in string units), so that this flow can only be accessed by the new dual description obtained upon taking 6 T-dualities (assuming again that the original space is simply a toroidal orientifold with O7's). In that case, the result would be  Type IIB with O5's, and the D7 would become a D1, so the corresponding perturbative regime would be described by its S-dual theory, becoming analogous to the Type I setup of Section \ref{s.SO32}. It is not surprising then that the arrangement of towers in Figures  \ref{f.IIB 1} and \ref{f.II2B} in this section are equivalent to those of Type I (Figures \ref{f.typeI1} and \ref{f.typeI2}) in Section \ref{s.SO32}, with both duality frames being valid at complementary regions of the moduli space. The boundary between the two polytopes corresponds to the loci where a divisor $D_a$ or the whole $X$ become small with respect to the string scale. We leave for future work a more detailed analysis of whether these polytopes can be glued together somehow in specific compactifications (to provide a global description of the entire saxionic cone); as well as an analysis of the gray region in a general CY. \\

On the other hand, if an effective curve $C_a$ gets small in string units, we can more easily follow the chain of dualities, which will depend on the microscopic nature of the EFT string arising from a D3-brane wrapping such curve. As discussed in \cite{Lee:2019tst,Klaewer:2020lfg,Cota:2022yjw}, such wrapping D3-brane is dual to a fundamental heterotic or a type IIB string, depending on whether the $C_a$ fiber is rational ($C_a\simeq \mathbb{P}^1$) or elliptic ($C_a\simeq\mathbb{T}^2$). The appearance of light modes coming from these strings will allow us to extend the perturbative descriptions to additional corners of the K\"ahler cone. In order to compute the genus of the $X$ fiber $C_a$ one needs more topological information about $X$ than simply the leading intersection number, such as the first Chern class or the anticanonical class of $X$. Hence, with only EFT data about the K\"ahler potential we cannot predict the type of emergent string limit appearing in this regime. In the following, we will describe both options, corresponding either to a $C_a\simeq\mathbb{T}^2$ or $C_a\simeq \mathbb{P}^1$ fibration over the K\"ahler base $X$.

\subsubsection*{$\mathbb{T}^2$-fibration on $X$}
We first consider that the shrinking $C_a$ curve is topologically $\mathbb{T}^2$.\footnote{\label{fn.kahler not CY}Note that now $X$ is not Calabi-Yau, but a K\"ahler manifold. If $X$ admits a curve $C_a$ as a fiber, then it must be either a genus-one (this is, topologically $\mathbb{T}^2$) or a rational curve (i.e., $\mathbb{P}^1$), see Section 2.1 from \cite{Klaewer:2020lfg}.} This results in an elliptic fiber over $X$, yielding the following fourfold from the F-theory perspective,
\begin{equation}
\mathbb{E}\hookrightarrow Y\to (C_a\simeq\mathbb{T}^2\hookrightarrow X \to D^a)\equiv (\mathbb{E}\hookrightarrow\mathbb{T}^4\to C_a\simeq\mathbb{T}^2)\hookrightarrow Y\to D^a\;.
\end{equation}
As commented above, when the $C_a\simeq \mathbb{T}^2$ becomes small, D3 branes wrapping such curve are dual to the fundamental type IIB string. This allows us to extend the perturbative regime through the following duality frames, represented in Figure \ref{f.IIB 2-IIB}:
	\begin{itemize}
	\item For $s^a\to\infty$ we follow the EFT string flow of the above wrapped D3-brane, which becomes a fundamental Type IIB string and sets the species scale in this perturbative regime (red region in Figure \ref{f.IIB 2-IIB}, with $s^a\equiv s^2$). This region is therefore described by \textbf{weakly coupled, large volume type IIB on an elliptically fibered orientifold $\mathbb{T}^2\hookrightarrow X'\to D^a$}, with O5-planes (not wrapping the $\mathbb{T}^2$ fiber) rather than O7. Since the string mass is controlled by $l_a$, we find the following map:
	 \begin{equation}\label{eq. dictionary}
		s^a\to {s^0}'=e^{-2\phi_{\rm IIB,4}}\;,\quad {\rm Im}\tau = e^{-\Phi}\to {s^a}'\;,
	\end{equation}
	with $\phi_{\rm IIB,4}$ the new type \textit{4d} IIB dilaton and ${s^a}'$ a new saxion measuring the volume in \textit{string units} of the new elliptic curve $\mathbb{E}\simeq C_a'\simeq \mathbb{T}^2$, which now is part of the base $X'$. The above correspondence is obtained by identifying the wrapped D3 brane with the dual fundamental type IIB string, and the D7 brane wrapped on the whole original type IIB base $X$ with the NS5 brane(S-dual to the D5 in the type IIB T-dual to the original frame) wrapped on the $D^a$ base of the new description (see similar arguments in \cite{Morrison:1996na, Lee:2018urn}). Since fibration-wise $X$ and $X'$ are analogous, they share the same description and tower arrangement (since $D_a$ remains unchanged and $C_a'\simeq C_a$). This can be inferred from Figure \ref{f.IIB 2-IIB}, since the triangle representing the convex hull of the towers in the F-theory regime (light orange region) is equivalent to the triangle representing the towers in this dual Type IIB regime (red region), albeit rotated. 
	\item The above IIB duality frame with O5-planes is joined with the original F-theory/type IIB description through an \emph{additional} \textbf{weakly coupled, large volume type IIB string theory} description \textbf{compactified on the same  $\mathbb{T}^2\hookrightarrow X'\to D^a$} orientifold with O5-planes (dark orange region in Figure \ref{f.IIB 2-IIB}). This can be accessed either from the original F-theory/type IIB description on $X$ through a double T-duality (thereby the O7-planes become O5-planes) when the curve $C_a\simeq \mathbb{T}^2$ becomes string-size,  
	or from the new one on $X'$ through an S-duality when the new 4d dilaton becomes large, ${s^0}'\ll \mathcal{V}_{X'}$. As before,  the tower structure in this duality frame is equivalent to the other two, since the fibration structure is qualitatively unchanged. In addition to the corresponding KK modes of $X'$, it also has a string tower given by the same fundamental string of the original type IIB on $X$+O7 description, setting the species scale.
	
	\end{itemize}
Even if the arrangement of towers and duality frames have been plotted  in Figure \ref{f.IIB 2-IIB} for $V_X\sim v_1^2v_2$, the above results also apply to $V_X\sim v_1v_2v_3$, as the base $D_a$ remains untouched in the limits where $C_a$ shrinks. In the latter case, we will have three copies of the above display of duality frames corresponding to each of three saxions getting parametrically larger than the others.

\subsubsection*{$\mathbb{P}^1$-fibration on $X$}

The other possibility (see Footnote \ref{fn.kahler not CY}) it that $C_a$ is topologically $\mathbb{P}^1$, we obtain a rational fiber over $X$, or equivalently the fourfold
\begin{equation}
 \mathbb{E}\hookrightarrow Y\to (C_a\simeq\mathbb{P}^1\hookrightarrow X \to D^a)\equiv (\mathbb{E}\hookrightarrow{\rm K3}\to\mathbb{P}^1)\hookrightarrow Y\to D^a
\end{equation}

The D3-branes wrapped over the shrinking $\mathbb{P}^1$ give rise to a fundamental heterotic string, with gauge group depending on the actual $\mathbb{E}\hookrightarrow{\rm K3}\to\mathbb{P}^1$, set by the overall $Y$ fourfold. In the same way as before, we can extend the perturbative regime through the following duality frames, represented in Figure \ref{f.IIB 2-het}:
\begin{itemize}
	\item For $s^a\to\infty$, we follow the EFT string flow of the above wrapping D3-brane, which becomes a heterotic string and sets the species scale.  This new perturbative regime (blue region in Figure \ref{f.IIB 2-het}) is then given by a \textbf{weakly coupled, large volume heterotic string theory compactified on an elliptically fibered Calabi-Yau $\mathbb{T}^2\hookrightarrow X'\to D^a$}. This can be seen from the mapping between the F-theory and heterotic moduli. The (relative) $\mathbb{P}^1$ volume and the imaginary part of the $\mathbb{E}$ complex structure $\tau=C_0+ie^{-\Phi}$ of the original F-theory frame get mapped to the heterotic 4d dilaton $\phi_{\rm h,4}$ and volume of the $\mathbb{T}^2$ (in string units). Similarly to the case with $\mathbb{T}^2$-fibration on $X$, this can be obtained by identifying the fundamental heterotic string with a D3 brane wrapping the shrinking $\mathbb{P}^1$ fiber, and the D7 wrapping the whole $X$ with an NS5 wrapping the $D^a$ base in the heterotic frame (see \cite{Morrison:1996na} and Section 2.4 from \cite{Lee:2018urn}):
	\begin{subequations}
		\begin{align}
			e^{2\phi_{\rm h,4}}M_{\rm Pl,4}^2= {\rm vol}(\mathbb{P}^1)M_{\rm Pl,10(IIB)}^4\ \Longrightarrow\ &  e^{2\phi_{\rm h,4}}=\frac{V_{\mathbb{P}^1}}{V_X}\sim V_{D^a}^{-1}\\
			e^{\Phi}{\rm vol}(X)M_{\rm Pl,10(IIB)}^8=e^{-2\Phi_{\rm h}}{\rm vol}(D^a)M_{\rm Pl,10(het)}^{6}\ \Longrightarrow\  &\mathcal{V}_{\mathbb{T}^2}\sim e^{-\Phi}\;.
		\end{align}
	\end{subequations}
	Here we denote by $\mathcal{V}_{\mathcal{C}}$ and $V_{\mathcal{C}}$ volumes in string and 10d Planck units, respectively, and by ${\rm vol} (\mathcal{C})$ the dimensionfull volume of a given manifold $\mathcal{C}$. The 10d and 4d heterotic dilatons are represented by $\Phi_{\rm h}$ and $\phi_{\rm h,4}$.
	
Therefore, sending $s^a\to\infty$ corresponds to the weak coupling limit in the heterotic frame. The blue region in Figure \ref{f.IIB 2-het} also contains limits for which ${\rm Im}\tau\to\infty$, which in the heterotic frame corresponds to the decompactification of the $\mathbb{T}^2$ fiber. This results in the same tower arrangement as in the $\mathbb{T}^2$-fibration or the original type IIB compactification.
	\item Unlike in the $\mathbb{T}^2$-fibration, the duality frame (green region in Figure \ref{f.IIB 2-het}) interpolating between the original F-theory/type IIB frame (orange region) and the new heterotic duality frame (blue region) is \emph{not unique}, and it will depend on the gauge group of the heterotic description (which in turn depends on the point of moduli space of the K3 elliptical fibration). From the heterotic perspective, this new interpolating duality frame (in green) will be its S-dual theory, which will then be related to the original F-theory frame by two T-dualities. Let us analyze the following two options:
	\begin{itemize}
		\item $SO(32)$: Heterotic $SO(32)$ is S-dual to type I string theory, so that the interpolating frame is given by \textbf{weakly coupled, large volume type I string theory on} the same \textbf{elliptically fibered Calabi-Yau $\mathbb{T}^2\hookrightarrow X'\to D^a$} \cite{Polchinski:1995df}. When the $\mathbb{T}^2$ fiber becomes small, under two T-dualities, type I string theory on $\mathbb{T}^2$ is dual to type IIB on the pillowcase orbifold $\mathbb{S}^2(2222)\simeq \mathbb{T}^2/\mathbb{Z}_2$ \cite{Thurston1979TheGA,eskin2005pillowcasesquasimodularforms} , with an O7-plane on each of the four fixed points of the involution $z\to -z$ on the torus. Since $\mathbb{S}^2(2222)$ has the same underlying topology as $\mathbb{P}^1$, we recover the original F-theory/type IIB description.
		\item $E_8\times E_8$: The strong coupling limit of heterotic $E_8\times E_8$ is given by M-theory with a large Ho\v{r}ava-Witten interval, and as such, the interpolating duality frame is better described by \textbf{M-theory on $\mathbb{S}^1/\mathbb{Z}_2\times(\mathbb{T}^2\hookrightarrow X'\to D^a)$}. In the limit in which the $\mathbb{T}^2$ fiber becomes small, one can perform double dimensional reduction down to type IIA on $(\mathbb{S}^1\times\mathbb{S}^1/\mathbb{Z}_2)\hookrightarrow X''\to D^a$, with O8-planes at the endpoints of $\mathbb{S}^1/\mathbb{Z}_2$ and small $\mathbb{S}^1$, and then further T-dualize to find type IIB on $\mathbb{S}^2(2222)$ (as in the $SO(32)$ case  above but with different D7 brane disposition). As discussed, this has the topology of $\mathbb{P}^1$, thus arriving to the original F-theory/type IIB description.
	\end{itemize}

The arrangement of towers and duality frame are depicted for $V_X\sim v_1^2v_2$ in Figure \ref{f.IIB 2-IIB}. As in the previous case, the generalization to $V_X\sim v_1v_2v_3$ is straightforward, since the base $D^a$ is not affected in these limits with shrinking $C_a$. The EFT string flows are represented as red arrows, as usual.
\end{itemize}

    \begin{figure}[h]
\begin{center}

\includegraphics[width=0.45\textwidth]{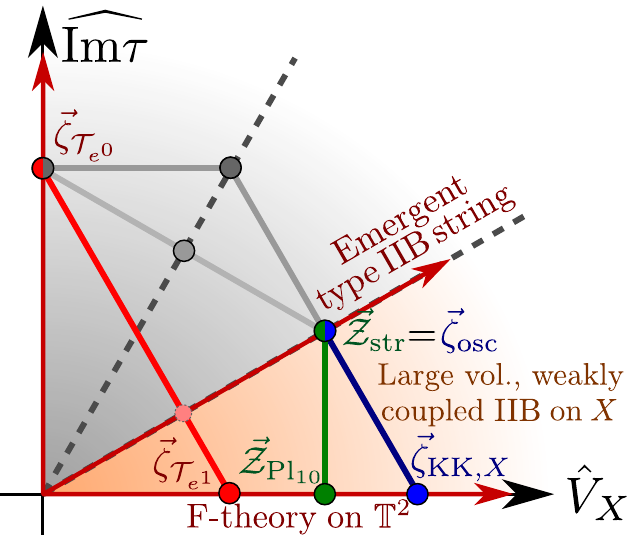}
\caption{Arrangement of perturbative duality frames, together with $\zeta$-vectors for towers (blue) and EFT strings (red), as well as $\mathcal{Z}$-vectors for  possible regimes of the species scale (green), in terms of the canonically normalized $\{\widehat{{\rm Im}\,\tau},\hat{V}_X\}$, wit $\tau$ the type IIB axio-dilaton and $V_X$ the overall volume in 10d Planck units. Depicted in gray are the regions which generally do not have a perturbative description reached through a series of dualities. Just for reference, the $\zeta$-vectors of the hypothetical towers also appear. In red arrows, we also depict the emergent dual theories along EFT string flows.}
\label{f.IIB 1}
\end{center}
\end{figure}

    \begin{figure}[h]
\begin{center}
\begin{subfigure}[b]{0.50\textwidth}
\center
\includegraphics[width=0.99\textwidth]{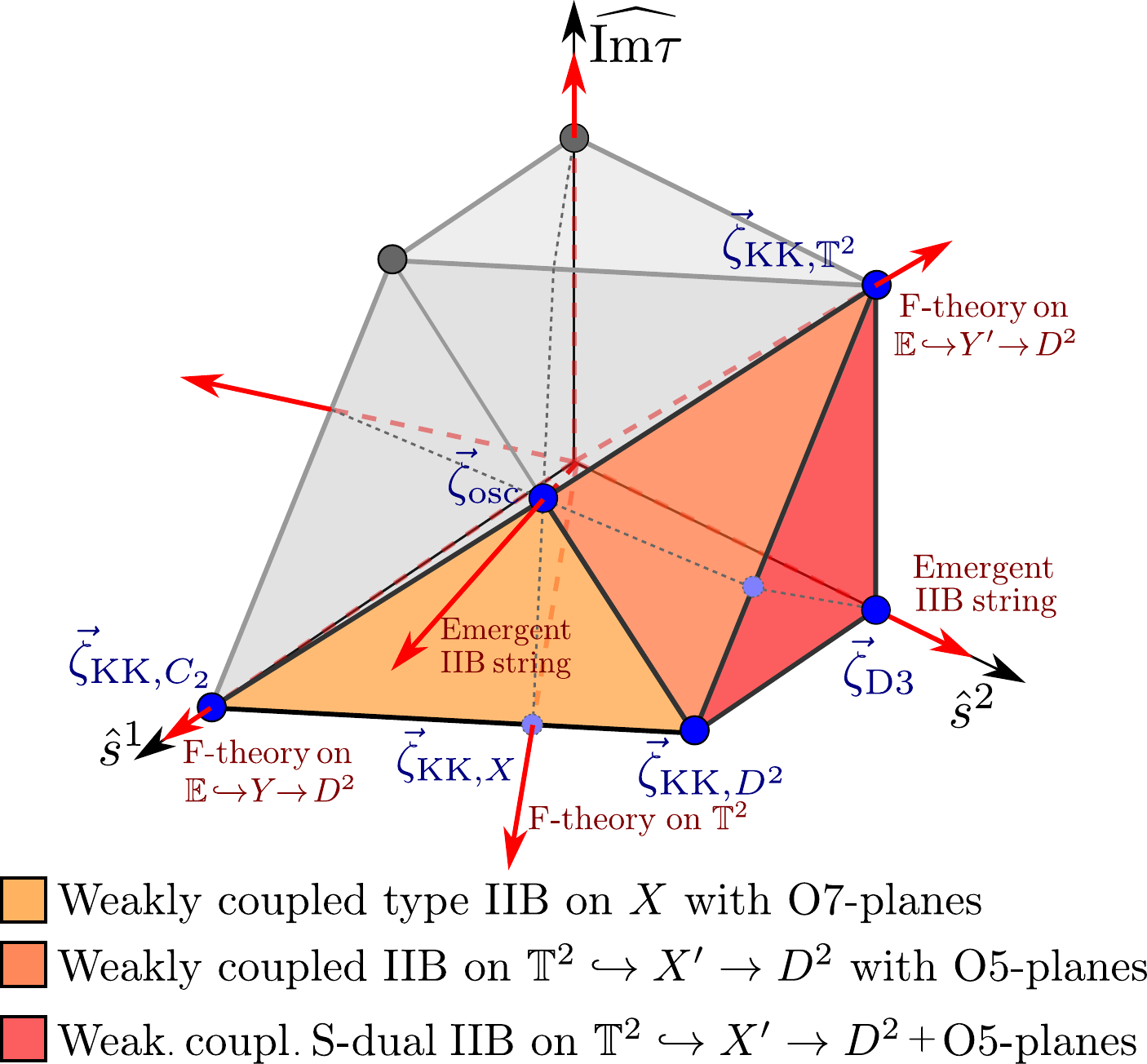}
\caption{\hspace{-0.3em} Base with $\mathbb{T}^2\hookrightarrow X\to D^a$ fibration.} \label{f.IIB 2-IIB}
\end{subfigure}\begin{subfigure}[b]{0.48\textwidth}
\center
\includegraphics[width=0.99\textwidth]{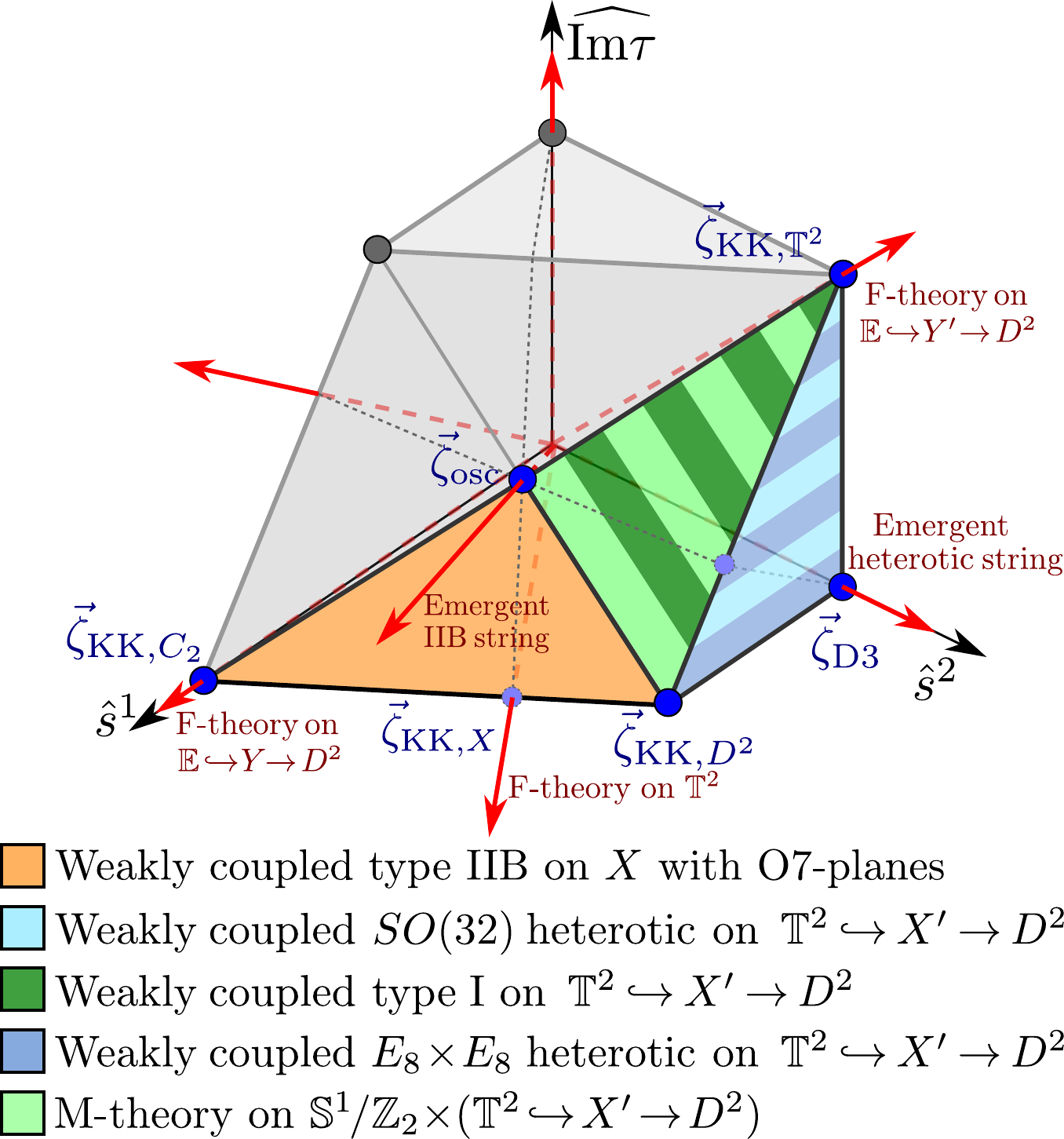}
\caption{\hspace{-0.32em} Base with $\mathbb{P}^1\hookrightarrow X\to D^a$ fibration.} \label{f.IIB 2-het}
\end{subfigure}
\caption{Arrangement of perturbative duality frames and  $\zeta$-vectors of light towers (blue), for F-theory compactifications on elliptically fibered Calabi-Yau fourfolds with base $X$ volume $V_X\sim v_1^2v_2$ in 10d Planck units and $\mathbb{T}^2$ (\ref{f.IIB 2-IIB}) and $\mathbb{P}^1$ (\ref{f.IIB 2-het}) asymptotic fibrations, in term of the canonically normalized saxions $\{\hat{s}^1,\hat{s}^2\}$, and the imaginary part of the axio-dilaton $\widehat{{\rm Im\,}\tau}$. Depicted in gray are the regions which would result through a series of dualities assuming a toroidal-like structure (but not necessarily in a general CY). Just for reference, the $\zeta$-vectors of the hypothetical towers also appear in gray. As red arrows, we show the  EFT string flows and the dual asymptotic descriptions emerging at infinite distance.
\label{f.II2B}}
\end{center}
\end{figure}

One important final consideration is that, due to the map \eqref{eq.saxion kahler} between the K\"ahler sector $\mathcal{K}(X)$ and the saxionic cone being injective but \emph{not necessarily surjective} \cite{fu2014relationskahlerconebalanced} (they are always in the strict asymptotic regime inside a given growth sector when $V_X$ is given by a monomial), then some saxion flows can move out of the K\"ahler cone $\mathcal{K}(X)$.  Crossing the $\mathcal{K}(X)$ boundary corresponds to a curve $C\subset X$ collapsing at finite distance (unlike in the cases studied above where a curve shrinks at infinite distance). However, since the divisors remain finite, we are still in the perturbative regime, and this crossing simply corresponds to a (finite distance) \emph{flop transition} to a new base $X'$, together with the blow-up  of a new curve $C'\subset X'$. See \cite{Witten:1996qb,Mayr:1996sh} for more details on possible quantum corrections to the EFT strings. One could glue together all the K\"ahler cones accessible from a given one through flop transitions to obtain an \emph{extended K\"ahler cone} $\mathcal{K}(X)_{\rm ext}$, which is now bijective with the saxionic cone $\Delta$. However, since under flop transitions the intersection numbers and the topology of the base change, the 4d $\mathcal{N}=1$ data, such as the K\"ahler potential, does too. The new K\"ahler cone after the flop will be again divided in a set of growth sectors, which can be understood and studied separately in an analogous way as the original one. In this sense, the tower polytopes studied here provide the building blocks that will then later have to be combined to produce the full figure associated to the extended K\"ahler  cone. We will refrain in this paper from understanding how the EFT string and tower $\zeta$-vectors, as well as the possible duality frames, fit together in a global way over $\mathcal{K}(X)_{\rm ext}$, and leave this for a future work \cite{TBAkahler}.\\

\section{M-theory compactifications\label{s.Mth}}
We now consider a different set of 4d $\mathcal{N}=1$ constructions, consisting in M-theory compactifications on $G_2$ manifolds. Unlike in the previous cases, where we had a quartic K\"ahler potential (with one saxion associated to the string coupling and a cubic volume), in 11d M-theory there are no moduli, and as such all the 4d saxions will be associated to geometric moduli of the compact 7-manifold $X_7$. In order to  obtain 4d $\mathcal{N}=1$ EFTs, we will reduce our 11d maximally supersymmetric theory on a compact $G_2$ manifold. Since the study of this type of spaces is way less developed than their 6-dimensional $SU(3)$ cousins, we will restrict to a subset known as \textbf{Joyce's compact manifolds} \cite{joyce1996a,joyce1996b,joyce2000compact}. This set of manifolds already have a rich variety of different asymptotic fibrations \cite{Liu:1998tha} (see Appendix \ref{app-Joyce} for more details), which are sufficiently different than those from CY constructions of the previous sections to allow us to learn important insights.
\vspace{0.5cm}

We will start with some generalities of M-theory compactifications on smooth $G_2$ 7-manifolds $X_7$ \cite{Beasley:2002db, Acharya:2004qe}. Given an associative 3-form $\Phi$, the volume of $X_7$ in $M_{\rm Pl,11}$ units is given by
\begin{equation}\label{e.G2vol}
	V_X=\frac{1}{7}\int_{X_7}\Phi\wedge\star\Phi\;.
\end{equation}
Such compactification results in the following 11d (Einstein frame) metric splitting as
\begin{equation}\label{e.11d metric}
	\dd s^2=\frac{l_{\rm M}^2M_{\rm Pl,4}^2}{2V_X}\dd s_4^2+\ell^2_{11}\dd s^2_{X}\;,
\end{equation}
where $\dd s_4^2$ is the 4d Einstein frame metric and $\ell_{11}=(4\pi\kappa_{11}^2)^{1/9}$ is the 11d Planck length. Given $C_3$ the (flat) M-theory potential under which M5-branes are magnetically charged, in the large volume regime, we can expand its (complex) combination with $\Phi$ in terms of the $\mathcal{N}=1$ chiral coordinates $t^b=a^b+is^b$ as
\begin{equation}\label{e.t exp G2}
	\mathbf{t}=\mathbf{a}+i\mathbf{s}=[C_3+i\Phi]=t^b[\Sigma_b]\in H^3(X_7;\mathbb{C})\;.
\end{equation}
Here, $\{\Sigma_a\}_{a=1}^{b_4(X_7)}$ are the basis of 4-cycles spanning the torsion-free part of $H_4(X_7;\mathbb{Z})$ and $\{[\Sigma_a]\}_{a=1}^{b_3(X_7)=b_4(X_7)}$ their Poincar\'e dual 3-forms. Up to numerical constants, the K\"ahler potential is given by 
\begin{equation}\label{.eq kahler Mth}
K=-3 \log V_X\;.
\end{equation}
Note that in the large volume regime, from \eqref{e.G2vol} and \eqref{e.t exp G2}, $\Phi$, $\star\Phi$ and $K$ depend only on the saxions $\{s^a\}_{a=1}^{b_3(X_7)}$ to leading order. The dual saxions $l_a$ defined in \eqref{TEFT}, measuring the relative volume of $\Sigma_a$, are given by
\begin{equation}
	{\boldsymbol l}=l_a[\tilde{\Sigma}^a]=-\frac{1}{2}\frac{\partial K}{\partial s^a}[\tilde{\Sigma}^a]=\frac{1}{2V_X}[\star\Phi]\in H^4(X_7;\mathbb{R})\;,
\end{equation}
with $\{\tilde\Sigma_a\}_{a=1}^{b_3(X_7)=b_4(X_7)}$ the 3-cycle basis dual to $\{\Sigma_a\}_{a=1}^{b_4(X_7)}$. The EFT strings are realized by M5-branes wrapped along a four cycle $C=e^a\Sigma_a$, such that their tension is then given by \ref{TEFT}
\begin{equation}\label{e.G2 tension}
	\mathcal{T}_{\mathbf{e}}=\frac{M^2_{\rm Pl,4}}{2V_{X}}\int_C\star\Phi=M^2_{\rm Pl,4}e^a l_a\;,
\end{equation}
with $\mathbf{e}$ the set of charges associated to the EFT string.

For simplicity and illustrative purposes, we will take our $G_2$-manifold to be an instance of Joyce's compact manifold of the first kind, see \cite{joyce1996a,joyce1996b,joyce2000compact,Liu:1998tha} and Appendix \ref{app-Joyce}, consisting in a toroidal orbifold $X_7=\mathbb{T}^7/\Gamma$, with the torus metric determined by the $\{e^i=R_i\dd x^i\}_{i=1}^7$ 7-bein, with $x^i\sim x^i+1$ and the radii $\{R_i\}_{i=1}^7$ in $M_{\rm Pl,11}$ units. Then $\Gamma$ is taken to be a finite group preserving the 3-form\footnote{As explained in Appendix \ref{app-Joyce}, we will use a different convention for the basis than in recent instance of $G_2$-manifolds in the literature, as this way there is a clearer interpretation of the different limits}
\begin{equation}\label{e.Phi Joyce}
	\Phi=\eta^{567}+\eta^{347}-\eta^{246}+\eta^{235}+\eta^{145}+\eta^{136}+\eta^{127}\;,\quad\text{where }\;\eta^{ijk}=e^i\wedge e^j\wedge e^k\;.
\end{equation}
We will take the manifold $J(0,\nicefrac12,\nicefrac12,\nicefrac12,0)$, such that $\Gamma=\mathbb{Z}_2\times\mathbb{Z}_2\times\mathbb{Z}_2$, generated by
\begin{equation}\label{e.group 1st BULK}
 	\begin{array}{l}
 		\alpha(\vec{x})=(-x^1,-x^2,-x^3,-x^4,x^5,x^6,x^7)\\
 		\beta(\vec{x})=(-x^1,\nicefrac12-x^2,x^3,x^4,-x^5,-x^6,x^7)\\
 		\gamma(\vec{x})=(\nicefrac12-x^1,x_2,\nicefrac12-x^3,x^4,-x^5,x^6,-x^7)\;,
 	\end{array}
 	\end{equation}
	following \cite{joyce1996a,joyce1996b,Liu:1998tha}.
It is easy to see that the top form $e^1\wedge\dots\wedge e^7$ in invariant under $\Gamma$, and thus $X_7$ is orientable. Furthermore, one can check that the $\eta^{a_Ia_Ja_K}$ 3-forms from \eqref{e.Phi Joyce} are linearly independent and invariant with respect to $\Gamma$, such that $b_3(X_7)=b_4(X_7)=7$, and dual to the seven 3-cycle basis $\{\tilde{C}_a\simeq\mathbb{T}^3\}_{a=1}^7$ of $H_3(X_7,\mathbb{Z})$, identified by\footnote{One can identify basis for $p$-forms on $X_7$ and, through Poincar\'e duality (recall that $X_7$ is both closed and oriented), the Betti numbers are $\vec{b}(X_7)=(1,0,0,7,7,0,0,1)$.} 
\begin{equation}\label{eq: table IJK}
	\begin{array}{c|ccccccc}
		a&1&2&3&4&5&6&7\\\hline
		a_I&5&3&2&2&1&1&1\\
		a_J&6&4&4&3&4&3&2\\
		a_K&7&7&6&5&5&6&7	
	\end{array}\;.
\end{equation} 
The volume of each of the 7 3-cycles is controlled by the saxions $\{s^a\}_{a=1}^7$, in such a way that $s^a=R_{a_I}R_{a_J}R_{a_K}$, following \eqref{eq: table IJK}. For instance, $s^1=R_5R_6R_7$, and similarly for the others. In the large volume limit, the contribution from the singularities resolution\footnote{The $X_7=\mathbb{T}^7/\Gamma$ singularities consist in the disjoint union of twelve $\mathbb{T}^3\subset \mathbb{T}^7$. Their resolution results in the introduction of $36=12\times 3$ extra 3-cycles $\{\tilde{C}_\alpha\simeq \mathbb{S}^2\times\mathbb{S}^1\}_{\alpha=1}^{36}$, with the associated 36 \emph{twisted saxions} $\{\tilde{s}^\alpha\}_{\alpha=1}^{36}$, \cite{Lanza:2021udy}. For fixed values of $\tilde{s}^\alpha$ the contribution of the resolutions to the overall volume is finite, and thus can be ignored in the large $V_X$ limit.} can be ignored, so that $V_{X}=R_1\dots R_7=(s^1\dots s^7)^{1/3}$ in $M_{\rm Pl,11}$ units. In the large volume limit the K\"ahler potential takes the simple form $K=-\log\left(s^1\dots s^7\right)$, which results in the following saxionic moduli space metric and dual saxions
\begin{equation}\label{e.saxions met G2}
	\mathsf{G}_{ab}=\frac{\delta_{ab}}{2(s^a)^2}\;,\qquad l_a=\frac{1}{2s^a}\;.
\end{equation}
This allows us to obtain the tension of the EFT strings and their $\zeta$-vectors
\begin{subequations}\label{e.zeta EFT string HET}
	\begin{align}
		\mathcal{T}_{\mathbf{e}}&=M_{\rm Pl,4}^2e^a l_a=\frac{M_{\rm Pl,4}^2}{2}\sum_{a=1}^7\frac{e^a}{s^a}\\
		\left.\zeta_{\mathcal{T}_{\mathbf{e}}}^i\right|_{\rm flow}&=-\frac{1}{2}\delta^{ij}\mathsf{e}^{a}_j\partial_a\log\frac{\mathcal{T}_{\mathbf{e}}}{M_{\rm Pl,4}^2}=\frac{\hat{e}^i}{\sqrt{2}(\hat{e}^1+\dots+\hat{e}^7)    }\;,
	\end{align}
\end{subequations}
where $\hat{e}^i=\sign e^i$. One can check that there are $127=7+21+35+35+21+7+1$ different $\left.\zeta_{\mathcal{T}}\right|_{\rm flow}$-vectors, of which 7 correspond to elementary EFT strings. 

As for the different light towers along the possible infinite distance limits, we first consider the critical \textbf{string oscillator modes} coming from elementary EFT strings above, with
\begin{equation}\label{e.zeta string HET}
	m_{{\rm osc},a}^2=\frac{M_{\rm Pl,4}^2}{2s^a}\;\Longrightarrow\;\zeta^i_{{\rm osc},a}=\frac{\delta^i_a}{\sqrt{2}}\;.
\end{equation}
As it will be discussed later, the topology of the 4-cycle will determine the type of fundamental string resulting from the wrapped M5-brane. 

Regarding the \textbf{KK towers} associated with the decompactification of different number of dimensions, similarly we have
\begin{subequations}
	\begin{align}\label{e.zeta KK HET}
		m_{{\rm KK},R_{i_1}\dots R_{i_n}}&=\frac{M_{\rm Pl,11}}{(R_{i_1}\dots R_{i_n})^{1/n}}=M_{\rm Pl,4}\left(\prod_{j=1}^ns^{a_I^j}s^{a_J^j}s^{a_K^j}\right)^{-1/2n}\\
		\zeta^i_{{\rm KK},R_{i_1}\dots R_{i_n}}&=\frac{\hat{a}^i}{\sqrt{2}n}\;,
	\end{align}
\end{subequations}
where $a_I^J$ are given by Table \ref{eq: table IJK} and $\hat{a}^i=a_I^i+a_J^i+a_K^i\in\{0,1,2,3\}$ is the number of radii associated to the saxion $s^i$ sent to infinity. It can be checked that $|\vec{\zeta}_{{\rm KK},R_{i_1}\dots R_{i_n}}|=\sqrt{\frac{n+2}{2n}}$ and $|\vec{\zeta}_{\rm osc}|=\frac{1}{\sqrt{2}}$, as expected from the taxonomy rules \eqref{e.taxonomy} from Section \ref{ss.review towers}. Same combinatorics as with the EFT strings show that there are a total of $127=7+21+35+35+21+7+1$ KK and 7 emergent string towers. 

Additionally, we must consider the states resulting from M2 branes wrapping internal cycles. As commented above, the Joyce manifold $X_7$ does not have globally defined 1- or 2-cycles, but,  as described in Appendix \ref{app-Joyce}, in some asymptotic limits one can have fibration structures with 1 or 2-dimensional bases or fibers that are becoming large or small. One can obtain light particles by wrapping M2 branes over shrinking 2-cycles (analogous to what occurred in the heterotic $E_8\times E_8$ and type IIA compactifications, dual to F-theory limits),
	\begin{align}
		m_{{\rm M2}}=M_{\rm Pl,11}V_2&\Longrightarrow \frac{m_{{\rm M2}}}{M_{\rm Pl,4}}=V_2 V_X^{-1/2}\;
	\end{align}
 and (non-BPS) strings\footnote{As we will see later, the cycle over which we wrap the M2 brane is the interval corresponding to the base of a Borcea-Voisin fibration. This would amount to a Ho\v{r}ava-Witten  interpretation of the heterotic $E_8\times E_8$ string. However, since the interval is non-calibrated, the resulting string is non-BPS at finite distance in moduli space.}  from M2-branes on 1-cycles becoming small:

\beq
		\mathcal{T}_{\rm osc,M2}=M_{\rm Pl,11}^2 R_a\Longrightarrow \frac{m_{\rm osc,M2}}{M_{\rm Pl,4}}=R_a^{\frac{1}{2}} V_X^{-1/2}
\eeq
Here, $R_a$ and $V_2=R_aR_b$ are the length and area of the 1- and 2-cycles. The associated $\zeta$-vectors are given by
\begin{equation}\label{e.zeta M2}
	\zeta^i_{{\rm osc\, M2},a}=\left\{
	\begin{array}{ll}
	0&\text{if }\frac{\partial s^a}{\partial R_i}\neq 0\\
	\frac{1}{2\sqrt{2}}&\text{otherwise}
	\end{array}
	\right.\qquad
	\zeta^i_{{\rm M2},ab}=\left\{
	\begin{array}{ll}
	-\frac{1}{\sqrt{2}}&\text{if }\frac{\partial s^a}{\partial R_i},\frac{\partial s^b}{\partial R_i}\neq 0\\
	0&\text{if }\frac{\partial s^a}{\partial R_i}\neq 0,\frac{\partial s^b}{\partial R_i}=0\text{ (or viceversa)}\\
	 \frac{1}{\sqrt{2}}&\text{otherwise}
	\end{array}
	\right.\;,
\end{equation}
where we used Table \ref{eq: table IJK} to read the saxion dependence on the radii. Note that $|\zeta^i_{{\rm osc\, M2},a}|=\frac{1}{\sqrt{2}}$ and $|\vec{\zeta}_{{\rm M2},ab}|=\sqrt{\frac{3}{2}}$, respectively corresponding to the expected length associated with the scaling of critical strings and KK-1 modes, with the later being explained through a duality with F-theory as in Section \ref{s.het}. As it was the case then, the $\vec{\zeta}_{{\rm M2},ab}$ vectors point outside the K\"ahler cone (note some of the components are negative), but their bounded states with other KK towers will have non-negative components. The $\vec{\zeta}_{\rm osc M2}$ vectors point to the interior of the saxionic cone, and thus do not act as generators of the convex hull of $\zeta$-vectors. In certain cases, they will still play the role of a co-leading tower, and will satisfy the integer scaling relation. However, there will be an instance in which this tower is subleading (in fact, setting the species scale) and, therefore, will not preserve the integer scaling relation, as shown in Table \ref{tab.w Mth}. The result is analogous to that of the non-BPS Type I string in sections \ref{s.SO32} and\ref{s.Fth}.

Clearly, the tower of critical oscillator modes will always have integer scaling $w=1$   for elementary EFT string flows since $\vec{\zeta}_{\mathcal{T}}=\vec{\zeta}_{\rm osc}$. Using \eqref{e.zeta EFT string HET}, \eqref{e.zeta string HET} and \eqref{e.zeta KK HET}, we can also check that the integer scaling (as in \eqref{lattice}) for the KK towers satisfies
\begin{equation}
w=\frac{\vec{\zeta}_{\mathcal{T}_{\mathbf{e}}}\cdot\vec{\zeta}_{{\rm KK},R_{i_1}\dots R_{i_n}}}{|\vec{\zeta}_{\mathcal{T}_{\mathbf{e}}}|^2}=\sum_{i=1}^{7}\frac{\hat{a}^i}{n}\in[0,3]\;.
\end{equation}
For elementary string  flows, this automatically results in $w\in\{0,1,2,3\}$ for all KK vectors associated with $n=1,\,2$, which we will see are precisely the \emph{generators} of the convex hull of tower $\zeta$-vectors. Something similar can be checked to occur for the KK-1 states resulting from wrapping M2 branes, see \eqref{e.zeta M2}. One can also check that for non-elementary EFT string limits, the integer scaling also holds for the (sub)leading tower(s) generating the convex hull.

In Table \ref{tab.w Mth} we show the scalings of the leading tower(s) and those subleading generating the $\zeta$-vector convex hull. Due to the involved fibration structure of the Joyce manifold (see Appendix \ref{app-Joyce} for more details), most towers do not exist in all limits.

\begin{table}[h]
	\centering
	\resizebox{\textwidth}{!}{\begin{tabular}{|c|cccccccccc|}
	\hline
	\# 	&$\vec{\zeta}_{\rm osc, M5}$	&$\vec{\zeta}_{\rm osc, M2}$	&$\vec{\zeta}_{{\rm KK},\mathbb{T}^3/\Gamma}$	&$\vec{\zeta}_{{\rm KK},\mathbb{T}^4/\Gamma_{(0)}}$	&$\vec{\zeta}_{{\rm KK},\mathbb{S}^1/\mathbb{Z}_2}$&$\vec{\zeta}_{{\rm KK},\mathbb{S}^1_{\rm IIB}}$&$\vec{\zeta}_{{\rm KK},\mathcal{Q}_3}$&$\vec{\zeta}_{{\rm KK},\mathbb{S}^2(2222)}$&$\vec{\zeta}_{{\rm KK},Y_6^{(')}}$&$\vec{\zeta}_{{\rm KK},X_7}$\\\hline
	1	&\textbf{1}	&-	&\textbf{1}	&-	&-	&-	&-	&-	&-	&-	\\
	2	&-	&-	&-	&1	&\textbf{2}	&\textbf{2}	&-	&-	&-	&-	\\
	3(a)&-	&${\color{gray}\nicefrac32}$&2	&-	&-	&-	&2	&-	&\textbf{2}	&-	\\
	3(b)&-	&-	&-	&1	&\textbf{3}	&-	&-	&1	&1	&-	\\
	4(a)&-	&\textbf{2}	&-	&2	&-	&-	&-	&2	&\textbf{2}	&-	\\
	4(b)&1	&-	&1	&-	&\textbf{3}	&-	&2	&-	&-	&-	\\
	5	&-	&2	&-	&2	&-	&-	&-	&\textbf{3}	&-	&-	\\
	6	&-	&-	&2	&\textbf{3}	&-	&-	&-	&-	&-	&-	\\
	7	&-	&-	&-	&-	&-	&-	&-	&-	&-	&\textbf{3}	\\
	\hline
	\end{tabular}
	}
	\caption{Integer scaling weight $w$ of the different towers of states along each EFT string limit, for M-theory on a $J(0,\nicefrac12,\nicefrac12,\nicefrac12,0)$ Joyce manifold.  Each tower with vector $\vec{\zeta}$ satisfies  $\vec{\zeta}_{\mathcal{T}_{\mathbf{e}}} \cdot \vec{\zeta} = w \  \abs{\vec{\zeta}_{\mathcal{T}_{\mathbf{e}}}}^2$ along the flow driven by an EFT string with vector $\vec{\zeta}_{\mathcal{T}_{\mathbf{e}}}$ and charge ${\mathbf{e}}$. The leading tower(s) for each EFT string limit are highlighted in bold face. As in Table \ref{t.Limits Mth}, the limits are labeled by the number \# of saxions sent to infinity. For simplicity, all the $\mathbb{T}^3/\Gamma$ and $\mathbb{T}^4/\Gamma_{(0)}$ (this one including K3) cycles/fibers are depicted together. Note that the string obtained by wrapping a M2-brane on the base interval does not have an integer scaling ($w=\frac{3}{2}$), but in such limit $\vec{\zeta}_{\rm osc, M2}$ is neither a leading tower nor a convex hull generator.\label{tab.w Mth}}
\end{table}

\vspace{0.25cm}

Depending on the number of saxions sent to infinity (or in other words, the number of non-vanishing EFT string charges), we reach the following string flow limits represented in Table \ref{t.Limits Mth}. 

\begin{table}[htp!]
	\begin{center}
			\resizebox{\textwidth}{!}{\begin{tabular}{|c|c|c|c|c|c|}
			\hline
$\#$ & $w$ & Leading tower(s) $m_\ast$ &$\Lambda_{\rm QG}$ & Sublead. towers& Emergent dual theory \\ 
\hline 
\rowcolor{blue!10!} 1&1&$m_{\rm M5},\,m_{{\rm KK},\mathbb{T}^3/\Gamma}$	& $m_{\rm str}$	&	$\setminus$&	\begin{tabular}{@{}c@{}}Emergent IIB ($\mathbb{T}^4$) or\\ HE (K3) string\end{tabular}\\
\rowcolor{green!10!}2&2& $m_{{\rm KK},\mathbb{S}^1/\mathbb{Z}_2},\, m_{{\rm KK},\mathbb{S}^1_{\rm IIB}}$&$M_{\rm Pl,6}$	& \begin{tabular}{@{}c@{}}$m_{\rm KK,K3}$\end{tabular}	&F-theory on $\mathbb{E}\hookrightarrow Y_6\to {\rm K3}$	\\
\rowcolor{green!10!}3(a)&2&	$m_{{\rm KK},Y_6'}$& $M_{\rm Pl,10}$& $m_{\rm M2}$	& Weakly coupled 10d HE	\\
\rowcolor{red!10!}3(b)&3&	$m_{{\rm KK},\mathbb{S}^1/\mathbb{Z}_2}$& $M_{\rm Pl,5}$& $m_{{\rm KK},Y_6}$	& M-theory on $Y_6$	\\
\rowcolor{green!10!}4(a)&2&$m_{\rm M2},m_{{\rm KK},Y_6}$	&$m_{\rm str}$	&	$\setminus$&	 Emergent HE string\\
\rowcolor{red!10!}4(b)&3&$m_{{\rm KK},\mathbb{S}^1/\mathbb{Z}_2}$& $M_{\rm Pl,11}$	&	$\setminus$&	M-theory on growing $\mathbb{T}^3$\\
\rowcolor{red!10!}5&3& $m_{{\rm KK},\mathbb{S}^2(2222)}$	&$m_{\rm str}$	& $m_{\rm M2},\, m_{\rm KK,K3}$	& Weak. coup. HE on K3	\\
\rowcolor{red!10!}6&3&$m_{{\rm KK},\mathbb{T}^4/\Gamma}$	&$M_{\rm Pl,8}$	&$m_{{\rm KK},\mathbb{T}^3}$	& M-theory on $\mathbb{T}^3$	\\
\rowcolor{red!10!}7&3&$m_{{\rm KK},X_7}$	&	$M_{\rm Pl,11}$&$\setminus$	&	11d M-theory\\
\hline 			
		\end{tabular}}
		\caption{Description of the different EFT string limits driven by the string charges ${\mathbf{e}}$  for M-theory on a $J(0,\nicefrac12,\nicefrac12,\nicefrac12,0)$ Joyce manifold. We provide the identification of the leading tower(s) $m_\ast$, their scaling weight $w$, the asymptotic species scale $\Lambda_{\rm QG}$, subleading towers with $m\lesssim \Lambda_{\rm QG}$ and the microscopic interpretation of the dual theory emerging at the infinite distance limit. We label each limit by the number $\#$ of the EFT non-zero charges or equivalently, by the number of saxions sent to infinity. Note that there are two different choices when sending three or four saxions to infinity.\label{t.Limits Mth}} 
	\end{center}
\end{table}

In the following, we explain in more detail each of this limits (referring to Appendix \ref{app-Joyce} for more details on the fibration structure of Joyce manifolds). Note that when writing fibration structures $\mathcal{F}_n^{{\color{gray}(a)}}\hookrightarrow X\to\mathcal{B}_{m}^{{\color{gray}(b)}}$, the gray superindices indicate the scaling of the characteristic length of fiber/base with the EFT string flow parameter $\sigma$. In other words, $\mathcal{F}_n^{{\color{gray}(a)}}$ means that its volume grows as $V_{\mathcal{F}}^{1/n}\sim \sigma^a$, while $\mathcal{B}_{m}^{{\color{gray}(b)}}$ that $V_{\mathcal{B}}^{1/m}\sim \sigma^b$.

Unlike in the 4d $\mathcal{N}=1$ dualities arising the previous string theory CY compactifications, much less is known for M-theory on $G_2$ manifolds. Our logic here is to use the scaling of the different light towers to infer its microscopical nature, and with this, the interpretation of the dual description. This is not a rigorous proof of such dualities, but rather serves as an \emph{informed guess} about such uncharted limits that can provide interesting targets for future research. 
\begin{enumerate}
	\item[\textbf{1.}] In this first limit a single saxion $s^i$ is sent to infinity (i.e., $s^i\sim e^i\sigma$ with $\sigma\rightarrow \infty$). Inverting \eqref{eq: table IJK}, we can compute  the scaling of the volumes in terms of the scaling of the saxions,  obtaining that the length of the associated 3-cycle (i.e. that with volume given by $R_{a_I}R_{a_J}R_{a_K}$) grows as $\sigma^{1/3}$. Analogously, the dual 4-cycle shrinks with length scale going as $\sigma^{-1/6}$. This way, the Joyce manifold adopts an a.c. fibration (see Appendix \ref{app-Joyce}), 
	\begin{equation}
	\mathcal{F}_4^{{\color{gray} (-\nicefrac{1}{6})}}\hookrightarrow X_7\to\mathbb{T}^3/\Gamma^{\color{gray} (\nicefrac{1}{3})}
	\end{equation}	
	with the growing base given by $\mathbb{T}^3/\Gamma$, with $\Gamma$ inherited from \eqref{e.group 1st BULK} and acting only on the $\mathbb{T}^3$ directions, and generic fiber $\mathcal{F}_4\simeq\mathbb{T}^4/\Gamma_0$, with $\Gamma_0$ the normal subgroup of $\Gamma$ leaving the base invariant. One can check that indeed this corresponds to $\mathbb{Z}_2\simeq \langle\alpha\rangle,\,\langle\beta\rangle$ and $\langle\gamma\rangle$, respectively for $s^1$, $s^2$ and $s^3$, while for the rest of saxion limits $\Gamma_0=\{0\}$. The generic fibers will correspond to singular K3$\simeq\mathbb{T}^4/\mathbb{Z}_2$ (which can later be resolved) for $i=1,2,3$ and $\mathbb{T}^4$ for $i=4,5,6,7$. Wrapped along these shrinking 4-cycles we can have M5 branes becoming light, resulting in the BPS elementary EFT strings, respectively dual to heterotic (with gauge group depending on the actual K3 geometry) or Type IIB fundamental strings, see \cite{Lee:2019wij}. These emergent string limits, where the leading tower and species scale are both given by the string scale, are accompanied by partial decompactification of the $\mathbb{T}^3/\Gamma$ base to $d=7$, with the associated KK modes becoming light at the same rate as $m_{\rm osc}\equiv m_{\rm str}$. For the asymptotic regime in which this $m_{\rm str}$ sets the species scale, the perturbative duality frame is better described by \textbf{weakly coupled, large volume heterotic/type IIB string theory on a dual Calabi-Yau $Y_6'$}, similar as in the M-theory/heterotic case in Section \ref{ss.het 2sec}. Since the $m_{{\rm KK},\mathbb{T}^3/\Gamma}$ tower lays at the scale of $\Lambda_{\rm QG}$, we expect such $Y_6'$ to have a limit with a $\mathbb{T}^3/\Gamma$ factor, with the volume of the additional cycles not controlled by the $\{s^i\}_{i=1}^7$ saxions.

	\item[\textbf{2.}] In this limit two saxions are sent to infinity, resulting in an asymptotic fibration structure 
	\begin{equation}
	({\rm K3}^{\color{gray} (\nicefrac{1}{6})}\hookrightarrow Y_6\to\mathbb{S}^2(2222)^{\color{gray} (-\nicefrac{1}{3})})\hookrightarrow X_7\to\mathbb{S}^1/\mathbb{Z}_2^{{\color{gray} (\nicefrac{2}{3})}}
	\end{equation}
	where $Y_6$ is a Calabi-Yau threefold known as the \emph{Borcea-Voisin manifold} and $\mathbb{S}^2(2222)\simeq\mathbb{T}^2/\mathbb{Z}_2$ is the ``pillowcase'' orbifold, defined as $\mathbb{T}^2$ orbifold under the involution $(x,y)\mapsto (-x,-y)$, topologically a 2-sphere or a $\mathbb{P}^1$ with four conical points of order 2, corresponding to the four fixed points. The characteristic lengths of K3, $\mathbb{S}^2(2222)$ and $\mathbb{S}^1/\mathbb{Z}_2$ scale as $\sigma^{1/6}$, $\sigma^{-1/3}$ and $\sigma^{2/3}$, respectively. As it was the case in Section \ref{s.het}, together with the decompactification of the M-theory interval $\mathbb{S}^1/\mathbb{Z}_2$, we have a candidate tower of states associated to M2-branes wrapping the shrinking $\mathbb{S}^2(2222)$ 2-cycle. The mass of all these states is given by
	\begin{equation}
		\frac{m_{\rm M2}}{M_{\rm Pl,4}}=\frac{V_{\mathbb{S}^2(2222)}}{\sqrt{V_{X_7}}},\quad\frac{m_{{\rm KK},\mathbb{S}^1/\mathbb{Z}_2}}{M_{\rm Pl,4}}=\frac{1}{R_{\mathbb{S}^1/\mathbb{Z}_2}\sqrt{V_{X_7}}}\;,
	\end{equation}
	which in this limit both scale as $\sigma^{-1}$. Potential bounded states between the two, scale as
	\begin{subequations}
	\begin{align}
	\frac{m_{{\rm KK},\mathbb{S}^1/\mathbb{Z}_2+\rm M2}}{M_{\rm Pl,4}}&\sim\sqrt{\frac{V_2}{R_{\mathbb{S}^1/\mathbb{Z}_2}V_X}}\\
	\vec{\zeta}_{{\rm KK},\mathbb{S}^1/\mathbb{Z}_2+\rm M2}&=\left(\frac{\delta^a_1+\delta^b_1}{\sqrt{2}},\frac{\delta^a_2+\delta^b_2}{\sqrt{2}},\frac{\delta^a_3+\delta^b_3}{\sqrt{2}},\frac{\delta^a_4+\delta^b_4}{\sqrt{2}},\frac{\delta^a_5+\delta^b_5}{\sqrt{2}},\frac{\delta^a_6+\delta^b_6}{\sqrt{2}},\frac{\delta^a_7+\delta^b_7}{\sqrt{2}}\right)\,,
	\end{align}
	\end{subequations}
	where $R_a$ and $R_b$ are the radii associated to the pillowcase orbifold $\mathbb{S}^2(2222)$, with $s^a$ and $s^b$ being the saxions sent to infinity. It is immediate to check that $|\vec{\zeta}_{{\rm KK},\mathbb{S}^1/\mathbb{Z}_2+\rm M2}|=1$, so it matches with the length expected for KK-2 towers.	This suggests that the limit corresponds to a double decompactification to a 6d theory (this is, $\Lambda_{\rm QG}=M_{\rm Pl,6}$), given by F-theory on $\mathbb{E}\hookrightarrow Y_6'\to $K3, although it would be nice to find further evidence for this result.
	The associated perturbative regime of the moduli space would then  be described by \textbf{F-theory on an elliptically fibered CY$_4$} with fibration structure $\mathbb{E}\hookrightarrow X'_8\to [{\rm K3}\hookrightarrow Y''_6\to(\mathbb{S}^1_{\rm IIB}\times\mathbb{S}^1/\mathbb{Z}_2)]$. Additional limits in the $\mathbb{E}$ complex structure are dual to heterotic $E_8\times E_8$ on $\mathbb{T}^4$, but in general these are not accessible from our saxionic sector. One can also compute that the EFT string resulting from wrapping a M5 brane on the (relatively shrinking) K3 has a mass scale that becomes light as the same rate as the species scale $M_{\rm Pl,6}\sim \sigma^{-1/2}$, with both
	\begin{equation}
		\vec{\zeta}_{\mathcal{T}_{M5}}=\vec{\mathcal{Z}}_{\rm Pl_6}=\left(\frac{\delta^a_1+\delta^b_1}{2\sqrt{2}},\frac{\delta^a_2+\delta^b_2}{2\sqrt{2}},\frac{\delta^a_3+\delta^b_3}{2\sqrt{2}},\frac{\delta^a_4+\delta^b_4}{2\sqrt{2}},\frac{\delta^a_5+\delta^b_5}{2\sqrt{2}},\frac{\delta^a_6+\delta^b_6}{2\sqrt{2}},\frac{\delta^a_7+\delta^b_7}{2\sqrt{2}}\right)\;.
	\end{equation}

	\item[\textbf{3.}] In this third limit, there are two possibilities for sending three saxions to infinity:
	\begin{itemize}
		\item[\textbf{(a)}] We can choose three saxions in such a way that there is a single radius $R_i$ that remains independent of them. For example, from Table \eqref{eq: table IJK}, having $s^1,\,s^2,\,s^3\to\infty$, implies that $R_1$ is independent of them. This results in $X_7$ adopting again a Borcea-Voisin fibration
\begin{equation}
	({\mathbb{T}^3}^{\color{gray} (0)}\hookrightarrow Y_6\to Q_3^{\color{gray} (\nicefrac{1}{2})})\hookrightarrow X_7\to \mathbb{S}^1/\mathbb{Z}_2^{{\color{gray} (-\nicefrac{1}{2})}}\;,
\end{equation}	
	where now the CY 3-fold $Y_6$ has a special Lagrangian $\mathbb{T}^3$-fibration with base a 3-orbifold $Q_3$ with a $\mathbb{S}^3$ topology. The scales of $\mathbb{T}^3$, $Q_3$ and $\mathbb{S}^1/\mathbb{Z}_2$ respectively grow as $\sigma^0$ (i.e., constant), $\sigma^{\frac{1}{2}}$ and $\sigma^{-1/2}$.  

    The $\mathbb{T}^3$ fiber is relatively shrinking with respect to the overall $\mathbb{T}^3{}^{\color{gray} (-\nicefrac{1}{4})}\hookrightarrow Y_6\to Q_3^{\color{gray} (\nicefrac{1}{4})}$ volume, doing so at the same rate at which $Q_3$ grows. While it is known that M-theory on a small 3-torus is U-dual to M-theory on a growing $\mathbb{T}^3$ (with $R_i\to \frac{R_i^{1/3}}{(R_i R_j R_k)^{2/3}}$ in $M_{\rm Pl,11}$ units) \cite{Obers:1998fb,Calderon-Infante:2023ler}, it is not obvious that this duality extends when considering non-trivial $\mathbb{T}^3$ fibrations. However, one can see that the KK modes of the $\mathbb{S}_i^1$ circle in $\mathbb{T}^3$ and the wrapped M2-branes on the Poincar\'e-dual $\mathbb{T}^2_{jk}$,
    \begin{subequations}
    	\begin{align}
    m_{{\rm KK},\mathbb{S}^1_i}&=M_{\rm Pl,11}R_i^{-1}=M_{\rm Pl,8}R_i^{-7/6}R_j^{-1/6}R_k^{-1/6}\\
    m_{{\rm M2},jk}&=\mathcal{T}_{\rm M2}\frac{R_j R_k}{M_{\rm Pl,11}^2}=M_{\rm Pl,8} R_i^{-1/6}R_j^{5/6}R_k^{5/6}
    	\end{align}
    \end{subequations}
    scale at the same rate and are mapped to each other under $R_i\to \frac{R_i^{1/3}}{(R_i R_j R_k)^{2/3}}$, with both $|\vec{\zeta}_{{\rm KK},\mathbb{S}^1_i}|=|\vec{\zeta}_{{\rm M2},jk}|=\sqrt{\frac{3}{2}}$, as expected from KK towers associated to the decompactification of 1 internal dimension (the scaling of $M_{\rm Pl,8}$ in terms of $M_{\rm Pl,8}$ and the volume of $Q_3$ and $\mathbb{S}^1/\mathbb{Z}_2$ is the same for both towers). Since the whole $\mathbb{T}^3$ is shrinking homogeneously, then the M2 branes wrapped along the 3 generators of $H_2(\mathbb{T}^3;\mathbb{Z})=\langle\{\mathbb{T}^2_{12},\mathbb{T}^2_{23},\mathbb{T}^2_{31}\}\rangle$ become light at the same rate, and their bounded states are equivalent to the KK modes from the decompactification of the whole $\mathbb{T}^3$. This serves as a self-consistency cross-check confirming our intuition coming from the 8d U-duality group of M-theory on $\mathbb{T}^3$.
    
    Based on the scaling of the towers becoming light, we then conclude that the emergent dual theory is given by M-theory on a homogeneously growing $\mathbb{T}^3{'}^{\color{gray} (\nicefrac{1}{4})}\hookrightarrow Y_6'\to Q_3^{\color{gray} (\nicefrac{1}{4})}$. The leading tower is given by $m_{{\rm KK},Y_6'}\sim M_{\rm Pl,4}\sigma^{-1}$, consisting on bounded KK states of the $Q_3$ and $\mathbb{T}^3{'}$, with the $\zeta$-vector of the later pointing outside the K\"ahler cone. The species scale is then given by the 10d Planck mass, $M_{\rm Pl,10}\sim M_{\rm Pl,4}\sigma^{-3/4}$. 
	
Furthermore, one finds the oscillation modes of the M2 brane wrapping the shrinking interval $\mathbb{S}^1/\mathbb{Z}_2\simeq[0,\frac{1}{2}]$ or $[0,\frac{1}{4}]$.\footnote{In the definition of the interval $\mathbb{S}^1/\mathbb{Z}_2$ from a circle in of $\mathbb{T}^7$ after quotienting by $\Gamma\simeq \mathbb{Z}_2\times\mathbb{Z}_2\times\mathbb{Z}_2$ \eqref{e.group 1st BULK}, one of the generators (note necessarily $\alpha$, $\beta$ or $\gamma$ but maybe a different set of generators) of $\Gamma$ leaves fixed the circle, while another breaks it into an interval, $x^i\sim -x^i$. Depending on the presence of $\frac{1}{2}$ terms, the third generator might leave it invariant or enhance the shift-symmetry to 
$x^i\sim x^i+\frac{1}{2}$, thus resulting in the interval having a total length $[0,\frac{1}{4}]$ rather than $[0,\frac{1}{2}]$.}  The oscillator modes scale as $m_{\rm osc,M2}\sim M_{ \rm Pl,4}\sigma^{-3/4}$, precisely at the same rate as the species scale. As seen above, this interval corresponds to the base	of a non-trivial Borcea-Voisin fibration $Y_6$ with singular fibers on the extremes of the interval, what breaks the calibration property as the base interval shrinks and the manifold looks less and less as a product. Hence, $\mathbb{S}^1/\mathbb{Z}_2$ is not a calibrated submanifold, so the resulting string is \textbf{non-BPS}, though it still sets the species scale $\Lambda_{\rm QG}=m_{\rm str}$.

 When decompactifying the orbifold along the infinite distance limit, supersymmetry is restored and the string becomes the usual heterotic $E_8\times E_8$ string. Hence, the theory emerging at the infinite distance limit is \textbf{weakly coupled 10d $E_8\times E_8$ heterotic string theory}.
From the 4d perspective, this duality frame is then described by \textbf{weakly coupled, large volume $E_8\times E_8$ heterotic string theory compactified on a $\mathbb{T}^3$-fibered Borcea-Voisin three-fold}.

\item [\textbf{(b)}] The second possibility is that all radii depend on the growing saxions, with six of them depending on a single saxion and one depending on the three. This occurs for example with $s^1,\,s^2,\,s^7\to\infty$, with $R_7$ being a function of the three, see Table \eqref{eq: table IJK}. Here the Joyce manifold is asymptotically fibered as
\begin{equation}
	Y_6^{{\color{gray}(0)}}\hookrightarrow X_7\to\mathbb{S}^1/\mathbb{Z}_2^{{\color{gray}(1)}}\;,
\end{equation}
	
	with $Y_6$ a Borcea-Voisin fiber that remains fixed as the base interval grows. It is immediate that the leading tower is given by the KK modes of $\mathbb{S}^1/\mathbb{Z}_2$ and the species scale is given by $M_{\rm Pl,5}$, respectively scaling as $m_{{\rm KK},\mathbb{S}^1/\mathbb{Z}_2}\sim M_{\rm Pl,4}\sigma^{-3/2}$ and $M_{\rm Pl,5}\sim M_{\rm Pl,4}\sigma^{-1/2}$. The emergent dual theory at infinite distance corresponds to M-theory on $Y_6$, so the 4d perturbative regime is described by \textbf{M-theory on $Y_6\hookrightarrow X_7\to\mathbb{S}^1/\mathbb{Z}_2$ fibration}. One can check that the pertinent EFT string results from wrapping a M5-brane on the internal K3 fiber of $Y_6$ and scales like $\mathcal{T}\sim M_{\rm Pl,5}^2\sigma^{-1}$, so this limit has $w=3$.
	\end{itemize}	
	\item[\textbf{4.}] As in the previous case, there are two possibilities for sending four saxions to infinity:
	\begin{itemize}
	\item[\textbf{(a)}] We can take four saxions in such a way that there is a radius that does not depend on them. From  Table \eqref{eq: table IJK}, this is exemplified by $s^1,\,s^2,\,s^3,\,s^4\to\infty$, with $R_1$ independent of them. This case is analogous to the previous \textbf{3(b)} one, with  a Borcea-Voisin fibration 
\begin{equation}
	{Y_6}^{\color{gray} (\nicefrac{1}{3})}\hookrightarrow X_7\to \mathbb{S}^1/\mathbb{Z}_2^{\color{gray} (-\nicefrac{2}{3})}\;,
\end{equation}	
	where now $Y_6$ homogeneously scales as $\sigma^{1/3}$ while the interval shrinks like $\sigma^{-2/3}$. Again, wrapping a M2-brane over the small $\mathbb{S}^1/\mathbb{Z}_2$ base results in a non-BPS HE string, whose oscillation modes now are the leading tower (also setting the species scale $\Lambda_{\rm QG}$). The KK tower associated to the growing $Y_6$ becomes light at the same rate as the strings, thus resulting in an emergent HE string that lives in 10d. The perturbative duality frame description is again \textbf{weakly coupled, large volume $E_8\times E_8$ heterotic string theory compactified on a Borcea-Voisin three-fold}, as above, but this time with a different scaling growth of the $Y_6$ cycles.
	
	Notice that this is an emergent string limit with $w>1$. As it was the case for the type I and type IIB strings (on an orientifold) from Sections \ref{s.SO32} and \ref{s.Fth}, this can occur because the string is not BPS in 4d, and thus is not a (BPS) EFT string.
	
	\item[\textbf{(b)}] The other possibility comes from taking four saxions  large such that one radius depends on 3 of these saxions, three radii on 2 saxions and the remaining three radii on only 1 of these saxions. This is achieved, for example, with $s^1,\,s^2,\,s^6\,s^7\to\infty$ in Table \eqref{eq: table IJK}. The fibration structure achieved is again Borcea-Voisin,
	\begin{equation}
	({\mathbb{T}^3}^{{\color{gray}(-\nicefrac{1}{6})}}\hookrightarrow Y_6\to \mathcal{Q}_3^{{\color{gray}(\nicefrac{1}{3})}})\hookrightarrow X_7\to\mathbb{S}^1/\mathbb{Z}_2^{{\color{gray}(\nicefrac{5}{6})}}\;,
	\end{equation}
	with  the interval base, $\mathbb{T}^3$ and $\mathcal{Q}_3$ cycles scaling as $\sigma^{5/6}$, $\sigma^{-1/6}$ and $\sigma^{1/3}$ respectively. The EFT string comes from wrapping an M5 brane over the $\mathbb{T}^3\times\mathbb{S}^1/\mathbb{Z}_2$ 4-cycle, which in this limit is a calibrated submanifold, resulting in $\sqrt{\mathcal{T}}\sim M_{\rm Pl,4}\sigma^{-1/2}$. On the other hand, the leading tower is given by the KK associated to the $\mathbb{S}^1/\mathbb{Z}_2$ interval, scaling as $m_{{\rm KK},\mathbb{S}^1/\mathbb{Z}_2}\sim M_{\rm Pl,4}\sigma^{-3/2}$ (so that $w=3$). There are additional towers given by $m_{{\rm KK},\mathcal{Q}_3}\sim M_{\rm Pl,4}\sigma^{-1}$ and $m_{{\rm KK},\mathbb{T}^3}\sim M_{\rm Pl,4}\sigma^{-1/2}$. The species scale is computed to be $M_{\rm Pl,8}\sim M_{\rm Pl,4}\sigma^{-5/6}$, signaling a decompactification to 8d. Since the KK modes associated to $\mathbb{T}^3$ are above the species scale, $\frac{m_{{\rm KK},\mathbb{T}^3}}{M_{\rm Pl,8}}\sim\sigma^{1/3}\to\infty$, the $\mathbb{T}^3$ manifold is shrinking in the asymptotic dual theory. However, as commented before, since M-theory on a small 3-torus is U-dual to M-theory on a growing $\mathbb{T}^3$ (with $R_i\to \frac{1}{R_i}$ in $M_{\rm Pl,11}$ units) \cite{Obers:1998fb,Calderon-Infante:2023ler}, the emergent dual theory is then given by 8d M-theory on a growing $\mathbb{T}^3$. Hence, the 4d perturbative description is described by \textbf{M-theory on a $({\mathbb{T}^3}\hookrightarrow Y_6\to \mathcal{Q}_3)\hookrightarrow X_7\to\mathbb{S}^1/\mathbb{Z}_2$ fibration}. 
		\end{itemize}

	\item[\textbf{5.}] The last chapter of the Borcea-Voisin fibration ``saga'' is achieved by sending 5 saxions to infinity, with our Joyce manifold adopting a fibration structure
	$$
	({\rm K3}^{\color{gray} (\nicefrac{1}{6})}\hookrightarrow Y_6\to\mathbb{S}^2(2222)^{\color{gray} (\nicefrac{2}{3})})\hookrightarrow X_7\to\mathbb{S}^1/\mathbb{Z}_2^{{\color{gray} (-\nicefrac{1}{3})}}\;,
	$$
	with the length scales of K3, $\mathbb{S}^2(2222)$ and $\mathbb{S}^1/\mathbb{Z}_2$ growing as $\sigma^{1/6}$, $\sigma^{2/3}$ and $\sigma^{-1/3}$. As before, M2-branes wrapping the shrinking interval result in non-BPS heterotic $E_8\times E_8$ strings, again setting the species scale with $\Lambda_{\rm QG}=m_{\rm M2}\sim M_{\rm Pl,4}\sigma^{-1}$. However, in this case the leading tower is given by $m_{{\rm KK},\mathbb{S}^2(2222)}\sim M_{\rm Pl,4}\sigma^{-3/2}$. The K3 fiber also grows, but its KK modes scale at the same rate as the M2 brane strings. This results in the same perturbative description as in the two above cases, a \textbf{weakly coupled, large volume $E_8\times E_8$ heterotic string theory compactified on a }K3\textbf{-fibered Borcea-Voisin three-fold}. The dual theory emerging in the infinite distance limit is weakly coupled BPS heterotic string theory on K3.

	\item[\textbf{6.}] When all the saxions but one are sent to infinity, the Joyce manifold adopts an a.a fibration (again, see Appendix \ref{app-Joyce})
	\begin{equation}
	{\mathbb{T}^3}^{{\color{gray} (0)}}\hookrightarrow X_7\to\mathbb{T}^4/\Gamma^{{\color{gray} (\nicefrac{1}{2})}}\;,
	\end{equation}
	with  $\mathbb{T}^3$ as generic fiber of fixed size and a base growing as $\sigma^{1/2}$. This is not a weakly coupled limit, with the leading tower given by the $\mathbb{T}^4/\Gamma$ KK modes, $m_{ {\rm KK},\mathbb{T}^4/\Gamma}\sim M_{\rm Pl,4}\sigma^{-3/2}$ and the species scale given by $M_{\rm Pl, 8}\sim M_{\rm Pl,4}\sigma^{-1}$. This way, this EFT string flow corresponds to a decompactification to M-theory on $\mathbb{T}^3$, and the 4d perturbative regime is better described by \textbf{M-theory on a $\mathbb{T}^3\hookrightarrow X_7\to \mathbb{T}^4/\Gamma$ fibration}.
	\item[\textbf{7.}] This final limit, with all the saxions sent to infinity, corresponds to full decompactification of $X_7$ (scaling as $\sigma^{1/3}$) to 11d M-theory. Since the destination theory does not feature any modulus that can be subsequently sent to infinity, there are no subleading towers becoming light at the same rate as $M_{\rm Pl,11}\sim \sigma^{-7/6}$. The asymptotic regions of moduli space where $M_{\rm Pl,11}$ sets the species scale are then described by \textbf{M-theory on $X_7$}.
\end{enumerate}
The above EFT string limits correspond to a discrete subset of asymptotic directions in moduli space. To have a complete description, we would need to divide the moduli space into different regions and analyze the global arrangement of towers and duality frames as done in the previous sections. The story here is more complicated due to the larger number of saxions, which makes impossible to plot it. However, the above EFT string limits are already representative of the possible duality frames that will emerge at the different regions of the moduli space. In any case, for illustration purposes, in Figures \ref{f.slices2dMth} and \ref {f.slices3dMth} we include the 2- and 3-dimensional slices of the 7-dimensional polytope constructed from the convex hull of all the towers and strings. The slices depicted here are those containing tower vectors directly on the 2- or 3-plane and generating the full quadrant/octant.

Note that some of the slices found here have tower arrangements already present in previous constructions, such as in heterotic $E_8\times E_8$ or type IIA compactifications (compare e.g. Figures \ref{f.slicesMth3-1} and \ref{f.slicesMth3-3} with Figures \ref{f.dual2} and \ref{f.slicess1s2s3-1}). There is however, an important difference. In those cases, the K\"ahler potential had degree 4 in the saxions, while here it has 7. This allows for slices in which some of the elementary EFT are associated to a saxion that appears in the K\"ahler potential with degree 4, in such a way that the associated tower is a fundamental string with scaling $w=2$. As commented along this section, this is consistent with the fact that in this compactifications we obtain non-BPS strings. Conversely, other tower arrangements found in previous string constructions are not found here, such as Figure \ref{f.slicess1s2s3-2} in type IIA/HE compactifications (since $X_7=\mathbb{T}^7/\Gamma$ does not have a $\mathbb{T}^6$ fibration), or Figures \ref{f.typeI1} and \ref{f.typeI2} in HO/type I (equivalently Figures \ref{f.IIB 1} and \ref{f.II2B} in F-theory compactifications), since these regimes do not have an M-theory limit.

Finally, notice that the different limits and duality descriptions above have been derived for a particular Joyce manifold of the first kind, $J(0,\nicefrac12,\nicefrac12,\nicefrac12,0)$. We expect that the results and lessons learned here also apply to a more general $J(b_1,b_2,c_1,c_3,c_5)$, as well as Joyce manifolds of the second kind. The reason is that all possible fibrations that have been classified in \cite{Liu:1998tha} (modulo blow ups of the singularities, whose volume we keep fixed) for this type of manifolds have already appeared in this simple example, which means that we expect to have analyzed all possible building blocks that can be later combined in different ways to reproduce different Joyce manifolds. It would be interesting, though, to perform a more thorough study of Joyce manifolds in this context.

\begin{figure}[htb!]
\begin{center}
\begin{subfigure}[b]{0.32\textwidth}
\captionsetup{width=.95\linewidth}
\center
\includegraphics[width=\textwidth]{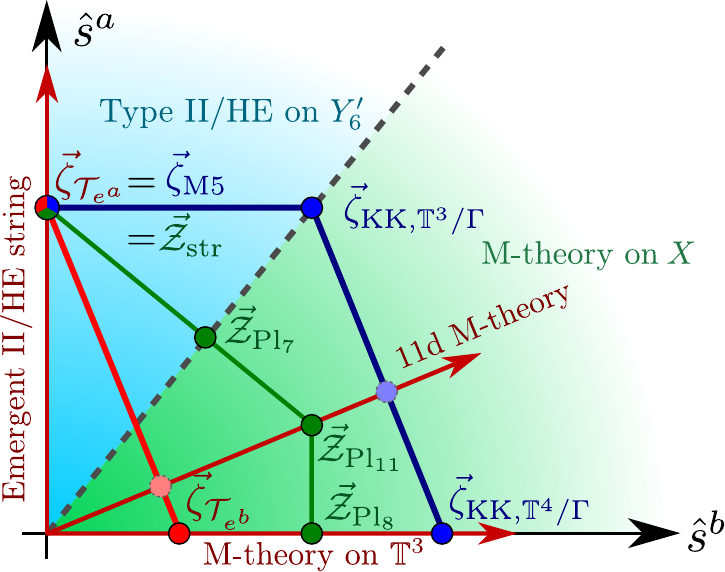}
\caption{\hspace{-0.3em}\scriptsize $(s^a,s^b)\sim(s^{a_1},s^{a_2}\dots s^{a_7})$.}\label{f.SlicesMth-1}
\end{subfigure}
\begin{subfigure}[b]{0.32\textwidth}
\captionsetup{width=.95\linewidth}
\center
\includegraphics[width=\textwidth]{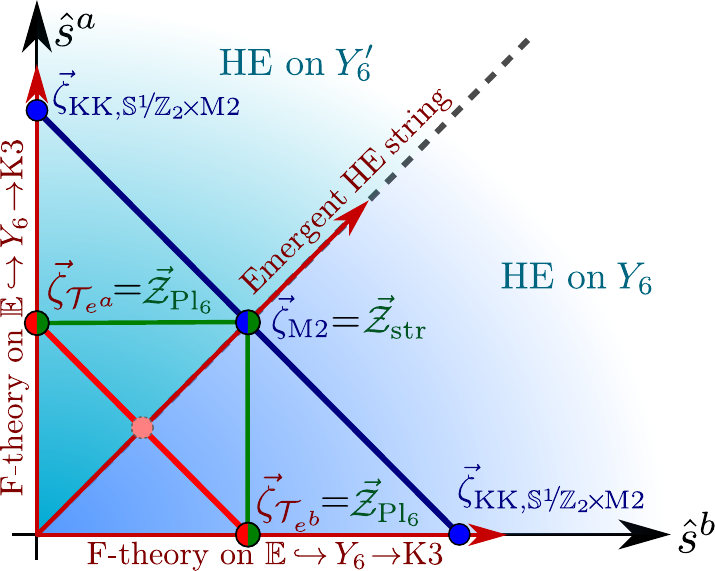}
\caption{\hspace{-0.3em} $(s^a,s^b)\sim(s^{a_1}s^{a_2},s^{a_3}s^{a_4})$.} \label{f.SlicesMth-2}
\end{subfigure}
\hfill
\begin{subfigure}[b]{0.32\textwidth}
\captionsetup{width=.95\linewidth}
\center
\includegraphics[width=\textwidth]{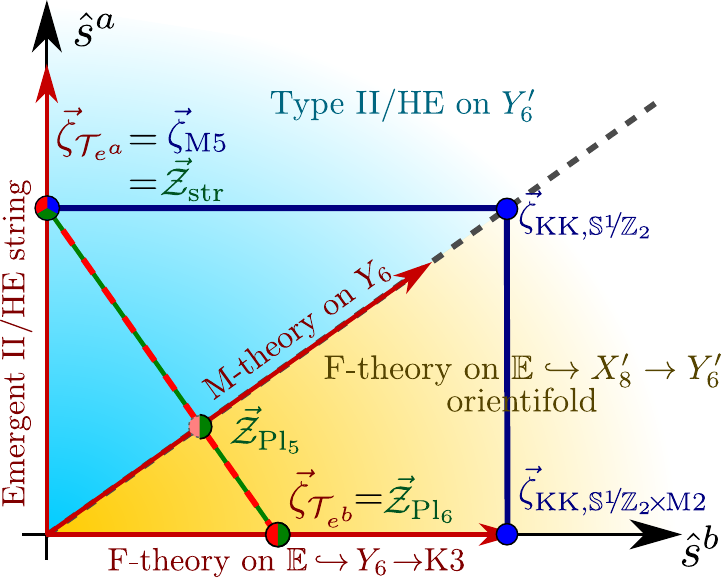}
\caption{\hspace{-0.3em} $(s^a,s^b)\sim(s^{a_1},s^{a_2}s^{a_3})$}\label{f.SlicesMth-3}
\end{subfigure}
\begin{subfigure}[b]{0.32\textwidth}
\captionsetup{width=.95\linewidth}
\center
\includegraphics[width=\textwidth]{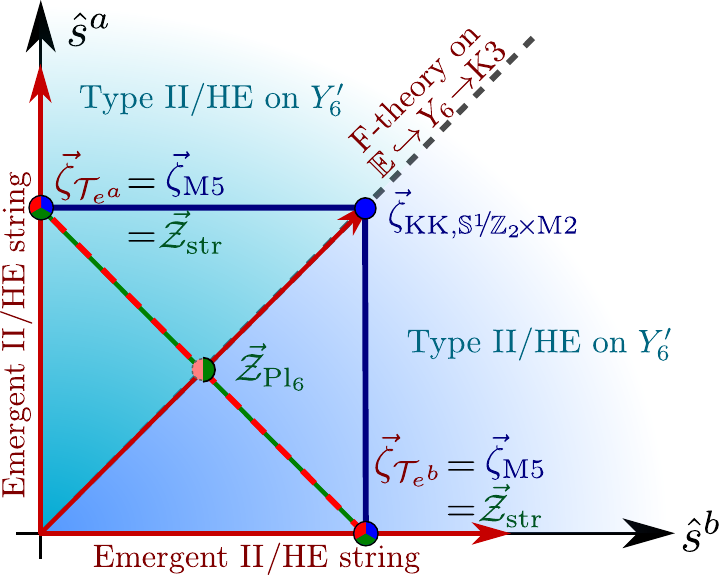}
\caption{\hspace{-0.3em} $(s^a,s^b)\sim(s^{a_1},s^{a_2})$.} \label{f.SlicesMth-4}
\end{subfigure}
\hfill
\begin{subfigure}[b]{0.32\textwidth}
\captionsetup{width=1.1\linewidth}
\center
\includegraphics[width=\textwidth]{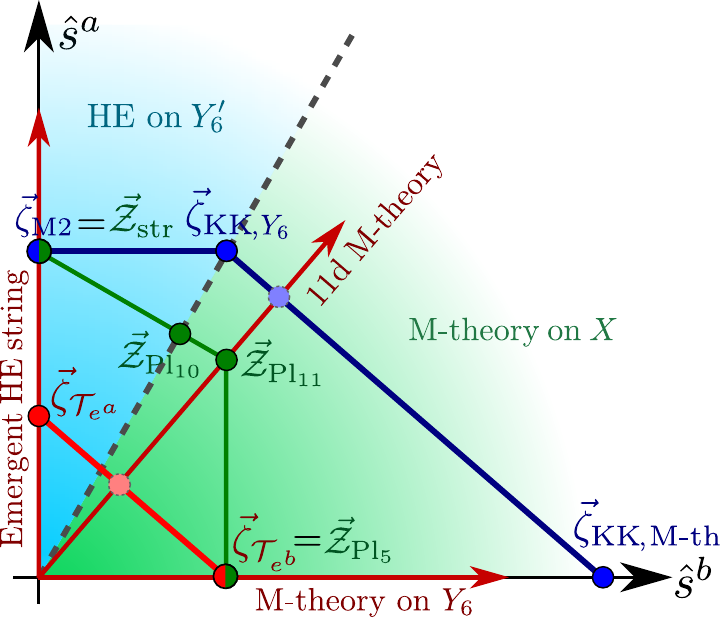}
\caption{\hspace{-0.3em} $(s^a,s^b)\sim(s^{a_1}\dots s^{a_4},s^{a_5}s^{a_6} s^{a_7})$}\label{f.SlicesMth-5}
\end{subfigure}
\begin{subfigure}[b]{0.32\textwidth}
\captionsetup{width=.95\linewidth}
\center
\includegraphics[width=\textwidth]{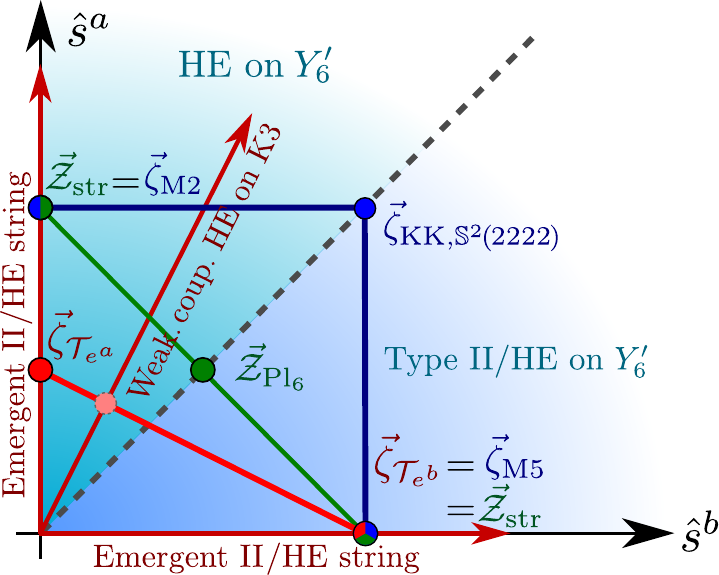}
\caption{\hspace{-0.3em} $(s^a,s^b)\sim (s^{a_1}\dots s^{a_4},s^{a_5})$} \label{f.SlicesMth-6}
\end{subfigure}
\hfill
\begin{subfigure}[b]{0.32\textwidth}
\captionsetup{width=1.1\linewidth}
\center
\includegraphics[width=\textwidth]{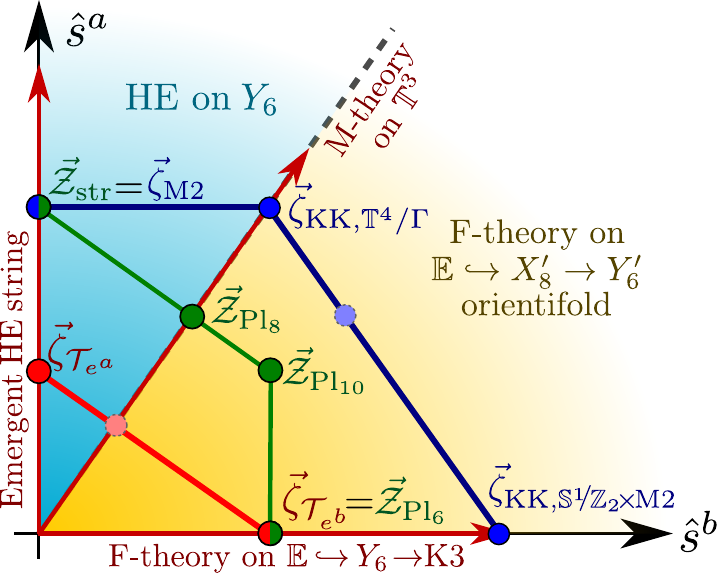}
\caption{\hspace{-0.3em} $(s^a,s^b)\sim (s^{a_1}\dots s^{a_4},s^{a_5}s^{a_6})$.} \label{f.SlicesMth-7}
\end{subfigure}
\begin{subfigure}[b]{0.32\textwidth}
\captionsetup{width=1.1\linewidth}
\center
\includegraphics[width=\textwidth]{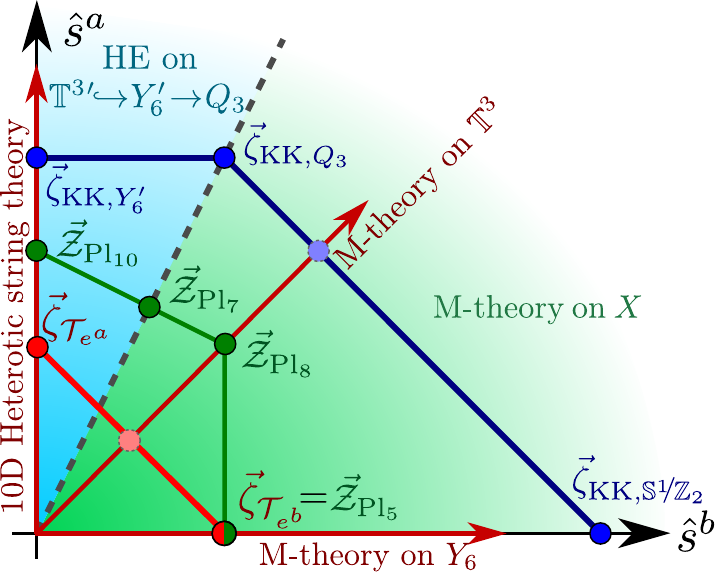}
\caption{\hspace{-0.3em} $(s^a,s^b)\sim (s^{a_1}s^{a_2} s^{a_3},s^{a_4}s^{a_5}s^{a_6})$.} \label{f.SlicesMth-8}
\end{subfigure}
\begin{subfigure}[b]{0.32\textwidth}
\captionsetup{width=1.1\linewidth}
\center
\includegraphics[width=\textwidth]{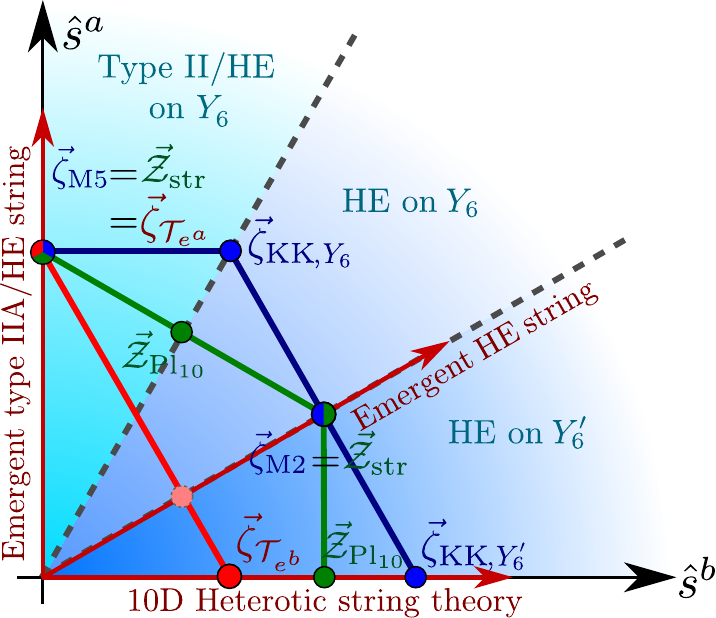}
\caption{\hspace{-0.3em} $(s^a,s^b)\sim (s^{a_1},s^{a_2} s^{a_3}s^{a_4})$.} \label{f.SlicesMth-9}
\end{subfigure}
\hfill
		\caption{All possible arrangements of the scaling vectors for towers and strings contained in 2-dimensional slices of the $\{s^1,\dots,s^7\}$ moduli space of M-theory on $\mathbb{T}^7/\Gamma$. Here $(s^a,s^b)\sim(s^{a_1}\dots s^{a_k},s^{a_{k+1}}\dots s^{a_{k+l}})$ denotes the slice of moduli space where $k$ and $l$ saxions are sent to infinity homogeneously, with the new saxionic coordinates $s^a$ and $s^b$ controlling the respective growth, while the remaining $7-k-l$ saxions are kept fixed. We include the $\zeta$-vectors for towers (blue),  EFT strings (red), and the $\mathcal{Z}$-vectors (green) for the species scale, as well as the different 4d perturbative descriptions and the emergent dual theories arising along EFT string flows (darker red arrows). All quantities are depicted in canonically normalized coordinates.}
		 
			\label{f.slices2dMth}
	\end{center}
\end{figure}

 \begin{figure}[htp!]
\begin{center}
\begin{subfigure}[b]{0.70\textwidth}
\captionsetup{width=.95\linewidth}
\center
\includegraphics[width=\textwidth]{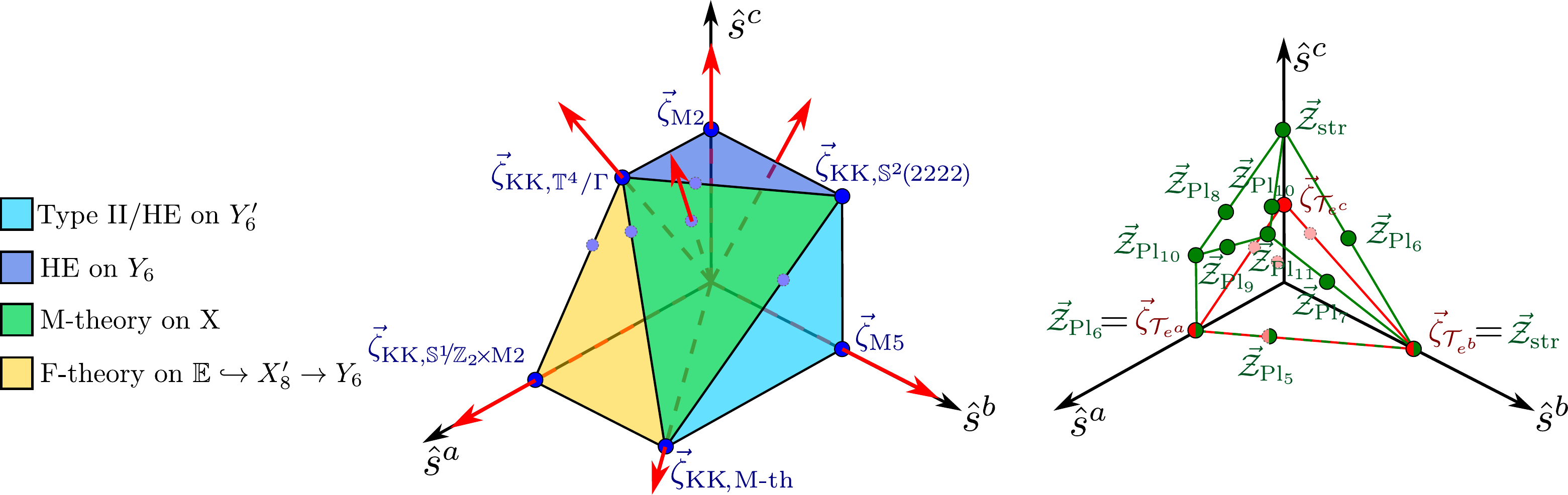}
\caption{\hspace{-0.3em} $(s^a,s^b,s^c)\sim(s^{a_1}s^{a_2},s^{a_3}\dots s^{a_6},s^{a_7})$.} \label{f.slicesMth3-1}
\end{subfigure}
\begin{subfigure}[b]{0.70\textwidth}
\captionsetup{width=.95\linewidth}
\center
\includegraphics[width=\textwidth]{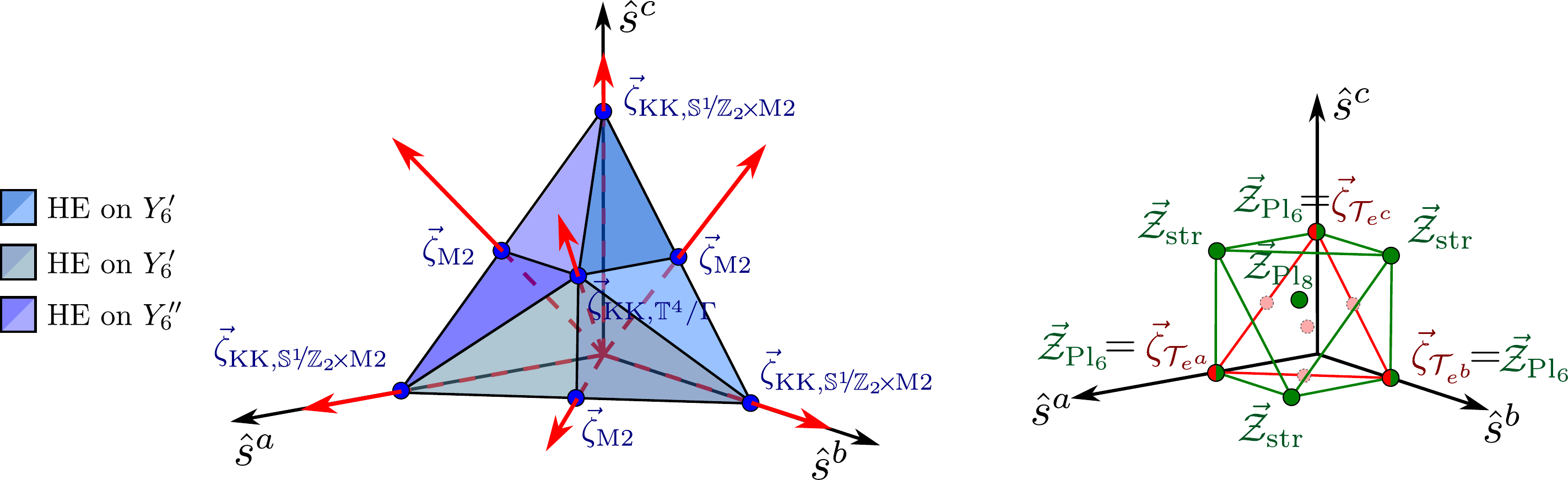}
\caption{\hspace{-0.3em} $(s^a,s^b,s^c)\sim(s^{a_1}s^{a_2},s^{a_3}s^{a_4},s^{a_5}s^{a_6})$.} \label{f.slicesMth3-2}
\end{subfigure}
\begin{subfigure}[b]{0.70\textwidth}
\captionsetup{width=.95\linewidth}
\center
\includegraphics[width=\textwidth]{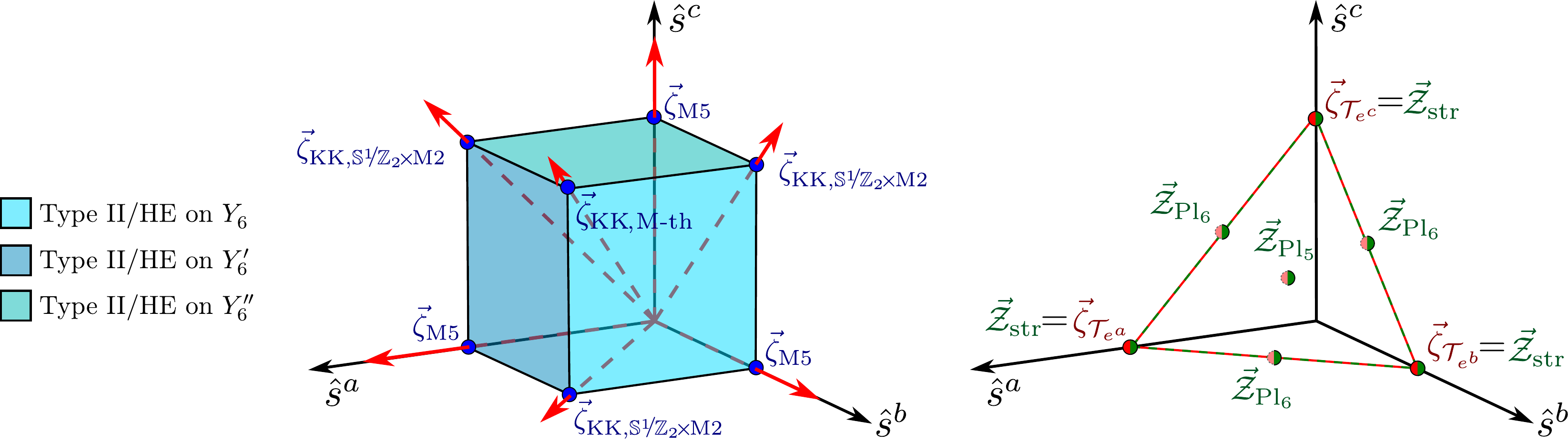}
\caption{\hspace{-0.3em} $(s^a,s^b,s^c)\sim(s^{a_1},s^{a_2},s^{a_3})$.} \label{f.slicesMth3-3}
\end{subfigure}
\begin{subfigure}[b]{0.70\textwidth}
\captionsetup{width=.95\linewidth}
\center
\includegraphics[width=\textwidth]{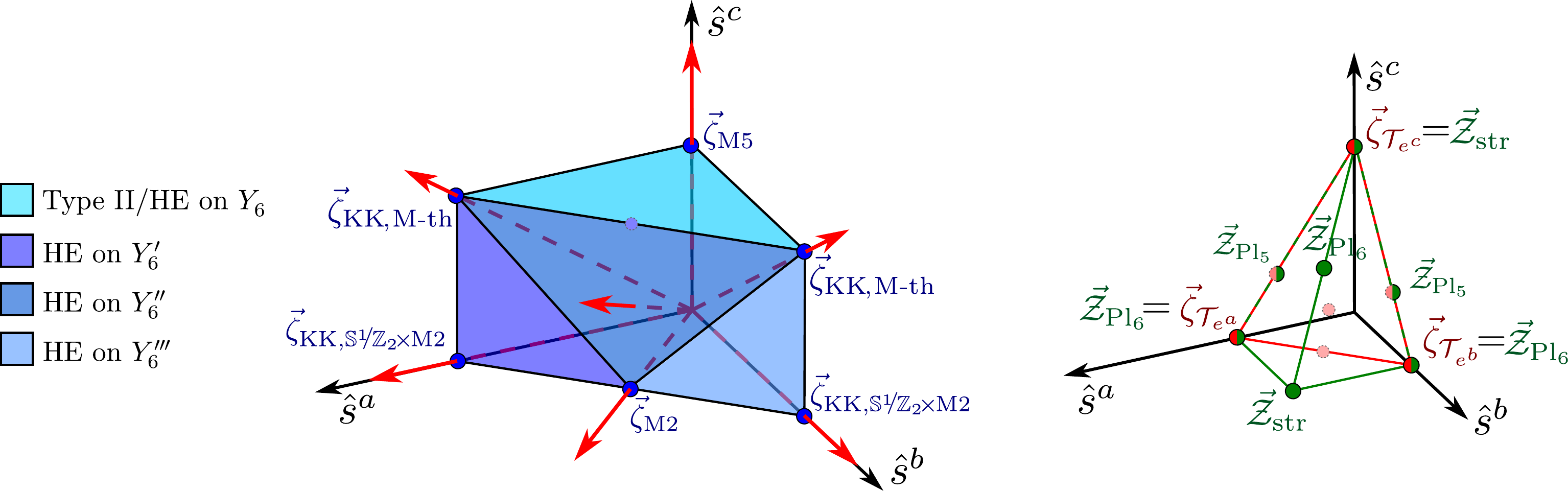}
\caption{\hspace{-0.3em} $(s^a,s^b,s^c)\sim(s^{a_1},s^{a_2}s^{a_3},s^{a_4}s^{a_5})$.} \label{f.slicesMth3-4}
\end{subfigure}
\begin{subfigure}[b]{0.70\textwidth}
\captionsetup{width=.95\linewidth}
\center
\includegraphics[width=\textwidth]{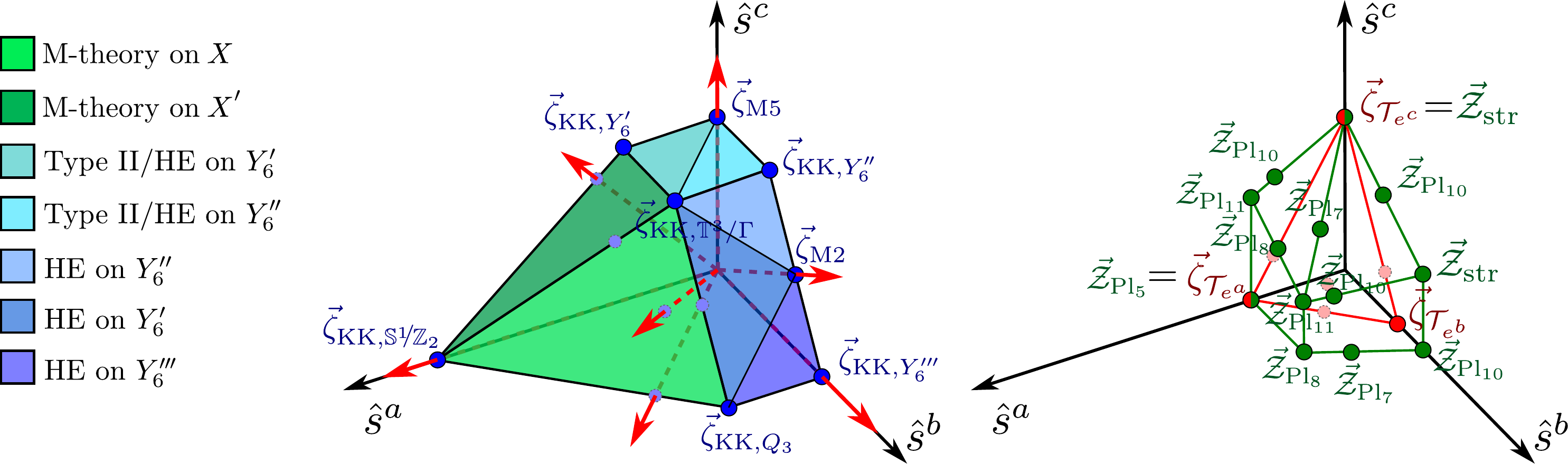}
\caption{\hspace{-0.3em} $(s^a,s^b,s^c)\sim(s^{a_1}s^{a_2}s^{a_3},s^{a_4}s^{a_5}s^{a_6},s^{a_7})$.} \label{f.slicesMth3-5}
\end{subfigure}
		\caption{All arrangements of the scaling vectors contained in 3-dimensional slices of the $\{s^1,\dots,s^7\}$ moduli space of M-theory on $\mathbb{T}^7/\Gamma$, namely the $\zeta$-vectors for towers (blue),  EFT strings (red), and the $\mathcal{Z}$-vectors (green) for the species scale, as well as the 4d perturbative descriptions. The emergent dual theories along EFT string flows (darker red arrows) are not labeled, as they appear already in Figure \ref{f.slices2dMth} for 2d slices. All quantities are depicted in canonically normalized coordinates, with $(s^a,s^b,s^c)\sim(s^{a_1}\dots s^{a_k},s^{a_{k+1}}\dots s^{a_{k+l}},s^{a_{k+l+1}}\dots s^{a_{k+l+h}})$ denoting the $k+l+h$ slice of moduli space with co-scaling saxions as in Figure \ref{f.slices2dMth}.}
			\label{f.slices3dMth}
	\end{center}
\end{figure}

\section{Conclusions\label{s.conc}}

EFT strings are a special class of BPS axionic strings that become weakly coupled and tensionless as approaching infinite distance limits in field space. They are magnetically charged under axionic shift symmetries that emerge asymptotically and characterize distinct perturbative regimes of the 4d effective theory. Their ubiquitous appearance in string compactifications can be understood from the absence of global symmetries in quantum gravity, and their tension can be derived directly from the Kähler potential. Since these perturbative regimes are at infinite distance in field space, they are a key player in the analysis of the Distance Conjecture and the asymptotic behavior of the infinite towers of states that become light in such limits.

In this paper we have investigated the global structure of the towers of particle states and EFT strings becoming light at infinite distance limits of 4d $\mathcal{N}=1$ string compactifications, as well as their interplay with the species scale and the dualities emerging at different perturbative corners of the field space. To this end, we have analyzed the structure of the convex hulls of the vectors $\vec\zeta=-\vec{\nabla} \log m$, where $m$ is the characteristic mass scale of the towers and the gradient is taken over the field space, as well as the analogous convex hulls for the EFT strings and the species scale. This provides a systematic description of the network of dualities arising in  4d $\mathcal{N}=1$ theories, offering organizing principles for the parametric hierarchies between all relevant physical scales in a given perturbative limit. It also yields constraints on how different duality frames can fit together in the moduli space. 

We should emphasize that, in order to compute the convex hulls of the towers, we simply study the scaling behaviour of the masses and use this to determine the implied structure of $\mathcal{N}=1$  dualities. However, it remains an open question to rigorously show that these towers of states are in fact populated by physical states which become stable asymptotically. Although this is more clear in the frame where the towers simply correspond to KK modes or oscillator modes of the critical string, it becomes a very challenging question in the duality frame in which they arise from e.g. wrapping branes (like in M-theory on G2 manifolds). We hope that our results can actually motivate future works to provide independent evidence for the structure of dualities implied by these towers in these uncharted corners of the $\mathcal{N}=1$ landscape.\\

The most important organizing principle in this work is the so-called Integer Scaling Conjecture first observed in \cite{Lanza:2021udy}. This relates the EFT strings tension to the mass scale of the light towers of states as follows,
\beq
m_*^2\simeq M_{\rm Pl,4}^2\left(\dfrac{\cT_{\bf e}}{M_{\rm Pl,4}^2}\right)^w \quad \text {with } w\in\mathbb{Z}_{\geq 0}
\eeq
if moving along the particular infinite distance trajectory associated to the EFT string flow. This equivalently implies a relation between their corresponding $\zeta$-vectors
\beq
\label{sc}
\vec\zeta_{*} \cdot \vec\zeta_{\cT_{\mathbf{e}}}=w|\vec\zeta_{\cT_{\mathbf{e}}}|^2 \ , \quad \text{ with } w\in\mathbb{Z}_{\geq 0}
\eeq
Strikingly, the scaling weight $w$ appears to take only the discrete value $w=1,2,3$ in all string theory examples that have been explored for the moment. In this work, we showed that this relation is not only satisfied by the leading tower of states, but also by all subleading towers that generate the tower convex hull. This implies that the \emph{$\zeta$-vector of all such towers lie at a lattice generated by the EFT string vectors, which span the 4d perturbative regime}. This constitutes an instance of a UV/IR interplay, since the EFT string tension is determined by IR data encoded in the Kähler potential, while the tower spectra are inherently UV. In Section \ref{sec.bounds}, we also provide some bounds on the possible values of $w$ in terms of the microscopic nature of the limit and the number of extra dimensions decompactifying. 

These results suggest a deeper connection between asymptotic towers and EFT strings, potentially governed by duality symmetries. From the EFT perspective, the EFT strings are the duality vortices implementing the axionic monodromies that emerge near infinite distance limits. It is tempting to speculate that the presence of additional towers of states—and the scaling of their masses—could be inferred from a completeness principle for the duality symmetries, in the spirit of \cite{Delgado:2024skw}. We hope that, by further investigating this integer scaling relation, we will eventually be able to derive the mass scale of the UV towers of states purely from bottom-up data, such as the K\"ahler potential, which determines the tension of the EFT strings. For now, all we can say is that the scaling weight appears to be related to the microscopic nature of the EFT string in the dual frame emerging at infinite distance, but why this in turn governs the scaling of the remaining towers of states is a question we leave for future work. Similar structures involving integer relations between the mass scale of towers of particles and tensions of extended objects have been found in \cite{Etheredge:2024amg} and will be further analysed in \cite{muldrowTBA}. It would be interesting to explore whether the lattice structure discussed in these papers is somehow related to the lattice structure studied here.\\

Remarkably, we have also found that only a handful of convex hull geometries arise across a wide array of 4d $\mathcal{N}=1$ string compactifications -- including heterotic, Type II and M-theory settings. While the microscopic interpretation of the dual quantum gravity theories may differ, the geometric structure of the convex hulls—and hence their identification in terms of emergent string limits or decompactifications of extra dimensions—is highly constrained and limited to a few cases.
In a follow-up paper, we will reverse the logic and ask the following question: given the K\"ahler potential and assuming \eqref{sc}, how much can we derive about the UV structure of the towers of states? Remarkably, we will find that the possible tower convex hulls compatible with \eqref{sc} and the Emergent String Conjecture, for a given Kähler potential, match precisely those arising in known string compactifications. This suggests that significant aspects of the UV physics -- such as the spectrum of light towers -- is  already encoded in the physics of EFT strings.\\

The interplay between asymptotic towers of states and extended objects at infinite distance limits has also been explored in \cite{Etheredge:2024amg}, where a generalization of the Distance Conjecture to branes was proposed. This Brane Distance Conjecture posits the existence of an extended $(p{-}1)$-brane whose tension scales as $\mathcal{T}_{p-1}\sim M_{{\rm Pl},d}^pe^{-\alpha_p\Delta\varphi}$, with exponent bounded below by $\alpha_p \geq \frac{1}{\sqrt{d - p_{\rm max} - 1}}$. This scaling is required to ensure that the bound $\alpha \geq \frac{1}{\sqrt{d-2}}$  \cite{Etheredge:2022opl} is satisfied upon toroidal compactification. 
Interestingly, some 4d $\mathcal{N}=1$ compactifications -- such as heterotic $SO(32)$ on Calabi-Yau threefolds and type IIB on orientifolds -- exhibit along certain 2d slices the same tower and string arrangement as the 4d maximal supergravity limit of type IIB on $\mathbb{T}^6$ discussed in \cite{Etheredge:2024amg}. These setups satisfy the Brane Distance Conjecture, but crucially only when the non-BPS fundamental Type I string is included; the (BPS) EFT strings alone are insufficient. In contrast, other setups -- such as heterotic $E_8 \times E_8$ on CY threefolds, type IIA on orientifolds, or M-theory on $G_2$ manifolds -- yield tower and string convex hulls that differ from those arising in toroidal compactifications, and do not obviously satisfy the conjecture. In these cases, the non-BPS Type I string is absent, and no alternative candidate is currently known to the best of our knowledge.

We leave a more detailed analysis of this comparison to a companion paper \cite{bottomUpTBA}, but we can already draw a preliminary conclusion: in general, the BPS EFT strings identified in this work do not suffice to satisfy the Brane Distance Conjecture. Their convex hulls span hyperplanes that typically cut the ball of radius $\alpha_p \geq \frac{1}{\sqrt{d - p_{\rm max} - 1}}$. If the Brane Distance Conjecture holds universally, it would imply the existence of additional non-BPS strings in 4d $\mathcal{N}=1$ compactifications that remain to be identified.\\

We also want to comment on the recent work \cite{Kaufmann:2024gqo}, which studied the asymptotic behavior of general 5d supergravity theories with eight supercharges from a bottom-up perspective. By requiring the consistency of the 5d analogues of our EFT strings—known as supergravity strings \cite{Katz:2020ewz}—the authors argued that every infinite-distance limit in the vector multiplet moduli space must correspond either to a decompactification to 6d or to an emergent string limit, without assuming an explicit UV completion. From our top-down perspective, these 5d $\mathcal{N}=1$ theories arise in strong coupling limits of heterotic $E_8 \times E_8$ on a threefold or type IIA on orientifolds that decompactify to 5d M-theory (see Sections \ref{s.het} and \ref{sIIA}). Only the states decaying at the same exponential rate as $M_{\rm Pl,5}$ will survive in the 5d theory upon decompactification from 4d. The EFT strings scaling as $\vec{\mathcal{Z}}_{\rm Pl_5}$ will then correspond to supergravity strings in 5d (see e.g. Figure \ref{f.slicess1s2s3-1sp}). As shown in Figures \ref{f.towers2} and \ref{f.slicess1s2s3-1}, the towers becoming light at the same rate as $M_{\rm Pl,5}$ along these particular limits are always either string or KK-2 towers (the latter corresponding to KK-1 towers in 5d),\footnote{Equivalently, in the species convex hull, the only $\vec{\mathcal{Z}}$-vectors lying at the same level as $\vec{\mathcal{Z}}_{\rm Pl_5}$ along such limits are $\vec{\mathcal{Z}}_{\rm Pl_6}$ and $\vec{\mathcal{Z}}_{\rm str}$. } so we get either emergent string limits or decompactifications to 6d, in agreement with the results of \cite{Kaufmann:2024gqo}.
While our conclusions are drawn from explicit top-down constructions, we will show in the companion paper \cite{bottomUpTBA} that the tower arrangements analyzed in this work are in fact the only ones compatible with the integer lattice structure \eqref{lattice} relating the tower and EFT string vectors—providing a bottom-up justification that aligns with the arguments of \cite{Kaufmann:2024gqo}.\\

A natural next step of our work is to incorporate a non-vanishing scalar potential. In the presence of fluxes, the asymptotic behavior of the flux potential is captured by the physics of EFT membranes, as described in \cite{Herraez:2020tih,Lanza:2020qmt}. It would therefore be compelling to explore whether a deeper structure emerges among EFT membranes, strings, and particles—potentially revealing new insights into the connection between the scalar potential and the towers of states (see also \cite{Casas:2024oak}). Since a non-trivial scalar potential typically leads to spontaneous supersymmetry breaking, this may bring the Distance Conjecture and related Swampland ideas closer to phenomenological relevance. Ultimately, understanding the underlying structure behind all these patterns and constraints may allow us to infer the cut-off scale at which quantum gravitational effects become significant -- and the responsible microscopic physics  -- purely from EFT data.\\

\textbf{Acknowledgements}: 
We are very grateful to Muldrow Etheredge, Ben Heidenreich, Luca Martucci, Luca Melotti, Miguel Montero, Tom Rudelius, Timo Weigand and Max Wiesner for illuminating discussions and comments. A.G. and I.R. also wish to acknowledge the hospitality of the Department of Theoretical Physics at CERN and the Department of Physics of Harvard University during different stages of this work. We also thank the Erwin Schr\"odinger International Institute for Mathematics and Physics of the University of Vienna for their hospitality during the programme ``The Landscape vs. the Swampland''. The authors thank the Spanish Agencia Estatal de Investigaci\'on through the grant ``IFT Centro de Excelencia Severo Ochoa'' CEX2020-001007-S and the grant PID2021-123017NB-I00, funded by MCIN/AEI/10.13039/ 501100011033 and by ERDF ``A way of making Europe''. The work of A.G. is also supported by the fellowship LCF/BQ/DI23/11990073 from ``La Caixa'' Foundation (ID 100010434). I.R. acknowledges the additional support of the Spanish FPI grant No. PRE2020-094163 and the ERC Starting Grant QGuide101042568 - StG 2021. The work of I.V.  was supported by the grant RYC2019-028512-I from the MCI (Spain), the ERC Starting Grant QGuide101042568 - StG 2021, and the Project ATR2023-145703 funded by MCIN/AEI/10.13039/501100011033.

\appendix
\section{Basic results on Joyce manifolds and their fibration structures\label{app-Joyce}}
In this appendix we discuss elementary results on \textbf{Joyce (compact) manifolds} $\mathbb{T}^7/\Gamma$ \cite{joyce1996a,joyce1996b,joyce2000compact} that are used in section \ref{s.Mth}, and the associated fibration structures that can be found. Most of the results appear in more detail in \cite{Liu:1998tha}, to which we refer for more details. We will be using the conventions of said reference, which differ from those in previous instances of Joyce manifolds in the string phenomenology literature \cite{Lanza:2021udy,Lanza:2022zyg}, but give way to analogous results.

Before introducing Joyce manifolds, consider the following 3-form and dual 4-form on standard $\mathbb{R}^7$,
\begin{subequations}
\begin{align}\label{e.Phi0}
	\Phi_0&=\eta^{567}+\eta^{347}-\eta^{246}+\eta^{235}+\eta^{145}+\eta^{136}+\eta^{127}\;,\\
	\star\Phi_0&=\eta^{1234}+\eta^{1256}-\eta^{1357}+\eta^{1467}+\eta^{2367}+\eta^{2457}+\eta^{3456}\;,
\end{align}
\end{subequations}
where $\eta^{abc}=\dd x^a\wedge\dd x^b\wedge \dd x^c$ and $\eta^{ijkl}=\dd x^i\wedge\dd x^j\wedge \dd x^k\wedge \dd x^l$. The group of oriented-preserving linear isomorphisms preserving $\Phi_0$ (and automatically $\star\Phi_0$) is the exceptional group $G_2<SO(7)$. An oriented 7-manifold $X_7$ endowed with a 3-form $\Phi$ that is point-wise isomorphic to $\Phi_0$ is called \emph{$G_2$ manifold}. If $\dd \Phi=\dd\star \Phi=0$, then the induced metric $g_\Phi$ is torsion-free and $(X_7,g_\Phi)$ is Ricci-flat, with $\mathsf{Hol}(g_\Phi)\leq G_2$ \cite{10.1007/BFb0084595}. This later type of $G_2$ manifolds are thus interesting in string and M-theory compactifications.

As there is no existence theorem theorem akin to that for Calabi-Yau manifolds \cite{Yau:1978cfy}, very few explicit examples of $G_2$ manifolds are known. We will focus in those found by Dominic Joyce
 \cite{joyce1996a,joyce1996b}, bearing his surname. These are constructed as toroidal orbifolds\footnote{Take $\mathbb{T}^n=\mathbb{R}^n/\Lambda$ to be the \emph{$n$-torus} determined by a lattice $\Lambda\subset\mathbb{R}^n$. A \emph{toroidal orbifold} is defined as $\mathcal{Q}_n=\mathbb{T}^n/\Gamma$, with $\Gamma$ a discrete group of isometries, acting effectively. For each $x\in \mathcal{Q}_n$ we associate the group $\Gamma_x$, isomorphic to the stabilizer of any preimage of $x$. The \emph{singular locus} of $\mathcal{Q}_n$ is then defined as $\Sigma_{\mathcal{Q}}=\{x\in \mathcal{Q}_n\,:\,\Gamma_x\neq\{1\}\}$.
 } $\mathcal{Q}_7$ starting from $(\mathbb{R}^7,\Phi_0)$, and are subdivided in two classes:
 \begin{itemize}
 	\item\textbf{Joyce manifolds of the first kind}: We construct $\mathcal{Q}_7$ from the 7-torus $\mathbb{T}^7=\mathbb{R}^7/\mathbb{Z}^7$, with coordinates $\{x^i\sim x^i+1\}_{i=1}^7$, and $\Gamma=\mathbb{Z}_2\oplus\mathbb{Z}_2\oplus\mathbb{Z}_2$, with generators
 	\begin{equation}\label{e.group 1st}
 	\begin{array}{l}
 		\alpha(\vec{x})=(-x^1,-x^2,-x^3,-x^4,x^5,x^6,x^7)\\
 		\beta(\vec{x})=(b_1-x^1,b_2-x^2,x^3,x^4,-x^5,-x^6,x^7)\\
 		\gamma(\vec{x})=(c_1-x^1,x_2,c_3-x^3,x^4,c^5-x^5,x^6,-x^7)\;,
 	\end{array}
 	\end{equation}
 	where $b_1,\,b_2,\,c_1,\,c_3,\,c_5\in\{0,\frac{1}{2}\}$.
 	\item\textbf{Joyce manifolds of the second kind}: We take the orbifold $\mathcal{Q}_7=(\mathbb{C}^3\times\mathbb{R})/\langle D_a,\Lambda\rangle$, with $\mathbb{C}^3\times\mathbb{R}$ parameterized by $(z^1,z^2,z^3,x)$, $\mathsf{D}_a$ the dihedral group of $2a$ elements, generated by
 	 \begin{equation}\label{e.group 2nd}
 	 	\begin{array}{l}
 	 		\alpha(z^1,z^2,z^3,x)=(Uz^1,Vz^2,\overline{UV}z^3,x+\frac{1}{a})\\
 	 		\beta(z^1,z^2,z^3,x)=(-\overline{z^1},-\overline{z^2},-\overline{z^3},-x)\;,
 	 	\end{array}
 	 \end{equation}
 	 with $U=\exp\left(\frac{2\pi i}{u}\right)$, $V=\exp\left(\frac{2\pi i}{v}\right)$, such that $a={\rm lcm}(u,v)>0$, and $\Lambda\subset \mathbb{C}^3\times\mathbb{R}$ a lattice invariant under $\mathsf{D}_a$.
 \end{itemize}
 Given $\mathcal{Q}_7$ such toroidal orbifold, $\Sigma_\mathcal{Q}$ is a collection of $\mathbb{T}^3$ fixed by some element of $\Gamma$, each of which can be resolved along a tubular neighborhood as $\mathbb{T}^3\times T^*\mathbb{P}^1$ or $\left(\mathbb{T}^3\times T^*\mathbb{P}^1\right)/\mathbb{Z}_2$. We will not pay much attention to the singular loci resolutions and the associated twisted saxions, which we will leave fixed and effectively ignore.

Now, on $(\mathbb{R}^7,\Phi)$ one can take take 7 decompositions in associative-coassociative\footnote{A calibrated submanifold of $(\mathcal{M}_ n,\omega_k)$, with $\omega_k$ a closed $k$-form (\emph{calibration}) is a $k$-dimensional submanifold $\mathcal{N}_k\subseteq \mathcal{M}_n$ such that $\omega_k|_{\mathcal{N}}$ is the induced volume-form on $\mathcal{N}$. For $G_2$-manifolds $(X_7,\Phi)$, calibrated submanifolds of $(X_7,\Phi)$ and $(X_7,\star\Phi)$ are respectively known as \emph{associative} and \emph{coassociative submanifolds}} products $\mathbb{R}^7=\mathbb{R}^3_{abc}\times \mathbb{R}^4_{ijkl}$, with ${abc}$ and $ijkl$ from the $\Phi_0$ and $\star\Phi_0$ expansion in \eqref{e.Phi0} such that $\{a,b,c\}\cup\{i,j,k,l\}=\{1,\dots,7\}$. Joyce manifolds of the first kind admit the seven decompositions in $\mathbb{T}^3_{abc}\times \mathbb{T}^4_{ijkl}$, all preserved by $\Gamma$, which induce a pair of fibrations on $\mathbb{T}^7/\Gamma$, respectively with associative and coassociative fibers. After resolving the singular set, the smooth manifold  still admits this kind of fibration, which (modulo an arbitrary small region) preserves the associative/coassociative structure. We call these \emph{asymptotically associative (a.a.)/asymptotically coassociative (a.c.) fibrations} of $\mathcal{Q}_7$.

For Joyce manifolds of the second kind, expanding the complex coordinates as $z^j=x^{2j-1}+i x^{2j}$, results in decomposition $\mathbb{C}^3\times\mathbb{R}\simeq \mathbb{R}^3_{127}\times \mathbb{R}^4_{3456}$, $\mathbb{C}^3\times\mathbb{R}\simeq \mathbb{R}^3_{347}\times \mathbb{R}^4_{1256}$ and $\mathbb{C}^3\times\mathbb{R}\simeq \mathbb{R}^3_{567}\times \mathbb{R}^4_{1234}$, preserved by $\mathsf{D}_a$ down to $\mathcal{Q}_7$. While in general it is not possible to construct a nice a.a./a.c. fibration after resolution, for most examples in \cite{joyce1996a,joyce1996b} the $\mathbb{T}^3_{abc}\times \mathbb{T}^4_{ijkl}$ fibration is kept after resolution.

Decomposing then $\mathbb{T}^7=\mathcal{F}\times\mathcal{B}$, $\mathcal{Q}_7=\mathbb{T}^7/\Gamma$, then we have the fibration structure $\mathcal{F}\hookrightarrow \mathcal{Q}_7\to \mathcal{B}/\Gamma$. Given $\Gamma_0\leq \Gamma$ the normal subgroup of $\Gamma$ given by the elements acting trivially on $\mathcal{B}$, then the generic fiber is $\mathcal{F}/\Gamma_0$. This allows us to consider the following fibrations case by case \cite{Liu:1998tha}
\begin{itemize}
	\item \textbf{a.a. fibr. for Joyce manifold of the 1$^\textbf{st}$ kind}: Here $\mathcal{B}=\mathbb{T}^4$ and $\mathcal{F}=\mathbb{T}^3$, and there are enough coordinates in the base such that $\Gamma_0$ is always trivial, in such a way that the generic fiber is always $\mathbb{T}^3$. There are some nuances on the topology of the singular point resolution depending on whether the tubular neighborhood of each of the singular locus components prior to the resolution look like $\mathbb{T}^3\times(\mathbb{C}^2/\langle-1\rangle)$ or $[\mathbb{T}^3\times(\mathbb{C}^2/\langle-1\rangle)]/\mathbb{Z}_2$. As this will not affect the considerations in Section \ref{s.Mth}, we will not comment further on this and refer to \cite{Liu:1998tha}.
	\item \textbf{a.c. fibr. for Joyce manifold of the 1$^\textbf{st}$ kind}: Now $\mathcal{B}=\mathbb{T}^3$ and $\mathcal{F}=\mathbb{T}^4$, and we can have either $\Gamma_0=\{0\}$ (having a $\mathbb{T}^4$ fibration), or $\Gamma=\langle g \rangle \simeq \mathbb{Z}_2$ for some $g\in\Gamma$. For this second case $\mathbb{T}^4/\mathbb{Z}_2$ corresponds to the singular K3 surface (whose isolated singularities can later be resolved).
	\item \textbf{Joyce manifolds of the 2$^\textbf{nd}$ kind}: Here $\Gamma_0$ is always the trivial group, and one always has $\mathbb{T}^3$ a.a. fibrations or $\mathbb{T}^4$ a.c. fibrations.
\end{itemize}

Finally, we will consider an additional fibration structure on Joyce manifolds of the first kind in terms of a class of Calabi-Yau manifolds known as \textbf{Borcea-Voisin threefolds} \cite{voisin,Borcea:1996mxz}, which we swiftly introduce. Consider $X_4$ a K3 surface and an involution $\imath$ implementing a $(-1)$ action on the holomorphic 2-form. The quotient $X_4/\imath$ is a smooth surface, with fixed points given by \emph{(a)} $\Sigma=\emptyset$, \emph{(b)} $\Sigma=\mathcal{C}_1\sqcup \mathcal{C}'_1$ or \emph{(c)} $\Sigma=\mathcal{C}_g\sqcup\mathcal{E}_1\sqcup\dots\sqcup\mathcal{E}_k$, with $\mathcal{C}_g$ a genus $g$ curve and $\mathcal{E}_i$ rational curves. On the other hand take $\mathbb{E}$ an additional elliptic curve and $j$ its involution by negation. Then $(X_4\times\mathbb{E})/(\imath,j)$ has four disjoint copies of $\Sigma$ as fixed points, and after resolution of these singularities the resulting manifold $Y_6$ is a Calabi-Yau known as \emph{Borcea-Voisin threefolds}. It is not complicated to see \cite{Liu:1998tha} that $Y_6$ can admit the following two fibrations:
\begin{itemize}
	\item A natural K3 fibration inherited from $(X_4\times\mathbb{E})/(\imath,j)$, with fiber K3 and base $\mathbb{E}/j\simeq\mathbb{S}^2(2222)$, a $\mathbb{T}^2$-orbifold by $(x,y)\sim -(x,y)$, with underlying topology of $\mathbb{S}^2$ or $\mathbb{P}^1$, and four fixed, conical points of order 2. After resolution, over each orbifold point the exceptional fiber looks like K3$/\imath\cup(\Sigma\times \mathbb{P}^1)$.  
	\item If the K3 manifold $X_4$ has an elliptic fiber invariant under $\imath$ which generically is a special Lagrangian submanifold with respect to the K\"ahler form on $X_4$, together with an $\mathbb{S}^1$ fibration from $\mathbb{E}$, this results in a $\mathbb{T}^3$ fibration. Here the base $\mathcal{Q}_3$ is a real 3-dimensional orbifold with the topology of $\mathbb{S}^3$.
\end{itemize}

Take then Joyce manifolds of the first kind and the group $\Gamma=\langle\alpha,\beta,\gamma\rangle\simeq\mathbb{Z}_2\oplus\mathbb{Z}_2\oplus\mathbb{Z}_2$ from \eqref{e.group 1st}. Splitting $\mathbb{T}^7=\mathbb{T}^6_{123456}\times\mathbb{S}^1_7$ and complexifying $z^j=x^{2j-1}+ix^{2j}$, we can consider $\mathcal{Q}_7=\mathbb{T}^7/\Gamma$ as a $\mathbb{T}^7/\langle\alpha\rangle$ orbifold by $\langle\beta,\gamma\rangle$. The resolution of $\mathbb{T}^7/\langle\alpha\rangle$ results in ${\rm K3}\times\mathbb{T}^2_{56}\times\mathbb{S}^1_7$ (since $\alpha$ negates $z^1$ and $z^2$, while leaving invariant $z^3$ and $x^7$), with $\beta$ and $\gamma$ acting holomorphically and antiholomorphically (respectively) on the three complex components. Precisely the action of $\beta$ on ${\rm K3}\times\mathbb{T}^2_{56}$ results, after resolution, in a Borcea-Voisin threefold $Y_6$, the trivially fibered over $\mathcal{S}^1_7$, $\mathcal{W}_7^{(\alpha,\beta)}=Y_6\times\mathbb{S}^1_7$. Finally, the action of $\gamma$ is a bundle automorphism on $\mathcal{W}_7^{(\alpha,\beta)}$, with two invariant fivers on $x_7=0,\frac{1}{2}$. This results in the fibration
\begin{equation}
	Y_6\hookrightarrow(\mathcal{Q}_7\simeq \mathcal{W}_7^{(\alpha,\beta)}/\langle\gamma\rangle)\to\mathbb{S}^1_7/\mathbb{Z}_2\;,
\end{equation}
with the Borcea-Voisin CY$_3$ $Y_6$ as generic fiber and exceptional fibers at $x_7=0,\,\frac{1}{2}$. See \cite{Liu:1998tha} for more details in the resolution of the exceptional fibers. As a curiosity, (see again \cite{Liu:1998tha}), $Y_6$ has as singular loci two copies of $\mathbb{T}^2_{34}$ (fixed under $\langle\beta\rangle$), resulting in Hodge numbers $h^{1,1}(Y_6)=h^{2,1}(Y_6)=19$.

Analogous constructions of Borcea-Voisin fibrations $Y_6\hookrightarrow \mathcal{Q}_/\to\mathbb{S}^1/\mathbb{Z}_2$ can be obtained in Joyce manifolds of the first kind through different complex combinations of the $\{x^i\}_{i=1}^7$ coordinates and generators of the $\Gamma=\mathbb{Z}_2\oplus\mathbb{Z}_2\oplus\mathbb{Z}_2$. The presence of $b_i$ or $c_i=\frac{1}{2}$ in \eqref{e.group 1st} only results in said direction having $\frac{1}{2}$ rather than $1$-shift symmetry, what does not affect any of the above descriptions.

For Joyce manifolds of the second kind, $J(u,v,\Lambda)$ from \eqref{e.group 2nd}, the existence of Borcea-Voisin fibrations is not as straightforward, though some constructions can be built, such as with  $u=1$, $v=2$ and $\Lambda=\mathbb{Z}^6$.

\bibliographystyle{JHEP}
\bibliography{ref}

\end{document}